\documentclass[dsingle]{Dissertate}

\usepackage{amsthm}
\usepackage{multirow}
\usepackage{float}
\usepackage{framed}
\usepackage{adjustbox}
\usepackage{longtable}

\usepackage{tikz}
\usepackage{tkz-fct}
\usepackage{tikz-cd}
\usetikzlibrary{matrix,arrows,decorations}

\makeatletter
\newcommand\thefontsize[2]{{#1 \noindent \f@size~pt: #2}}
\makeatother

\makeatletter
\def\nobreakhline{%
  \noalign{\ifnum0=`}\fi
    \penalty\@M
    \futurelet\@let@token\LT@@nobreakhline}
\def\LT@@nobreakhline{%
  \ifx\@let@token\hline
    \global\let\@gtempa\@gobble
    \gdef\LT@sep{\penalty\@M\vskip\doublerulesep}
  \else
    \global\let\@gtempa\@empty
    \gdef\LT@sep{\penalty\@M\vskip-\arrayrulewidth}
  \fi
  \ifnum0=`{\fi}%
  \multispan\LT@cols
     \unskip\leaders\hrule\@height\arrayrulewidth\hfill\cr
  \noalign{\LT@sep}%
  \multispan\LT@cols
     \unskip\leaders\hrule\@height\arrayrulewidth\hfill\cr
  \noalign{\penalty\@M}%
  \@gtempa}
\makeatother

\newtheoremstyle{definitionstyle}	
	{0.2cm}				
	{0.2cm}				
	{\it}						
	{}							
	{\it\bfseries}			
	{:}						
	{ }						
	{\thmname{#1}\thmnumber{~#2}\thmnote{~(#3)}}	

\newtheoremstyle{nameddefinitionstyle}	
	{\baselineskip\@plus.2\baselineskip\@minus.2\baselineskip}	
	{\baselineskip\@plus.2\baselineskip\@minus.2\baselineskip}	
	{}							
	{}							
	{\bfseries}			
	{:}						
	{ }						
	{\thmnote{#3}}	

\newtheoremstyle{framednameddefinitionstyle}	
	{0.2cm}				
	{0.2cm}				
	{\it}						
	{}							
	{\it\bfseries}			
	{:}						
	{ }						
	{\thmnote{#3}}	

\newtheoremstyle{theoremstyle}	
	{0.2cm}				
	{0.2cm}				
	{}							
	{}							
	{\bfseries}			
	{:}						
	{ }						
	{\thmname{#1}\thmnumber{~#2}\thmnote{~(#3)}}	

\newtheoremstyle{framedtheoremstyle}	
	{\baselineskip\@plus.2\baselineskip\@minus.2\baselineskip}	
	{\baselineskip\@plus.2\baselineskip\@minus.2\baselineskip}	
	{\sl}						
	{}							
	{\bfseries}			
	{:}						
	{ }						
	{\thmname{#1}\thmnumber{~#2}\thmnote{~(#3)}}	

\newtheoremstyle{proofstyle}
	{\baselineskip\@plus.2\baselineskip\@minus.2\baselineskip}	
	{\baselineskip\@plus.2\baselineskip\@minus.2\baselineskip}	
	{}							
	{}							
	{}							
	{:}						
	{ }						
	{\textsc{\thmname{#1}\thmnote{~#3}}}	

\theoremstyle{theoremstyle}

\theoremstyle{framedtheoremstyle}

\theoremstyle{nameddefinitionstyle}

\theoremstyle{framednameddefinitionstyle}

\theoremstyle{proofstyle}

\theoremstyle{definitionstyle}

\newcommand{\Z}{\mathbb{Z}}
\newcommand{\R}{\mathbb{Z}}

\DeclareMathOperator{\Tr}{Tr}

\newcommand{\bra}[1]{\left\langle #1\right|}
\newcommand{\ket}[1]{\left|#1\right\rangle}
\newcommand{\braket}[2]{\left\langle#1\middle|#2\right\rangle}

\def\ind\hspace{0.2in}

\newcommand{\up}{\uparrow}
\newcommand{\down}{\downarrow}
\newcommand{\abs}[1]{\left|#1\right|}

\newcommand{\bpm}{\begin{pmatrix}}
\newcommand{\epm}{\end{pmatrix}}

\newcommand{\bal}{\begin{align}}

\newcommand{\diag}{\mathrm{diag}}

\newcommand{\avg}[1]{\overline{#1}} 

\newcommand{\proj}[1]{\ket{#1}\bra{#1}}
\newcommand{\matrixel}[3]{\big\langle #1 \vphantom{#2#3} \big|
	#2 \big| #3 \vphantom{#1#2} \big\rangle} 
\usepackage{wasysym}
\newcommand{\gae}{\lower 2pt \hbox{$\,
    \buildrel{\scriptstyle >}\over {\scriptstyle \sim}\,$}}
\newcommand{\lae}{\lower 2pt \hbox{$\,
    \buildrel{\scriptstyle <}\over {\scriptstyle \sim}\,$}}
    \newcommand{\pd}[2]{\frac{\partial #1}{\partial #2}} 
\newcommand{\expectation}[1]{\left\langle #1 \right\rangle}
\newcommand{\disavg}[1]{\overline{#1}} 
\newcommand{\eps}{\varepsilon} 
\let\adot=\dot 
\newcommand{\bea}{\begin{equation}\begin{aligned}}
\newcommand{\eea}{\end{aligned}\end{equation}}
\newcommand{\Li}{\mathrm{Li}}
\newcommand{\sinc}{\mathrm{sinc}}

\renewcommand{\Im}{\mathrm{Im} \ }

\usepackage{dsfont}
\newcommand{\1}{\mathds{1}}

\usepackage{xcolor}

\def\mod{{\rm\ mod\ }}

\def\widebar{\accentset{{\cc@style\underline{\mskip10mu}}}} 
\def\wideubar{\underaccent{{\cc@style\underline{\mskip10mu}}}} 

\begin{document}

\title{Universality in Non-Equilibrium Quantum Systems}
\author{William Virgil Berdanier}

\advisor{Joel Moore}
\committeeInternalOne{Ehud Altman}

\committeeExternal{K. Birgitta Whaley}

\degree{Doctor of Philosophy}
\field{Physics}
\degreeyear{2020}
\degreeterm{Summer}
\degreemonth{August}
\department{Physics}

\pdOneName{B.S.}
\pdOneSchool{University of Texas at Austin}
\pdOneYear{2013}

\pdTwoName{M.A.St.}
\pdTwoSchool{University of Cambridge}
\pdTwoYear{2014}

\pdThreeName{M.St.}
\pdThreeSchool{University of Oxford}
\pdThreeYear{2015}
\frontmatter
\setstretch{\dnormalspacing}

\singlespacing
\setcounter{chapter}{0}  
\begin{savequote}[75mm]
To see a World in a Grain of Sand \\
And a Heaven in a Wild Flower \\
Hold Infinity in the palm of your hand \\
And Eternity in an hour\ldots \\
\qauthor{William Blake, \textit{Auguries of Innocence}}
\end{savequote}

\chapter{Universality and Equilibrium}
\label{introduction}

    \setcounter{page}{1}
    \pagenumbering{arabic}

A common rallying cry for the field of condensed matter physics is Anderson's\cite{Anderson393} ``more is different''\footnote{Or, as those of us in Joel's group like to joke, ``Moore is different.''} -- that a large number of relatively simple constituents can behave in completely new and unexpected ways. There are a few archetypical examples of this amazing phenomenon. One is the phenomenon of chaos, commonly known as the butterfly effect. In chaotic systems, even a small difference in initial conditions can lead to vastly different outcomes, as the apocryphal flap of a butterfly's wings may end in a far-off tornado.\footnote{Chaotic systems are highly prevalent, and generally occur whenever the underlying equations of motion are sufficiently non-linear. The fact that atmospheric systems are chaotic is what makes it so difficult to reliably predict the weather!} Another example is that of superconductivity. As a chunk of metal is cooled, its electrical resistance drops, until suddenly, at about 3 or 4 Kelvin -- far below the freezing point of liquid nitrogen -- it drops to zero. The metal becomes a `super'-conductor, with whirlpools of spinning electric currents flowing endlessly without dissipation, and it expels magnetic fields so strongly that it can float.\footnote{This is the technological basis of magnetic levitation (`maglev') trains, first proposed by physicists Gordon Danby and James R. Powell working at Brookhaven National Laboratory. The invention won them the Franklin Medal in 2000, and maglev has taken root in high speed train systems around the world.} Both of these examples arise from the interactions between many constituent particles, and are emergent effects that only manifest from the whole. `More' behaves differently from how any particular part would if taken by itself.

Sometimes, however, more is really \emph{the same}. Systems with totally different microscopic constituents, such as drops of water or magnetic spins, can nonetheless appear completely identical. This is the phenomenon of \emph{universality}, where, under certain conditions, Nature seems to forget the building blocks of which it is composed, and behaves in only a few different ways. 

We are all exposed to the idea of universal theories early on in our scientific education. Probably the simplest is the Ideal Gas Law, 

\begin{equation}
\label{eq:IdealGasLaw}
PV = N k_B T.
\end{equation}

Here, $P$ is the pressure of the gas, $V$ its volume, $N$ the number of particles, and $T$ the temperature, with $k_B$ the universal Boltzmann constant that holds across myriad varieties of gases. One might first think that a gas of hydrogen, a gas of CO$_2$ and a gas of vaporized lead would all behave rather differently, and show just as much variation as their atomic counterparts. Perhaps there might be some structure, such as gases with similar densities or numbers of free electrons behaving in similar ways, just as elements of the periodic table in the same column share similar properties. But reality turns out to be much more uniform, as \emph{all} gases are (approximately) described by this one thermodynamic law. How can this be?

One reason that equation~\ref{eq:IdealGasLaw} has such broad purchase is that its central approximation -- namely, that particles of the gas don't interact all that strongly -- is widely valid. (Indeed, gases that violate this assumption, such as those that have been ionized to form a plasma, don't obey equation~\ref{eq:IdealGasLaw} whatsoever.) This is something of a quirk of the gaseous state, where densities are generally low and temperatures high, such that particle collisions are not as dominant as one might expect. There is no Ideal Liquid Law or Ideal Solid Law for the simple reason that particles in those states are far closer together and hence interact far more strongly; we cannot ignore the differences in building blocks. 

Another, deeper reason why equation~\ref{eq:IdealGasLaw} broadly holds is its central assumption of thermodynamic equilibrium. Once we zap a gas out of its equilibrium state, concepts like pressure and temperature no longer make sense, and so the ideal gas law is meaningless, let alone true. Even in less extreme cases where temperature and pressure vary slowly and can be thought of as local quantities, $T(r)$ and $P(r)$, the ideal gas law cannot hold globally and must be amended to hold in a more restricted, local sense. (This is often exactly what is done to derive hydrodynamic theories, which study the spatial variations of thermodynamic variables.) Once we are far from equilibrium, though, all bets on universal relationships are off. 

\begin{figure}
	\includegraphics[width=0.45\columnwidth]{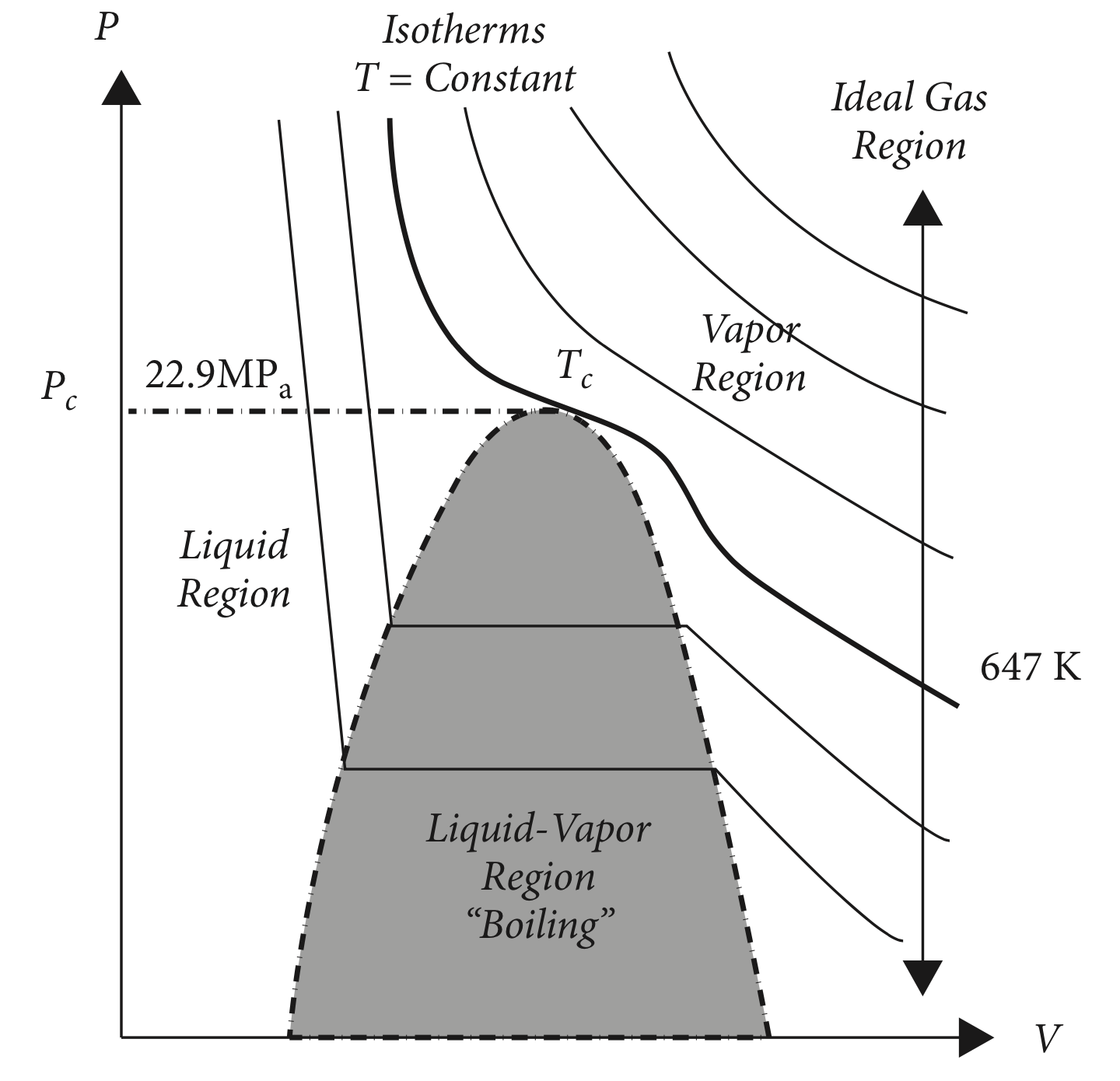}
	\includegraphics[width=0.55\columnwidth]{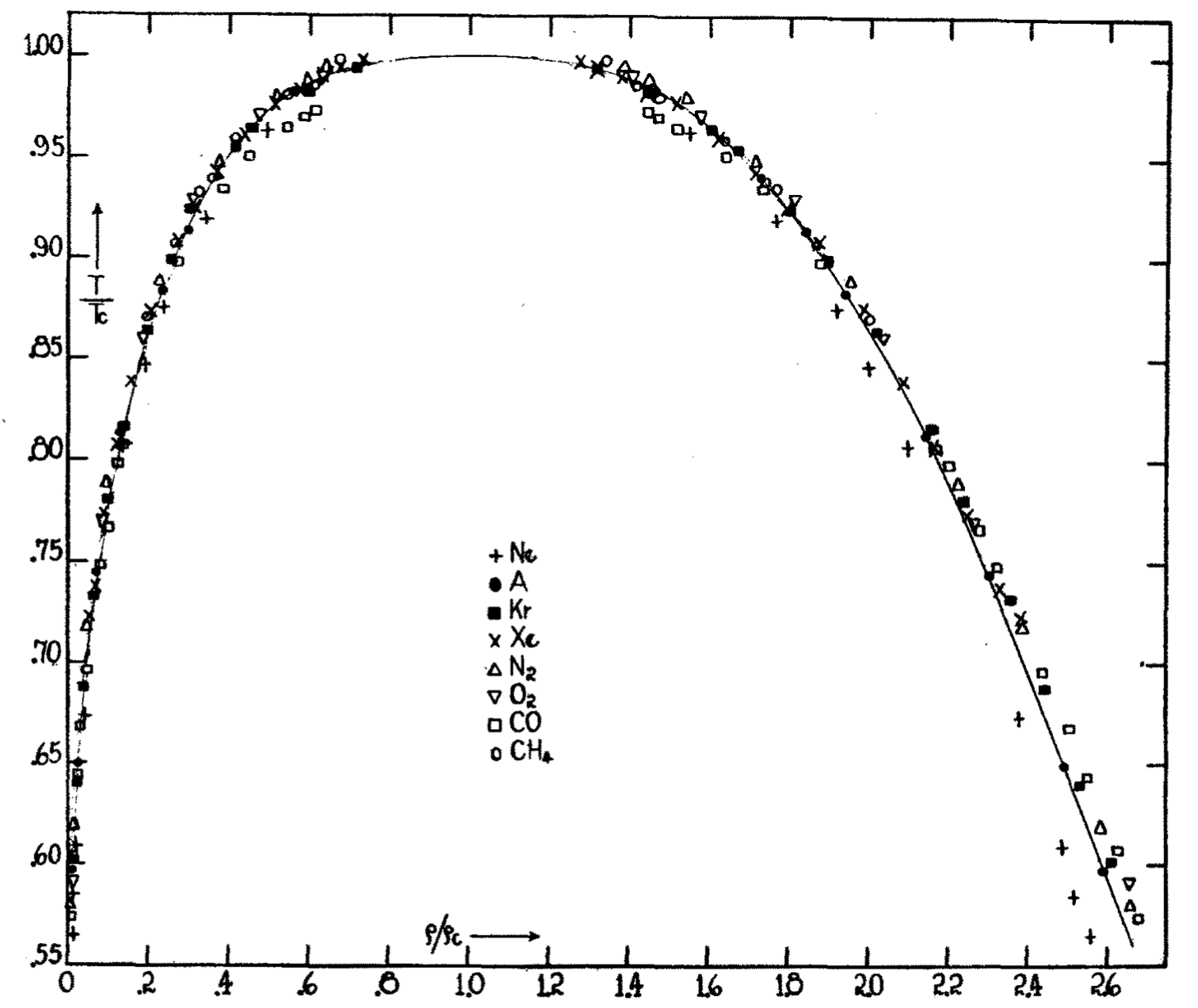}
	\caption{\label{fig:gases}Universality of the liquid-vapor transition. Left: schematic phase diagram~\cite{TheoriesofMatterInfinitiesandRenormalization}. At a critical value of the pressure $P_c$, volume $V_c$ and temperature $T_c$, the liquid undergoes a second-order transition to vapor~. Right: Remarkably, though these critical parameters depend on the microscopic details, when plotted in universal coordinates $\rho/\rho_c$ and $T/T_c$, many different substances collapse onto the same universal curve~\cite{doi:10.1063/1.1724033}. Taking this further still, the ferromagnetic transition shows the same quantitative behavior, just with different variables -- gases and magnets behave the same way!}
\end{figure}

Though there is no one ideal law for the other states of matter, these states can nonetheless sometimes behave universally, generally when undergoing a \emph{phase transition} from one type of order to another. A simple example is the liquid-vapor transition, or boiling, familiar to all coffee-lovers. The universality of this transition was shown by Guggenheim~\cite{doi:10.1063/1.1724033}. As one varies the pressure, volume and temperature of a liquid, at critical values of these parameters the liquid spontaneously changes to a gas -- a second-order, or continuous, phase transition (Fig.~\ref{fig:gases}). Though these parameter values $P_c, V_c, T_c$ depend strongly on the substance in question, when we form universal ratios $\rho/\rho_c$ (with $\rho$ the density) and $T/T_c$, many different substances collapse onto precisely one universal curve. This phenomenon of \emph{universal collapse} is a hallmark of second-order phase transitions. 

\section{Landau's Theory of Phase Transitions}

A seminal result in the history of condensed matter physics was Landau's theory of phases and phase transitions~\cite{Landau_theory}. The defining concept is that of an \emph{order parameter}, namely a thermodynamic variable that takes on a finite value when the substance is ordered, and is zero when the substance is disordered. Returning to the liquid-vapor transition, one may form the order parameter

\begin{equation}
\tilde{\rho}= \frac{\rho_l - \rho_g}{\rho_c} \propto \left( 1 - \frac{T}{T_c}\right)^{\beta},
\end{equation}

with $\rho_l$ the liquid density and $\rho_g$ the vapor density, and empirically $\beta \approx 1/3$.~\cite{doi:10.1063/1.1724033} One may similarly define an order parameter for the ferromagnetic transition by $M$, the average magnetization. We can define this in the simplest possible model of magnetism (and one that will appear multitudinous times in this dissertation!), the (classical) Ising model~\cite{Ising:1925aa},\footnote{Interestingly, Ernst Ising, after receiving his PhD from the University of Hamburg under Lenz for solving this model in one dimension, essentially left physics research. He first worked in the patent office AEG Berlin, then as a teacher in a boarding school, and finally as a professor at Bradley University in Illinois, having fled Nazi Germany. Despite introducing perhaps the most important model in statistical physics, he never published again.~\cite{Kobe:1997aa}}

\begin{equation}
\label{eq:Ising}
H(\{s_i\}) = -J \sum_{\langle ij \rangle} s_i s_j - h \sum_i s_i.
\end{equation}

Here, $s_i = \pm 1$ is a binary variable, and $H(\{s_i\})$ is the energy (also called the Hamiltonian) associated with a collection of these variables, or \emph{configuration} $\{s_i\}$. Notationally, $\langle ij \rangle$ means nearest-neighbor pairs $i,j$, and the parameters $J$ and $h$ tune the strength of the interaction and magnetic field, respectively. (We assume for simplicity that we are working on a square lattice.\footnote{The case of, for instance, a triangular lattice in 2D makes this problem much harder, as the triangular lattice shape causes geometric frustration, leading to spin glass order.}) We then form the \emph{thermal average} magnetization of site $i$ and take a spatial average of this over the sample to form the \emph{magnetization} $M$:

\begin{equation}
M(T) = \frac{1}{N} \sum_i m_{i}(T) = \sum_{\{s_i\}} \frac{e^{-H(\{s_i\}) / k_B T}}{Z(T)} \frac{1}{N} \sum_i s_i ,
\end{equation}

with $Z(T) = \sum_{\{s_i\}} e^{-H(\{s_i\}) / k_B T}$ the partition function. This $M$ is the order parameter for the ferromagnetic transition. A central object of interest is the free energy density $f(M)$, which we assume can be written as a function of the order parameter. In Landau theory, we postulate that (1) the free energy has all the symmetries of the Hamiltonian, and (2) that the free energy is analytic. Let us set $h=0$. In this case, there is a global $\Z_2$ symmetry associated to flipping the sign of all spins: the product $s_i s_j \to (-s_i)(-s_j) = s_i s_j$ is preserved, so the Hamiltonian is identical. The action of this symmetry on the order parameter is to send $M \to -M$. Because of this, only even powers of $M$ are allowed in the free energy!\footnote{We exclude absolute values like $\abs{M}$ because they are not analytic at $M=0$.} That is, we write

\begin{equation}
f(M) = f_0 + f_2 M^2 + f_4 M^4 + \ldots,
\end{equation}

expanding the free energy in Taylor series in $M^2$. In the vicinity of the phase transition, $M$ will be small -- it is zero to the one side, after all, and is smooth since the transition is second-order -- and so we can drop higher order terms. The minimum of $f(M)$ can be readily calculated by taking the derivative: $M(f_2 + 2f_4 M^2) = 0$, which has solutions $M=0$ and $M = \pm \sqrt{- f_2 / 2f_4}$. Obviously, the magnetization must be real, so if the ratio $f_2 / f_4$ is positive, the \emph{only} solution is $M=0$. This is the disordered phase. However, once $f_2/f_4$ turns negative, the solutions $M = \pm \sqrt{- f_2 / 2f_4}$ are the two minima, and $M=0$ becomes a local maximum. This is the ordered phase: the free energy is minimized by a \emph{finite} value of the order parameter. We know that this transition is tuned by temperature, so let's assume that we have $-f_2/2f_4 \propto T-T_c$.\footnote{One may wonder why we assume that the dependence here is linear. Indeed, in general we should assume it to be some unknown function, $-f_2/2f_4 = G(T-T_c)$. Near the critical point, we can expand in Taylor series, keeping only the linear term $G(T-T_c) \propto T-T_c$; any constant term is 0, as we know that $G$ changes sign at $T_c$, and we have no reason to impose that the first derivative vanishes. So applying Occam's razor, we get linear behavior. In a situation that would surely be pleasing to Friar William of Ockham, Landau theory feels like a divine miracle.} Then the magnetization behaves like 

\begin{equation}
M \propto \sqrt{T-T_c}. 
\end{equation}

This is remarkable: using nothing but symmetry and analyticity, we have deduced the presence of two phases, and that at the critical point, we should have the exponent $M \propto (T-T_c)^\beta$ with $\beta=1/2$. We have further deduced that in the phase with $M\not=0$, the Ising symmetry is \emph{spontaneously broken}; despite the fact that the Hamiltonian is symmetric under $s_i \to -s_i$, the order parameter itself is not symmetric (the only symmetric possibility being $M=0$), and sending $s_i \to -s_i$ toggles between two degenerate states with equal free energy.\footnote{For this reason, Landau theory is a general theory of symmetry-breaking transitions; transitions that do not involve symmetry breaking, such as topological transitions, are outside the Landau paradigm.} 

We may rightly wonder if this solution is correct. In fact, it is not correct, or at least not until we go to a high enough dimension. In Ising's exact solution in 1D, we know that there is in fact \emph{no transition}~\cite{Ising:1925aa}, and the model is always disordered ($M=0$).\footnote{The one-dimensional classical Ising model is easily solved with, e.g., transfer matrices.} In 2D, we again have an exact solution due to Onsager~\cite{PhysRev.65.117}, which shows that $\beta = 1/8$. In 3D, we have no exact solution, but state-of-the-art numerics~\cite{El-Showk:2014aa} pin the exponent to $\beta=0.326419(3)$, rather close to $\beta=1/3$. Finally, in dimension $D\geq 4$, the Landau theory solution is correct. This is also known as the \emph{mean-field} solution, since we essentially replaced the fluctuating order parameter $M(\{s_i\}) = \frac{1}{N}\sum_i s_i$ -- which changes from configuration to configuration, even at the same energy -- by its thermal mean, $M$, in the free energy. The failure of mean-field theory underscores the importance of fluctuations in low dimensions.\footnote{We can understand why, in high dimensions, mean field theory works by noting that the number of nearest neighbors scales as $2d$, with $d$ the dimensionality. Thus, in high $d$, there are many nearest neighbors, so each spin feels something close to the mean of all the other spins. For the Ising model, it turns out, the \emph{lower critical dimension}, below which mean field theory completely fails, is $d_l=1$, and the \emph{upper critical dimension}, above which mean field theory is exact, is $d_c=4$. In between, mean field theory gets some aspects right and some aspects wrong. Every universality class has its own $d_l$ and $d_c$.}

Incredibly, the values of $\beta$ for the 3-dimensional zero-field Ising model and the liquid-vapor transition are the same, at nearly 1/3. That is, this value of $\beta$ is characteristic of the \emph{Ising universality class}. In general, there are only a small number of possible universality classes, each of which hold a collection of sometimes seemingly unrelated models. Quoting Leo Kadanoff's ``principle of universality,''

\begin{quotation}
	All phase transition problems can be divided into a small number of different classes depending upon the dimensionality of the system and the symmetries of the order state. Within each class, all phase transitions have identical behavior in the critical region, only the names of the variables are changed.~\cite{TheoriesofMatterInfinitiesandRenormalization}
\end{quotation}

The 3-dimensional superfluid Helium transition is also in this universality class. 

\section{The Renormalization Group}

The puzzle of how to understand the origins of these critical exponents, and crucially why there are only a few different classes -- the phenomenon of universality -- was tackled in the equilibrium case by a set of tools collectively called the \emph{renormalization group}, or the RG.\footnote{See, e.g., Cardy~\cite{Cardy:1996xt} for an elegant treatment of this topic.} In broad strokes, the main idea of the RG is that at a critical point, we expect \emph{scale invariance}: if we ``zoom out'', or rescale the system, the form of the laws (the Hamiltonian) is precisely the same. To understand phase transitions, then, we can perform a ``blocking'' transformation that manually enacts this zooming out process, often referred to as the real-space renormalization group (RSRG).\footnote{This is in contrast to the complementary momentum-space picture, which systematically derives a low-energy (i.e. long length-scale) approximation to the starting theory. Rather than zoom out in space, that picture zooms out in energy.}

Let's make this concrete by considering the 1-dimensional Ising model, Eq.~\ref{eq:Ising}.\footnote{The 1D classical Ising model is analytically solvable, and here the RG procedure may in fact be carried out exactly. This is exceedingly rare; almost all RGs involve some kind of approximation, which places caveats on the predictions. We will see some of this in the real-space RG for disordered systems in Chapter~\ref{ch:RSRG}.} The partition function for this model is 

\begin{equation}
\label{eq:IsingZ}
Z = \Tr e^{-\beta H} = \Tr e^{-H_0} = \prod_{i=1}^N e^{-\beta J s_i s_{i+1}} = \prod_{i=1}^N e^{-K s_i s_{i+1}},
\end{equation}

absorbing $\beta=1/k_B T$ into the definition of the Hamiltonian. We write $H_0 = -\beta J \sum_{i=1}^N s_i s_{i+1} = -K \sum_{i=1}^N s_i s_{i+1}$, with $K=\beta J$. Let's perform a blocking transformation:\footnote{The first paper to propose a blocking transformation for the Ising model was Kadanoff's groundbreaking 1966 paper, ``Scaling laws for Ising models near $T_c$.''~\cite{PhysicsPhysiqueFizika.2.263}} for every three consecutive spins (non-overlapping triplets), we simply assign the group to have the spin of the middle element.\footnote{A more democratic RG rule would be majority vote, but it's a little more analytically complicated. Reflecting the stability of RG fixed points, it doesn't really matter how we block, so long as it is physical, and many different RG schemes will lead to the same predictions provided they make compatible approximations. The insight behind this RG rule is that in the ordered phase, all the spins tend to point the same way anyway, so picking the middle one and picking the majority vote give the same answer.} We then trace over the spins at the edges of each block, leaving the middle one untouched. This is called a \emph{decimation}. 

Say that we have two blocks, composed of spins $s_1$ to $s_6$ in the raw variables, and we decimate once. This means we sum over spins $s_3$ and $s_4$, which are at the edges of the blocks, and keep $s_1' = s_2$, $s_2' = s_5$. The partition function (Eq.~\ref{eq:IsingZ}) factors involved in the decimation are 

$$
e^{K s_1' s_3} e^{K s_3 s_4} e^{K s_4 s_2'}.
$$

We can analytically perform the trace, by noting that $e^{K s_3 s_4} = \cosh K (1 + s_3 s_4 \tanh K)$, and performing the sum over $s_3, s_4 = \pm 1$ gives $4 \cosh^3 K (1 + s_1' s_2' \tanh^3 K)$. Remarkably, this has the same form as before the decimation, but now in the new primed variables! In particular, this is

\begin{equation}
\label{eq:RGRule}
e^{\tilde K s_1' s_2'}, \qquad \tilde K = \tanh^{-1} \left( \tanh^3 K \right).
\end{equation}

This is known as an \emph{RG rule}. Once we've derived this rule, we can imagine performing this decimation procedure many times on the chain, and track how the couplings flow. The full form of the Hamiltonian after the blocking is 

$$
\tilde H = N' g(K)  - \tilde{K} \sum_i s_i' s_{i+1}', \qquad g(K) = -\frac{1}{3} \log \frac{\cosh^3 K}{\cosh \tilde{K}} - \frac{2}{3} \ln 2,
$$

with $N'=N/3$ the number of blocks. As we iterate the RG rule Eq.~\ref{eq:RGRule}, which in terms of $x=\tanh K$ is $\tilde x = x^3$, we see that $x$ simply shrinks to 0.\footnote{$x$ lies between 0 and 1, so unless $x=1$, it will eventually shrink to 0.} Since $K$ has a factor of $1/T$ in it, this means that we flow to an effectively infinite temperature state. Infinite temperature does not host order, so is paramagnetic. We have therefore deduced that, no matter our initial condition (unless we are precisely at $T=0$), our state is paramagnetic at long lengths; there is no order! There are two fixed points, $x=0$ (paramagnetic) and $x=1$ (ferromagnetic, it turns out), with $x=0$ being `stable', i.e. an attractor, and $x=1$ `unstable'. The \emph{RG flow} is from $x=1$ to $x=0$.  

\begin{figure}
	\includegraphics[width=0.45\columnwidth]{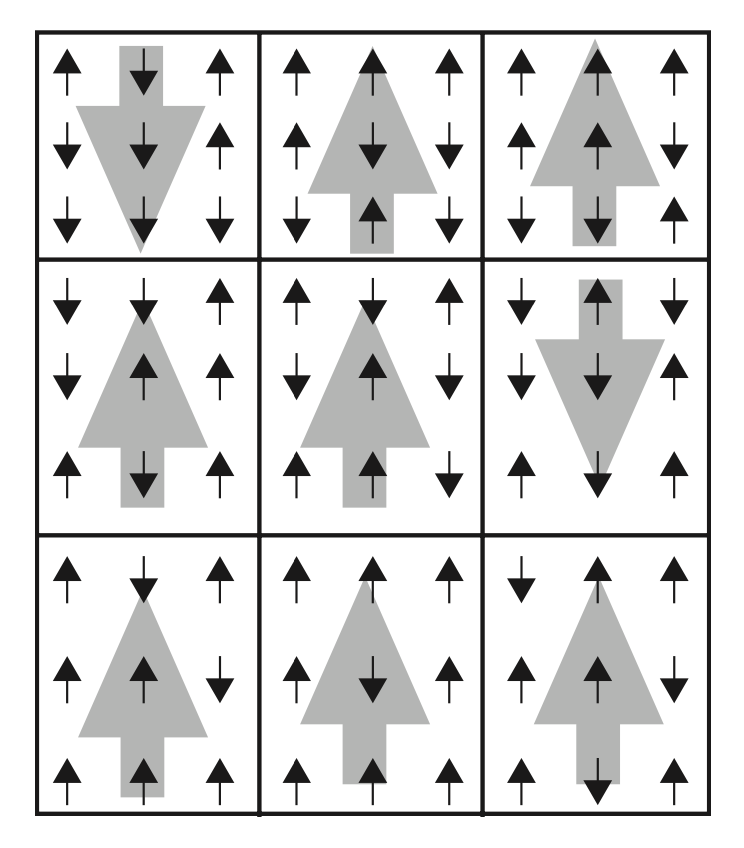}
	\includegraphics[width=0.55\columnwidth]{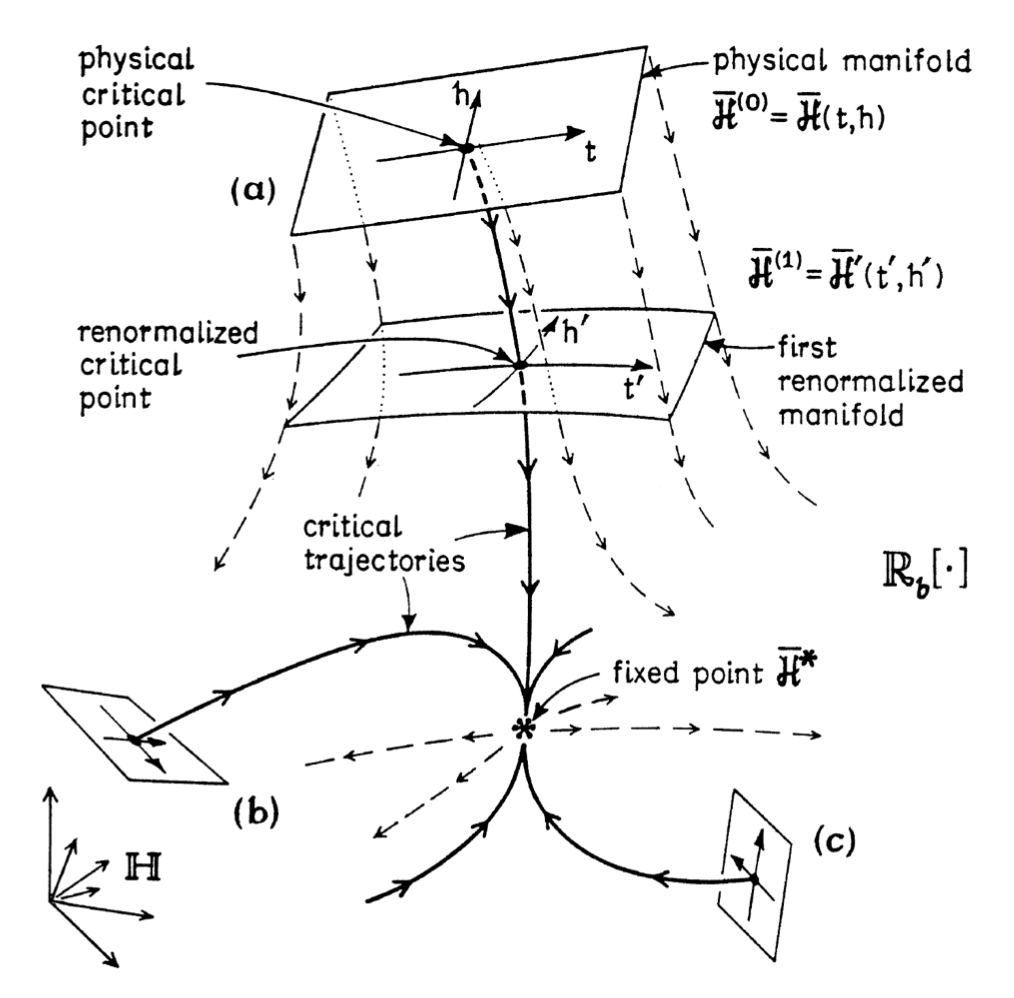}
	\caption{\label{fig:RG} Left: an example ``blocking'' transformation on the classical Ising model (from Ref.~\citenum{TheoriesofMatterInfinitiesandRenormalization}). The `super'-spin of the block is the majority vote of the microscopic spins, with couplings between them gotten from adding up the interactions among the microscopic spins. Repeatedly blocking and following the flow of these couplings is the central idea of Kadanoff's real-space renormalization group. A similar picture exists where blocking occurs in $k$-space, where low-energy modes are integrated out. Right: a schematic diagram of RG flow given by Michael Fisher\cite{RevModPhys.70.653}. As we perform infinitesimal blocking transformations, we flow along trajectories in coupling space, ultimately arriving at (or diverging away from) a fixed point. These fixed points are the canonical models of the universality classes describing critical points.}
\end{figure}

We can view this blocking procedure as an infinitesimal transformation, once we have coarse grained so much that we no longer can make out the individual spins. This continuum limit is, in general, very revealing. Every time we block, we zoom out by a factor of $b=3$. This means that whatever lengths are in the system, in particular the correlation length $\xi$, must change by 

$$
\xi(x') = \frac{1}{b} \xi(x), \qquad x' = x^b.
$$

This has the solution $\xi \sim 1/\log \tanh K$, which is finite but growing extremely long as $T\to 0$. 

\begin{table}
	\caption{\label{table:Ising} Some critical exponents of the Ising universality classes, and operator dimensions from the RG. $\beta$ is the exponent $M \propto (T-T_c)^\beta$. Defining a susceptibility $\chi$ in response to a source field $h$, we have $\gamma$ is the exponent $\chi \propto (T-T_c)^\gamma$. The correlation length $\xi$ is also a power-law with exponent $\nu$, in $\xi \propto (T-T_c)^\nu$. Finally, at the critical point the (thermal average) two-point correlation function decays as a power law with exponent related to $\eta$, in $\langle s_0 s_r\rangle \sim r^{-d+2-\eta}$. The quantities $\Delta_\sigma$ and $\Delta_\epsilon$ are the renormalization group (RG) scaling dimensions of the $\sigma$ and $\epsilon$ fields, respectively.}
	\begin{center}
		\begin{tabular}{c|ccccc}
			& \multicolumn{5}{c}{The Ising Universality Class: a snapshot} \\
			\hline
			& \multicolumn{4}{c}{dimension $d$} & \\
			\cline{2-5}
			& 1~\cite{Ising:1925aa} & 2~\cite{PhysRev.65.117}  & 3~\cite{El-Showk:2014aa} & 4 (mean field) & RG expression~\cite{Cardy:1996xt}\\
			\hline
			$\beta$ &  none  & 1/8   & 0.326419(3)  &  1/2 & $\Delta_\sigma / (d-\Delta_\epsilon)$ \\
			$\gamma$ & none    & 7/4     & 1.237075(10) & 1  &  $(d - 2\Delta_\sigma) / (d - \Delta_\epsilon)$ \\
			$\nu$ & none   &  1   & 0.629971(4) & 1/2 & $1/(d - \Delta_\epsilon)$ \\
			$\eta$ & none & 1/4 & 0.036298(2) & 0 & $2\Delta_\sigma - d + 2$ \\
			\cline{1-5} 
			$\Delta_\sigma$ & & 1/8 & 0.5181489(10) & 1 & \\
			$\Delta_\epsilon$ & & 1 & 1.412625(10)  & 2 & \\
		\end{tabular}
	\end{center}
	\vspace{2in}
\end{table}

This simple example reveals some key aspects of RG transformations in general. Given microscopic variables $K_i$, under a decimation, we form new couplings $\tilde{K}_i = R(\{K_j\})$, where $R$ is some function that we often will linearize about the fixed point $K_*$. We seek to form \emph{scaling variables} $u_i$, which are combinations of the microscopic variables $K_i$, that transform multiplicatively under the RG flow: $\tilde u_i = \lambda^{i} u_i = b^{\Delta_i} u_i$.\footnote{Formally, $\lambda_i$ is an eigenvalue of the Jacobian for the RG rule, $\partial K'/\partial K$, at the fixed point $K_*$. If we linearize about the fixed point, then the eigenvector $u_i$ indeed transforms multiplicatively under the RG rule.} The variable $u_i$ is considered to fall into three possible classes, depending on $\Delta_i$: if $\Delta_i > 0$, it is \emph{relevant}, as it will grow arbitrarily large under the RG flow; if $\Delta_i < 0$ it is \emph{irrelevant}, as it will shrink to 0 under the RG flow; and if $\Delta_i=0$ it is \emph{marginal}, and we cannot immediately tell what will happen to it, needing to go beyond the linear approximation. $\Delta_i$ is the \emph{scaling dimension} of scaling variable $u_i$. The surface spanned by the $u_i$ variables is the \emph{renormalized manifold}, which flows with the RG; an illustration is in Figure~\ref{fig:RG}.

Calculating the $\Delta_i$ is one of the main goals of any RG procedure. This is because their physical importance is paramount, as they fully determine the critical exponents. For instance, for the Ising universality class, the field-theoretic variables are $\sigma$ and $\epsilon$; the relationships between $\Delta_\sigma$, $\Delta_\epsilon$, and the dimension $d$ precisely give the critical exponents of measurable quantities like the specific heat and susceptibility. A snapshot of the Ising class is given in Table~\ref{table:Ising}. The breathtaking success of the renormalization group in understanding the origin of these exponents, in a quantitatively sharp way, is one of the triumphs of twentieth century physics and the provenance of Ken Wilson's 1982 Nobel Prize.\footnote{Wilson was one of the true pioneers of the RG, greatly expanding on Kadanoff's earlier work. In 1971, he introduced the differential form of the RG equations and explored its consequences in a pair of works (Refs.~\citenum{PhysRevB.4.3174,PhysRevB.4.3184}). His 1972 paper with Michael Fisher, cutely titled ``Critical exponents in 3.99 dimensions,''~\cite{PhysRevLett.28.240} introduced the $\epsilon$-expansion -- a radical idea of expanding in the number of spatial dimensions, treated as a smooth parameter -- and the celebrated Wilson-Fisher fixed point. Finally, in 1975 Wilson gave a resolution to the notorious Kondo problem in terms of the RG~\cite{RevModPhys.47.773}, cementing his Nobel prospects. Incidentally, Michael Fisher is the father of Daniel S. Fisher, whose work on RGs in disordered systems is covered in Chapter~\ref{ch:RSRG}, and noted physicist Matthew P. A. Fisher.}

\section{Non-Equilibrium Dynamics and Floquet Theory}

The previous analysis of universality has relied on the tacit assumption of thermal equilibrium -- there is no notion of thermodynamic variables, free energies, etc., without it. However, one would like to move beyond this rather restricted space, and try to understand to what extent \emph{non}-equilibrium systems can behave universally. 

The full non-equilibrium problem is hard, and perhaps intractable. One special case where we have some amount of control is that of \emph{periodic driving} or \emph{Floquet driving}. In Floquet systems, the Hamiltonian is a periodic function of time, 

$$
H(t+T) = H(t)
$$

for some period $T$. This means that the evolution operator over one cycle, $$F = U(T) = \exp\left(-i \mathcal T \int_0^T H(t) dt\right)$$ with $\mathcal T$ ensuring time-ordering, is of central importance.\footnote{Throughout this dissertation, we assume we are working in units where $\hbar = 1$. We also often set $T=1$, absorbing it into $H_F$, such that quasienergies lie between 0 and $2\pi$.} Since it is a unitary operator, we can always write it as 

\begin{equation}
F = e^{-i T H_F} \quad \Leftrightarrow \quad H_F = \frac{i}{T} \log F,
\end{equation}

which defines the \emph{Floquet Hamiltonian}  $H_F$. Here we can glimpse the power of Floquet systems; despite being strongly non-equilibrium systems, continuously driven, they admit a time-independent Hamiltonian description when probed at multiples of the drive frequency $T$, as $U(NT) = F^N = e^{-i NT H_F}$. In other words, every cycle, the system appears as though it was simply evolved under a time-independent operator $H_F$. By tuning the types of driving, we may make $H_F$ as complicated, simple, or exotic as we want, in a process known as \emph{Floquet engineering}. 

A final fundamental consequence of periodic driving is that the eigenstates of $F$, which are also those of $H_F$, must take a special form. This is known as \emph{Floquet's theorem}, analogous to Bloch's theorem for Hamiltonians with discrete spatial translation symmetry (on a lattice), and reflects the discrete time translation symmetry of Floquet systems.\footnote{Gaston Floquet was a French mathematician in the 1800s who mainly worked on differential equations. He published his eponymous theorem across three works in 1881-83 (all with the same title)~\cite{ASENS_1883_2_12__47_0}, shortly after finishing his doctorate at the \'Ecole Normale Sup\'erieure. He spent most of his life as a professor in Nancy, helping make it one of the leading research centers outside Paris. His theorem was generalized to the quantum setting by Shirley in 1965~\cite{PhysRev.138.B979}.} It states,

\begin{equation}
\label{eq:FloquetTheorem}
\ket{\Psi(t)} = e^{-i \epsilon t} \ket{\psi(t)}, \qquad \ket{\psi(t+T)} = \ket{\psi(t)}.
\end{equation}

The quantity $\epsilon$, defined on the circle modulo $2 \pi / T$, is called the \emph{quasienergy} of the state $\ket{\Psi}$. These are \emph{not} energies in that they are periodic, and only conserved up to multiples of $2 \pi / T$. The periodic nature of quasienergy space is one of the defining features of Floquet systems, allowing them to escape many no-go theorems for static Hamiltonians.\footnote{One simple example is the possibility of chiral one-dimensional Floquet systems. Bands of a Hamiltonian must be periodic in $k$-space, which means that any band of a static Hamiltonian must have both left- and right-moving pieces; a purely right-moving band $\epsilon = v k$ could not be periodic. In contrast, in a Floquet system a purely right-moving band is perfectly fine, as we identify the quasienergies 0 and $2\pi$ with one another.} 

Floquet systems also raise the tantalizing prospect of dynamical \emph{phases of matter}. We may wonder, can the Floquet Hamiltonian host types of order that are impossible in traditional Hamiltonians, or not present in any of the drives used to produce $H_F$? Can we obtain fundamentally new types of quantum orders in driven systems, and if so, what are they? Do these orders also lead to new types of transitions, and new notions of driven, non-equilibrium universality? These striking questions are at the heart of this dissertation. 

\section{Universality out of Equilibrium}

There have been several non-equilibrium universality classes discovered in the past several decades. Some are biologically inspired: the growth of surfaces obeys what is known as Kardar-Parisi-Zhang (KPZ) universality~\cite{PhysRevLett.56.889}, where $x \sim t^{2/3}$. This is, quite remarkably, different from almost all equilibrium classes, which have the ballistic scaling of $x\sim t$.\footnote{The astute reader may rightly object that equilibrium systems have no notion of time, $t$. This is of course correct, but we may relate time in a $d$-dimensional system to an equilibrium $d+1$-dimensional system through analytic continuation: we write $\tau = i t$, and treat $\tau$ as a new Euclidean dimension.} This is also distinct from diffusion, which obeys $x \sim \sqrt{t}$. Other models of surface growth, such TASEP, are also in the KPZ class~\cite{CORWIN:2012aa}. Other universality classes include reaction-diffusion, percolation and directed percolation, the Edwards-Wilkinson class, and various universality classes of cellular automata~\cite{RevModPhys.76.663}. Finally, a fascinating dynamical universality class is that of aging in spin glasses, such as the Sherrington-Kirkpatrick model, the Edwards-Anderson model, and the random energy model~\cite{spin_glass_aging}.\footnote{The author thanks Leticia Cugliando for introducing him to this subject at the Les Houches school, `Dynamics and Disorder in Quantum Many Body Systems Far from Equilibrium', Summer 2019.}

The focus of this dissertation is on universality in non-equilibrium \emph{quantum} systems, in particular, those that are driven by some kind of external perturbation. This generally implies that the system does not conserve energy, as the Hamiltonian is time-dependent. We write 

\begin{equation}
H = H_0 + H_D(t),
\end{equation}

where $H_D(t)$ is the drive term. In this dissertation, we consider two generic types of drive: (1) $H_D(t)$ is stochastic, in particular, a Poisson process, and (2) $H_D(t) = H_D(t+T)$ for some period $T$, also known as Floquet driving. A central issue with considering order and criticality in driven systems is that of heating. If a system heats to infinite temperature, we expect no order, as every configuration becomes equally likely ($e^{-E/k_B T} \to 1$) and any order simply averages out. How then can we observe universal responses in driven systems?

In this dissertation, we engineer two ways around this problem. The first, explored in Chapter~\ref{ch:CFT}, is to consider \emph{boundary} driving in clean systems. By boundary driving, we mean that $H_D(t)$ has support only on the boundary degrees of freedom of the system, while $H_0$ has support everywhere. While indeed boundary-driven systems may heat to infinite temperature at infinitely long times, the problem is less severe than in the bulk-driven case. First, any heating must be local; at a finite time $t$, the light cone of excitations can only have extended to a finite region $x \sim t$, and any part of the system outside of this will not have heated at all.\footnote{Strictly speaking, the light cone in a non-relativistic quantum system is not absolute, and correlations are instead bounded by the Lieb-Robinson inequality (provided interactions are short-ranged enough)~\cite{Lieb:1972aa, PhysRevX.9.031006}. Correlations outside the light cone decay exponentially with distance, though, so heating there is negligible.} Second, at each time step, we are only inputting a sub-extensive amount of energy into the system, rather than an extensive amount. This means that any heating will occur on a much slower timescale than in the bulk-driven case. In the non-interacting limit in particular, excitations will move from the boundary outward in a lightlike fashion, leaving no trace once they have passed out of the region of interest. We generally consider starting with a bulk-critical system and driving its boundary, leading the dynamics to inherit universality from the critical bulk. 

The second way around the issue of heating is through the consideration of disorder. A seminal discovery of the twentieth century was that of Anderson localization -- explained in detail in Chapter~\ref{ch:RSRG} -- whereby a quantum wavefunction ceases to be extended, and instead becomes exponentially decaying, in a disordered potential. This has quite recently been generalized to the many-body case, where interactions are strong. Importantly, such many-body localized systems do not reach thermal equilibrium in the traditional sense, and can host quantum orders at arbitrarily large energy densities. Further, such localized phases are robust to periodic driving, and have been shown to host new dynamical types of order, such as time-crystalline order. Within this context, we examine what happens as we transition between different many-body-localized phases. 

Finally, in Chapter~\ref{ch:Hydro} we consider hydrodynamics, which is in some sense the most universal non-equilibrium theory. All thermalizing systems are expected to be described by hydrodynamics in their late-time limit, and only a few coarse-grained parameters, such as viscosities and scattering times, need go into the hydrodynamic description. Part of the power of this description is that equilibrium is close at hand, as the system is still in \emph{local} equilibrium on a sufficiently short scale. By considering the Coulomb drag phenomenon, whereby a current in one plate can pull along a reciprocal current in another nearby plate via quantum fluctuations, we investigate a quantum limit of hydrodynamics. In particular, we study the drag between thermal currents, showing that they behave in remarkably different ways to charge currents.  
\begin{savequote}[75mm]
Mulla had lost his ring in the living room. He searched for it for a while, but since he could not find it, he went out into the yard and began to look there. His wife, who saw what he was doing, asked: ``Mulla, you lost your ring in the room, why are you looking for it in the yard?''

Mulla stroked his beard and said: ``The room is too dark and I can't see very well. I came out to the courtyard to look for my ring because there is much more light out here.''
\qauthor{Nasreddin}
\end{savequote} \vspace{12mm}

\chapter{Boundary Driving and Conformal Field Theory}
\label{ch:CFT}

We would like to understand how quantum systems behave when out of equilibrium, and in particular, if they can behave in a universal way. A natural starting point, then, is an equilibrium quantum system at criticality; we can then gently push it away from equilibrium, and see if the resulting phenomena are still universal. It may seem obvious that a critical system, gently pushed, may remain universal; however, there are good reasons to expect the opposite. As a system is pushed away from equilibrium by the injection of energy, the system should heat.\footnote{The process of heating in quantum systems is itself a fascinating topic that has seen much progress in recent years, and one that will be discussed at length throughout this dissertation. Suffice it to say that a ``generic'' interacting many-body quantum system (obeying the Eigenstate Thermalization Hypothesis~\cite{PhysRevE.50.888,0034-4885-81-8-082001,PhysRevA.43.2046}) should eventually ``heat'', meaning equilibrate to a thermal density matrix with temperature equal to the initial state's average energy density. Exceptions include free (non-interacting) systems, integrable systems (which instead equilibrate to a Generalized Gibbs Ensemble~\cite{Vidmar_2016}), and many-body localized systems; more on the last can be found in Chapter~\ref{ch:RSRG}.} In many cases, temperature is a relevant perturbation in a quantum critical system, meaning that even small increases in temperature will have large effects at the longest length scales.\footnote{This is an abuse of terminology, though not of my initial doing. Strictly speaking, all terms in the Hamiltonian (that survive the RG flow) are \emph{marginal}, with RG-dimension 0. Sometimes operators with even irrelevant character can be ``dangerously irrelevant'' in that they can lead to heating, which in turn destroys the criticality. Temperature is `relevant' in spirit, though, for the reason that it has a large effect on critical phenomena.} Truly quantum phase transitions are often defined only at \emph{absolute zero}, and as one moves away from the $T=0$ limit, the quantum critical point broadens into a `quantum critical fan'~\cite{sachdev2011quantum} with a mix of quantum and classical critical phenomena.\footnote{Quantum critical fans are their own vibrant area of both experimental and theoretical research, e.g., in the cuprates (high-temperature superconductors).} 

A central problem with studying non-equilibrium physics is the absence of a basic physical framework. In equilibrium, we have the framework of thermodynamics, which allows us to make sense of quantum systems in terms of temperature, free energies, susceptibilities, specific heats, and the like. Depending on how strongly we are out of equilibrium, some or none of this framework makes sense. So long as we move adiabatically between equilibrium states, thermodynamics is totally valid -- the definition of adiabaticity being, roughly, that motion between two equilibrium thermodynamic states is so slow that all intervening states are also equilibrium thermodynamic states.\footnote{Note that this is slightly different than the usual definition of adiabaticity as no exchange of heat, i.e. $dQ=0$. This generalizes better to the quantum setting, though, where we care about moving slowly enough to stay in the system's ground state as we change the parameters in the Hamiltonian.} Once we allow for non-adiabaticity, we can still sometimes use local notions of thermodynamics, where, e.g., temperature becomes a spatially varying function $T(r)$; on a microscopic scale, each small region of the system is in equilibrium, with flows between these regions reflecting the non-equilibrium nature of the problem. This is often what is done in fluid mechanics; for more, see Chapter~\ref{ch:Hydro}.  

The spirit of the hydrodynamic approach above is to start with what we do know how to handle -- namely equilibrium systems -- and patch them together to get a special case of what we don't know how to handle, namely non-equilibrium systems. In this vein, one may start with something we know well -- perhaps a certain class of quantum critical systems -- and then weakly drive them away from equilibrium. This is the motivation for this chapter's approach to non-equilibrium universality. In particular, we consider quantum critical systems described by \emph{conformal field theories}, or CFTs, which are quantum field theories obeying a particularly powerful symmetry called conformal symmetry.\footnote{Per usual, there are also classical field theories with conformal symmetry -- with a general relation between them being the quantum-classical mapping -- but we will concern ourselves with the quantum kind.} Roughly, conformal symmetry means scale invariance, and this is why it is so common at quantum critical points. Critical points are the fixed points of an RG-flow, and hence rescaling them -- or flowing with the RG -- doesn't do anything! So, they are scale invariant.\footnote{One has to be quite careful with this logic in disordered systems, which are not scale invariant in this sense at criticality; rather the distribution of their randomness is invariant.} Strictly, conformal invariance is more than this: it is the symmetry group of the metric tensor itself, in that a conformal transformation is an isomorphism of the metric up to a rescaling: $g_{\mu\nu} \to \Omega(r) g_{\mu\nu}$. In layman's terms, conformal transformations preserve angles, but not lengths, with simple rescaling being a special case.\footnote{It is much less obvious why criticality should imply conformal invariance, and not just scale invariance. This is actually a question I have asked myself many times over the past five years! To the best of my knowledge, criticality does NOT in general imply conformal invariance. A nice discussion is provided by Nakayama~\cite{nakayama2013scale}, the upshot being that in 2 dimensions, one can prove that scale invariance + quantum field theory + short-enough-ranged interactions implies conformal invariance. Quoting him, ``As of January 2014, our consensus is that there is no known example of scale invariant but non-conformal field theories in d=4 dimension under the assumptions of (1) unitarity, (2) Poincar\'e invariance (causality), (3) discrete spectrum in scaling dimension, (4) existence of scale current and (5) unbroken scale invariance in the vacuum.'' (There is also no rigorous proof, either.) Remarkably, the (quite physical and not pathological) theory of elasticity in two dimensions displays scale but NOT conformal invariance, as shown by Cardy and Riva~\cite{RIVA2005339}.} Sadly, powerful as conformal invariance is, its effect on quantum field theory is only truly understood in two dimensions. In $d>2$, the conformal group simply consists of (1) translations, (2) dilations, (3) rotations, and (4) `special conformal transformations'.\footnote{Namely, $r^\mu \to r^\mu + \alpha^\mu (r)$ with $\alpha^\mu(r) = b^\mu r^2 - 2 b^\lambda r_\lambda r^\mu$, which in plain English is a normal conformal transformation followed by an inversion $r^\mu \to r^\mu / r^2$, a translation by $b^\mu$, and another inversion.} In $d=2$, the conformal group is much larger, and consequently much more illuminating. This is because two-dimensional conformal transformations are simply analytic functions of a complex variable $z=x+i\tau$ in disguise, and we can thus bring the power of complex analysis to bear. 

In this chapter, we focus on the case of a 1+1-dimensional quantum critical system, described by a conformal field theory, driven at its boundary. We focus on boundary driving for the reason that boundary-driven systems are expected to be less sensitive to heating, as we only input a sub-extensive amount of energy with each cycle of the drive. The focus on 1+1-dimensional systems is not as restrictive as it might sound, as some symmetrical three-dimensional problems, such as the Kondo problem, reduce to a one-dimensional problem in the coordinate $r$; nonetheless the main application would be towards understanding quantum wires, and gleaning insight into the possible universal behavior of non-equilibrium quantum systems generally. 

We begin with a CFT primer that introduces the reader to the main concepts of conformal field theory, borrowed heavily from Cardy~\cite{Cardy:1996xt, cardy_boundary_2006}.\footnote{This would be an opportune time to thank John for his help throughout my PhD; his coming to Berkeley at the time that I started was extremely fortuitous, and I learned a lot from him.} The bible is the `yellow book' of Di Francesco, Mathieu and S\'en\'echal~\cite{di_francesco_conformal_2011}, which is extremely useful as a reference, but in the author's opinion lacks the physical insight of Cardy. The author also recommends Michael Flohr's Conformal Field Theory Survival Kit\footnote{\url{https://www.itp.uni-hannover.de/fileadmin/arbeitsgruppen/ag_flohr/papers/w-cft-survivalkit.pdf}}. Paul Ginsparg's notes from Les Houches~\cite{ginsparg_CFT} are also a classic (and in David Tong's opinion~\cite{tong2009lectures}, the canonical way to learn CFT). CFT with boundaries is more specialized, and the best entr\'ee to the literature is Cardy's article~\cite{cardy_boundary_2006}. For more on the use of boundary-condition-changing (BCC) operators in quantum impurity problems (such as the Kondo problem), Affleck's article is very nice~\cite{Affleck:1996mm}. 

We then move on to include slightly amended versions of the author's works `Floquet Dynamics of Boundary-Driven Systems at Criticality'~\cite{PhysRevLett.118.260602} and its appendices, and `Universal Dynamics of Stochastically Driven Quantum Impurities'~\cite{PhysRevLett.123.230604}. 

\section{A Conformal Field Theory Primer}

In this section, we give a brief introduction to the main tools and techniques of two-dimensional conformal field theory. This will be quite rapid and non-rigorous; for more detailed explanations, please refer to the above resources. 

\subsection{Conformal symmetry and the plane}

Conformal symmetry is invariance under conformal transformations, or transformations that preserve angles. More specifically, a conformal transformation is one that rescales the metric $g_{\mu\nu} \to \Omega(r) g_{\mu\nu}$. In two dimensions, we can move to complex coordinates by defining the variables $z = x + i \tau$, $\bar z = x - i \tau$, which are treated as independent of one another. We can at this stage remain agnostic as to the physical significance of $x$ and $\tau$, and simply treat them as Euclidean coordinates, with the theory defined on this Euclidean space. In practice, a spin chain or other quantum system tends to map onto such a statistical mechanics model only through the use of a Wick rotation to imaginary time, $\tau = i t$ with $t$ the physical time,\footnote{This comes with its own host of problems; one must make sure that, when the imaginary time answer is analytically continued back to real time, no divergences or non-analyticities are encountered. A simple failure mode is the analytic continuation of a sinusoid: perhaps some answer gives $e^{-\omega \tau}$, which is decaying and negligible for large $\tau$, but is simply oscillating with no decay in real time, as $e^{-i \omega t}$. Often issues arise when there are poles in the imaginary time answer, preventing a smooth analytic continuation to real time. Worries about analytic continuation can generally be put to rest with numerical evidence, in the case that the analytic continuation cannot be proven to be safe.} or when $\tau$ is treated as an inverse temperature $\tau = \beta$. Once we've made the move to the complex plane, it can be shown that \emph{any analytic function} is a conformal transformation. This amazing fact is what empowers CFT in two dimensions, as first elucidated by Belavin, Polyakov and A. B. Zamolodchikov~\cite{belavin_infinite_1984}.\footnote{Note that this is Alexander Borisovich Zamolodchikov, not his twin brother Alexei Borisovich Zamolodchikov -- also a noted theoretical physicist in his own right. Having two prominent A. B. Zamolodchikov's in the same field at the same time must have made for a publisher's nightmare (and it's still difficult to tell sometimes which work is by whom).}

A central object of interest in CFT is what is known as a \emph{primary operator}. Note that operators can be viewed either as functions of position $r = (x, \tau)$ as $\phi(r)$, or in complex notation as $\phi(z, \bar{z})$. Primary operators transform simply under conformal transformations; each primary operator $\phi$ has \emph{conformal weights} $h_\phi$ and $\bar{h}_\phi$, which are generally distinct real numbers, and in a unitary CFT, $h,\bar{h} \geq 0$. These are usually combined to form the quantities $\Delta = h+\bar{h}$, called the \emph{conformal dimension}, and $s=h - \bar{h}$, called the \emph{spin}. Under a conformal transformation $z \to w(z)$, $\phi(z, \bar z)$ transforms as

\begin{equation}
\phi(z, \bar z) = \left( \pd{w}{z} \right)^{h_\phi} \left( \pd{\bar w}{\bar z} \right)^{\bar{h}_\phi} \phi(w(z), \bar w(\bar{z})).
\end{equation}

This expression is valid when $\phi$ is inside of an expectation, $\langle \phi \ldots \rangle$, and the factors out front are just the Jacobian of the transformation raised to conformal dimension of the operator. This little formula has extremely broad implications; often in CFT we don't know how to compute correlation functions of an operator in some complicated geometry, like a multi-sheeted Riemann surface with cuts\footnote{As is common in entanglement entropy calculations with the replica trick; see Cardy and Calabrese~\cite{1751-8121-42-50-504005,calabrese_entanglement_2004}.} or even a simple disk, but we can use this relation to conformally map onto a surface where we do know how to compute correlation functions. That said, let us examine correlation functions on the plane, with the understanding that this gives us all correlation functions on geometries with the same topology. 

Conformal invariance completely fixes the form of the two-point function on the plane:\footnote{The one-point function vanishes by symmetry.}

\begin{equation}
\label{eq:twpoint}
\langle \phi_i(r_1) \phi_j(r_2) \rangle  = \frac{\delta_{ij} C_i}{|z_1 - z_2|^{2h_i} |\bar{z}_1 - \bar{z}_2|^{2\bar{h}_i}} = \frac{\delta_{ij} C_i}{|z_1 - z_2|^{2(h_i + \bar{h}_i)}} = \frac{\delta_{ij} C_i}{|r_1 - r_2|^{2\Delta_i}}.
\end{equation}

The $\delta_{ij}$ is because any two-point function of operators with different weights must vanish.\footnote{A simple argument goes as follows: make a conformal transformation on the plane, and consider the two quantities $\langle \phi_i(z_1, \bar z_1) \phi_j(z_2, \bar z_2)  \rangle$ and $\langle \phi_j(z_1, \bar z_1) \phi_i(z_2, \bar z_2)  \rangle$, switching the operators. In fact, these must be equal (!), since the two point function has just been rotated by $\pi$. After the transformation, the only way they may be equal is if their Jacobian prefactors $a^{h_i} b^{h_j} c^{\bar{h}_i} d^{\bar{h}_j}$ and $b^{h_i} a^{h_j} d^{\bar{h}_i} c^{\bar{h}_j}$ are equal, which entails either $h_i = h_j$ and $\bar{h}_i = \bar{h}_j$, meaning that two operators are really the same; or the two-point function is 0. $\qed$} $C_{i}$ is a constant that depends on the operator in question; in general, we cannot deduce its value from CFT alone, and it needs to be determined from a microscopic calculation in the particular model (it is not a universal number). Similarly, the form of the three-point function is also fixed by conformal symmetry alone, namely

\begin{equation}
\langle \phi_i(r_1) \phi_j(r_2) \phi_k(r_3) \rangle = \frac{C_{ijk}}{ |r_1 - r_2|^{\Delta_i + \Delta_j - \Delta_k} |r_2 - r_3|^{\Delta_j + \Delta_k - \Delta_i} |r_3 - r_1|^{\Delta_k + \Delta_i - \Delta_j} },
\end{equation}

which is quite elegant. Our luck runs out for the four point function, though, due to the presence of the \emph{cross-ratios} or \emph{anharmonic ratios}, like $\frac{|r_1 - r_2| |r_3 - r_4|}{|r_1 - r_3| |r_2 - r_4|}$, which are invariant under a conformal transformation. Since these ratios are invariant, the 4-point function could be any function of these ratios and still satisfy conformal symmetry, so we cannot fix the form; obviously, higher point functions have this issue as well. Nonetheless, knowing the 2- and 3-point functions is quite powerful, and in some cases, such as the free fermion $\psi$, we know the full $N$-point function due to Wick's theorem (which relates $N$-point functions to sums of products of 2-point functions). The Coulomb gas formalism can give $N$-point functions for some other operators (such as the $\sigma$ field in the Ising model with $\Delta = 1/8$), but generic $N$-point functions are not known. Finally, a central concept in CFT is of the \emph{operator product expansion} or OPE, which states that as we bring two operators close together, we may replace them by a sum of operators (rather than a product). The intuition for this comes from, essentially, counting poles; bringing two poles close together creates a single pole of higher order, in the same sense that bringing two operators of some dimension close together creates a new operator of a different conformal dimension, if we don't care about the non-singular part of the function. In math, the OPE is

\begin{equation}
\phi_i(z_1, \bar{z}_1) \phi_j(z_2, \bar{z}_2) = \sum_k C_{ij}^k \phi_k(z_1, \bar{z}_1).
\end{equation}

This is again only valid inside correlation functions $\langle \phi_i(z_1, \bar{z}_1) \phi_j(z_2, \bar{z}_2) \ldots \rangle$ in the limit $z_1 \approx z_2$, $\bar{z}_1 \approx \bar{z}_2$. The $C_{ij}^k$ are known as \emph{structure constants}, with the above formula also called a \emph{fusion rule}.\footnote{There is a deep connection between CFTs and the braiding of anyons, hence the similarity in terminology.} OPEs are not restricted to primary operators, and exist for all pairs of operators; but, the primary fields being the atoms of any CFT, we are generally most interested in their OPEs.\footnote{With the exception of OPEs of primaries against the stress-energy tensor $T$.}

\subsection{Central charges and Virasoro algebras}

The last bit of CFT we will need is the concept of a \emph{central charge}, $c$. Earlier, we introduced primary operators, which transform simply under conformal maps. Luckily, many physical operators are primary, such as the Ising field $\psi$ in the Ising CFT. However, it is easily shown that their derivatives (also called \emph{descendants}) like $\partial_\mu \phi$ are not primary, transforming in more complicated ways. The most notable non-primary field is the \emph{stress-energy tensor} $T_{\mu \nu}$. Now, the stress energy tensor may be expanded in Laurent series as

\begin{equation}
T(z) = \sum_{m=-\infty}^{\infty} \frac{L_m}{z^{m+2}} \Leftrightarrow L_m = \frac{1}{2\pi i} \oint dz \ z^{m+1} T(z),
\end{equation}

with similar expressions for $\bar{T}(\bar z)$. The significance of the $L_m$'s (and their cousins, the $\bar{L}_m$'s) are that they generate the group of conformal transformations, in the sense of Lie algebras; alternatively, they are the Noether charges associated to the symmetry transformations $\delta z = z^{n+1}$. In particular, $L_{-1}$ generates translations, and $L_0$ generates dilations and rotations. These $L_m$ operators form the \emph{Virasoro algebra}, perhaps the most fundamental aspect of CFT.\footnote{This is reminiscent of the algebra of angular momentum operators, but now with an infinite tower of $L$'s and a term with $c$ in it.} In particular,

\begin{equation}
[L_m, L_n] = (m-n)L_{m+n} + \frac{c}{12} m(m^2-1) \delta_{n+m,0}.
\end{equation}

The quantity $c$ is the central charge. (This is the formal definition, and is pretty dry.) Note that we have two copies of this algebra, one for $L_m$'s and one for $\bar{L}_m$'s; the $\bar{L}_m$ have their own central charge $\bar{c}$, which may differ from $c$.

The central charge is the defining characteristic of a CFT.\footnote{We'll concern ourselves here with only the ``minimal models'', with a finite number of primary fields, plus the free boson at $c=1$ with an infinite number.} The Ising universality class is defined by $c=1/2$, the free boson by $c=1$, the tricritical Ising model by $c=7/10$, and the 3-state Potts model by $c=4/5$, to name a few. Unitary theories have $c>0$, and minimal models (i.e. the models we generally care about in condensed matter physics) have $c\leq 1$.\footnote{In this range, $c$ may only take discrete values $c=1 - 6/m(m+1)$, where $m=3,4,5\ldots$.} The central charge is universal, in some sense the ur-universal object, and appears extensively. 

The modern interpretation of $c$ is as the prefactor of the growth of entanglement; without getting unnecessarily derailed, since CFTs are gapless, the entanglement entropy, defined as $S_L = -\Tr \rho_L \log \rho_L$ for a subsystem of size $L$, grows logarithmically\footnote{In a gapped theory, we expect an \emph{area law} for entanglement in the ground state, $S_L$ $\sim$ $\partial L$, i.e. a constant in 1+1 dimensions, and a \emph{volume law} for generic states with $S_L \sim L$. The area law is a remarkable conjecture that has only been proven in the case of one-dimensional gapped Hamiltonians, by Hastings~\cite{Hastings_2007}.} as

\begin{equation}
S_L \sim \frac{c + \bar{c}}{6} \log L.
\end{equation}

The central charge also appears in the specific heat, in Cardy's formula $C \sim c \pi k_B / 3\beta$, and pops up in many places where $T_{\mu\nu}$ makes an appearance. Finally, $c$ satisfies a remarkable theorem known, straightforwardly, as the $c$-theorem~\cite{zamo_cTheorem}, that states that $c$ is monotonically decreasing along RG flows.\footnote{More accurately, there is a function $C(H, \lambda)$, with $H$ the Hamiltonian and $\lambda$ the RG scale, that is monotonically decreasing along flow, and is fixed only at the RG fixed points, where it equals the central charge $c$.} We remark that in higher dimensions, while there is no $c$, there are analogues of the $c$-theorem, including the $A$-theorem in 3+1 dimensions~\cite{CARDY1988749, OSBORN198997} and the $F$-``theorem'' in 2+1 dimensions~\cite{Klebanov:2011aa}.\footnote{Conjectured theorem, i.e. a conjecture. Special cases have been, to my understanding, proven only using supersymmetry in particular theories like $\mathcal{N}$=2 supersymmetric Yang-Mills.}

\subsection{Boundary conformal field theory}

We should finally spend a few words on boundary CFT, which is usually defined on the half-plane $\Im{z} \geq 0$. Boundary CFT is actually easier than bulk CFT; there is only \emph{one} copy of the Virasoro algebra, which we can think of as being $z$ for $\Im{z} \geq 0$ and $\bar z$ for $\Im z \leq 0$. Consequently there is only one central charge $c$, and, when the region $L$ is taken from the boundary inward, we have $S_L \sim (c/6) \log L$.\footnote{This is the formula for open boundary conditions; for periodic boundary conditions or for regions in the bulk of an OBC system, we again use $(c/3) \log L$, assuming $c=\bar{c}$ and no finite-size effects. Incidentally, finite size effects can be exactly corrected for, as shown by Cardy and Calabrese, by mapping the strip to a cylinder with radius $L$.} Boundary CFT has its own peculiarities, however.

There are, for any CFT, only a finite number of boundary conditions that are consistent with conformal symmetry.\footnote{The seminal papers here are Cardy 1984~\cite{cardy_conformal_1984} and 1989~\cite{cardy_boundary_1989}, and Ishibashi 1989~\cite{doi:10.1142/S0217732389000320}.} In the case of the Ising model, calling the boundary field $h_B$,\footnote{More correctly, the value of the field $\psi$ is what is fixed to have expectation 0 or $\pm 1$, but this corresponds to the stated values of the coupling $h_B$.} these are $h_B = 0$ (``free'', which is always conformal for any theory) and $h_B = \pm \infty$ (``fixed''); any other values would change upon rescaling. We can thus flow boundary conditions under the RG, leading to some boundary conditions being stable, and some unstable, fixed points. For Ising, $h_B=0$ is unstable, and arbitrarily small $h_B$ will flow off to infinity. Whenever there is sharp change in boundary conditions, correlation functions in this geometry can be related to those in the geometry with a free boundary via the insertion of a \emph{boundary-condition-changing operator}, or BCC operator.\footnote{The reason why this works is deep, and related to the \emph{state-operator correspondence}. In general in CFT, there is actually an isomorphism between states and operators -- which is certainly not true for general quantum field theories! The hand-waving way to see this correspondence is to imagine the far past state at the end of a cylinder, which under a conformal transformation gets mapped to a disturbance at the origin -- an operator. Similarly, boundary conditions, viewed as states, can be mapped to operators, and the mixed boundary condition case is a BCC operator.} That is, if we have a sharp change from boundary condition $a$ to boundary condition $b$ at a point $x$ along the boundary, then $\langle \ldots \rangle_{\text{ab}} = \langle \phi_{ab}(x) \ldots \rangle_{\text{free}}$. Finally, even the bulk is slightly different: due to the presence of the boundary, any $N$-point function in the half-plane is related to a $2N$-point function in the full plane, via the method of images: $\langle \phi(x,\tau) \ldots \rangle_{\text{half plane}} = \langle \phi(-x,\tau)\phi(x,\tau) \ldots \rangle_{\text{full plane}}$.\footnote{The reason why this works is interesting. When I first learned it, I wondered whether there was an equivalent to Laplace's equation, as for the usual method of images: solutions are unique given fixed boundary conditions, so we arrange charges so as to mimic a particular set of isocontours of the field (such as a plane conductor or spherical conductor). The reason the method of images works here is because we can ``unfold'' the CFT from the half plane to the full plane by mapping the left-moving sector to $x<0$ and the right-moving sector to $x > 0$. So long as we are not on the boundary of $x=0$, we get two copies of the operator in the plane after we do this, at $x$ and $-x$.} This means that the one-point function with a boundary does not generally vanish (and its form is Eq.~\ref{eq:twpoint}), while the two-point function is not fixed by symmetry due to the cross-ratios.

This concludes our lightning review of the key concepts of CFT; we now move on to apply these techniques to a critical quantum spin chain driven at its boundary. 

\section{Floquet Dynamics of Boundary-Driven Systems at Criticality}

\subsection{Introduction}

Recent years have witnessed substantial progress in understanding the dynamics of periodically driven (Floquet) systems. Such driving has traditionally been used for engineering non-trivial effective Hamiltonians~\cite{doi:10.1080/00018732.2015.1055918,PhysRevLett.116.205301,0953-4075-49-1-013001,PhysRevX.4.031027}, but recent research has shown that these dynamics can differ drastically from their static counterparts. Examples include the recently observed Floquet time crystals~\cite{khemani_prl_2016,pi-spin-glass,else_floquet_2016,PhysRevX.7.011026,Zhang:2017aa}, the emergence of topological quasiparticles protected by driving~\cite{PhysRevLett.106.220402,PhysRevB.94.045127}, Floquet topological insulators~\cite{PhysRevB.79.081406,PhysRevB.82.235114,Lindner:2011aa,rechtsman_photonic_2013,cayssol_floquet_2013,titum_disorder-induced_2015}, and Floquet symmetry-protected topological phases~\cite{PhysRevB.92.125107,PhysRevB.93.245145,PhysRevB.93.201103,potter_classification_2016}. More broadly, periodically driven systems touch on fundamental issues in statistical and condensed-matter physics such as thermalization~\cite{lazarides_periodic_2014,abanin_effective_2015,PhysRevLett.115.256803,kuwahara_floquetmagnus_2016,mori_rigorous_2016} and phase structure~\cite{khemani_prl_2016}.

However, relatively little attention has been paid to driven systems at criticality, whose low-energy dynamics are often described by a conformal field theory (CFT). Such quantum critical systems are a natural setting in which to study Floquet dynamics, as many insights into the non-equilibrium dynamics of many-body systems have come from the study of CFTs in 1+1d~\cite{PhysRevLett.96.136801,1742-5468-2005-04-P04010,1742-5468-2007-10-P10004,1742-5468-2016-6-064003}. A na\"ive expectation is that such a driven critical system would simply heat up. However, in the presence of a boundary drive, the energy injected per cycle is not extensive in system size, and there are multiple possible behaviors in an arbitrarily long period prior to thermalization. Moreover, as CFTs are integrable, it is natural to expect they can escape heating even at low frequencies provided the scaling limit is taken before the long time limit. This opens the door to using scaling theory combined with the analytical toolkit of boundary CFT~\cite{cardy_boundary_2006,cardy_conformal_1984,cardy_boundary_1989,Affleck:1996mm} to characterize multiple regimes of universal dynamics in such boundary driven quantum critical points.

In this Letter, we study the dynamics of entanglement entropy $S_l(t)$ and Loschmidt echo ${\cal L}(t) = \abs{\braket{\psi(0)}{\psi(t)}}^2$ in conformally-invariant quantum critical systems subject to a periodic boundary drive. We find two distinct regimes in which boundary conformal field theory provides an excellent description of the dynamics. For suitably slow drives, the system behaves almost as though subject to a series of independent quantum quenches but with amplitude corrections related to multiple-point correlation functions, while for fast drives, the boundary drive can be averaged out, and the system responds as though subject to a single quench at an averaged value of the field. For intermediate driving frequency, we find universal heating which crosses over from a perturbative regime at weak drive to non-perturbative boundary CFT regime at strong drive. The dynamics in all driving regimes are universal and can be described using field-theoretic tools. We numerically confirm that the dynamics remain robust against adding integrability-breaking interactions up to the finite times that may be simulated. 

\subsection{Model} 

While our results apply to arbitrary boundary-driven CFTs, for concreteness we will focus on the archetypical transverse-field Ising (TFI) model on the half-line with a time-dependent symmetry-breaking boundary field
\begin{equation}
H = - \sum_{i \geq 0} \left( J  \sigma_i^z \sigma_{i+1}^z + h  \sigma_i^x  +\Gamma \sigma_i^x \sigma_{i+1}^x \right) -  h_b(t) \sigma_0^z, 
\label{eq:TFI}
\end{equation}
with $\Gamma$ an integrability-breaking perturbation and $h \sim J$ tuned to the critical point. This model has a convenient description in terms of free fermions when $\Gamma = 0$, seen by performing a Jordan-Wigner transformation~\cite{sachdev2011quantum} and is thus an ideal numerical test-bed for our model-independent analytical arguments. We initially prepare the system in the ground state at fixed boundary field $h_b(t<0)$ then quench on a periodic boundary drive, $h_{b}(t+T) = h_b(t)$, for $t \geq 0$. In equilibrium, the low energy description of this spin chain at criticality is well-understood in terms of gapless left- and right-moving Majorana fields satisfying $\lbrace \eta_{R/L} (x), \eta_{R/L} (y)\rbrace = \delta(x-y)$, with Hamiltonian 
\begin{equation}
H = - \frac{i v }{2} \int_{0}^{\infty} dx \left( \eta_R \partial_x \eta_R  -  \eta_L \partial_x \eta_L\right) - \lambda(t) \sigma_b(0),
\label{eq:TFIFieldTheory}
\end{equation}
where we dropped irrelevant terms. Here, $v$ is a non-universal velocity ($v=2 J$ for $\Gamma=0$)  and $\lambda \propto h_b$. In this Majorana formulation, the boundary spin can be represented as $\sigma_b(0) = i (\eta_R + \eta_L) \gamma$~\cite{doi:10.1142/S0217751X94001552}, where $\gamma^\dagger = \gamma$ is an ancilla Majorana satisfying $\gamma^2=1$ that anticommutes with all fields. In the following, we will assume that the drive is characterized by a single scale $\| h_b(t) \| \sim h_b $ which we take to be much smaller than the single particle bandwidth $h_b \ll \Lambda \equiv 2J=2$ (setting $J=1$), for which field theory is a good equilibrium description. The boundary field is a relevant perturbation with scaling dimension $\Delta=\frac{1}{2} < 1$ in the renormalization group (RG) language, with characteristic time scale $t_b \sim |h_b|^{-\nu_b} $, $\nu_b=(1-\Delta)^{-1}=2$.

There are three energy scales in this problem: the driving frequency $\omega=2\pi/T$, the bandwidth $\Lambda$, and the scale of the boundary perturbation $t_b^{-1} \sim h_b^{\nu_b} \ll \Lambda$. We will now consider various orderings of these scales and argue that essentially all regimes can be understood using a combination of field theory and scaling arguments, even though the drive is continuously injecting energy into the system. While the Hamiltonian~\eqref{eq:TFI} for $\Gamma=0$ can be mapped onto free fermions for numerical convenience (see Sec.~\ref{sec:numerics}), we note that our main conclusions follow from general field theory arguments and therefore continue to hold in the non-integrable case. We emphasize that although we choose to focus on the Ising field theory~\eqref{eq:TFIFieldTheory} as an example, our field-theoretic arguments are model-independent, so our results carry over immediately to any boundary driven CFT, such as a driven quantum impurity problem with $t_b^{-1} \to T_K$, the Kondo temperature.

\subsection{Slow driving regime: step drive} 

\begin{figure}
	\includegraphics[width = \columnwidth]{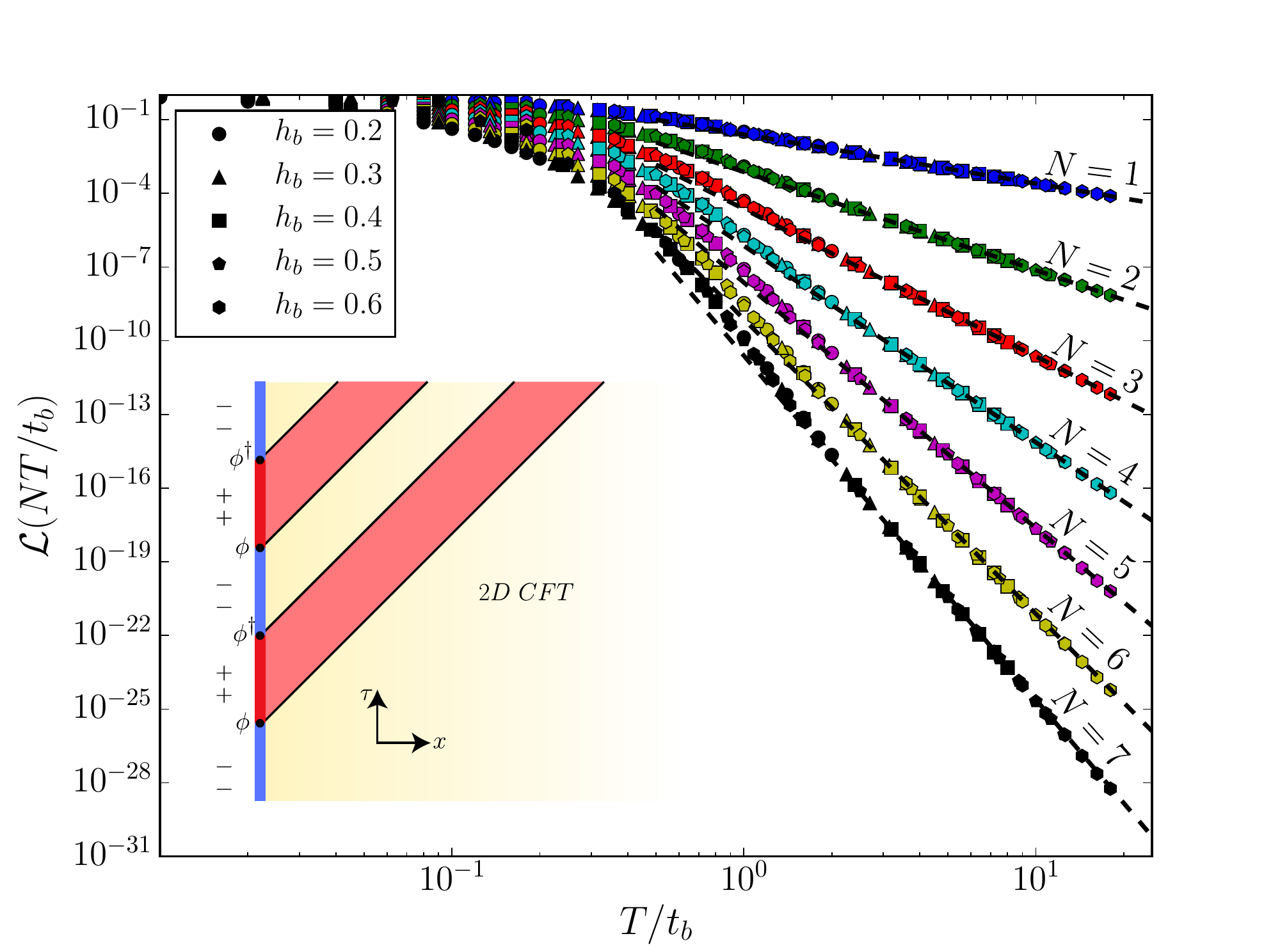}
	\caption{Slow driving regime $\omega \ll t_b^{-1} \sim h_b^2 \ll \Lambda$ for a step drive alternating between $-h_b$ and $+h_b$ for systems up to $L \sim 3200$ sites ($\Gamma=0$). For large $T$, we see clear power-law scaling of the Loschmidt echo with slope $-2 N $ as predicted from boundary CFT. The agreement between the CFT $2N$-point function prediction (dashed lines) and numerical data is excellent, where we stress that the only fit parameter is the non-universal offset $c_1$. Note also the universal collapse of the Loschmidt echo as a function of universal parameter $T / t_b \sim h_b^2 T$. Inset: sketch of the imaginary time picture where the step drive corresponds to inserting BCC operators.}
	\label{figslow}
\end{figure}

We start by considering the slow driving regime $\omega \ll t_b^{-1} \ll \Lambda$ for a step drive starting from the initial field $h_b(t<0) = -h_b$ with $h_b (t) = +h_b$ for $0 \leq t \leq T/2$ (Hamiltonian $H_1$) and $h_b (t) = -h_b$ if $T/2 \leq t \leq T$  (Hamiltonian $H_0$) for $t\geq 0$. Intuitively, this drive looks like independent local quenches. Focusing on the Loschmidt echo (return probability) ${\cal L} = \left| \langle \psi_0   \ket{\psi(t)}  \right|^2$~\cite{Goussev:2012}, this behavior is best understood by Wick rotating to imaginary time $\tau = i t$, where the spin-chain Loschmidt echo can be mapped onto a CFT correlation function. After computing this correlation function, we Wick rotate back to real time to obtain the dynamical echo. In imaginary time, the initial state can be generated by an infinite imaginary time evolution $\lim_{\tau\to\infty} e^{-\tau H_0}\ket{0} \propto \ket{\psi_0}$ from arbitrary initial state $\ket{0}$. In imaginary time, $\exp(-\tau H)$ acts as a projector onto the ground state of $H$, so for large $T \gg t_b$ we essentially oscillate between the ground states of $H_0$ and $H_1$, for which $\sigma_0^z$ is locked in the direction of the boundary field $\pm h_b$. In the CFT language, a sharp change in boundary conditions can be treated by inserting a boundary-condition changing (BCC) operator~\cite{cardy_boundary_2006}, as diagrammed in Fig.~\ref{figslow}. This means that the Loschmidt echo $\mathcal L(NT)$ after $N$ periods of drive corresponds to the $2N$-point correlation function of a BCC operator $\phi_{\rm BCC}$ changing the boundary condition from fixed $\sigma_0^z= \pm 1$ to $\sigma_0^z= \mp 1$. 

Analytically continuing to real time, we expect the Loschmidt echo to be a universal function ${\cal L} (T /t_b,N)$ in the field theory regime. In the limit $T \gg t_b$, this reduces to the $2N$-point function
\begin{equation}
{\cal L}(NT) \underset{T \gg t_b}{\sim} \abs{\expectation{\prod_{n=0}^{2N-1}\phi_{{\rm BCC}}(n T/2)}}^2 = c_N  \left(\frac{T}{t_b}\right)^{- \gamma N} ,
\label{eqLoschQuench}
\end{equation}
whose form is fixed by scale invariance. The universal exponent $\gamma = 4 h_{+-}=2$ is given by the scaling dimension $h_{+-}=\frac{1}{2}$ of the BCC operator $\phi_{\rm BCC}$~\cite{cardy_conformal_1984,cardy_boundary_1989}.  Other step drives can be dealt with in a similar fashion; for example, a step drive from $h_b=0$ to $h_b \neq 0$ corresponds to the insertion of a BCC field with scaling dimension $h_{\rm BCC}=\frac{1}{16}$. We emphasize that Eq.~\eqref{eqLoschQuench} holds for arbitrary boundary step drives in more general CFTs with the appropriate choice of BCC operator.

Note that although the Loschmidt echo decays exponentially with $N$, consistent with the independent quenches picture, the fact that the quenches are not fully independent is encoded in the non-trivial $N$ dependence of the coefficients $c_N$. The ratio $c_N/(c_1)^N$ is universal and can be computed exactly for this specific drive, since the BCC operator $\phi_{\rm BCC}$ corresponds to a chiral fermionic field $\psi$ in the Ising field theory with $2N$-point correlator given by a Pfaffian: ${\cal L}(NT) \sim \abs{\expectation{\psi(0) \psi(T/2) \psi (T) \dots }}^2 \sim \left| {\rm Pf}( 1/(t_i-t_j)) \right|^2 $ with $t_i=0,T/2,T, \dots, (N-\frac{1}{2})T$. For step drives in general CFTs, such universal ratios can be computed within the Coulomb gas (bosonization) framework.  These analytical expressions are in excellent agreement with numerical simulations for $\Gamma=0$ (Fig.~\ref{figslow}), where the only non-universal fit parameter is $c_1$. Since these predictions rely solely on field theory, they apply equally well to the non-integrable case $\Gamma \neq 0$; the interactions $\Gamma \sigma_i^x \sigma_{i+1}^x$ are irrelevant in the RG sense and therefore do not change the universality class. We confirm this numerically by locating the new critical point for $\Gamma \neq 0$ using exact diagonalization, obtaining the ground state using standard density matrix renormalization group (DMRG) techniques~\cite{PhysRevLett.69.2863,Schollwock201196}, and simulating the dynamics of this driven interacting chain using time-evolving block decimation (TEBD)~\cite{PhysRevLett.93.040502}. We find excellent agreement with our field-theoretic argument, as shown in Sec.~\ref{sec:ints2}.

\begin{figure}
	\includegraphics[width = \columnwidth]{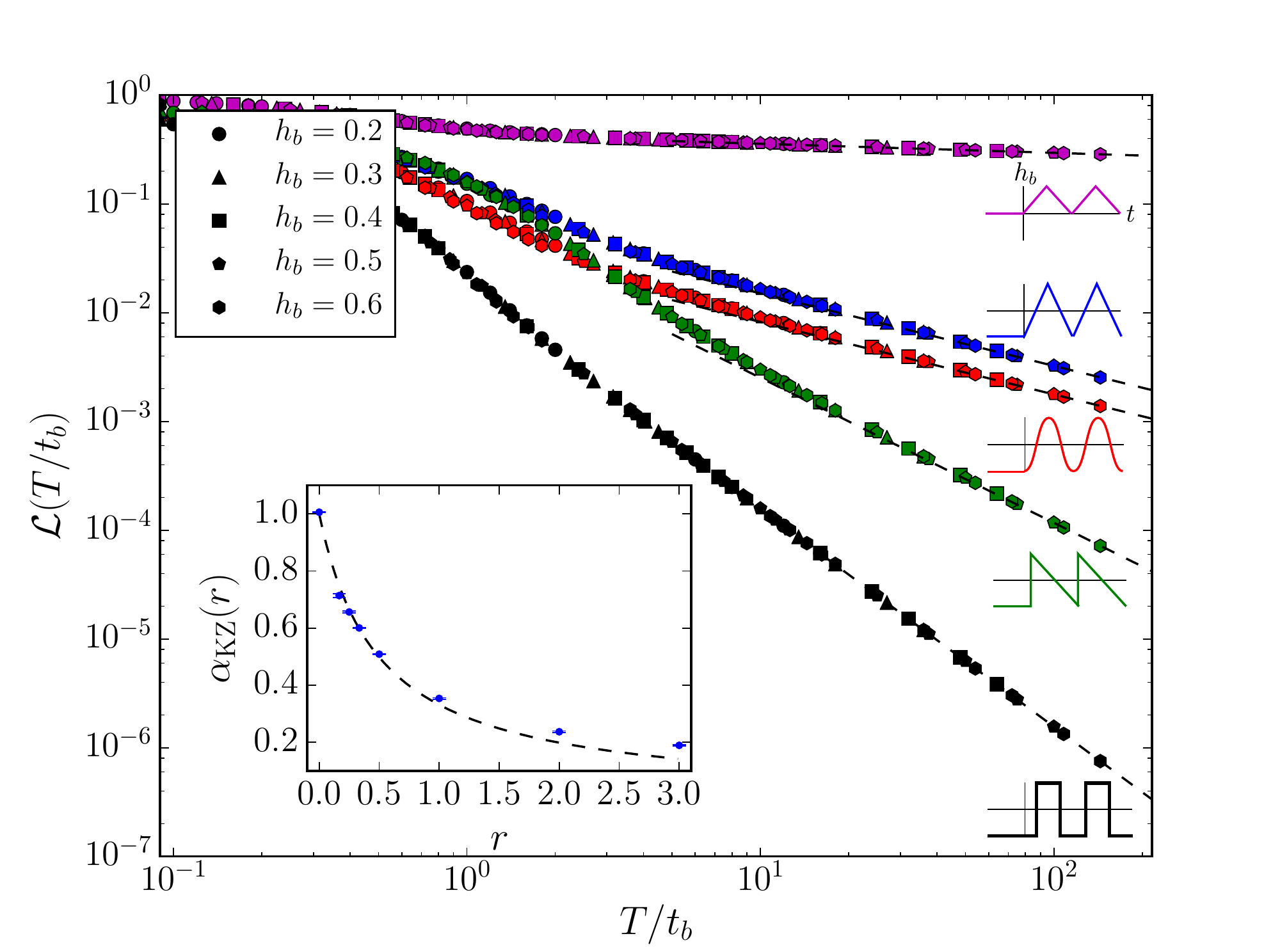}
	\caption{Loschmidt echo for $\Gamma=0$ over a single cycle ($N=1$) in the slow regime for various drive geometries showing renormalized power laws and universal collapse as a function of $T/t_b = T h_b^2$. The dashed lines correspond to the analytic prediction~\eqref{eqKZexp} from boundary CFT and KZ arguments. Inset: KZ renormalization factor $\alpha_{\rm KZ}$ of the BCC exponents for a boundary field scaling as $h_b(t) = h_b  (\frac{t}{T})^r$ compared to the KZ prediction $\alpha_{\rm KZ}= (1+\nu_b r)^{-1}$ with $\nu_b=2$. The dashed line is a fit of the numerical data for small $r$ giving $\nu_b \approx 2.02 \pm 0.08$. }
	\label{FigKZ}
\end{figure}

\subsection{Slow driving regime: general drives} 

Consider now a more general drive such as $h_b(t>0) = -h_b \cos(\pi t / T)$ with $h_b(t<0)=-h_b$. In the large $T$ limit $h_b(t)$ crosses the critical value slowly rather than suddenly, yet the BCC picture suggests that the field should quickly flow to infinity. We find, however, that the vanishing (but finite) crossing speed is strongly relevant, changing the power law entirely (Fig.~\ref{FigKZ}). To understand this difference, we use the concept of Kibble-Zurek (KZ) scaling, which is frequently applied to bulk drives crossing a bulk quantum critical point~\cite{PhysRevLett.95.105701,PhysRevB.72.161201,PhysRevLett.95.245701,2016arXiv161202259L}  but has not been studied for such boundary drives to our knowledge.

Let us imagine that the drive crosses $h_b=0$ as a power-law $h_b(t) = h_b  |\frac{t}{T}|^r \mathrm{sgn}(t)$ with $r=1$ in the cosine drive considered above and $r = 0$ for a quench~\cite{PhysRevB.81.012303}. The effective time scale $t_b(t) \sim \left[ h_b(t) \right]^{-\nu_b} $ now becomes time-dependent, and we expect the dynamics to be controlled by an emergent time scale
\begin{equation}
t_{\rm KZ} \sim T^{\frac{r \nu_b}{1+r \nu_b}} h_b^{-\frac{ \nu_b}{1+r \nu_b}},
\label{eqKZscale}
\end{equation}
given by $t_{\rm KZ} \sim t_b(t_{\rm KZ})$. Though our system is always gapless so that there is no adiabatic limit, it is straightforward to show that this dynamical scale emerges directly from the equations of motion of Eq.~\eqref{eq:TFIFieldTheory} \cite{mike_prl}. It is natural to expect that the slow driving limit $T \gg t_{\rm KZ}$ should still be described by boundary CFT, suggesting that the Loschmidt echo would scale as~\eqref{eqLoschQuench} with $t_b$ replaced $t_{\rm KZ}$. We therefore see that the effect of the slow driving amounts to renormalizing the dimension $h_{\rm BCC}$ of the BCC operator by a factor $\alpha_{\rm KZ} = 1/(1+r \nu_b)$ with $\nu_b=2$ in our case. More generally, for a drive where $h_b(t)$ crosses or touches the critical value $n$ times within a single cycle, we predict that the universal exponent $\gamma$ controlling the exponential decay of the Loschmidt echo is given by 
\begin{equation}
\gamma = 2 \sum_{i=1}^n \frac{h^i_{\rm BCC}}{1+r_i \nu_b},
\label{eqKZexp}
\end{equation}
where $r_i$ is the power of $|h_b(t)| \sim |t-t^i_c|^{r_i}$ near the critical time $t^i_c$. For our model, $h^i_{\rm BCC} =\frac{1}{2}$  if $h_b(t)$ crosses zero and $h^i_{\rm BCC} =\frac{1}{16}$ if it touches zero without changing sign~\cite{cardy_conformal_1984,cardy_boundary_1989,cardy_boundary_2006}. For example, a cosine or triangle drive oscillating between $\pm h_b$ has $n=2$, $r_1=r_2=1$ so that $\gamma = 2/3$, while a sawtooth drive combines slow ($r_1=1$) and fast ($r_2=0$) crossings to give $\gamma=4/3$. 

These predictions give good agreement with numerics (Fig.~\ref{FigKZ}).\footnote{We note that the agreement is systematically worse for larger $r$ due to finite system size and Trotter step errors.} Furthermore, the only effect of the slow driving is to renormalize the scaling dimensions of the BCC operators while keeping the structure of the $2N$-point function unchanged. In particular, we find that the universal numbers $c_N/(c_1)^N$ in Eq.~\eqref{eqLoschQuench} are still given by the boundary CFT predictions for a step drive (see Sec.~\ref{sec:KZ}).

\begin{figure}
	\includegraphics[width = \columnwidth]{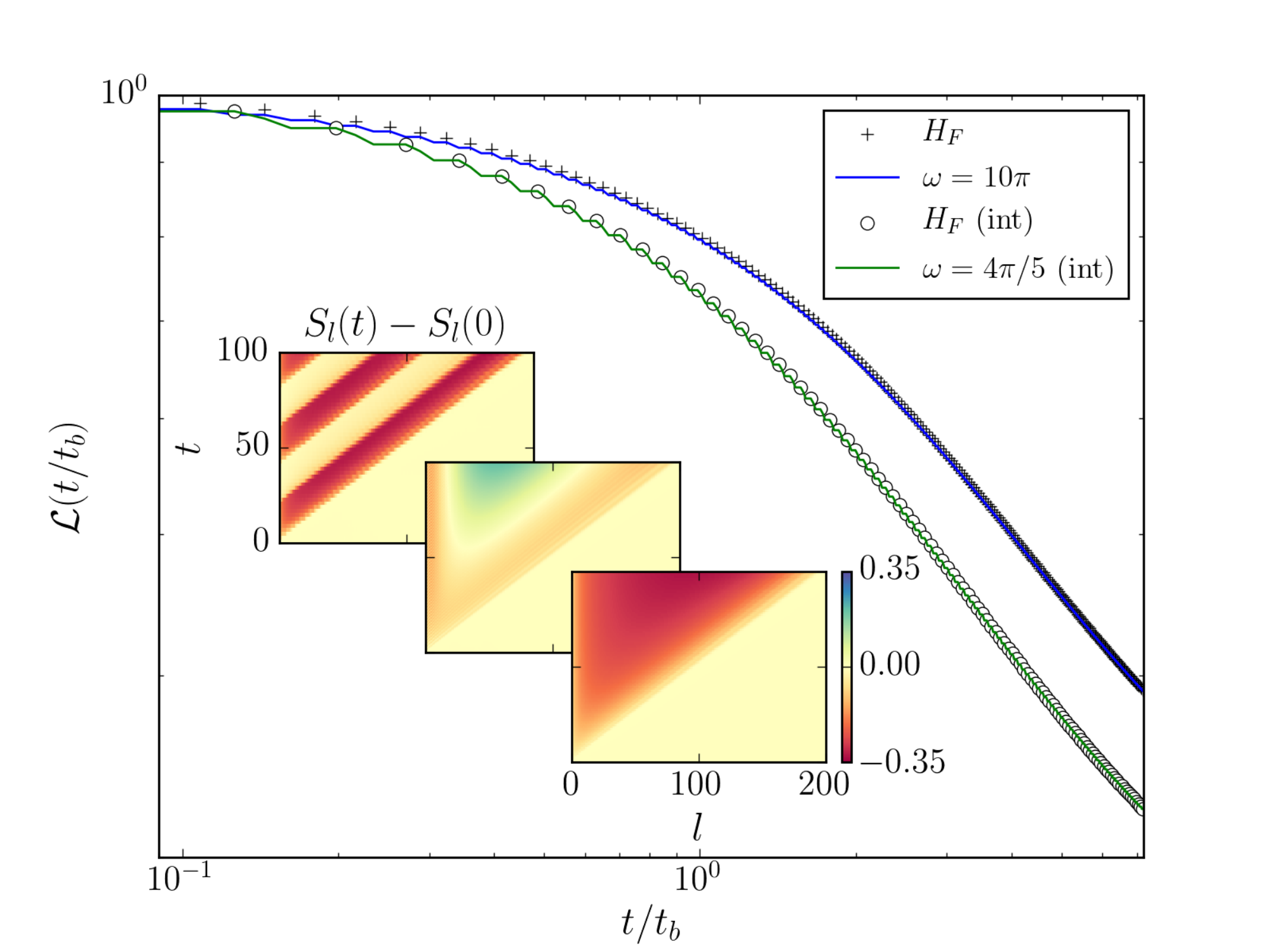}
	\caption{Fast regime: the Loschmidt echo at frequencies $\omega > \Lambda$ for a step drive oscillating between $0$ and $h_b$ coincides with the echo after a single local quench with effective field $h_b/2$ with Floquet Hamiltonian $H_F$ (black crosses). This result also holds when interactions are added with $\Gamma=0.25$ (white circles, green line). Insets: entanglement entropy difference $S_l(t) - S_l(0)$ for $\Gamma = 0$ as the drive frequency crosses over from the intermediate to the fast regime.}
	\label{FigFast}
\end{figure}

\subsection{Fast driving regime} 

We now consider the high-frequency regime $t_b^{-1} \ll \Lambda \ll \omega$. This is na\"ively outside the regime where field theory results should apply, but we can take advantage of standard Floquet machinery to write a Floquet-Magnus high-frequency expansion for the Floquet Hamiltonian $H_F$ defined by $U(T) ={\cal T}  {\rm e}^{- i \int_{0}^T dt H(t)}= {\rm e}^{-i T H_F}$~\cite{CPA:CPA3160070404}. For example, $H_F = \frac{1}{2}(H_0 + H_1) - \frac{i }{4 \omega} [H_0,H_1] + \mathcal O (\omega^{-2}) $ for a step drive. While higher order terms in this expansion are suppressed by powers of $\omega^{-1}$ as for any high-frequency Floquet system, we note here that the Floquet Hamiltonian $H_F$ itself corresponds to a CFT subject to an effective boundary field $\overline{h}_b = (1/T) \int_0^T h_b(t) dt$ with higher order terms in the high frequency expansion being RG irrelevant. This is most easily seen using the field theory Hamiltonian~\eqref{eq:TFIFieldTheory} where the small parameter controlling the expansion is $v/\omega \ll 1$ with $v \sim \Lambda = 2J$. While the first boundary term has scaling dimension $\Delta=1/2$ and corresponds to the averaged field $\overline{h}_b$, dimensional analysis immediately implies that terms of order $\omega^{-n}$ have scaling dimension of  at least $n+1/2$ due to terms such as $\partial^{n} \eta (0)$ and are thus irrelevant for $n>0$ (see Sec.~\ref{sec:highfreq}). Therefore at late times, the system behaves as though subject to a single local quantum quench with effective boundary field $\overline{h}_b$ (Fig.~\ref{FigFast}), a problem whose universal dynamics has been studied extensively~\cite{PhysRevLett.110.240601,PhysRevX.4.041007}. We remark that though RG techniques may be in general ill-defined in a Floquet system which, for instance, lacks a notion of ground state, in this high-frequency limit the Floquet evolution is well-controlled by an effective static Hamiltonian. Since our initial state is a conformally invariant ground state and the effective Hamiltonian implements a local quench, the notion of RG flow is well-defined~\cite{PhysRevLett.110.240601} and provides a powerful tool of analysis. Additionally, for the non-interacting (free fermions) case with $\Gamma=0$ in Eq.~\eqref{eq:TFI}, one may prove that the high-frequency expansion is convergent for $\omega \gae \Lambda$ by bounding the spectral width of the single-particle Hamiltonian (see Sec.~\ref{sec:HFEheating}). More generally, this effective single quench picture will survive even in the presence of integrability-breaking interactions controlled by $\Gamma$ up to exponentially long time scales $\tau_{\rm th} \sim {\rm e}^{C \omega/ \Lambda}$ \cite{abanin_effective_2015,PhysRevLett.115.256803,mori_rigorous_2016,kuwahara_floquetmagnus_2016}. We simulated the dynamics of this interacting chain subject to the same drive using TEBD and found excellent agreement with the single effective quench picture even at moderate frequencies (Fig.~\ref{FigFast}).

\begin{figure}
	\includegraphics[width = \columnwidth]{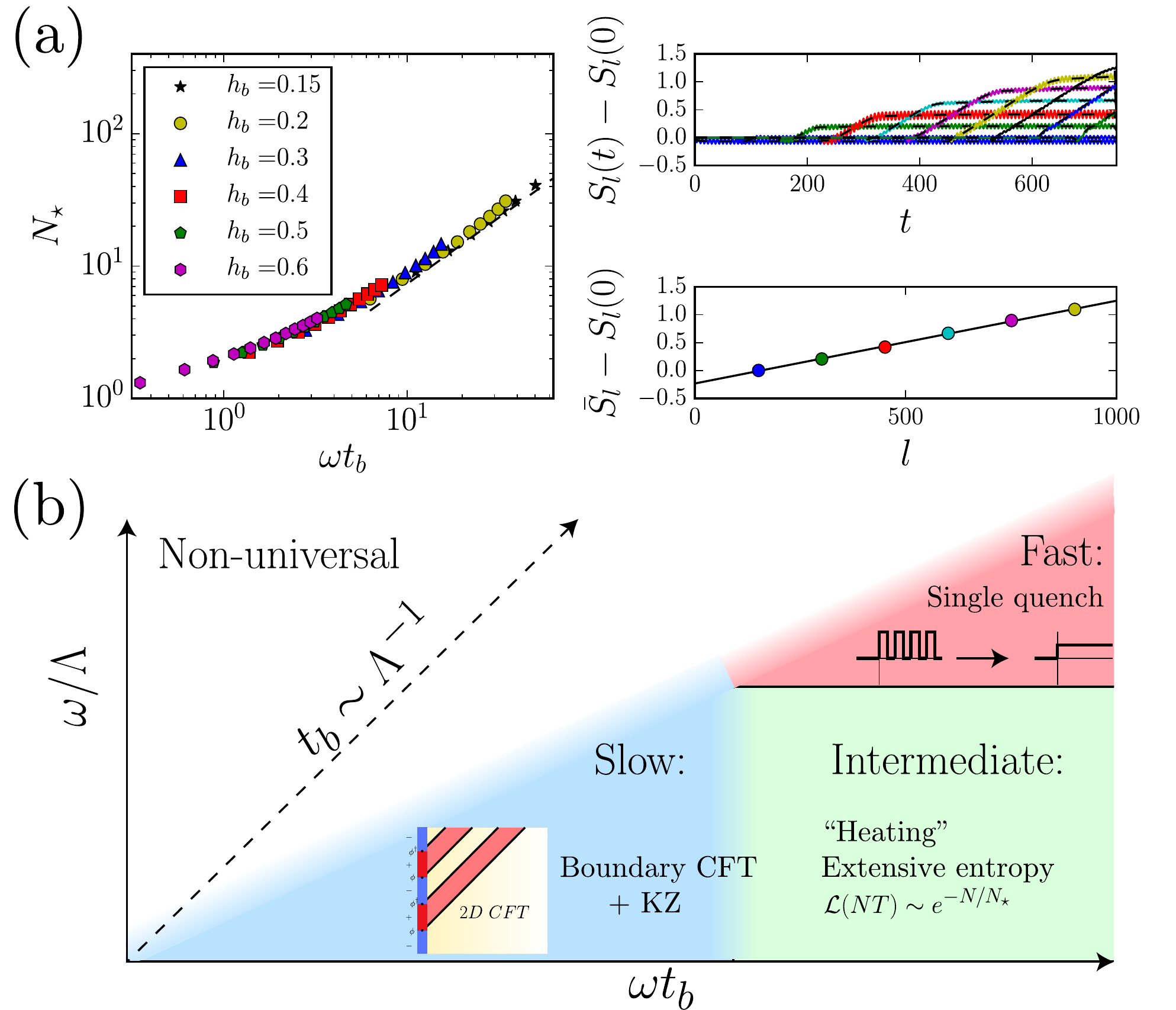}
	\caption{Intermediate regime $t_b^{-1} \sim \omega \ll \Lambda $ for a step drive from $h_b = 0$ to $h_b \not=0$. (a) Left panel: universal scaling function $N_\star( \omega t_b)$ characterizing the exponential decay of the Loschmidt echo ${\cal L} (N T) \sim {\rm e}^{-N/N_\star}$, with the dashed line showing the linear behavior expected from Fermi's golden rule. Right panel: volume-law scaling of the entanglement entropy in the long-time limit for $\omega t_b = 10.5 \gg 1$. An overbar denotes the value of the late-time plateau. (b) Sketch of the three universal driving regimes analyzed in this Letter.}
	\label{FigIntermediate}
\end{figure}

\subsection{Crossover regime} 

Finally, we discuss the intermediate crossover regime $t_b^{-1} \sim \omega \ll \Lambda$. We focus on a free-to-fixed step drive from $h_b = 0$ to $h_b \not=0$ with $\Gamma=0$ for simplicity. In this regime, we expect the system to absorb energy (``heat'') via resonant processes within the single-particle bandwidth. This leads to exponential decay of the Loschmidt echo,
\begin{equation}
{\cal L} (N T) \underset{t_b^{-1}, \omega \ll \Lambda} {\sim}{\rm e}^{-N/N_\star( \omega t_b)},
\end{equation}
with $N_\star( \omega t_b)$ a universal function (Fig.~\ref{FigIntermediate}a). For weak drive ($\omega t_b \gg 1$), resonant heating occurs with a rate $\tau^{-1} \sim h_b^2/J $ given by Fermi's golden rule, so that  $N_\star \sim \tau /T \sim \omega t_b$. For strong drive ($\omega t_b \ll 1$), we recover the boundary CFT prediction $N_\star \sim -1/(\gamma \log \omega t_b)$. We also find that entanglement entropy of boundary intervals of size $\ell$, relative to the ground state entropy, saturates to a volume law behavior $S_\ell \sim \ell$ at long times in the regime $ \omega t_b \gg 1$, consistent with heating.\footnote{We note that by heating we simply mean absorption of energy, not thermalization. Since this is an integrable model, the reduced density matrix should more accurately be described by a generalized Gibbs ensemble~\cite{PhysRevLett.98.050405} which in some observables can look highly athermal} At low frequencies, the entropy simply oscillates between ground state values\footnote{In the absence of boundary field, at large distances the ground state entanglement entropy is given by the well-known result $S_l(h_b=0)=\frac{c}{3} \ln \frac{l}{\tilde a}$\cite{1751-8121-42-50-504005}, where $c=1/2$ is the central charge of the CFT and $\tilde a$ is a non-universal constant. Pinning the boundary spin causes a universal reduction of the entanglement entropy for $h_b \neq 0$ of $\Delta S = -\log \sqrt 2$, known as the Affleck-Ludwig entropy \cite{affleck_universal_1991}.} though it may become extensive at much later times. We leave a detailed analysis of the role of interactions in this intermediate regime for future work. 

\subsection{Discussion} 

We have investigated CFTs subject to a Floquet boundary drive. Despite the na\"ive expectation that such gapless systems should absorb energy and simply heat up, we have identified three distinct regimes summarized in Fig.~\ref{FigIntermediate}b in which the system shows universal features that can be understood using tools of field theory and scaling theory. We expect our main conclusions to apply to a broad class of systems, and it will be especially interesting to investigate the consequences of our results for the physics of driven quantum dots and the non-equilibrium signatures of topological edge modes~\cite{PhysRevLett.106.220402}. In general, our results represent an analytically tractable model of a driven gapless system, an active area of research increasingly relevant to experiments.


\section{Universal Dynamics of Stochastically Driven Quantum Impurities}

\subsection{Introduction}

Universality lies at the core of our understanding of equilibrium critical phenomena and is successfully captured by the renormalization group framework~\cite{Cardy:1996xt,sachdev2011quantum}. 
This program has been extended to non-equilibrium classical systems, leading to the discovery of new dynamical universality classes, including coarsening, reaction-diffusion, and surface growth, among several others~\cite{Tauberbook2014}. Recent developments in experiments with quantum many-body systems call for a further extension of the program to universal phenomena in quantum dynamics. For example, systems of ultracold atoms and ions exhibit new dynamical transitions~\cite{PhysRevLett.119.080501, 1806.11044, zhang}, as well as new forms of dynamical scaling~\cite{PhysRevLett.115.245301, ober, schmied, cor}. Other classes of universal phenomena are seen in driven open quantum systems. These include experiments with  non-equilibrium Bose-Einstein condensation of polaritons~\cite{Kasp2006}, dissipative phase transitions  in cavity QED circuits~\cite{PhysRevX.7.011016}, and dynamical phase diagrams of condensates trapped in optical cavities~\cite{PhysRevA.99.053605, klinder2015dynamical}.
The common wisdom is that  driven-dissipative quantum  systems exhibit emergent classical dynamics  because the coupling to the environment  washes out the delicate quantum coherences. For instance, the occurrence of effective Langevin dynamics is common to many quantum systems coupled to a bath, with examples ranging from cold atoms to
solid state platforms~\cite{sieb, mitra06, PhysRevB.85.184302}. In certain cases an intermediate regime of universal quantum scaling can be identified~\cite{PhysRevB.85.184302,PhysRevB.94.085150}, but it remains an open question whether such quantum scaling can persist to all scales in a driven-dissipative system.

In this Letter,  we show that universal, inherently quantum scaling can emerge in a conformally invariant system driven out of equilibrium by a stochastic boundary field.  We consider microscopic models with Hamiltonian of the form

\begin{equation}\label{cftham}
\hat{H}=\hat{H}_{\mathrm{CFT}}+{h}_b(t)\hat{O}_b,
\end{equation}

where $\hat{H}_\mathrm{CFT}$ is a one-dimensional bulk  critical Hamiltonian driven by a stochastic noise field $h_b(t)$, weighted by a relevant operator $\hat{O}_b$ that lives on the boundary of the system. 
Generically, $\hat{H}_\mathrm{CFT}$ can include irrelevant terms that break the conformal symmetry, and only emergent conformal invariance in the infrared limit is required. 
Previous work investigated the coupling of quantum systems to different types of boundary drives, which lead to eventual thermalization~\cite{prosen,PhysRevB.85.184302} or to non-universal relaxation~\cite{PhysRevLett.122.040604}; in contrast, we show that the dynamics induced by a conformal boundary drive are universal in a certain limit and inherently quantum. 

Before proceeding, we note that the problem of a CFT driven by a periodic (Floquet) boundary drive, considered by one of us \cite{PhysRevLett.118.260602},  does lead to universal relaxation. In this work we find that universality persists even with a more generic stochastic drive. Furthermore, we show that the behavior of the Loschmidt echo is richer than in the periodically driven case: one may have a different class of universal relaxation when looking at the typical decay in a single realization of the noise compared to the average echo over many noise realizations. 
We corroborate these results with a direct numerical calculation of a boundary driven transverse field Ising model at its critical point.

\begin{figure}
    \centering
    \includegraphics[width=\columnwidth]{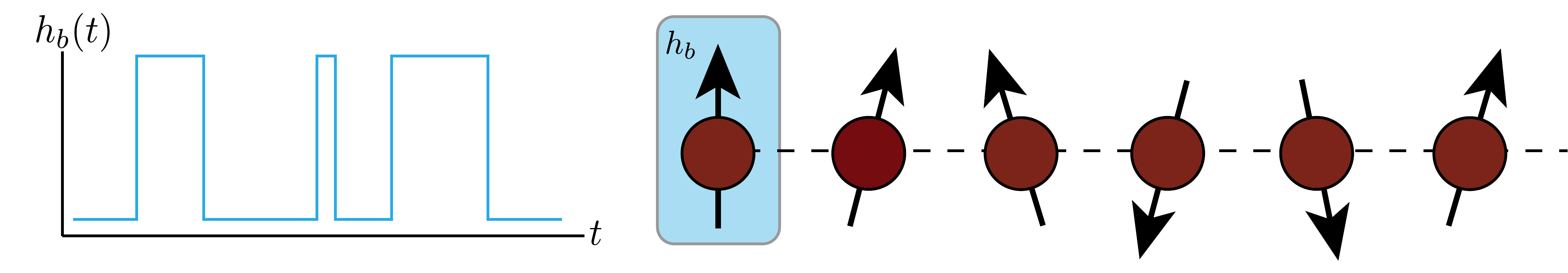}
    \caption{Sketch of the class of systems under study in this work. We consider a quantum critical spin chain (red spins) subjected to a stochastic boundary drive (blue line).}
    \label{fig:model}
\end{figure}

\subsection{Stochastically driven boundary in CFT} 

For concreteness, let us consider the Poisson process whereby the boundary coupling $h_b$ stochastically jumps between two values with some fixed probability $p$ over an interval of time $\delta t$ as illustrated schematically in Fig. \ref{fig:model}. We note that generically any type of sufficiently weak Markovian noise will flow to Poisson noise under the renormalization group (RG), since events that flip the boundary conditions are more relevant than those that do not. To define our scaling variables, we have an average time between flips $T = \delta t/p$, with a Poisson parameter of the shot noise after total time $t$ of $\lambda=pt/\delta t$. Finally, the strength of the boundary field  $h_b \sim ||h_b(t)||$ sets the timescale $t_b = h_b^{-\nu_b}$. Here, $\nu_b = 1/(1-\Delta_b)$ with $\Delta_b$ the scaling dimension of the boundary operator $\mathcal O_b$. Application of the boundary CFT framework is valid while the time between flips of boundary conditions is much larger than the timescale $t_b$, hence the latter serves as a short time cutoff for our theory.

In what follows we focus on the Loschmidt echo, or return-probability amplitude of the wavefunction,
\begin{equation}\label{eq:los}
\mathcal{L}(t)=|\langle \psi_0|\psi(t)\rangle|^2,
\end{equation}
{which in recent years has become an important quantity for the study of universal properties of quantum many-body systems~\cite{dpt, PhysRevLett.119.080501,PhysRevB.99.134301,Silva08} and can be measured via spectroscopic techniques~\cite{adilet, Latta11}.}
We consider the behavior of this function in a typical realization of the stochastic drive field as well as its expected average over all possible realizations of the noise. 

For each realization of the stochastic field $h_b(t)$, $\mathcal{L}(t)$ can be mapped to a partition function of a field theory, which flows to a conformally invariant one in the scaling limit where the time between flips is much larger than $t_b$.\footnote{This generalizes the constructions of Ref.~\citenum{PhysRevX.4.041007} and Ref.~\citenum{PhysRevLett.118.260602}.} After a Wick rotation to imaginary time, the ground state $\ket{\psi_0}$ is determined as the asymptotic evolution $\lim_{\tau \to \infty} e^{-\tau H_0} \ket{\Omega}$,  with $\ket{\Omega}$ a generic state and the operator $e^{-\tau H_0}$ acting as a projector onto the ground state of $H_0$ in the limit $\tau\to\infty$. The boundary field flips between different fixed values at random times; therefore, in any given realization of the flips, the unitary time evolution operator takes the form of a succession of imaginary time evolutions, given by the Hamiltonian~\eqref{cftham} with different fixed boundary fields over the intervals between flips. Thus, we have $\mathcal{L}(t) \propto |\expectation{e^{-\tau_0 H_0} \ldots e^{-\tau_2 H_2} e^{-\tau_1 H_1} e^{-\tau_0 H_0}}|^2$, with $\tau_0\to\infty$. Since the Hamiltonians $H_i$ differ only by a relevant boundary operator, we see that this maps exactly onto a partition function in a two-dimensional conformal field theory with mixed boundary conditions along the imaginary time direction. 

Now let us focus on the case of $T \gg t_b$, that is, the average time between flips being much greater than the timescale induced by the finite boundary field. This is to ensure that the dynamics enters into a universal regime where it can exhibit scaling.  It is also important that we impose $\delta t \gtrsim t_b$, since we only expect universal physics on timescales longer than $t_b$, and $\delta t$ is the minimal spacing between flips. These limits allow us to use the technique of boundary condition changing operators, generic to any two dimensional conformal field theory, in which sharp changes in the boundary condition may be replaced inside all correlation functions by a particular type of primary operator, often referred to as a boundary-condition changing (BCC) operator, inserted at the location of the change~\cite{cardy_conformal_1984, cardy_boundary_1989, cardy_boundary_2006}. We can therefore identify the Loschmidt echo with a $2n$-point function of primary operators $\phi_{\mathrm{BCC}}$. Analytically continuing to real time, for any realization of the noise with flips at times within some configuration $S=\{t_i \}$, the Loschmidt echo is then

\begin{equation}
\mathcal{L}(t|\{t_i\}) \sim \Big| \Big\langle \prod_{t_i \in S}\phi_{\mathrm{BCC}}^{(i)}(t_i) \Big\rangle \Big|^2.
\end{equation}

For simplicity, let us now assume that we have a binary drive between two Hamiltonians $H_0$ and $H_1$, and hence only one type of BCC operator, $\phi$, per drive, though we note that the argument follows for more complicated drives as well. The specific examples that we consider below are boundary drives in the critical Ising model. One class of drive in this case is given by a boundary condition that jumps back and forth between fixing the boundary spin up/down. We call this the ``fixed-fixed'' drive, and it corresponds to insertions of a fermion BCC operator with scaling dimension $\Delta_{\mathrm{BCC}}=1/2$. Another class of drive is given by a field that jumps between a free and fixed (say, spin up) boundary condition. This drive corresponds to inserting a BCC operator with a scaling dimension $\Delta_{\mathrm{BCC}}=1/16$.

\subsection{Typical echo} We first calculate the typical echo $\mathcal{L}_{\mathrm{typ}} \equiv e^{\overline{\log \mathcal L}}$. We have

\begin{equation}\label{eq:typ}
\overline{\log \mathcal L} = \sum_{n=0}^\infty P(n) \frac{1}{t^n/n!} \int \prod_i dt_i \log \abs{\mathcal{C}(t_1, ..., t_n) }^2,
\end{equation}

with $\mathcal{C}(t_1, ..., t_n)=\expectation{\phi(t_n) \ldots \phi(t_1)}$ the time-ordered correlation function associated to $n$ insertions of the BCC operators, $P(n) = e^{-\lambda} \lambda^n / n!$ for a Poisson process, and we note that in Eq.~\eqref{eq:typ} only $2n$-point functions enter the expectation value. In fact, because of the ket in the echo, both the one-flip process and the two-flip process are controlled by the two-point function of BCCs, and similarly for higher orders: the $(2n-1)$- and $2n$-flip processes are controlled by the $2n$-point function of BCCs. In taking the average over the BCC insertions, we normalize by $\int \prod_i dt_i = t^n/n!$, where the $n!$ factor is due to the time-ordering. 

Now, for average flipping times $T$ much larger than the microscopic timescale $t_b$ ($T\gg t_b$), we can utilize the finite-size scaling relation for primary operators at the bulk critical point~\cite{Cardy:1996xt}, i.e. $|\mathcal{C}(t_1, ..., t_n)|^2 =  (T/t_b)^{-4n\Delta_{\mathrm{BCC}}} \mathcal F(t_1/T,\ldots, t_n/T)$, with $\mathcal F$ a universal scaling function. We therefore expect the typical Loschmidt echo to be a universal scaling function $\mathcal L_{\mathrm{typ}} = \mathcal L_{\mathrm{typ}}(T/t_b, \lambda)$, and after explicit evaluation of the sum we arrive at

\begin{equation}\label{eq:scaling2}
\overline{\log \mathcal L} \simeq -4 \Delta_{\mathrm{BCC}} \sum_{n=1}^\infty (P(2n-1) + P(2n))n \log(T/t_b)= -2\Delta_{\mathrm{BCC}} (\lambda + e^{-\lambda} \sinh \lambda) \log(T/t_b),
\end{equation}

up to an additive universal average amplitude term $\overline{\log \mathcal F}$ that may be neglected in the large $T$ limit. We note that averaging the logarithm is crucial, as the amplitude itself may in general diverge. For large $\lambda \gg 1$ we expand this result to obtain $\overline{\log L} \approx_{\lambda \gg 1} -2\Delta_{\mathrm{BCC}} \lambda  \log(T/t_b)$. Thus, we predict a universal power-law form of the typical echo

\begin{equation}\label{scaling}
\mathcal{L}_{\mathrm{typ}}  \underset{\lambda\gg 1}{\sim} \left(\frac{T}{t_b}\right)^{-2\Delta_{\mathrm{BCC}} \lambda},
\end{equation}
which is in good agreement with the numerical data on the Ising model shown in Fig.~\ref{fig:data}. 

\subsection{Mean echo} 

Having argued for universal behavior of the Loschmidt echo in a typical realization of the boundary stochastic field, we now turn to the calculation of the mean echo. 
In many cases, the mean echo should follow the same universal scaling form as the typical. However, as we argue below, for certain types of drives the mean and typical echo may differ drastically.

The general form of the mean echo is given by 

\begin{equation}
\overline{\mathcal L(t)} = \sum_n P(n) \frac{1}{t^n/n!}\int \prod_{k=1}^n dt_k \abs{\expectation{\phi(t_n) \ldots \phi(t_1)}}^2 .
\label{eq:avL}
\end{equation}

As noted previously, finite-size scaling implies that for $T\gg t_b$ the $n$-point function should be a power law in $T$, with an exponent determined by the scaling dimension of the BCC operator.
If $\Delta_{\mathrm{BCC}} \geq 1/4$, the power-law can produce a divergence in (\ref{eq:avL}) when integrating over the insertions of the BCCs. In the divergent case, rare configurations where the insertions are all closely spaced can give a dominant large contribution to the mean echo, while they do not affect the typical echo because the integral is over the logarithm. 

Let us show this explicitly. Consider first the case $\Delta_{\mathrm{BCC}} < 1/4$, where the integrals are non divergent. An example is the fixed-free drive of the Ising model with $\Delta_{\mathrm{BCC}}=1/16$.
Using the aforementioned finite-size scaling relation $|\mathcal{C}(t_1, ..., t_n)|^2 =  (T/t_b)^{-4n\Delta_{\mathrm{BCC}}} \mathcal F(t_1/T,\ldots, t_n/T)$, we have $\overline{\mathcal L(t)} = \sum_n P(n) (T/t_b)^{-4n \Delta_{\mathrm{BCC}}} \overline{\mathcal F(n)}$, where $\overline{\mathcal F(n)} = \int \prod dt_i \mathcal F(t_1/T,\ldots, t_n/T)$ is finite and independent of the lower cutoff. This sum can be evaluated using the saddle point approximation. One finds that, under the assumption $\lambda \gg T/t_b$, the sum is dominated simply by the term $n_* = \lambda / 2$, recalling that the sum runs only over $n$ even. Therefore, one obtains $$\overline{\mathcal L(t)} \sim e^{-\lambda/2} \overline{\mathcal F(\lambda/2)}  (T/t_b)^{-2\Delta_{\mathrm{BCC}} \lambda}.$$ This gives the same power law dependence on $T$ as the typical echo, and hence the same scaling form. 

Now consider the divergent case $\Delta_{\mathrm{BCC}}\ge 1/4$, which is realized, for example, by the fixed-fixed drive of the Ising model ($\Delta_{\mathrm{BCC}}=1/2$). 
In this case the integral over $\mathcal F(t_1/T,\ldots, t_n/T)$ depends sensitively on the lower cutoff $\delta t/T$. In order to estimate of the scaling form, we replace the averaged correlation function of BCCs by the largest contribution in the limit $\delta t \to 0$. Namely, we take $\int \prod dx_k {\mathcal F}(x_1, ..., x_n) \approx (\delta t/T)^{(1-4\Delta_{\mathrm{BCC}})n/2}$, where $\delta t^{1-4\Delta_{\mathrm{BCC}}}$ is the divergent part of the two-point function. Substituting this into the sum and taking $\delta t \approx t_b$ gives $$\overline{\mathcal L} \sim \sum_n P(n) \frac{n!}{\lambda^n} (T/t_b)^{-4n\Delta_{\mathrm{BCC}}} (t_b/T)^{{n\over 2}(1-4\Delta_{\mathrm{BCC}})}.$$ Finally, using the saddle point method with the sum dominated by $n_* = \lambda / 2$, we obtain the power law $\overline{\mathcal L} \sim (T/t_b)^{-\lambda(\Delta+{1\over 4})}$, which is different than power law governing the typical echo. In particular, for the fixed-fixed Ising drive we get $\overline{\mathcal L}\sim (T/t_b)^{-3\lambda/4}$, which should be compared with ${\mathcal L}_{\mathrm{typ}}\sim (T/t_b)^{-\lambda}$.

\begin{figure*}
	\includegraphics[width=0.5\columnwidth]{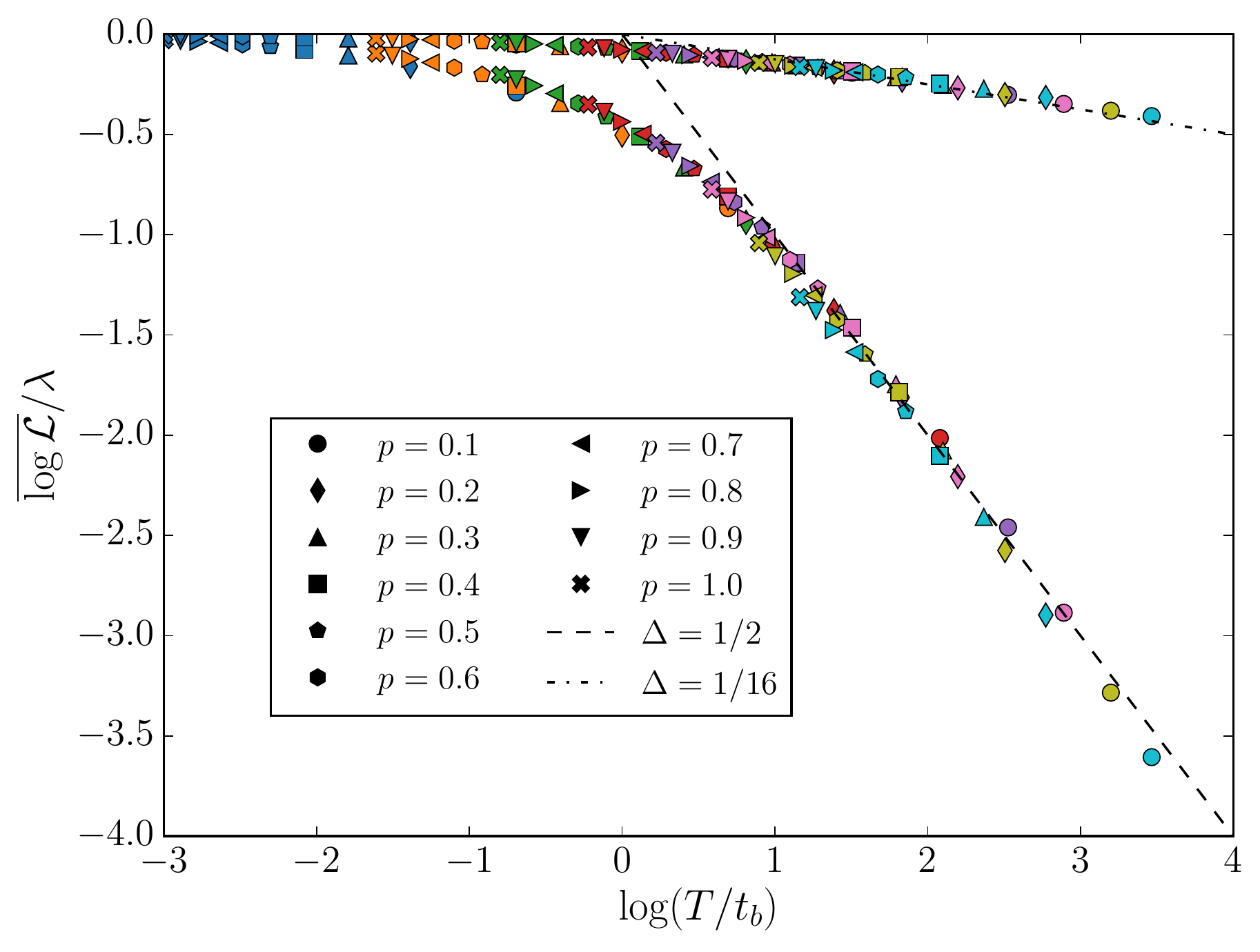}
	\includegraphics[width=0.5\columnwidth]{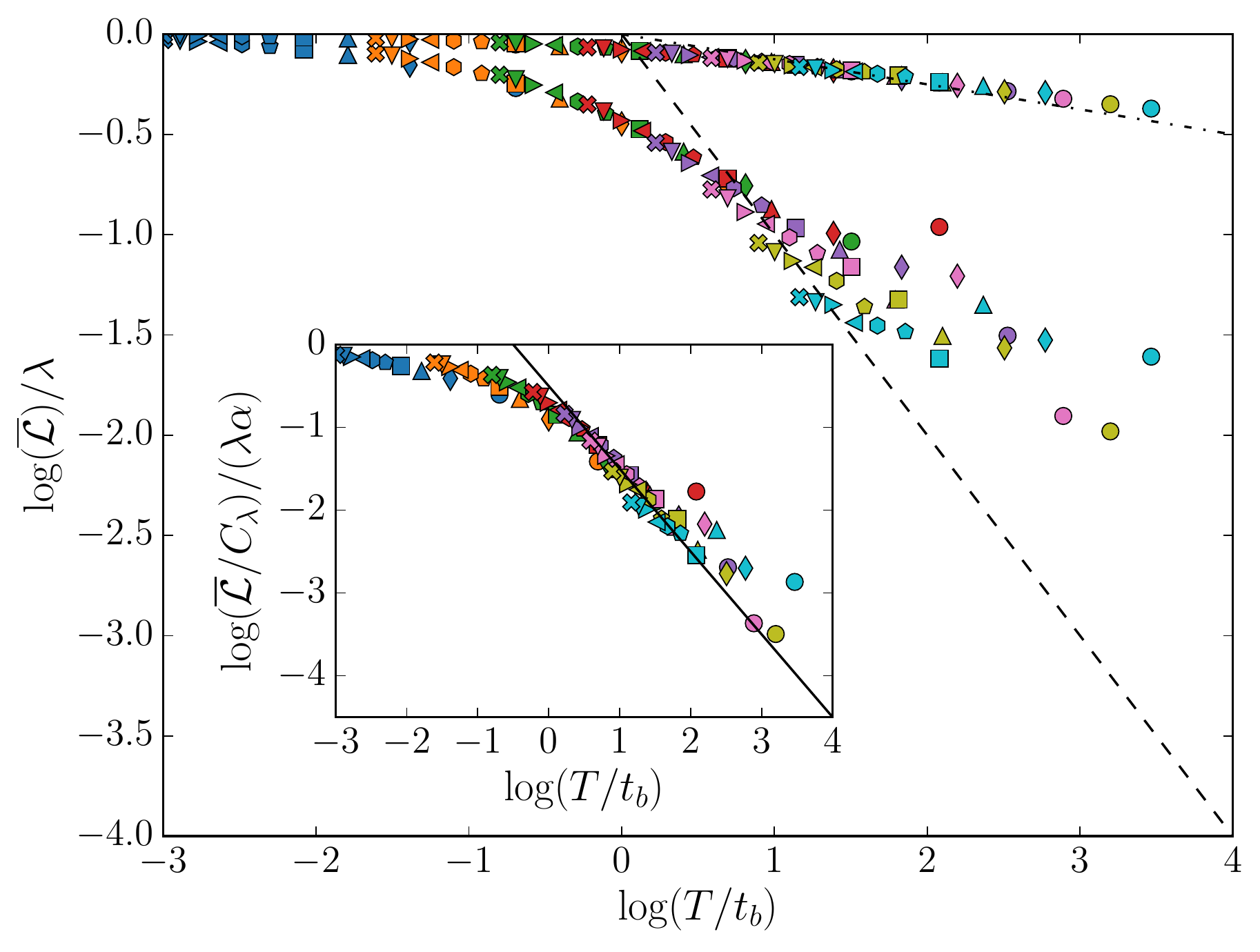}
	\caption{\label{fig:data} Left: The typical Loschmidt echo averaged over $r=1000$ realizations and for  system sizes up to $L=1000$, for different values of the boundary field $h_b$ and flipping probabilities $p$. The boundary field takes the values $h_b = 0.1$ (blue), 0.2 (orange), 0.3 (green), 0.4 (red), 0.5 (purple), 0.6 (pink), 0.7 (yellow), 0.8 (teal), and the probability $p$ varies as marked in the legend. The Poisson parameter $\lambda$ takes values $\lambda \geq 10$ throughout. The dashed lines are the prediction from boundary CFT (Eq.~\eqref{scaling}): for the fixed-fixed drive $\Delta = 1/2$, and for the free-fixed drive $\Delta = 1/16$, with both showing excellent agreement. Right: The mean echo of the same data. For the free-fixed drive, the mean and typical (black dashed lines) are very similar, but, strikingly, for the fixed-fixed drive the mean lies far above the typical. This is due to rare events that dominate the average and give a renormalized scaling form (inset), where $\alpha = 0.71\pm 0.03$, in good agreement with the estimate of $\alpha = 0.75$ in the main text.  
	}
	\label{fig:numerics}
\end{figure*}

\subsection{Numerical results} 

Having expounded our arguments in generality for stochastically boundary-driven CFTs, let us now validate them in an explicit model. Consider a one dimensional integrable quantum Ising chain in a transverse field, $g$, tuned to criticality, $g\to g_c$, and driven by a stochastic time-dependent noise coupled to the longitudinal spin field at the boundary of the chain, 

\begin{equation}\label{eq:ham}
H(t) = - J\sum^L_{i=1} ( \sigma_i^z \sigma_{i+1}^z + g \sigma_i^x ) - h_b(t) \sigma_1^z.
\end{equation}

This Hamiltonian falls in the class given by~\eqref{cftham}, as its low-energy excitations are described in equilibrium by the Ising conformal field theory. We note that the critical Ising model with a spatially disordered boundary field was studied in Ref.~\citenum{Cardy_1991}.

After a Jordan-Wigner transformation~\cite{sachdev2011quantum}, the model (\ref{eq:ham})  maps onto a chain of free Majorana fermions 
\begin{equation}
    H(t)=-J\sum_{n=1}^{2L}i\eta_n\eta_{n+1}-h_b(t)\,i\gamma\eta_1,
    \label{eq:majorana}
\end{equation}
where $\eta_{2i-1}$, and $\eta_{2i}$ are Majorana operators located on site $i$ of the Ising chain. 
Note that expressing the boundary coupling to the edge operator $\sigma_1^z$, which breaks the Ising-symmetry, requires an additional ancilla Majorana operator $\gamma$ that anticommutes with all fields and satisfies $\gamma^2 = 1$~\cite{doi:10.1142/S0217751X94001552}. The quadratic Hamiltonian \eqref{eq:majorana} can be easily diagonalized numerically on large systems, {and is thus an ideal testbed for our earlier analytical arguments, which require large system sizes, late times and extensive disorder averaging to numerically observe.}

The system is endowed with three characteristic time scales:  the inverse bandwidth, $t_J\sim1/J$, which is the ultra-violet scale in the problem and controls the onset of non-universal effects in dynamics; the time-scale associated to the  boundary field $t_b = h_b^{-2}$;  and the intrinsic time of a stochastic Poisson flip, $\delta t$.  To ensure universal scaling, we choose  $t_J \ll t_b$, equivalent to the condition $h_b^2 \ll J$ (the boundary CFT limit). 
We note that if we were to integrate over the stochastic boundary field from the start, we would obtain an effective non-unitary evolution of a density matrix. However, because the Poisson switching process cannot be represented by a Gaussian white noise field, this not in general described by a quantum master equation in Lindblad form~\cite{PhysRevA.95.012115, Lucz}. Thus, the results presented here are distinct from previous works on driven-dissipative impurities, which used Lindblad equations to represent the drive~\cite{PhysRevLett.122.040604,PhysRevLett.122.040402, PhysRevB.85.184302, dries}. 

In our exact numerical calculations,\footnote{For details of the free-fermion numerical procedure, see the supplemental material of e.g. ~\cite{PhysRevLett.118.260602}, or the original references in~\cite{peschel_calculation_2003,eisler_evolution_2007}.}
we prepare the ground state of the chain and then  compute the time-dependent Loschmidt echo for at least 1000 realizations of the noise, on system sizes up to $L=1000$ and with $J=2$. At any given time step, we randomly select whether or not to flip the boundary field, corresponding to a Markovian process. We then scan over many values of the boundary field $h_b$ and the probability of flipping $p$ for two different types of drives: 1) a ``fixed-fixed'' drive, where the boundary field takes values $\pm h_b$ (with the system prepared in the ground state of $-h_b$), and 2) a ``free-fixed'' drive, where the boundary field takes values $+h_b$ and 0 (with the system in the ground state of $h_b=0$). Note that at very long-times we generally expect to see decay of the Loschmidt echo in any finite system as it heats up under the action of the incoherent drive, $h_b(t)$~\cite{Marinolong2012,cai}. However, this occurs on time scales of at least $t_{*}\propto L$~\cite{marko,PhysRevA.91.052107}, while in our simulations we keep $t < L/2$ to reduce finite-size effects, ensuring $t \lesssim t_*$.

Fig. ~\ref{fig:numerics} (left panel) shows the decay of the typical echo $\overline{\log\mathcal{L}}$ for different instances of the boundary field. The universal collapse, the asymptotic power law and the specific exponents obtained for both types of drive (fixed-fixed and free-fixed) are in excellent agreement with the CFT predictions. 

The right panel of Fig. ~\ref{fig:numerics} shows the results for the mean echo. As expected from the discussion of the previous section we see that the mean echo is identical to the typical echo in the case of the free-fixed drive. This is because the BCC operator has dimension $\Delta_{\mathrm{BCC}} = 1/16 < 1/4$ in this case. Again, as expected, the mean and typical echos differ substantially in the case of the fixed-fixed drive, for which the BCC operator $\Delta_{\mathrm{BCC}} = 1/2>1/4$. Furthermore, the inset shows reasonable data collapse with the ansatz 
$\overline{\mathcal L} \sim C_\lambda (T/t_b)^{-\lambda \alpha}$, where $\alpha=0.71\pm 0.03$ and $C_\lambda$ is a constant prefactor dependent on the Poisson rate of flipping. 
This should be compared to the analytical prediction of $\alpha=0.75$ obtained from our approximation above, taking into account only the leading divergences in the average over BCC insertions. Notice, however, that there is a larger statistical error in the average echo compared to the typical one; therefore, the imperfect collapse could either be due to  statistical errors or from actual small corrections to the  scaling exponent predicted from the bCFT analysis above.

\subsection{Discussion} 

The scaling exponents that control the dynamics of the Loschmidt echo in the critical transverse field Ising model are those of the boundary Ising CFT;  we therefore expect our results to hold upon adding integrability breaking perturbations $V$ to the Hamiltonian in~\eqref{eq:ham}, provided they are irrelevant operators under renormalization group flow (for instance, $V=\Gamma\sum_i\sigma^x_i\sigma^x_{i+m}$, with $m>0$).  
Furthermore, other critical points with central charge $c=1/2$ will give the same  dynamical scaling exponents. While we have demonstrated the scaling numerically for the Ising CFT, we emphasize that the mechanism for universality outlined here is model-independent. Any boundary-driven CFT  will display similar universal collapse when driven by appropriate boundary perturbations, with  exponents that depend on the particular form of the drive and driving operator. 
{We remark that the stochastic boundary Ising problem solved here does not map onto a Kondo problem (as done  in Ref.~\citenum{PhysRevLett.100.165706}), since the average echo and the
mean echo studied in our work are not expressible as the statistical partition
function of a Coulomb gas.}

 An important general question is under what conditions one should expect to find universal behavior of a driven impurity. The problem of a quantum critical Ising chain driven by noise acting on a local transverse spin operator $h_x(t)\sigma^x_1$ was studied by one of us in Ref.~\citenum{PhysRevLett.122.040604}. 
In that study, crucially, the critical Ising chain was driven by a \textit{marginal} boundary operator, $\sigma^x_1$, rather than by a \textit{relevant} boundary operator, $\sigma^z_1$.
 Despite this seemingly small difference, driving by a marginal spin operator yielded a decaying Loschmidt echo $\mathcal{L}(t)\propto e^{-\gamma t}t^\theta$, with a non-universal exponent $\theta$.
 This is in sharp contrast to the universal scaling collapse found in this work, and suggests that the  RG relevance of the driving operators can play an important role in dictating the universality (or lack thereof) of the dynamical response to dissipative impurities. 
Further, whether other classes of noise, such as $1/f$ noise or non-Markovian noise, can lead to novel dynamical universal scaling is an intriguing open question. Answering such questions would hopefully serve as stepping stones towards the goal of a systematic categorization of the universality classes of driven-dissipative impurities.

\section{Appendices}

\subsection{Numerical methods: free fermions and interactions} \label{sec:numerics}

In this appendix, we provide details on the numerical simulation of the transverse-field Ising (TFI) chain with longitudinal field, and the extraction of the entanglement entropy and the Loschmidt echo. To numerically study the time evolution of the TFI chain ground state subject to the Floquet drive described above, we utilize the fact that the non-interacting TFI chain can be efficiently described as a system of free Majorana fermions. This fact will prove useful in calculating the entanglement entropy of subsections of the chain as a function of time~\cite{peschel_calculation_2003,eisler_evolution_2007}. We cover the free case first, then move on to the interacting one in subsection C. For completeness, the TFI Hamiltonian in terms of spins is

\begin{equation}
H = - \sum_{j = 0}^{L-1} (J\sigma_i^z \sigma_{i+1}^z + h \sigma_i^x + \Gamma \sigma_i^x \sigma_{i+1}^x) - h_b(t) \sigma_0^z 
\end{equation}

with interactions controlled by $\Gamma$. 

\subsubsection{Entanglement entropy}\label{sec:entropy}

First, let us apply a Jordan-Wigner transformation to the non-interacting TFI chain $\sigma_j^x = i \gamma_{B,j} \gamma_{A,j}$, 
$\sigma_j^z = -i \gamma \left( \prod_{l=0}^{j-1} i \gamma_{A,l} \gamma_{B,l} \right) \gamma_{B,j}$ where the $\gamma$ operators obey the Majorana algebra $\{ \gamma_{\alpha,i},\gamma_{\beta,j} \} = 2\delta_{\alpha \beta}\delta_{ij}$, $\gamma_{\alpha,i}^2 = 1$. We include in our Jordan-Wigner transformation an ancilla Majorana operator $\gamma$ that plays no dynamical role. In the Majorana language, then, the TFI Hamiltonian is 

\begin{equation}
H = \sum_{j=0}^{L-2} i \gamma_{A,j} \gamma_{B,j+1} + \sum_{j=0}^{L-1} i \gamma_{A,j} \gamma_{B,j} + h_0 i \gamma \gamma_{B,0}.
\end{equation}

at the critical point $h=J=1$. This doubles the size of the Hilbert space, making each original level doubly degenerate. Now, if we are in a state satisfying Wick's theorem, everything is essentially determined by the two-point correlator $C_{ij} = \expectation{\gamma_i \gamma_j}$.
The Majorana anti-commutation relation implies that $C_{ij} = 2 \delta_{ij} - C_{ji}$, so $C_{ij} = \delta_{ij} + a_{ij}$, where $a$ is some anti-symmetric matrix. Let us now diagonalize $a = q^T \sigma q$, where $q$ is orthogonal and $\sigma$ has form  $\sigma = \diag\begin{pmatrix}
0 & \lambda_i \\ -\lambda_i & 0 
\end{pmatrix}_{i=1}^L$. This form has the eigenvalues arranged such that $\sigma_{i,i+1} = \lambda_i$, and satisfies $\sigma^T = - \sigma$. Now define $\gamma' = q \gamma$. Then 

\begin{equation}
\expectation{ \gamma'_{2k'-1} \gamma'_{2k }} = q_{2k'-1,i}q_{2k,j} (\delta_{ij} + \lambda_{k''} (q_{2k''-1,i}q_{2k'',j} - q_{2k'',i}q_{2k'',j}) ).
\end{equation}

From the orthogonality of $q$, $q_{\alpha i}q_{\beta i} = \delta_{\alpha\beta}$, so the only non-vanishing term is $$q_{2k'-1,i} q_{2k,j} \lambda_{k''} q_{2k''-1,i} q_{2k'',j} = \lambda_k \delta_{kk'}\delta_{k'k''}.$$ Thus the only non-vanishing two-point function is

\begin{equation}
\expectation{\gamma_{2k-1}' \gamma_{2k}} = - \expectation{\gamma_{2k}\gamma_{2k-1}'} = \lambda_k.
\end{equation}

We can write this correlation function as arising from a single particle density matrix $\rho = \frac{1}{Z}\prod_k e^{ i \epsilon_k \gamma_{2k-1}' \gamma_{2k}' }$. 
Now, we can construct a complex fermionic operator from Majorana operators via $c_k = \frac{\gamma_{2k-1}' + i \gamma_{2k}'}{2}$. This gives $\gamma_{2k-1}' \gamma_{2k}' = -i(2 c_k^\dagger c_k - 1)$. This gives the density matrix as

\[
\rho = \prod_k \frac{e^{\epsilon_k (2 n_k - 1)}}{e^{\epsilon_k} + e^{-\epsilon_k}}.
\]

Thus the two-point function is $\expectation{\gamma_{2k-1}'\gamma_{2k}} = -i \expectation{2 n_k - 1} = \lambda_k = -i \frac{e^{\epsilon_k}(+1) + e^{-\epsilon_k}(-1)}{e^{\epsilon_k} + e^{-\epsilon_k}} = -i \tanh \epsilon_k$. Thus we have the non-trivial relation $i \lambda_k = \tanh \epsilon_k$. Now define $\mu_k = \abs{\lambda_k}$. Then $\epsilon_k = \tanh^{-1}(\mu_k)$. To find the entanglement entropy, write the density matrix as

\begin{equation}
\rho = \prod_k  \Big[ p_k \proj{0_k} + (1-p_k) \proj{1_k} \Big], \qquad p_k = \frac{e^{-\epsilon_k}}{e^{\epsilon_k} + e^{-\epsilon_k}}.
\end{equation}

Then the entanglement entropy is $S =- \Tr \rho \log \rho = -\sum_{k=1}^{L} p_k \log p_k + (1-p_k) \log (1-p_k)$. To time evolve the correlation function, in the Heisenberg picture $C_{ij}(t) = \expectation{\gamma_i(t) \gamma_j(t)}$. Now, for $A$ and $B$ Majorana operators, $e^{\alpha A B} = \cos \alpha + AB \sin \alpha$. Thus, $e^{\alpha AB} A e^{-\alpha AB} = A \cos 2 \alpha - B \sin 2 \alpha$, and $e^{\alpha AB} B e^{-\alpha AB} = B \cos 2 \alpha + A \sin 2\alpha$. Thus, defining the diagonal matrix 

\begin{equation}
D(t) = \diag\left\{\begin{pmatrix}
\cos (2\epsilon_k t) & -\sin (2\epsilon_k t) \\
\sin (2\epsilon_k t) & \cos (2\epsilon_k t)
\end{pmatrix}_{k=1}^{2L}\right\},
\end{equation}

we find that $\gamma_i' (t) = D_{ij}(t) \gamma_j'(0)$. Defining $\Gamma (t) \equiv Q^T D(t) Q $, we see that the correlation function evolves particularly simply as $C(t) = \Gamma(t) C(0) \Gamma(t)^T$. From the time-evolved correlation function we can dynamically calculate the entanglement entropy as above. \\

\subsubsection{Loschmidt echo}\label{sec:echo}

Calculation of the Loschmidt echo directly from the Majorana operator two-point function above is a bit more challenging due to the fact that the TFI Hamiltonian does not conserve particle number when written in terms of complex fermion operators. This is not an issue for a single quench, as has been explored in Refs.~\cite{peschel_calculation_2003,eisler_evolution_2007,stephan_local_2011} and many others, but becomes quite complicated even for three quenches. Instead, it is convenient to use the fact that the $XX$ model -- which does conserve particle number -- can be decomposed into two independent copies of the TFI chain. The mapping proceeds as follows. Take the $XX$ chain with an ancilla fermion at ``site'' 0:

\begin{equation}
H_{XX} = - J \sum_{i=1}^{L} (\sigma_i^x \sigma_{i+1}^x + \sigma_i^y \sigma_{i+1}^y) - J' \sigma_0^x \sigma_1^x.
\end{equation}

The total length of the chain is $L+1$ in this notation. Via a Jordan-Wigner transformation, namely $c_i^\dagger = \sigma_i^+ \prod_{j < i}\sigma_j^z$, $c_i = \left(\prod_{j < i}\sigma_j^z\right) \sigma_i^-$ with $\sigma^\pm = (\sigma^x \pm i \sigma^y)/2$,we get that the $XX$ Hamiltonian is $H_{XX} = - \frac{J}{2} \sum_{i = 1}^{L} c_i^\dagger c_{i+1} - \frac{J'}{2} c_0^\dagger c_1 + h.c.$ Now we can decompose each fermion operator into two Majorana operators, via $c_i^\dagger = (\gamma_{A,i} - i \gamma_{B,i})/2$,  $c_i  =(\gamma_{A,i} + i\gamma_{B,i})/2$, where $\gamma^2 = 1$, $\gamma^\dagger = \gamma$ and $\{ \gamma_{\alpha,i},\gamma_{\beta,j} \} = 2\delta_{ij}\delta_{\alpha\beta}$. Thus,

\begin{equation}
H_{XX} = - \frac{J}{4} \sum_{i=1}^{L} i (\gamma_{A,i} \gamma_{B,i+1}) - \frac{J'}{4} i \gamma_{A,0} \gamma_{B,1} 
-\frac{J}{4}\sum_{i=1}^{L} i (\gamma_{A,i+1} \gamma_{B,i})- \frac{J'}{4} i \gamma_{A,1} \gamma_{B,0}. 
\end{equation}

We now see that the $XX$ chain is simply two uncoupled copies of the TFI chain. Thus, the Loschmidt echoes will be related by $\mathcal L_{XX}(t) = \mathcal L_{TFI}(t)^2$. To get the Loschmidt echo of the $XX$ chain with a driven first link, we generalize the methods of Refs.~\cite{PhysRevA.72.013604,kennes_universal_2014} to handle multiple quenches between two $XX$ Hamiltonians $H_0$ and $H_1$. We first write the ground state as a filled Fermi sea,

\begin{equation}
\ket{\psi(0)} = \prod_{m=1}^{N_f} \left( \sum_{j=1}^{L} P_{jm} c_j^\dagger \right) \ket{0}
\end{equation}

where $N_f$ is the total number of negative eigenvalues of the initial Hamiltonian $H_0$ and the $L \times N_f$ matrix $P$ is the (sorted) matrix of corresponding eigenvectors. Now, under time evolution, $\ket{\psi (t)} = \ldots e^{-i H_0 T/2}e^{-i H_1 T/2} \ket{\psi(0)} = \hat U(t) \ket{\psi(0)}$, so $\ket{\psi(t)} = \prod_{m=1}^{N_f} \left( \sum_{j=1}^{L} P_{jm}(t) c_j^\dagger \right) \ket{0}$ with $P(t) = U(t) P$ where $U(t)$ is an $L \times L$ matrix. This straightforwardly gives the Loschmidt echo as

\begin{align}
\mathcal L_{XX}(t) &= \abs{ \matrixel{\psi(0)}{\ldots e^{-i H_0 T/2}e^{-i H_1 T/2}}{\psi(0)} }^2 \nonumber\\
&= \abs{ \matrixel{0}{\prod_{m=1}^{N_f} \sum_{j=1}^{L} P_{jm}^* c_j \prod_{n=1}^{N_f} \sum_{i=1}^{L} P_{in}(t) c_i^\dagger }{0} }^2 \nonumber\\
&= \abs{\det( P^\dagger P(t) )}^2
\end{align}

We then simply compute $\mathcal L_{TFI}(t) = \sqrt{\mathcal L_{XX}(t)}$ to get the desired echo.

\subsubsection{Interactions}\label{sec:ints2}

In the presence of nonzero $\Gamma$, the integrability of the TFI chain is broken, and a description in terms of free fermion operators no longer holds, as the $\sigma_i^x \sigma_{i+1}^x$ terms produce four-fermion interaction terms after a Jordan-Wigner transformation. While interactions break integrability and ultimately lead to late-time thermalization among other effects, they crucially have no effect on the underlying CFT because they are irrelevant in the renormalization-group (RG) sense. We therefore expect the late-time Loschmidt echo to display the same power law behavior as in the free case, albeit with minor deviations at finite times due to the presence of an irrelevant operator. \\

We first note that interactions will shift the critical point away from the self-dual point $h=J=1$. We determine the location of this new critical point numerically by exact diagonalization by looking at the scaling of the gap for systems sizes $L=18,\dots,24$.  We then simulate the dynamics using matrix product states (MPS) techniques~\cite{Schollwock201196}. The initial state is determined using standard density matrix renormalization group (DMRG) methods~\cite{PhysRevLett.69.2863,Schollwock201196}, and the Floquet dynamics is simulated using the time-evolving block decimation (TEBD) algorithm based on a fourth-order Trotter decomposition with Trotter time step $dt=0.2$. We adapt the bond dimension of the MPS in order to keep the discarded weight below $10^{-8}$ throughout the whole time evolution. The Loschmidt echo for a step drive from $h_b = 0$ to $h_b \not = 0$ of an interacting Ising chain $\Gamma=0.25$ with $L=400$ sites is shown in Fig.~\ref{fig:ints}. Despite the shorter time scales and smaller system sizes accessible using TEBD compared to the non-interacting case, we find that our field theory predictions agree well with TEBD simulations in the low-frequency regime, confirming the universality of our results.

\begin{figure}
	\includegraphics[width = \columnwidth]{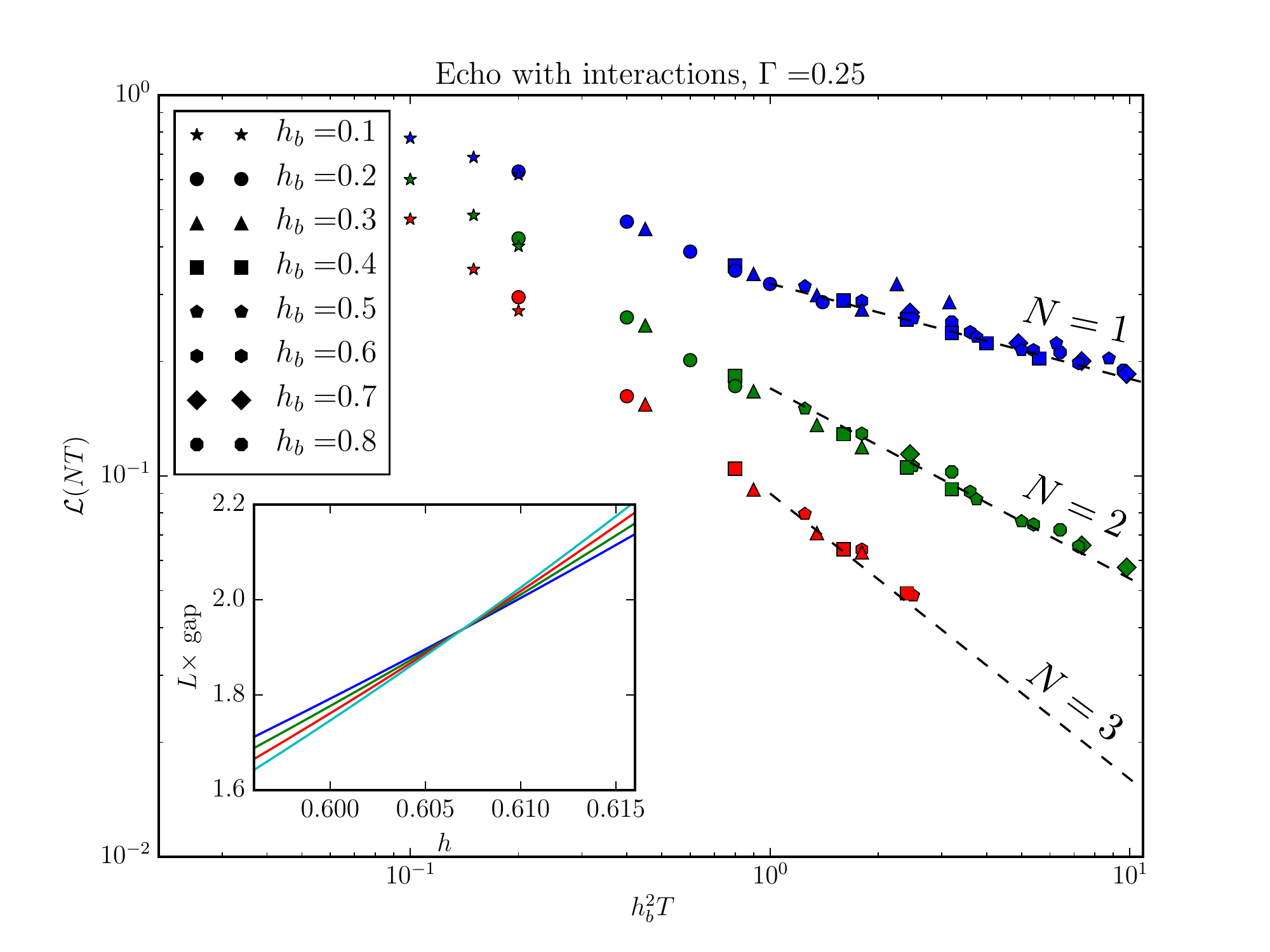}
	\caption{Loschmidt echo as a function of time in the interacting case, with $\Gamma = 0.25$, for a step drive from $h_b = 0$ to $h_b \not = 0$. Arguments in the main text show that the echo should decay as $\mathcal L(NT) \sim T^{-N/4}$ and should be a universal function of $h_b^2 T$, in good agreement with TEBD simulations on $L=400$ sites. Inset: determination of the new critical point with interactions from exact diagonalization on systems of size $L=18,20,22,24$. For $\Gamma = 0.25$, the new critical point is at $h = 0.6066(2)$ with $J=1$. }
	\label{fig:ints}
\end{figure}

\subsection{Structure of the $N$-point correlation functions and Kibble-Zurek Scaling}\label{sec:KZ}

\begin{figure}
	\includegraphics[width = \columnwidth]{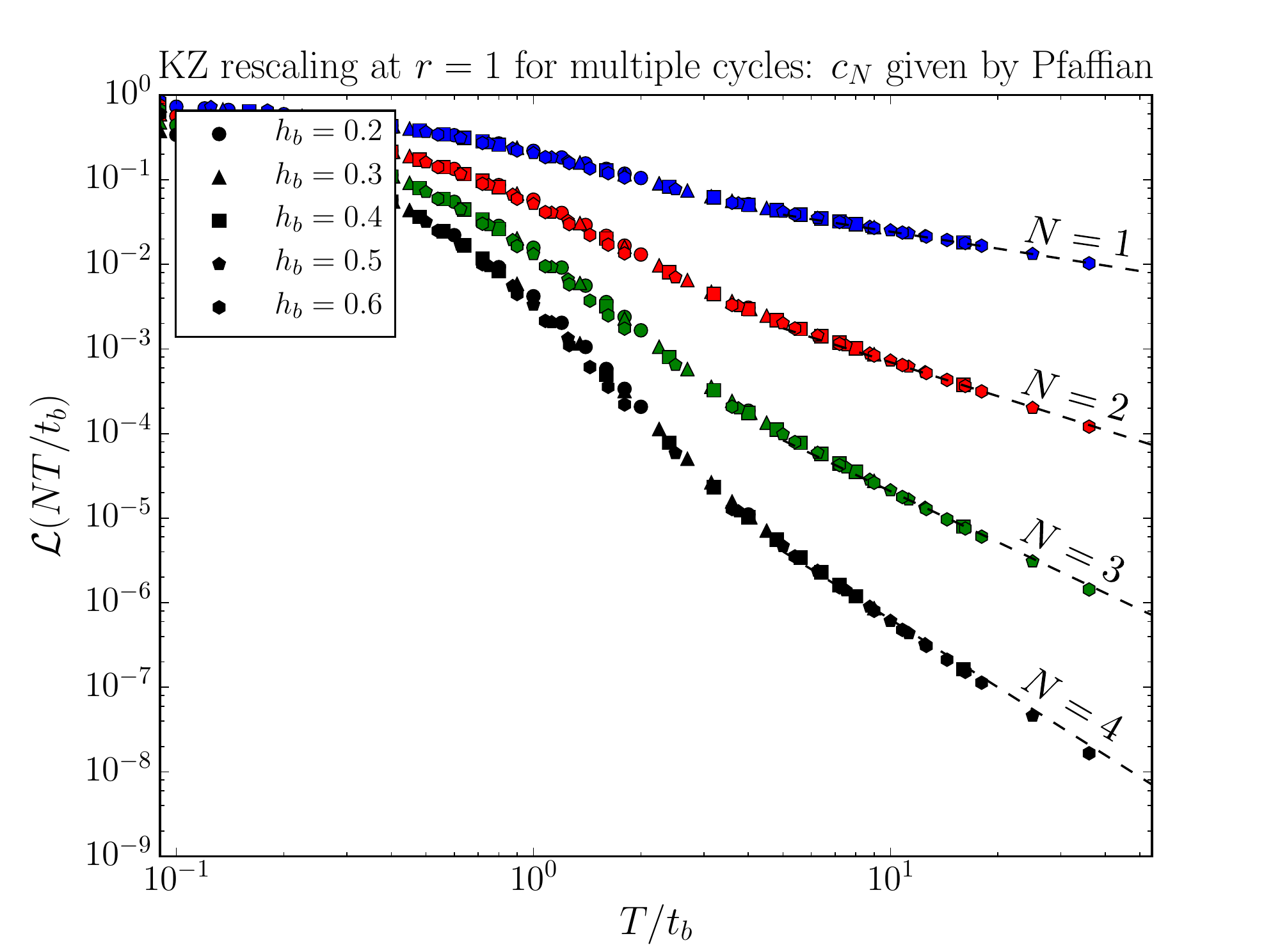}
	\caption{The boundary Kibble-Zurek scaling mechanism does not affect the $N$-point function structure of the Loschmidt echo. Here we consider a triangle drive from minus to plus, so the power law exponent is rescaled by a factor 1/3 due to crossing $h_b = 0$ with $r=1$. The $c_N$ coefficients are still given by a Pfaffian, which is due to the fact that the underlying BCC operator is a chiral fermion with $h_{\rm BCC} = 1/2$.}
	\label{fig:KZcN}
\end{figure}

In the main text we argued that for a step-drive in the low frequency limit, the system is essentially subject to almost independent local quenches, implying a simple exponential decay for the Loschmidt echo (return probability) 

\begin{equation}
{\cal L}(NT) \underset{T \gg t_b}{\sim} \abs{\expectation{\prod_{n=0}^{2N-1}\phi_{{\rm BCC}}(n T/2)}}^2 = c_N  \left(\frac{T}{t_b}\right)^{- \gamma N},
\end{equation}

given by a $2N$-point boundary CFT correlation function. While dependence on the period $T$ is completely fixed by scale invariance with the exponent $\gamma$ being related to the scaling dimension of the operator $\phi_{\rm BCC}$, the behavior of the coefficients $c_N$ is a bit more subtle. In particular, we note that $c_N \neq (c_1)^N$, indicating that the successive local quenches are of course not exactly independent. In fact, the coefficients $c_N/(c_1)^N$ are universal numbers that can be computed using CFT techniques. In the case of a step drive between positive and negative boundary field in the Ising model, this is especially simple since the BCC operator happens to be a fermionic field $\psi$ with dimension $h_{\rm BCC}=\frac{1}{2}$ in the Ising CFT. The $2N$ point function is therefore given by a Pfaffian of an anti-symmetric Toeplitz matrix ${\cal L}(NT) \sim \abs{\expectation{\psi(0) \psi(T/2) \psi (T) \dots }}^2 \sim \left| {\rm Pf}( 1/(t_i-t_j)) \right|^2 $ with $t_i=0,T/2,T, \dots, (N-\frac{1}{2})T$. This is consistent with $\gamma=2$, and yields

\begin{equation}
\frac{c_N}{(c_1)^N} =  \det \begin{pmatrix}
0 & 1 & \frac{1}{2} & \ddots & \frac{1}{2N-1} \\
-1 & 0 & 1 & \frac{1}{2} & \ddots \\
-\frac{1}{2} & -1 & 0 & 1 & \ddots \\
\ddots & \ddots& \ddots& \ddots& \ddots\\
-\frac{1}{2N-1} & \ddots & \ddots& \ddots& 0\\
\end{pmatrix}.
\end{equation}

We emphasize that the normalization with the (non-universal) coefficient of the two-point function $c_1$ is necessary to compare to numerical results. 

For more general CFTs and arbitrary step drives, these coefficients can be computed using the Coulomb Gas formalism~\cite{di_francesco_conformal_2011}, or in some cases using bosonization. As a simple example, we consider a step drive oscillating between 0 and $h_b$ for which the Loschmidt echo is given by the (chiral) $2N$-point function of the spin operator $\sigma$ with conformal weight $h=\frac{1}{16}$ in the Ising CFT. In order to compute this correlation function, one can  ``double'' the Ising CFT to obtain a free boson theory with central charge $c=1=\frac{1}{2}+\frac{1}{2}$ (we already used this trick in Sec.~\ref{sec:numerics} by expressing the $XX$ chain as two independent copies of the Ising model), and compute the (square of the) $2N$-point function of the spin operator as simple free boson correlator~\cite{di_francesco_conformal_2011}. In principle, more complicated multi-point correlators can also be computed using the Coulomb gas formalism~\cite{di_francesco_conformal_2011}.

Remarkably, we note that the boundary Kibble-Zurek (KZ) scaling mechanism at play for more complicated (non-step) drives does not seem to affect the $2N$-point function structure of the Loschmidt echo (Fig.~\ref{fig:KZcN}). In other words, while the BCC scaling dimensions are renormalized by the KZ mechanism as explained in the main text, the universal ratios $c_N/(c_1)^N$ are still given by the CFT expressions above.  

\subsection{RG analysis of the High Frequency Expansion}\label{sec:highfreq}

In this appendix, we detail the high-frequency expansion and renormalization-group (RG) argument described in the main text. We use the so-called Floquet-Magnus (FM) high frequency expansion, a perturbative scheme in the driving period $T$ to compute the Floquet Hamiltonian $H_F[t_0]$ defined from Floquet's theorem by $U(t_0 + T, t_0 ) = \exp(-i H_F[t_0] T)$. For simplicity we will consider a step drive, though as we shall see a simple scaling argument ensures that our results hold in general. Say that we initially prepare a quantum system in the ground state of some Hamiltonian $H_0$, at time $t=0$ quench on a different Hamiltonian $H_1$, and at $t = T/2$ again apply $H_0$. Then the Floquet-Magnus high-frequency expansion, fixing the Floquet gauge $t_0 = 0$, takes a particularly simple form, and is in fact just equivalent to the Baker-Campbell-Hausdorff series:

\begin{equation}
\begin{split}
H_F = &\frac{1}{2}(H_0 + H_1) - \frac{iT}{4} [ H_0, H_1] +\frac{(-iT)^2}{24} ( [H_0,[H_0,H_1]] + [H_1,[H_1,H_0]] ) \\&-\frac{(-iT)^3}{48}[H_1,[H_0,[H_0,[H_1]]]] + \mathcal O (T^4). 
\end{split}
\end{equation}

Applying this expansion to the transverse-field Ising (TFI) spin chain described in the main text is straightforward; however, more insight can be gleaned from applying the FM expansion directly to the Ising conformal field theory (CFT) itself. Let us first apply an ``unfolding'' procedure to the Ising CFT~\cite{Giamarchi:743140}. In order to halve the number of fields, we remap $x$ from the half-line to the whole line. We define a new chiral (say, without loss of generality, right-moving) field $\eta(x)=\eta_R(x)$ for $ x \geq 0$ and  $\eta(x)=\eta_L(-x)$ for $ x < 0$ with anti-commutation relation $\{ \eta(x),\eta(y) \} = \delta(x-y)$. This gives the unfolded Hamiltonian as

\begin{equation}
H = -i v \int_\R dx \ \eta(x) \partial_x \eta(x) - i \lambda(t) \gamma \eta(0) , 
\end{equation}

with $v = \Lambda = 2J$ and $\lambda \propto h_b$ from the lattice model in the main text, and $\gamma = \gamma^\dagger$ an ancilla Majorana fermion with $\gamma^2 = 1$ that anticommutes with all fields. $H_0$ is then the above Hamiltonian with $\lambda = 0$, and $H_1$ with $\lambda \not= 0$ but constant. At first order, then, $[H_0,H_1] = - \lambda a \int_\R dx \ ( \eta \{\partial_x \eta, \eta(0) \} - \{ \eta, \eta(0) \} \partial_x \eta ) = 2 \lambda \gamma \int_{\R} dx \ \delta(x) \eta'   = 2 \lambda \gamma \eta'(0)$. We first note that $[\lambda] = {1-\Delta} = 1/\nu_b = 1/2$ in our case, and $[\eta] = 1/2$ and $[\gamma] = 0$. We can now confirm that the RG dimension of this term is $[\gamma \partial_x \eta] = 3/2 > 1$, so it is RG-irrelevant as claimed in the main text. The second order commutators are likewise 

\begin{align}
[H_0,[H_0,H_1]] &= 4i \lambda \gamma \eta''(0), \\
[H_1,[H_1,H_0]] &= -4 i \lambda^2 \eta(0) \eta'(0) - i \lambda^2 \delta'(0),
\end{align}

where we have used the fact that $\{\eta(x),\eta'(y) \} = \partial_x \delta(x-y)$. These operators have RG dimension $[\gamma \partial_x^2 \eta] = 5/2$ and $[\eta \partial_x \eta] = [\partial_x \delta ] = 2$, so we see that all terms but the lowest order term in the FM high-frequency expansion are irrelevant, getting progressively more irrelevant with higher order in $T$. \\

This is in fact a feature of any high-frequency expansion applied to this field theory. The zeroth order term will always be just the time-averaged Hamiltonian $\frac{1}{T} \int_{0}^{T} dt \ H(t)$ which has RG dimension $1$, so factoring out a $\lambda$, the operator itself has dimension 1/2 and is thus relevant. Now, the $n$th order term, for $n\geq 1$, in any high-frequency expansion will be of the form $\lambda^{m} T^n \hat{\mathcal O}$ with $1 \leq m \leq n$ and $\hat{\mathcal O}$ some operator. Since this term has to have units of energy, its overall RG dimension must be 1, so we find

\begin{equation}
[\hat{\mathcal O}] = 1 - (1 - \Delta) m + n \geq 1 + \Delta n > 1.
\end{equation}

In any unitary CFT, $0 < \Delta < 1$ for any relevant perturbation~\cite{di_francesco_conformal_2011}, and here $\Delta = 1/2$. Thus, any higher order term must always be irrelevant, regardless of the particular drive and particular expansion considered. 

\subsection{Convergence of the HFE and heating in the high-frequency limit}\label{sec:HFEheating}

In this appendix, we address a couple of questions about the boundary-driven CFT, namely: 1) for the integrable model with $\Gamma=0$,
can we prove convergence of the HFE/Magnus expansion for some region of phase space and 2) for the CFT with interactions at $\omega>\Lambda$, is there are regime where the physics is described by the CFT quench results before heating takes over? We will argue that the answer to
both is yes.

First, consider the integrable, non-interacting TFI model with $\Gamma=0$. If $C(t)\equiv\langle\eta(t)\eta(t)^{T}\rangle$
is the Majorana correlation matrix, then we showed above (see Sec.~\ref{sec:numerics}) that 

\begin{equation*}
C(t) = U(t)C(0)U(t)^{T},
\end{equation*}

\begin{equation*}
	U(t) = \begin{cases}
		Q_{1}^{T}D_{1}(t)Q_{1}, & 0<t<T/2\\
		Q_{2}^{T}D_{2}(t-T/2)Q_{2}Q_{1}^{T}D_{1}(T/2)Q_{1}, & T/2<t<T
	\end{cases},
\end{equation*}

\begin{equation*}
	D_{1,2}(t) = \begin{pmatrix}
		\cos(2\epsilon_{1}t) & \sin(2\epsilon_{1}t)\\
		-\sin(2\epsilon_{1}t) & \cos(2\epsilon_{1}t)\\
		&  & \cos(2\epsilon_{2}t) & \sin(2\epsilon_{2}t)\\
		&  & -\sin(2\epsilon_{2}t) & \cos(2\epsilon_{2}t)\\
		&  &  &  & \ddots
	\end{pmatrix},
\end{equation*}
where $Q_{1,2}$ are fixed orthogonal matrices that diagonalize the
single-particle Hamiltonian with (positive) eigenenergies $\epsilon_{j}$.
Note that the single-particle bandwidth is $\Lambda=2\max\epsilon$. Since all
physical properties depend solely on the correlation matrix via Wick's
theorem, we are interested in doing the Magnus expansion on the single-particle
unitary $U(T)=e^{-iH_{F}T}$. To prove convergence, we will use Theorem
9 of Ref.~\citenum{Blanes2009151}, which states that if
$Y^{\prime}=A(t)Y$ with $Y(0)=1$, then the Magnus expansion converges
if $\int_{0}^{T}||A(s)||ds<\pi$ where $||\cdot||$ denotes the 2-norm.
We therefore must massage the above expression into $Y^{\prime}=AY$.
To do so, let us first consider $0<t<T/2$. Then

\begin{align*}
	U^{\prime}(t) &= Q_{1}^{T}D_{1}^{\prime}(t)Q_{1}\\
	&= Q_{1}^{T}D_{1}^{\prime}(t)Q_{1}U^{T}(t)U(t)\\
	&= \underbrace{Q_{1}^{T}D_{1}^{\prime}(t)D_{1}(t)Q_{1}}_{A(t)}U(t),
\end{align*}

\begin{align*}
	D_{1}^{\prime}D_{1} &= \begin{pmatrix}
		-2\epsilon_{1}\sin(2\epsilon_{1}t) & 2\epsilon_{1}\cos(2\epsilon_{1}t)\\
		-2\epsilon_{1}\cos(2\epsilon_{1}t) & -2\epsilon_{1}\sin(2\epsilon_{1}t)\\
		&  & \ddots
	\end{pmatrix} \begin{pmatrix}
		\cos(2\epsilon_{1}t) & \sin(2\epsilon_{1}t)\\
		-\sin(2\epsilon_{1}t) & \cos(2\epsilon_{1}t)\\
		&  & \ddots
	\end{pmatrix} \\
	&= \begin{pmatrix}{ccc}
		-\epsilon_{1}\sin(4\epsilon_{1}t) & \epsilon_{1}\cos(4\epsilon_{1}t)\\
		-\epsilon_{1}\cos(4\epsilon_{1}t) & -\epsilon_{1}\sin(4\epsilon_{1}t)\\
		&  & \ddots
	\end{pmatrix},
\end{align*}

so that

\begin{align*}
	||A||=||D_{1}^{\prime}D_{1}|| &= \max_{i}\epsilon_{i}\left|\sin(4\epsilon_{i}t)\pm\cos(4\epsilon_{i}t)\right|\\
	&\leq \sqrt{2}\max_{i}\epsilon_{i}=\frac{\Lambda}{\sqrt{2}}.
\end{align*}

So convergence is guaranteed if

\begin{align*}
	\int_{0}^{T}||A(s)||ds\leq\frac{T\Lambda}{\sqrt{2}} <  \pi,
\end{align*}

that is, $\Lambda  <  \frac{\omega}{\sqrt{2}}$. Thus we can prove that for frequencies slightly above the single-particle bandwidth, convergence is guaranteed. Note that this bound is not tight in general,
so this is still consistent with the possibility of convergence all
the way down to $\Lambda=\omega$. Note also that, up to prefactors,
similar proofs should hold for all non-interacting CFTs, meaning that
our high-frequency regime is a well-defined dynamical phase.

The second question is whether HFE CFT behavior can be observed before
heating occurs in a generic interacting model. To address this, let
us specifically ask about perturbative heating rates as discussed
in Ref.~\citenum{PhysRevLett.115.256803}. Specifically, consider quenching
a monochromatic drive on an interacting CFT,

\[
H(t)=\underbrace{H_{CFT}(U\neq0)+h_{B}\sigma_{0}^{z}\Theta(t)}_{H_{f}}+h_{B}\sigma_{0}^{z}\Theta(t)\sin(\omega t).
\]

At leading order, the system will absorb energy due to the drive at
a rate 

\[
\frac{dE}{dt}\equiv\frac{d\langle H_{f}\rangle}{dt}\approx2h_{B}^{2}\omega\sigma(\omega);\;\sigma(\omega)=\frac{1}{2}\int_{-\infty}^{\infty}\langle[\sigma_{0}^{z}(t),\sigma_{0}^{z}(0)]\rangle.
\]

Note that this expectation value is intended to be taken in the ground
state of the post-quench Hamiltonian $H_{f}$ which, in the case where
heating is slower than CFT dynamics, should be a valid description of
heating due to drive near the boundary (where the system looks like
it's in the ground state). The question is whether the time scale
of this heating is larger than the time scale $t_{B}\sim h_{B}^{-\nu_{B}}$
for the CFT quench physics to take place. For local drive, Abanin
et al \cite{PhysRevLett.115.256803} bound the susceptibility as

\[
\sigma(\omega)\apprle e^{-C\omega/\Lambda},
\]

where $C$ is a constant of order $1$. An important subtlety is that
this involves integrating over a windows of width $\delta\omega$
which is left out of the final expression. From Fermi's golden
rule, the relevant width should be the inverse density of states at
energy $\omega$ above the ground state, since the argument involves
bounding excitation rates from the ground state to states at energy
$E_{gs}+\omega$. Therefore, we expect this bound could be tightened, but for now we can substitute this result into the heating rate in order to lower-bound the time scale $\tau_{heat}$ for heating processes to occur.

The most relevant energy scale is the single-particle bandwidth, so let us define $\tau_{\rm heat} = \Lambda/ (dE/dt)$. Then 

\begin{align*}
	\frac{\tau_{\rm heat}}{t_{B}} \apprge \frac{\Lambda}{h_{B}^{2-\nu_{B}}\omega}e^{C\omega/\Lambda}.
\end{align*}

For our case, $\nu_{B}=2$ and for sufficiently large $\omega/\Lambda$
heating will occur exponentially slower than CFT dynamics. However,
for models with sufficiently weak boundary perturbations ($\nu_{B}>2$),
this ratio will vanish in the scaling limit and the CFT dynamics becomes irrelevant.
\begin{savequote}[75mm]
`Order and disorder', said the speaker, `they each have their beauty.'
\qauthor{Orson Scott Card, \textit{Speaker for the Dead}}
\end{savequote}

\chapter{Driving and Disorder}
\label{ch:RSRG}

Can a generic driven quantum system avoid the doom of infinite temperature? After all, an infinite temperature state must be completely bland, or so the lore goes; the infinite temperature state is an equal superposition of all eigenstates of the Hamiltonian $\rho \propto \1$, and any symmetry breaking must therefore cancel out. Take the case of the 1-dimensional quantum Ising model. The symmetry operator $\prod_i X_i$ preserves the Hamiltonian, but sends the order parameter $M = (1/N) \sum_i \langle Z_i \rangle \to -M$. Therefore, for every eigenstate with some $ M = a$, there is a partner eigenstate with $ M  = -a$, so an equal superposition of all eigenstates has $M = 0$. There are no magnets nor superconductors at infinite temperature -- no order at all. But is this really true?

In the previous chapter, we considered interacting quantum critical systems driven by an external source at the boundary. Despite the universal signatures we found -- even strongly out of equilibrium! -- the inescapable fate of such systems is heating.\footnote{Strictly speaking, this may not be true in certain limits, as investigated by Wen and collaborators~\cite{wen2018floquet, fan2019emergent}. In their case, a CFT is driven in the bulk, and they observe some ``non-heating'' regimes depending on the drive parameters. Presumably, this is due to integrability, where in CFT, the integrals of motion can be written in terms of the $L_m$ operators~\cite{Bazhanov:1996aa}. Even so, there is a bit of a bone to pick with this result, namely that a spin chain or other quantum system is only described by a CFT in its low-energy limit. Driving the CFT itself and examining heating ignores the contributions of irrelevant terms, which do not change the universal properties but lead definitively to heating. So, the driving considered therein is somewhat academic and unphysical; a generic driven spin chain at criticality will heat up.} This renders our previous results \emph{prethermal} in nature. This is not as bad as it sounds; even though the eventual fate is heating, the timescale for heating can be exponentially long,\footnote{Specifically, as shown in Appendix~\ref{sec:HFEheating}, it is $\tau \sim \Lambda e^{\omega/\Lambda}$ with $\Lambda$ the single-particle bandwidth, which in a system with extensive local interactions scales as $\Lambda \sim \max\{J, U\}$.} which may quickly grow beyond any experimentally reasonable timescale (such as a heating time longer than the age of the universe). In other words, the heating is observable only in principle, and not in practice. Nonetheless, we would still like to know whether true driven phases and phase transitions can exist.\footnote{One promising tactic for engineering interesting driven systems is to allow them to dissipate energy to their environment in a controlled fashion via a thermal bath. This goes by the name of `driven-dissipative systems', both quantum and classical, and has been extensively studied; Ref.~\citenum{PhysRevB.85.184302} and Ref.~\citenum{sieb}, among many others, investigate universality in such systems. The challenge there is that the coupling to the environment generically leads to decoherence, and so although an infinite-temperature state is avoided, delicate quantum coherences are usually absent.}

One escape route from thermalization to infinite temperature is \emph{integrability}. In an integrable system, we have an extensive (infinite) number of conserved quantities, rather than simply energy and momentum. In the late-time description, each conserved quantity keeps the initial value with which it was seeded, leading to more structure than in an infinite temperature state. The steady state is athermal, with each conserved quantity appearing with a conjugate variable in the generalized Gibbs ensemble description~\cite{Vidmar_2016}. While fascinating, the main caveat is that integrable systems are unstable, in the sense that adding a weak integrability-breaking perturbation causes the integrals of motion to be only approximately conserved. But approximately conserved is not the same as exactly conserved: given an infinitely long time, the initial seeds of the quasi-integrals of motion will decay away, and the late-time state will again be an infinite temperature one. So, once we break integrability, we inevitably flow into the basin of the infinite temperature state.\footnote{We are tacitly assuming that the system is driven. If not, then the system need not thermalize to infinite temperature, but it will nonetheless thermalize to a finite-temperature Gibbs ensemble.}

Integrability provides a way to sidestep thermalization, but it is not robust to perturbations. Recently, though, a second route around thermalization was found, namely that of \emph{many body localization}, which relies on strong disorder to generically endow systems with an emergent form of integrability that is, in fact, robust. 

\section{Disorder, Criticality and Thermalization}

In this section, we give a lightning overview of the current and extremely active field of many-body localization (MBL). Again, we skip over detailed explanations and proofs -- and actually, in the MBL field, rigorous arguments are few and far between anyway\footnote{As stated by Anderson in his 1977 Nobel Prize Lecture regarding many-body localization, ``one has to resort to the indignity of numerical simulations to settle even the simplest questions about it.''~\cite{AndersonNobel} Not much has changed in that regard since, but resort many did, to great effect.} -- and simply hit the highlights, giving the intuition behind such results. In general, the field of MBL relies heavily on numerical evidence, as the twin complexities of disorder and interactions leave few analytic tools at our disposal.\footnote{One notable exception is the real-space renormalization group, covered in the second part of this introduction.} As the field is still rapidly evolving, there are no firm, established textbooks on the subject nor canonical ways to learn the material. Instead, I would direct the reader to the excellent review article of Nandkishore and Huse~\cite{doi:10.1146/annurev-conmatphys-031214-014726}, and to that of Abanin, Altman, Bloch and Serbyn~\cite{2018arXiv180411065A} for more recent developments. 

\subsection{Many-body localization}

Many-body localization (MBL) is the many-body generalization of the celebrated \emph{Anderson localization}~\cite{PhysRev.109.1492, PhysRevLett.42.673}, which goes as follows. A single particle wavefunction in a translation-invariant system is generally \emph{extended}, namely some form of a plane wave that has weight across the entire system. For instance, in a crystal, the wavefunction is a Bloch wave, which is a simple plane wave times a periodic function. In the presence of disorder, however, the wavefunction \emph{localizes}, turning from an oscillatory function to one which is sharply peaked around a particular point in space. This leads to a total absence of diffusion, as originally noted by Anderson. Rather than $\langle r(t) \rangle \sim D \sqrt t$, as expected on general grounds, we have $\langle r(t) \rangle \sim \text{constant}$ -- wavepackets do not spread. How much disorder is required for this phenomenon to manifest depends on the dimension: for three dimensional systems, there is a critical value of disorder $W_c$ above which we have localization, while in two- and one-dimensional systems, \emph{any} amount of disorder leads to Anderson localization. Incidentally, this means that Anderson-localized systems do not thermalize, since local injections of energy do not diffuse throughout the system. 

We might wonder what happens when we introduce interactions into an Anderson-localized system. At first blush, they seem like they should destroy the localization. When we turn on a small interaction term, at lowest order in perturbation theory we see hybridization between single-particle states of similar energy. Now, there is no relation between energy and localization center $r$ in an Anderson-localized state in general; two states of similar energy may well correspond to localization centers that are far apart. Turning up the interactions further, we hybridize more and more levels, all of which are roughly randomly distributed in space across the sample. This seemingly would make new eigenstates that are extended, rather than localized, and allow for interaction-mediated diffusion processes. Thus Anderson localization appears unstable, and perhaps just a quirk of single-particle quantum mechanics.\footnote{Even if this had been the case, Anderson localization would have still been quite physical, as many non-quantum waves (such as classical light waves and water waves) Anderson localize. Further, photons are non-interacting, so would Anderson localize in a disordered medium.}

It was therefore shocking when Basko, Aleiner and Altschuler~\cite{BASKO20061126} (BAA) showed, from an extended perturbation theory argument,\footnote{Pun intended.} that localization survives the introduction of interactions.\footnote{The history here is a little more complicated than I am implying; the possibility of MBL at weak interactions was raised by Anderson in his original article~\cite{PhysRev.109.1492}, with the argument that any transport process must conserve energy and hence has to be virtual (i.e. a particle hops to a new energy level but then hops back), leading to no conductivity even with interactions. Fleischman and Anderson showed that localization was robust to lowest order in perturbation theory in 1980~\cite{PhysRevB.21.2366}, and in 1997 Altschuler and collaborators~\cite{PhysRevLett.78.2803} showed that it was stable to all orders in a quantum dot. BAA was nonetheless the breakthrough paper that kicked off MBL as a sub-field, though.} Specifically, they showed that there was a metal-insulator transition in such systems at a critical temperature $T_c$, and showed that in the insulating phase, the probability of an electron escaping vanished to all orders in Feynman diagrams. Interestingly, BAA's arguments are agnostic as to dimension. However, subsequent works have only rigorously proven MBL for one-dimensional systems~\cite{PhysRevLett.117.027201}, and the proof is known to break down for other dimensions~\cite{doi:10.1098/rsta.2016.0422}.\footnote{This has proven quite contentious in recent years, as Imbrie's proof relies on the assumption of `limited level attraction'. We generally expect level \emph{repulsion} in quantum systems as we turn on interactions, also called \emph{avoided crossings}. There is ample physical evidence for limited level attraction in generic systems, but it has not been proven for a particular model to date.} In fact, there is a non-perturbative argument, due to de Roeck and Huveneers\cite{2017arXiv170609338T,PhysRevB.95.155129}, that many-body localization does not exist in dimension greater than one. Called the `avalanche argument',\footnote{As David Huse has remarked, it's a bit of a misnomer, as the `avalanches' are ridiculously slowly moving and move even slower as they grow larger.} one assumes that there is a small ergodic bubble embedded in an insulating bulk, and asks whether the bubble grows. In $d=1$, such ergodic grains are stable, but for $d \geq 2$, the bubbles appear to grow without bound, presumably eventually thermalizing the system.\footnote{These bubbles are nearly certain to exist in a large enough system, since if the disorder is random, there will be some probability of getting a sizable region of low disorder (called a `rare region') that tends to 1 when $L\to \infty$.} The argument is not a rigorous proof, though, and the existence of MBL in $d \geq 2$ remains a topic of debate.

The canonical picture of the MBL phase is that of an emergent type of integrability, where the disorder leads to the emergence of extensively many conserved quantities. The integrals of motion are referred to as $\ell$-bits~\cite{PhysRevLett.111.127201, PhysRevB.90.174202}, with $\ell$ meaning localized. More specifically, the Hamiltonian can be written as a diagonal operator in the set of $\ell$-bits $\{\tau\}$, a set of Pauli operators, which (for a spin chain) are a local superposition of the actual bits $\{\sigma \}$. Concretely, there exists a finite-depth (local) unitary transformation $U$ such that

\begin{equation}
\label{eq:lbits}
U H U^\dagger = H_{\text{MBL}} = \sum_i h_i \tau_i^z + \sum_{ij} J_{ij} \tau_i^z \tau_j^z + \ldots . 
\end{equation}

The existence of $\ell$-bits is an ansatz for general MBL systems, but they were constructed explicitly in Imbrie's one-dimensional proof~\cite{PhysRevLett.117.027201}. Each $\tau_i^z$ commutes with $H$ and is hence a conserved operator, and there are extensively many of them. In this basis, there is no entanglement, and eigenstates are just products of up and down states $\ket{\up \up \down \ldots}$ with respect to the $\tau$'s. In contrast to the usual notion of integrable quantum systems, such as the Heisenberg chain or the $XXZ$ model where the integrals of motion are non-local or quasi-local, in MBL the $\ell$-bits are strictly (exponentially) local. In terms of the original spins, the $\tau$ operators have weight exponentially decaying with distance (an `exponential tail'). 

The $\ell$-bit picture explains much of the phenomenology of the MBL phase~\cite{PhysRevB.90.174202}. In the MBL phase, \emph{all} eigenstates are like ground states, in the sense that the entanglement entropy scales with the area of the system rather than the volume: $S_L = \Tr \rho_L\log \rho_L \sim \text{constant}$, for a one-dimensional system. This is quite remarkable, as in generic systems, we have excited state entanglement entropy that is volume law, $S_L \sim L$.\footnote{This is a challenge for numerical methods based on entanglement, such as the density-matrix renormalization group or DMRG. DMRG has been quite successful in simulating ground states, but not excited states, which makes time-dependent simulations quite limited; in an MBL system, we can hope to explore to much longer lengths and later times.} We can see this from the $\ell$-bit picture by noting that each eigenstate is a product state of $\ell$-bits, which are themselves local superpositions of the physical bits. So, in the physical bit frame, there can only be the entanglement generated by a finite-depth unitary transformation, which is a constant that does not scale with the volume of the system. The entanglement entropy is non-zero (eigenstates are not product states in the physical basis), but it is area law. 

A second characteristic of the MBL phase is the growth of entanglement following a quantum quench, which is not $S_L(t) \sim t$ as one expects in a generic quantum system, but rather $S_L(t) \sim \log t$.~\cite{PhysRevB.77.064426, PhysRevLett.109.017202} Since the $\ell$-bits have exponential tails in the physical basis, interactions between far away physical spins are exponentially suppressed with distance. This means that the light cone of excitations spreads in a logarithmic fashion. Since the $\ell$-bits only couple through their $z$-components, they can only get entangled through these exponentially decaying interactions, implying that entanglement entropy can grow no faster than logarithmically. 

The final canonical signature of MBL is that of \emph{Poisson level statistics}. Considering two consecutive energy levels, one can form the quantities $\delta_n = E_{n+1} - E_{n}$ and the resultant $r$\emph{-ratio} $$r_n = \frac{ \min\{\delta_n, \delta_{n+1}\} }{ \max\{\delta_n, \delta_{n+1}\} }.$$ We are usually interested in the mean $r$-ratio, $\overline{r}$, which is obtained by first averaging over some window size of levels in a particular sample to form $r_S$, then averaging that over disorder realizations to form $\overline{r}$. In a generic quantum system, i.e. one satisfying the Eigenstate Thermalization Hypothesis, the $r$-ratio would be distributed according to a Gaussian Orthogonal Ensemble (GOE)\footnote{The three classical ensembles of random matrix theory relevant for Hamiltonians (also called the Wigner-Dyson distributions) are: (1) the Gaussian Orthogonal Ensemble (GOE) for hermitian operators with real entries, (2) the Gaussian Unitary Ensemble (GUE) for hermitian operators with complex entries, and (3) the Gaussian Symplectic Ensemble (GSE) for hermitian operators with quaternionic entries. We assume all entries are independent and identically distributed (IID) random variables.} with $\overline{r} = 4-2\sqrt{3} \approx 0.536$. In an integrable system such as MBL, however, it is distributed according to a Poisson distribution~\cite{doi:10.1098/rspa.1977.0140}, with $\overline{r} = 2\ln 2 - 1 \approx 0.386$.\footnote{For members of the GUE we have $\overline{r} = 2\sqrt{3} / \pi - 1/2 \approx 0.603$, and for the GSE we have $\overline{r} = 32\sqrt{3} / 15\pi - 1/2 \approx 0.676$. The $r$-ratio is a powerful and numerically simple way to distinguish between different classes of ergodic behavior; for a catalog of these values and their derivations, see Ref.~\citenum{PhysRevLett.110.084101}.} This reflects the deep fact that in general quantum levels repel, but in an MBL system, they behave as if they had no knowledge of each other. 

Many-body-localized systems have the remarkable potential to host quantum orders \emph{in their excited states}. This is in stark contrast to most condensed matter systems, which are only e.g. magnetically ordered in the ground state, and is due to the special structure of the excited states just discussed. For instance, the entire spectrum could display ferromagnetic order, or more appropriately spin-glass order due to the disorder in the couplings. Then, even at infinite temperature, an MBL system could retain its ordering and still display phase structure! This is a true escape from the usual bland infinite temperature state, and this phenomenon goes by the name of \emph{eigenstate order} or \emph{localization protected quantum order}~\cite{PhysRevB.88.014206}.\footnote{This has also been shown to hold for symmetry-protected topological order in Ref.~\citenum{PhysRevB.89.144201} and Ref.~\citenum{BahriMBLSPT}.} The onset of an eigenstate order happens \emph{simultaneously} across the entire spectrum. This makes criticality in the MBL setting quite different from conventional criticality, which concerns the onset of order in the ground state only. As we will see, these are generally a special kind of critical point called an infinite randomness fixed point (IRFP), with every eigenstate displaying the properties of an IRFP at the transition. 

Finally, we may ask whether or not MBL is stable to periodic driving. This was indeed shown to be true for fast enough drives~\cite{PhysRevLett.115.030402, PhysRevLett.114.140401, PhysRevLett.114.140401}. Essentially, if the drive is very high energy, it cannot excite the many-body resonances associated with flipping an $\ell$-bit. The Floquet Hamiltonian can be constructed using the Floquet-Magnus expansion, and shown to itself be a fully-MBL Hamiltonian. The $\ell$-bit picture survives, and much of the same phenomenology goes through to the driven case. In particular, driven MBL systems can also host quantum order, leading to the exciting discovery of intrinsically dynamical, driven phases~\cite{khemani_prl_2016}. Foremost among these are `time-crystalline' phases,\footnote{A better name for these phases is `discrete time crystals' or `time-density waves', in analogy to charge density wave states. This is certainly a weaker notion of time-crystal behavior than was originally developed by Wilczek~\cite{PhysRevLett.109.160401}, which proposed that \emph{continuous} time-translation symmetry could be spontaneously broken. After several criticisms~\cite{PhysRevLett.110.118901, PhysRevLett.111.070402}, Wilczek's proposal was proven impossible in ground states by Watanabe and Oshikawa~\cite{PhysRevLett.114.251603}. These discrete time crystals, however, break a discrete time translation symmetry in their entire spectrum, sidestepping all of these arguments.} which exhibit spontaneous symmetry breaking of the discrete time translation symmetry, and form the subject of much of this chapter. 

\subsection{The real-space renormalization group}

In many body localization, analytical results are few and far between.\footnote{One might say they form a `rare region' in the space of all papers.} This is due to the inherent difficulty of disorder and interactions. Disordered systems lack any kind of symmetry, such as the conformal symmetry of the previous chapter, which greatly hurts their prospects for analytic solubility. Occasionally, though, we get lucky, and disorder can help us: we can take there to be so much disorder that, on long length scales, we essentially have an ensemble of individually disordered systems over which we can average. This is the idea behind a technique known as the `real space renormalization group', or RSRG. Originally developed by Dasgupta and Ma~\cite{PhysRevB.22.1305} to study the ground state of the disordered Heisenberg chain, the RSRG was greatly expanded by seminal work of Daniel Fisher in the 1990s\cite{PhysRevLett.69.534, PhysRevB.50.3799, PhysRevB.51.6411}, and has been extended to access the entire spectrum of many-body-localized systems (`RSRG-X'~\cite{PhysRevX.4.011052}). The purpose of this chapter is to extend the technique to driven (Floquet) systems, which is the author's main contribution to the field. 

The starting point for the RSRG is the assumption of strong disorder. Consider the probability distribution of the couplings throughout a disordered spin chain; for concreteness, let us consider the disordered Ising model in the Majorana language,

\begin{equation}
\label{eq:tfiMajoranaMBL}
H = i \sum_{n} J_n \gamma_{2n} \gamma_{2n+1},
\end{equation}

and consider the probability distribution $P(J)$. These Majorana operators are fermionic $\{\gamma_n, \gamma_m\} = 2 \delta_{nm}$ and are their own hermitian conjugates, $\gamma_n^\dagger = \gamma_n$.\footnote{Interestingly, shortly after discovering his eponymous fermions, in 1938 Ettore Majorana -- called by Fermi as `one of the geniuses, like Galilei and Newton' -- sent a farewell note to his colleagues and mysteriously vanished. His disappearance has fueled much speculation over the years, from proposals of his suicide, kidnapping, becoming a beggar or even becoming a monk. In 2015, the Rome Attorney's office found sufficient evidence to conclude that Majorana had voluntarily emigrated to Venezuela and lived there until at least 1955-9, closing the case on his disappearance~\cite{majoranaFate}.} 

We begin with an assumption of \emph{strong disorder}, namely that the distribution $P(J)$ is long-tailed. While we cannot justify it now, we will see that it becomes more and more justified as the RG proceeds. This long-tailed assumption means that if we pick the strongest bond in the chain $\Omega = \max\{J_n\}$, that its nearest neighbors $J_L$ and $J_R$ will be much weaker than it: $J_L, J_R \ll \Omega$. So, we can do perturbation theory locally, treating $J_L$ and $J_R$ as perturbative with respect to the strong piece of the Hamiltonian, $\Omega$.\footnote{Strictly speaking, the entire Hamiltonian except for the $\Omega$ term is the perturbation, but by truncating our perturbation theory to second order, we only need to consider $J_L$ and $J_R$.} 

The canonical method of doing perturbation theory at the operator level is to use a \emph{Schrieffer-Wolff transformation}~\cite{PhysRev.149.491}. The idea is to first divide the Hamiltonian operator $H$ into a strong piece $H_0$ and a weak piece $V$ as $H=H_0+V$, with $|H_0| \gg |V|$, as is usual in perturbation theory. We then seek a unitary rotation $U = e^{iS}$, with $S$ a hermitian operator, that diagonalizes $H$ with respect to $H_0$. That is, we have $[U H U^\dagger, H_0] = 0$. Formally, we expand this in series in $S$, and truncate at a particular order in $V$. For our purposes, we expand to second order, seeking $S = S_0 + S_1 + S_2$ such that $[e^{iS} H e^{-iS}, H_0] = \mathcal O(V^2)$. We then project onto the lowest eigenstate of the strong piece $H_0$, generating a virtual coupling between the adjacent sites and forming a new Hamiltonian $\tilde H$. 

Returning to the above setup, we find that, if the strong piece is $H_0 = i \Omega \gamma_{n} \gamma_{n+1}$, then after projecting onto the $i \gamma_{n} \gamma_{n+1} = -1$ subspace, the virtual coupling generated between Majoranas $\gamma_{n-1}$ and $\gamma_{n+2}$ is $$\tilde J = \frac{J_L J_R}{\Omega} \quad \Leftrightarrow \quad \log \tilde J = \log J_L + \log J_R - \log \Omega.$$ One of the great insights of Fisher was to cast this \emph{renormalization rule} in terms of logarithmic variables, since, as we shall see, in these variables the problem becomes exactly solvable. 

With our renormalization rule in hand, we can run the RG. Define the RG scale $\Gamma$ in terms of the initial strongest bond $\Omega_0$ and the strongest bond at the current scale $\Omega$ as $\Gamma = \log \Omega_0 / \Omega$. This RG scale starts at 0 and is always monotonically increasing (since successive `strongest bonds' $\Omega$ are always smaller). We consider the probability distribution of couplings at a given scale $P^\Gamma_J(J)$, and ask how it is changing under the flow. 

At this point it is convenient to define separate quantities for the even and odd sub-lattice couplings, namely $J_i = J_{2n}$ and $h_i = J_{2n+1}$. (These are the bonds and fields of the original, spin-language Ising chain.) In logarithmic variables, we can define $\zeta = \log(\Omega / J)$ and $\beta = \log(\Omega / h)$, each with their own distributions $P_\zeta$ and $P_\beta$. The renormalization rule splits into two (and slightly simplifies) as $\tilde{\beta} = \zeta_L + \zeta_R$, $\tilde{\zeta} = \beta_L + \beta_R$, with $\log \Omega$ dropping out. This ultimately leads to the coupled integro-differential equations

\begin{align*}
\pd{P_\zeta}{\Gamma} &= \pd{P_\zeta}{\zeta} + P_\beta(0) \int_{0}^{\infty} d\zeta' P_\zeta(\zeta') P_\zeta(\zeta-\zeta') + P_\zeta(\zeta) \left[P_\zeta(0) - P_\beta(0)\right] \\
\pd{P_\beta}{\Gamma} &= \pd{P_\beta}{\beta} + P_\zeta(0) \int_{0}^{\infty} d\beta' P_\beta(\beta') P_\beta(\beta-\beta') + P_\beta(\beta) \left[P_\beta(0) - P_\zeta(0)\right] .
\end{align*}

These look extremely nasty, but miraculously, they are exactly solvable. The fixed point solution is the distribution

$$
P_*(\chi \equiv \log x) = \lim_{\Gamma \to \infty} \frac{1}{\Gamma} e^{-\frac{\chi}{ \Gamma}} \quad \Leftrightarrow \quad P_*(x) = \lim_{\Gamma \to \infty} \frac{1}{\Omega_0 \Gamma} \left(\frac{\Omega_0}{x}\right)^{1 - 1/\Gamma}.
$$

This is known as the \emph{infinite randomness fixed point} (IRFP) distribution, and it is a universal attractor. For the Ising chain above, both $P_\beta$ and $P_\zeta$ flow to this distribution at criticality, where criticality is given by $\overline \zeta = \overline \beta$ (by symmetry), or $\overline{\log J} = \overline{\log h}$, with overline meaning disorder average. The IRFP also underlies the random singlet phase of the disordered Heisenberg chain, where the ground state looks like a collection of singlets of various ranges. A remarkable property of this fixed point is that, as we flow towards it, our distribution $P^\Gamma_{\beta, \zeta}$ becomes longer- and longer-tailed, terminating in the longest possible tailed distribution of $1/x$ (which is not a true distribution as it is not normalizable). This justifies our initial assumption of strong disorder \emph{post hoc}, and we find that any amount of disorder flows to an `infinite' amount of disorder under the RG -- hence, infinite randomness. 

The RSRG is an extremely powerful technique, and allows for the exact calculation of critical exponents, correlation functions, and other quantities. Among the most important is, as with clean (conformal) critical points, the scaling of the entanglement entropy. Seminal work of Refael and Moore~\cite{PhysRevLett.93.260602} found that, at an infinite randomness fixed point, the entanglement entropy $S_L = \Tr \rho_L \log \rho_L$ scales as 

\begin{equation}
S_L \sim \frac{\tilde c}{6} \log L
\end{equation}

for open boundary conditions (and twice this for periodic boundary conditions). This is remarkably close to the CFT result, and the quantity $\tilde c$ is known as the \emph{disordered central charge}. For several canonical systems, $\tilde c$ is related to the clean central charge $c$ by a simple factor of $\ln 2$; for the Ising model, we have $\tilde c = \ln 2 / 2$, and for the disordered Heisenberg chain we have $\tilde c = \ln 2$. Here, though, the similarities end. IRFPs are $z=\infty$ theories,\footnote{More precisely, we have excitation energy (i.e. singlet energy) scaling with length as $E \sim e^{-\sqrt{\ell}}$, in contrast to excitation energy $E\sim \ell^{-z}$ at usual quantum critical points. Off criticality, we have $E \sim \ell^{-1/2|\delta|}$, where $\delta$ is the detuning $\delta = (1/\sigma_J \sigma_h) (\overline{\log h} - \overline{\log J})$, with $\sigma$ the standard deviation. This slow scaling is characteristic of the \emph{Griffiths phase} surrounding the IRFP, and is due to the effect of rare regions of weak and strong disorder. This leads to stark differences between mean ($\overline{\mathcal{O}}$) and typical ($e^{\overline{\log \mathcal O}}$) quantities.} in stark contrast to the $z=1$ of a CFT, and feature logarithmically slow dynamics, among many other differences. 

There are many kinds of IRFP\footnote{Much recent work has attempted to deduce the nature of the transition into the MBL phase from the ergodic (ETH) phase as a function of disorder strength, postulating that it is an IRFP of some kind and building an RSRG procedure to tackle it. These procedures are much less rigorous than the ones shown here, as they start with an already coarse grained model (rather than a microscopic model), based on certain physical assumptions of ergodic regions coupled to insulating regions, to seed the RSRG.} resulting from various RSRG procedures, and like with any RG, we must perform it on a specific model, case-by-case, to find its universality class.\footnote{That said, the universality class is generally determined by the renormalization rule, and the characteristic of the Ising class is the rule $\tilde J = J_L J_R / \Omega$. We generally expect any system with this rule, even in new variables, to be in the Ising class, for example.} The main adjustment of this procedure to tackle MBL systems was to allow for projections into excited states (hence `RSRG-X'), such as $i \Omega \gamma_{n} \gamma_{n+1} = +1$. Now, a limitation of the RSRG is its rather extreme locality, and it will inevitably miss long-ranged resonances that could lead to thermalization. For instance, there may be a second-order process associated to flipping spin 0 and spin 100 simultaneously that is low-energy, but due to the large separation between these spins, it will not be captured by the RSRG-X. (This is not an issue for the ground state, but only for the excited states in the middle of the spectrum.) Thus, we must simply assume that these processes are not important, in the sense that they do not lead to thermalization. Numerical evidence is therefore needed to support any RSRG-X procedure.

The final warning we should issue before delving into the RSRG generalized to periodically driven systems is the stability of IRFPs under periodic driving. While Floquet MBL is on a solid footing due to the $\ell$-bit picture, and is robust to rapid enough driving, IRFPs are inherently more long-ranged. At a transition between different ordered MBL phases, such as between a paramagnet and a spin glass, the $\ell$-bits become algebraically delocalized (rather than exponentially localized), and the stability under periodic driving is no longer known. It is therefore somewhat of a philosophical question whether these IRFPs thermalize under infinitely long periodic driving. This question is unable to be settled with numerics nor analytics at the time of this writing, and we have only intuitions to guide us. 

\section{Floquet Quantum Criticality}

\subsection{Introduction}

The assignment of robust phase structure to periodically driven quantum many-body systems is among the most striking results in the study of non-equilibrium dynamics~\cite{khemani_prl_2016}. There has been dramatic  progress in understanding such `Floquet' systems, ranging from proposals to  engineer new states of matter via the 
\\ \noindent drive~\cite{RevModPhys.89.011004,doi:10.1080/00018732.2015.1055918,PhysRevLett.116.205301,0953-4075-49-1-013001,PhysRevX.4.031027,PhysRevB.82.235114,Lindner:2011aa,PhysRevX.3.031005,1367-2630-17-12-125014,PhysRevB.79.081406,Gorg:2018aa} to the classification of driven analogs of symmetry-protected topological  phases (`Floquet SPTs')~\cite{PhysRevB.92.125107,PhysRevB.93.245145,PhysRevB.93.245146,PhysRevB.93.201103,potter_classification_2016,PhysRevB.94.125105,PhysRevB.94.214203,PhysRevB.95.195128}. These typically require that the  system under investigation possess one or more microscopic global symmetries. In addition, {\it all} Floquet systems share an invariance under time translations by an integer multiple of their drive period. Unlike the continuous time translational symmetry characteristic of undriven Hamiltonian systems~\cite{PhysRevLett.109.160401,PhysRevLett.111.070402,PhysRevLett.114.251603}, this discrete symmetry may be spontaneously broken, 
leading to a distinctive dynamical response at rational fractions of the drive period --- a phenomenon dubbed `
time crystallinity'~\cite{pi-spin-glass,else_floquet_2016,PhysRevX.7.011026,PhysRevLett.118.030401,PhysRevB.96.115127,PhysRevLett.119.010602}. The time translation symmetry breaking (TTSB) exhibited by Floquet time crystals is stable against perturbations that preserve the periodicity of the drive, permitting generalizations of notions such as broken symmetry and phase rigidity to the temporal setting. Experiments have begun to probe these predictions in well-isolated systems such as ultracold gases, ion traps~\cite{Zhang:2017aa}, nitrogen-vacancy centers in diamond~\cite{Choi:2017aa}, and even spatially ordered crystals~\cite{PhysRevB.97.184301,PhysRevLett.120.180603}. 

In light of these developments, 
it is desirable to construct a theory of Floquet (multi-)critical points between distinct Floquet phases. Ideally, this should emerge as the fixed point of a coarse-graining/renormalization group procedure, enable us to identify critical  degrees of freedom, especially those responsible for TTSB, and allow us to compute the critical scaling behavior.

Here, we  develop such a theory for a prototypical Floquet system: the  driven random quantum Ising chain.  Extensive analysis has shown that this model hosts four phases~\cite{khemani_prl_2016,pi-spin-glass}. Two of these, the paramagnet (PM) and the spin glass (SG), are present already in the static problem~\cite{PhysRevB.88.014206,PhysRevLett.112.217204,PhysRevX.4.011052}. A third, the $\pi$ spin glass/
time crystal,  has spatiotemporal long-range order and subharmonic bulk response at half-integer multiples of the drive frequency. This phase, and its Ising dual --- the $0\pi$ paramagnet, which also exhibits TTSB but only at the boundaries of a finite sample --- are unique to the driven setting. A precise understanding of the (multi)critical points  between these distinct Floquet phases accessed by tuning drive parameters is the subject of this work.

Our approach relies on the presence of quenched disorder, required for a generic {periodically-driven} system to have Floquet phase structure rather than thermalize to a featureless infinite-temperature state~\cite{PONTE2015196,PhysRevLett.114.140401,PhysRevLett.115.030402,ABANIN20161}. 
We argue that transitions between distinct one-dimensional Floquet phases are then best described in terms of an infinite-randomness fixed point accessed via a strong-disorder real space renormalization group procedure. In the non-equilibrium setting, the stability of infinite-randomness fixed points against thermalization via long-range resonances remains a topic of debate~\cite{PhysRevB.95.155129,PhysRevLett.119.150602,Ponte20160428}. However, even if unstable, we expect that they will control the dynamics of prethermalization relevant to all reasonably accessible experimental timescales~\cite{PhysRevLett.115.256803,kuwahara_floquetmagnus_2016}.

The universality of our analysis turns on the fact that, in the vicinity of such infinite-randomness critical points, a typical configuration of the system can be viewed as being composed of domains deep in one of two proximate phases~\cite{PhysRevLett.69.534,PhysRevB.50.3799,PhysRevB.51.6411,PhysRevLett.89.277203,PhysRevB.61.1160,PhysRevX.5.031032}. Transitions that do {\it not}  involve TTSB  (i.e., the SG/PM or $0\pi$PM/$\pi$SG transitions) map 
to the static (random) Ising critical point and can be understood in similar terms. 
In contrast, transitions that involve the onset of TTSB in the bulk (PM to $\pi$SG) or at the boundary (SG to $0\pi$PM) can be understood in terms of 
a new class of domain wall special to driven systems, that 
separate regions  driven at a frequency primarily near  0 or near $\pi$ --- a picture we verify numerically. When the Ising model is rewritten as a fermion problem, this picture yields a simple description of Floquet criticality in terms of domain walls that bind Majorana states at quasienergy $0$ or $\pi$, allowing us to further study the multicritical point where all four phases meet.

\subsection{Model} 

\begin{figure}
\includegraphics[width = \columnwidth]{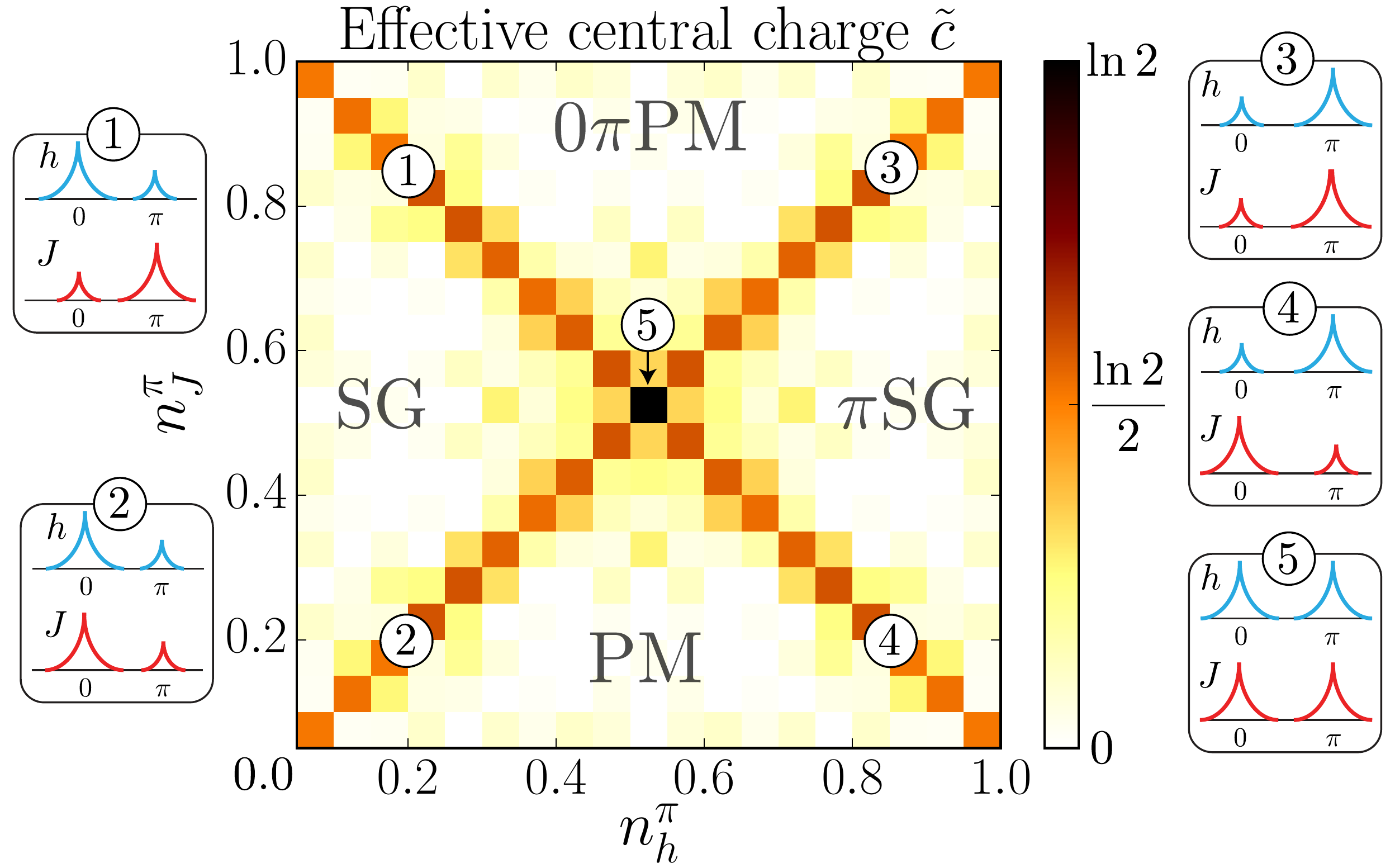}
\caption{\label{fig:phase_diagram} {Phase diagram deduced by fitting ``effective central charge'' from entanglement scaling (see Fig. 3 for details). Insets: sketches of infinite-randomness coupling distributions along the critical lines (1-4) and at the multicritical point (5).}} 
\end{figure}

Floquet systems are defined by a time-periodic Hamiltonian $H(t) = H(t+T)$.  For reasons similar to Bloch's theorem, eigenstates satisfy
$\ket{\psi_\alpha (t)} = e^{-i E_\alpha t} \ket{\phi_\alpha(t)}$, where $\ket{\phi_\alpha(t+T)} = \ket{\phi_\alpha(t)}$ and we set $\hbar =1$~\cite{PhysRev.138.B979,PhysRevA.7.2203}. In contrast to the case of static Hamiltonians, the quasi-energies $E_\alpha$ are only defined modulo $2\pi/T$, voiding the notion of a `ground state'.

 An object of fundamental interest is the single-period evolution operator or Floquet operator 
$F \equiv U(T)$.
If disorder is strong enough,  $F$ can have an extensive set of local conserved quantities. 
 This implies area-law scaling of entanglement in Floquet eigenstates, and consequently the absence of thermalization~\cite{doi:10.1146/annurev-conmatphys-031214-014726}.

Unlike  generic (thermalizing) Floquet systems, such many-body localized (MBL) Floquet systems retain a notion of phase structure to infinitely long times. For concreteness, we focus on the driven quantum Ising chain, the simplest Floquet system that hosts uniquely dynamical phases. The corresponding Floquet operator is
\begin{equation}
F = {\rm e}^{-i \frac{T}{2} \sum_i J_i  \sigma_i^z \sigma_{i+1}^z + U \sigma^z_i \sigma^z_{i+2} } {\rm e}^{-i \frac{T}{2} \sum_i h_i \sigma_i^x + U \sigma^x_i \sigma^x_{i+1} },\!\!\! 
\label{eqModel}
\end{equation} 
where $\sigma^\alpha_i$ are Pauli operators. Here $J_i$ and $h_i$ are uncorrelated  random variables, 
and $U$ corresponds to small interaction terms that respect the ${\mathbb Z}_2$ symmetry of the model generated by $G_{\text{Ising}} = \prod_i \sigma_i^x$. {For specificity, 
we  draw couplings $h,J$ randomly with probability $n_\pi^{h,J}$ from a box distribution of maximal width about $\pi$, namely $[\pi/2,3\pi/2]$, and with probability $n_0^{h,J} = 1-n_\pi^{h,J}$ from a box distribution of maximal width about $0$, namely $[-\pi/2,\pi/2]$. The reasons for this parametrization will become evident below.}
$F$ corresponds to an interacting transverse-field Ising model where for $U=0$  we stroboscopically alternate between field and bond terms. It is helpful to perform a Jordan-Wigner transformation to map bond and field terms to Majorana fermion hopping terms, yielding a $p$-wave free fermion superconductor with density-density interactions given by $U$. In the high-frequency limit $T \to 0$, we can rewrite $F =  e^{-i H_FT}$ by expanding and re-exponentiating order-by order in $T$ 
and the Floquet Hamiltonian $H_F$ recovers a static Ising model.  We work far from this limit, setting $T=1$. 

\subsection{Phases and duality}

\begin{figure}[t!]
\includegraphics[width = \columnwidth]{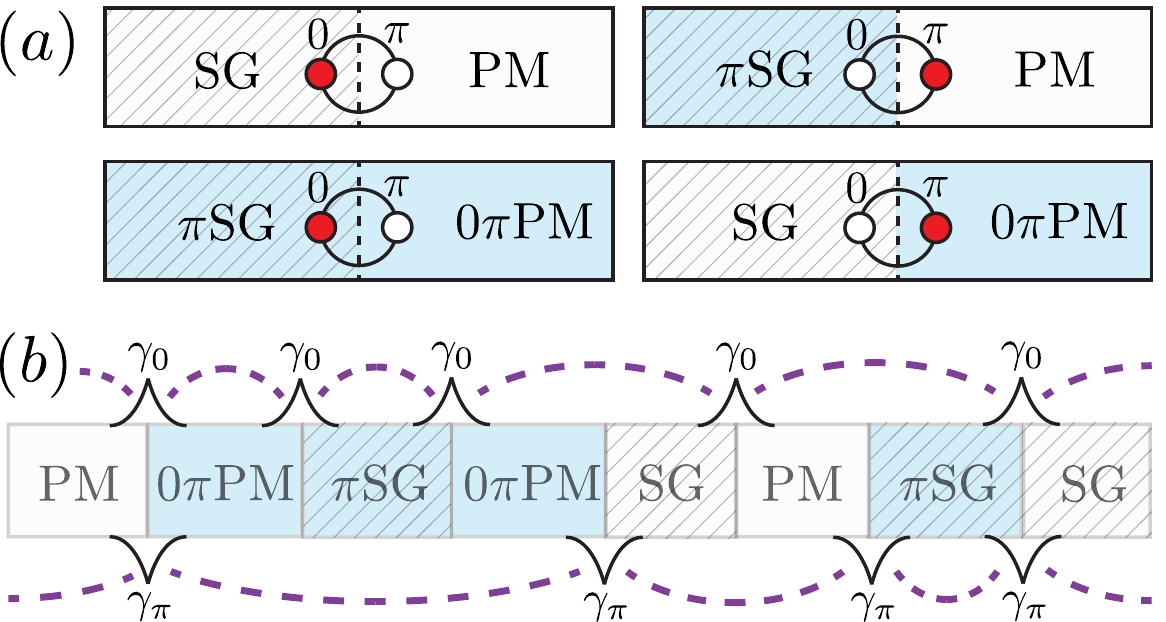}
\caption{\label{fig:cartoons} (a) Domain walls (DWs) between proximate phases of the driven Ising model.  In fermionic language, these host topological edge states at either 0 or $\pi$ quasienergy (red). Blue regions exhibit bulk/boundary  time-translational symmetry breaking (TTSB), and hatched regions have bulk spin glass order. (b) A typical multicritical configuration. Tunneling between DW states $\gamma_{0,\pi}$ yields two independent chains around $0$ and $\pi$ quasienergy.}
\end{figure}

{Observe that $(n_\pi^h, n_\pi^J) =\frac{1}{\pi}({\overline{h_i}},{\overline{J_i}})$, where the bars denote disorder averages, and hence tune between phases of model \eqref{eqModel} analogously to $h, J$ in the clean case.} 
The four phases are {summarized in the phase diagram in Figure~\ref{fig:phase_diagram}}. 
The trivial Floquet paramagnet (PM) breaks no symmetries and has short range spin-spin correlations. The spin glass (SG) spontaneously breaks Ising symmetry with long-range spin correlations in time, or equivalently localized edge modes at 0 quasienergy in the fermion language. These two phases are connected to the undriven paramagnet and ferromagnet/spin glass phases of the random Ising model~\cite{PhysRevB.88.014206,PhysRevLett.112.217204,PhysRevX.4.011052}. Unique to the Floquet system are the $\pi$-spin glass ($\pi$SG) and the $0\pi$ paramagnet ($0\pi$PM). The $\pi$SG  spontaneously breaks both Ising and time translation symmetry in the bulk. Often referred to as a ``
time crystal''~\cite{khemani_prl_2016,else_floquet_2016,PhysRevLett.118.030401}, it maps to a fermion phase with localized Majorana edge modes at $\pi$ quasienergy~\cite{PhysRevLett.106.220402}.
Finally, the $0\pi$PM has short range bulk correlations but also boundary TTSB; its fermion dual has both $0$ and $\pi$ Majorana edge modes and is a simple example of a Floquet SPT. In the fermion language, domain walls between these different phases host either $0$ or $\pi$ Majorana bound states (Fig.~\ref{fig:cartoons}a)  central to the infinite-randomness criticality discussed below.  

The absence of energy conservation in the Floquet setting admits two new  {eigenstate-preserving} changes of parameter to \eqref{eqModel}. The transformations $J_j \mapsto J_j + \pi$ and $h_j \mapsto h_j + \pi$ both separately map $F$ onto another interacting Ising-like Floquet operator with precisely the same eigenstates (see Sec.~\ref{sec:duality}), but possibly distinct quasienergies: $J_j \mapsto J_j+\pi$ preserves $F$ exactly (up to boundary terms), while $h_j \mapsto h_j + \pi$ sends $F \mapsto FG_{\rm Ising}=G_{\rm Ising}F$. {Note that, despite not changing bulk properties of the eigenstates, these transformations map the PM to the $0\pi$PM and the SG to the $\pi$SG respectively.} Additionally,  a global rotation about the $y$ axis takes $h_j\mapsto -h_j$. Below, we fix phase transition lines by combining
 these {\it Floquet symmetries} with the usual Ising bond-field duality that exchanges $h$ and $J$ (and hence SG and PM in the static random case).
  
\subsection{Infinite-randomness structure} 

In analogy with the critical point between PM and SG phases in the static random Ising model (both at zero temperature and in highly excited states), we expect that the dynamical Floquet transitions of~\eqref{eqModel} are controlled by an infinite-randomness fixed point  (IRFP) of a real space renormalization group (RSRG) procedure. At a static IRFP,  the distribution of the effective couplings broadens without bound under renormalization, so the effective disorder strength diverges with the RG scale. A typical configuration of the system in the vicinity of such a transition can be viewed as being composed of puddles deep in one of the two proximate phases, in contrast with  continuous phase transitions in clean systems~\cite{PhysRevB.61.1160,PhysRevX.5.031032}. 
 
 {In order to generalize this picture to the Floquet Ising setting we must identify appropriate scaling variables. For $J_i, h_i \ll \pi$ we recover the criticality of the static model controlled by an IRFP  if $J_i$ and $h_i$ are drawn from the same distribution.  In this case, the relevant operator at the critical point controls the asymmetry between the $J_i$, $h_i$ distributions. At static IRFPs, critical couplings are power-law distributed near $0$.  The absence of energy conservation in the Floquet setting  complicates this picture since there is no longer a clear notion of `low' energies. However, a natural resolution is to allow for fixed-point couplings to be symmetrically and power-law distributed around {\it both} $0, \pi$ quasienergy (or more generally, all quasienergies that can be mapped to $0$ by applying Floquet symmetries of the drive). This introduces a new parameter for Floquet-Ising IRFPs, namely the fractions $n_0$ and $n_\pi$ of couplings near $0$ and $\pi$, respectively. Evidently, we have $n_0=1 - n_\pi$. We will show  
 that there is a new type of IRFP specific to the Floquet setting for $n_\pi = 1/2$, where the criticality is tuned by the asymmetry between the distributions at $0$ and $\pi$ quasienergy, at {\it fixed} values of the  $J_i$ -
 $h_i$ distribution asymmetry.
  
\subsection{Emergent $\pi$-criticality} 

For $J_i, h_i$ near $0$ ($n_\pi \ll 1$), the IRFP distribution is similar to the static case, and the critical point can be understood in terms of domain walls (DWs) between regions where $J_i\gg h_i$ and those where $J_i\ll h_i$. Standard results show that in the fermionic language each DW binds a Majorana  state ${\tilde \gamma}^0_i$ at zero quasienergy, and the transition can be understood in terms of these. For $n_\pi\sim 1$, we again have a single IRFP distribution, but now centered at $\pi$. However, following~\cite{khemani_prl_2016} we may factor a global $\pi$ pulse from both terms of the drive, to recover the previous DW structure. Although still at zero quasienergy, here the DW Majoranas drive a transition between $\pi$SG-$0\pi$PM, owing to the global $\pi$-pulse. Again, the relevant parameter tuning the transition is the asymmetry between the distributions of $J_i$ and $h_i$ so the physics is essentially the same. 

Quite different physics arises for $n_\pi \sim 1/2$ where the couplings exhibit strong quenched spatial fluctuations between $0$ and $\pi$. 
This follows from the fact that there are {\it distinct} IRFP distributions for couplings near $0$ and $\pi$, such that the relevant critical physics is captured by a new class of ``$0\pi$-DWs'' unique to the Floquet setting. If $J_i$ is small and $h_i \sim \pi$ (consistent with $n_\pi \sim 1/2$), these correspond to DWs between $\pi$SG and PM regions, whereas if $h_i$ is small and $J_i \sim \pi$, the critical behavior can be understood in terms of DWs between SG and $0\pi$PM. In the fermion language
 each such $0\pi$ DW traps a Majorana bound state ${\tilde \gamma}^\pi_i$ at quasienergy $\pi$. This may also be deduced by comparing the edge modes of the adjacent phases (Fig.~\ref{fig:cartoons}a).  Since they are topological edge modes, a given $\pi$-Majorana trapped at a $0\pi$ DW can only couple to other $\pi$-Majoranas bound to $0\pi$ DWs, leading to a second emergent Majorana fermion chain whose dynamics are 
 independent from the initial chain (Fig.~\ref{fig:cartoons}b). If the intervening {puddles} are MBL, the tunneling between $\pi$-Majoranas is exponentially suppressed as $\sim {\rm e}^{-\ell}$, with the size $\ell$ of the puddles. Even if we start from a configuration where $J$ and $h$ are drawn from the same distribution, {there are still $\pi$-Majoranas bound to DWs separating infinite-randomness quantum critical regions where the couplings are near $0$ or $\pi$ (see Sec.~\ref{sec:decoupling}), and} the typical tunnel 
  coupling is stretched exponential $\sim {\rm e}^{- \sqrt{\ell}}$~\cite{PhysRevB.51.6411,PhysRevLett.69.534}. Thus, the tunneling terms between the $\pi$-Majoranas remain short-ranged. Crucially, the criticality of this emergent $\pi$-Majorana chain is tuned by  $n_\pi$ (with $n_\pi=\frac{1}{2}$ at criticality), independently of the field-bond asymmetry that tunes the usual Ising transition. {We emphasize that although the universality class of this transition is still random Ising, it is described by flow towards an IRFP at $\pi$ quasienergy, and hence the spectral properties of the transition are distinct.} 

Observe that the  PM-$\pi$SG transition 
 involves the onset of TTSB, since the $\pi$SG is the prototypical example of a discrete 
 time crystal. Similarly, the SG-$0\pi$PM transition involves the onset of TTSB at the ends of an open system. Therefore, we identify the $0\pi$ DWs as the degrees of freedom that are responsible for changes in TTSB.

\subsection{RG treatment} 

The above infinite randomness hypothesis suggests that the critical behavior at the dynamical Floquet transitions can be understood in terms of two  effectively static Majorana chains, one near quasienergy $0$ (${\tilde \gamma}^0_i$) and the other near quasienergy $\pi$ (${\tilde \gamma}^\pi_i$). While the criticality of the $0$ chain is driven by the asymmetry between $J$ and $h$ as in the usual Ising chain, the $\pi$ chain is critical for $n_\pi=\frac{1}{2}$ where there is a symmetry between $0$ and $\pi$ {couplings}. This picture can be confirmed explicitly (see Sec.~\ref{sec:decoupling}) by considering instead the criticality of $F^2$, which should have couplings only near 0 and is described by an effectively static Hamiltonian $F^2 = {\rm e}^{-i 2 H_F}$. The dynamical properties of these two Majorana chains can be analyzed using standard RG techniques designed for static MBL Hamiltonians~\cite{PhysRevX.4.011052,PhysRevLett.112.217204,PhysRevLett.114.217201}. We decimate stronger couplings before weaker ones, putting the pair of Majoranas involved in the strongest coupling in a local eigenstate. Iterating this process leads to an IRFP which self-consistently justifies the strong disorder perturbative treatment. The resulting RG equations match those for the static random Ising model, except crucially, we can now have renormalization towards 0 or towards $\pi$ quasienergies in $F$ reflecting the decoupling of the two effective Majorana chains. This effective decoupling also persists in the presence of interactions ($U\neq 0$). Interactions within the $0$ and $\pi$ Ising chains flow towards 0 under RG much faster than the other couplings and are therefore irrelevant~\cite{PhysRevB.50.3799,PhysRevLett.112.217204}. While interactions also 
permit  Floquet-umklapp terms $\tilde \gamma^0_i \tilde \gamma^0_j \tilde \gamma^\pi_k \tilde \gamma^\pi_l$ that would couple the critical $0$ and $\pi$ chains at the multicritical point, such terms are also irrelevant, and so can be ignored as long as interactions are relatively weak~\cite{PhysRevB.45.2167,PhysRevB.50.3799,1402-4896-2015-T165-014040,PhysRevB.94.014205}. While weak interactions are irrelevant at the multicritical point, and very strong interactions are likely to drive thermalization, we leave open the possibility that intermediate interactions might drive the system to a new infinite-randomness critical point in the universality class of the random Ashkin-Teller model~\cite{1402-4896-2015-T165-014040}.

Therefore, for sufficiently weak interactions, the critical lines are always in the random Ising universality class. The four-phase multicritical point --- at which all four distributions are symmetric --- is in the Ising $\times$ Ising universality class.
This  picture of Floquet (multi) criticality  extends both symmetry-based reasoning used when all $h_i$ couplings are near $\pi$~\cite{PhysRevLett.118.030401}, and the analysis of the essentially static  $J_i,h_i \ll 1$~\cite{1742-5468-2017-7-073301} case.

\subsection{Floquet (multi)criticality} 
 
Combining this reasoning with standard IRFP results, 
we conclude that all the transitions show infinite-randomness Ising scaling:  
 the correlation length diverges as $\xi\sim |\Delta|^{-\nu}$ 
  where $\Delta$ characterizes the deviation from the critical lines, and $\nu=2$ or $1$ 
for average or typical quantities, respectively~\cite{PhysRevB.51.6411,PhysRevLett.69.534}.
This scaling should have universal signatures in dynamical (or eigenstate) correlation functions~\cite{PhysRevLett.69.534,PhysRevB.51.6411,PhysRevLett.112.217204,PhysRevLett.118.030401}, and in particular in the eigenstate entanglement entropy~\cite{PhysRevLett.93.260602,PhysRevB.90.220202,PhysRevLett.114.217201}. Assuming a system of size $L$ and open boundary conditions, the half-system entanglement entropy should scale with system size 
as $S_L \sim (\tilde c/6) \ln L$, up to nonuniversal additive contributions, with ``effective central charge'' $\tilde c=\ln 2/2$~\cite{PhysRevLett.93.260602}. At the multicritical point, we predict  $\tilde c = \ln2$ due to the criticality of the $0$ and $\pi$ Majorana chains. Our picture also predicts an emergent $\Z_2 \times \Z_2$ symmetry at the multicritical point, where the additional $\Z_2$ symmetry can be constructed explicitly as $D = F \sqrt{F^2}^\dagger$~\cite{pi-spin-glass,PhysRevLett.118.030401,PhysRevX.7.011026}. For a multicritical configuration with couplings near 0 or $\pi$, we find that $D$ is distinct from the original Ising symmetry of the model, and coincides with the fermion parity of the emergent $\pi$-Majorana chain,
\begin{equation}
\label{eq:majorana_D}
D = \prod_{j \in \{\rm 0\pi \ {\rm DWs} \}} \tilde \gamma^\pi_j.
\end{equation}
}

\subsection{Numerics}  

\begin{figure}
\includegraphics[width = \columnwidth]{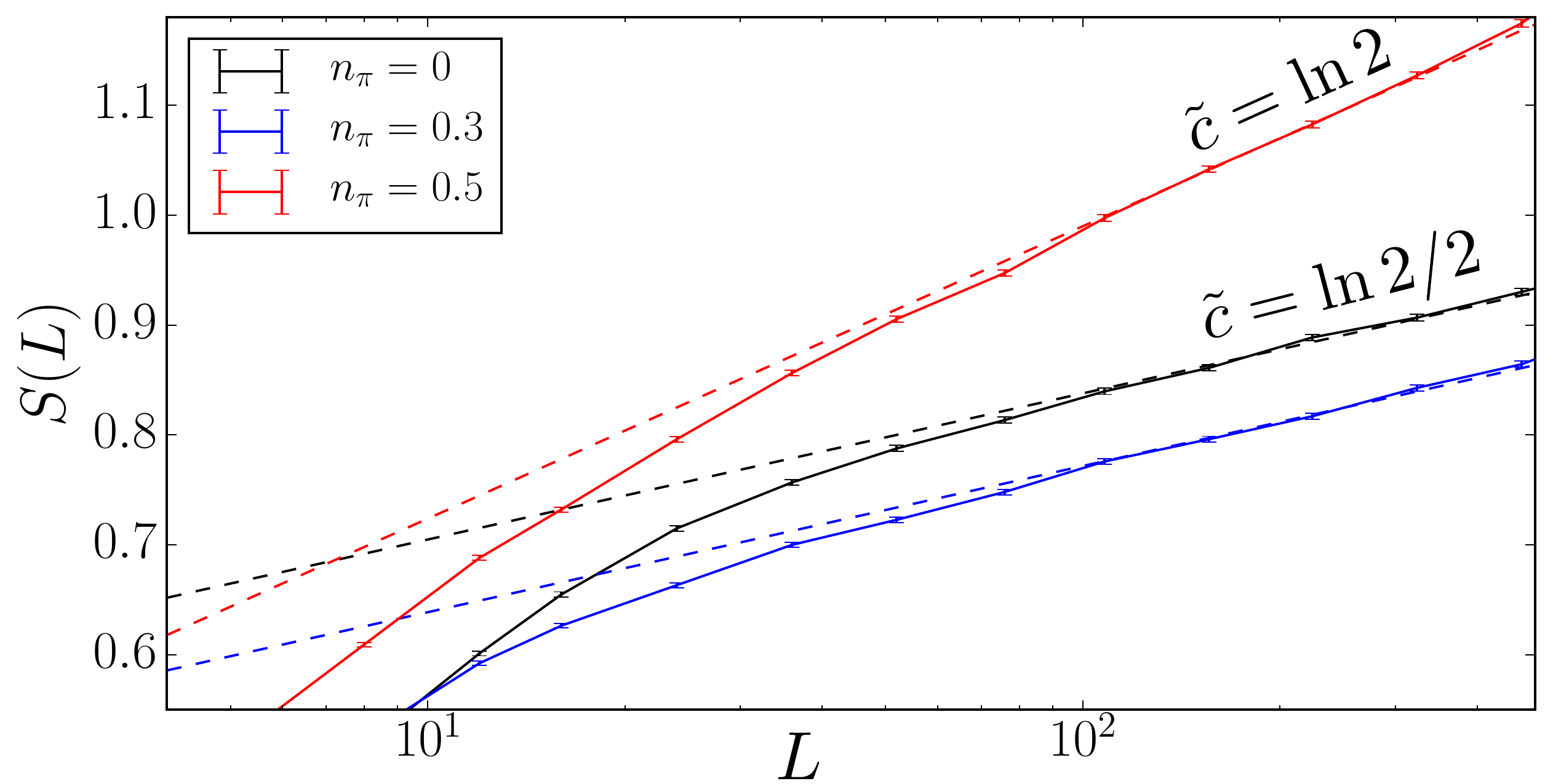}
\caption{\label{fig:entropy}{Scaling with system size of disorder- and eigenstate-averaged entanglement entropy $S$ for a cut at $L/2$. Dashed lines show predicted slopes for strong-disorder Ising criticality along the transition lines (blue, black); this doubles at the multicritical point (red).} }
\end{figure}

As stressed above, our picture of these transitions relies on the infinite randomness assumption. To justify this and to confirm our analytical predictions, we have performed extensive numerical simulations on the non-interacting model, leveraging its  free-fermion representation to access the full single-particle spectrum and to calculate the entanglement entropy of arbitrary eigenstates. We average over 20,000 disorder realizations (with open boundary conditions), randomly choosing a Floquet eigenstate in each.

 {Given our parametrization of disorder, the combination $\frac{1}{2}(n_\pi^h - n_\pi^J)$  provides a measure of the {asymmetry between $J$ and $h$ couplings}, while $\frac{1}{2}(n_\pi^h + n_\pi^J)$ measures the average probability of a $\pi$ coupling. Therefore, from our reasoning above and using}
  the usual Ising duality, we expect a critical line for $n^\pi_J=n^\pi_h$. 
  {Combining} the Ising duality with Floquet symmetries leads to another critical line $n^\pi_J+n^\pi_h=1$ where we expect $0\pi$ infinite randomness behavior. Note that the bare disorder distributions are far from the infinite randomness fixed point expected to emerge at criticality. Nonetheless, as shown in Figure~\ref{fig:entropy}, we observe clear logarithmic scaling of entanglement along the self-dual lines $n^\pi_J=n^\pi_h$ and $n^\pi_J+n^\pi_h=1$ of Eq.~\eqref{eqModel}. We find $\tilde c \approx \ln 2 / 2$, consistent with the prediction that the lines are in the random Ising universality class. Deep in the phases, we find $\tilde c \approx 0$ consistent with the area-law scaling expected for Floquet MBL phases~\cite{1742-5468-2013-09-P09005,PhysRevLett.111.127201}. At the multicritical point $n_\pi^h = n_\pi^J = 1/2$, we find $\tilde c \approx \ln 2$, consistent with our expectation of two decoupled critical Ising chains. Although stability to quartic interchain couplings cannot be addressed in this noninteracting limit, we expect it on general grounds~\cite{PhysRevB.45.2167,PhysRevB.50.3799,1402-4896-2015-T165-014040}, modulo usual caveats on thermalization. Fig.~\ref{fig:phase_diagram}, showing the entanglement scaling across the entire phase diagram, summarizes these results. Finally, we have also numerically calculated the relative number of single particle quasienergies near 0 and near $\pi$, finding good agreement with a simple prediction from the infinite-randomness domain wall picture. Moreover, Fig.~\ref{fig:phase_diagram} clearly shows that changing $n^\pi_J\pm n^\pi_h$ tunes across the critical lines, confirming that these parameters control  distribution asymmetries as in the IRFP picture (Fig.~\ref{fig:phase_diagram}, insets).
  
\subsection{Experimental consequences}

Let us now turn to some experimental consequences of the above predictions. Recent advances in the control of ultracold atomic arrays have brought models such as Eq.~\ref{eqModel} into the realm of experimental realizability~\cite{Kim:2010aa,Blatt:2012aa,Smith:2016aa}. The model hosts a time-crystal phase (the $\pi$ spin glass), the phenomenology of which has recently been directly observed~\cite{Choi:2017aa,Zhang:2017aa}. Even though, as mentioned earlier, these critical lines may eventually thermalize due to long-range resonances~\cite{PhysRevB.95.155129,PhysRevLett.119.150602,Ponte20160428}, the dynamics of the Ising universality class should persist through a prethermalization regime relevant to all reasonably accessible experimental timescales~\cite{PhysRevLett.115.256803,kuwahara_floquetmagnus_2016}. Thus, the dynamical signatures of the transitions we have identified should be readily experimentally observable. 

One prominent experimental signature of this physics is the scaling behavior of the dynamical spin-spin autocorrelation function in Fourier space $C(\omega,t) \equiv \int_0^\infty d\tau e^{-i \omega \tau} \overline{\expectation{\sigma_i^z (t+\tau) \sigma_i^z (\tau)}}$, with the overline representing a disorder average~\cite{PhysRevLett.118.030401}. In accordance with the random Ising universality class, the spin-spin autocorrelation function will scale as $\disavg{\expectation{\sigma_i^z(t) \sigma_{i}^z(0)} } \sim {1}/{\log^{2-\phi} t}$~\cite{PhysRevLett.112.217204}, with the overline representing a disorder average and $\phi = (1 + \sqrt{5} ) / 2$ the golden ratio. Performing the Fourier transform, our analysis then predicts that along the $n^\pi_h = n^\pi_J$ critical line of the model, the Fourier peak at 0 quasienergy will decay as $C(0,t) \sim 1/\log^{2-\phi} t$; along the $n^\pi_h = 1 - n^\pi_J$ critical line the peak at $\pi$ quasienergy will decay the same way as $C(\omega/2,t) \sim 1/\log^{2-\phi} t$; and at the multicritical point, both peaks will decay in this way simultaneously, giving
\begin{equation}
C(0,t) \sim C(\omega/2,t) \sim \frac{1}{\log^{2-\phi} t}.
\end{equation} 
This slow, logarithmic decay, independently for the decoupled chains at $0$ and $\omega/2$, serves as an unambiguous signature of the universal multicritical physics we describe. The fact that the two decays are independent is highly nontrivial, since generic $\Z_2\times \Z_2$ multicritical points would have distinct scaling from either $\Z_2$ individually.

\subsection{Discussion} 

We have presented a generic picture of the transitions between MBL Floquet phases, and applied it to study the criticality of the periodically driven interacting random Ising chain. Our work can be generalized to more intricate Floquet systems, under the (reasonable) assumption that they flow to infinite randomness under coarse-graning. The resulting IRFP is enriched in the  Floquet setting: each distinct invariant Floquet quasienergy hosts an independent set of fixed-point coupling distributions. (For instance the $\Z_n$ model has $n$ such invariant quasienergies, $2\pi k/n$, with $k=1,\ldots,n$.)  
Systems at conventional IRFPs are tuned across criticality by adjusting the imbalance between distributions of  distinct couplings at the {\it same} quasienergy. At Floquet IRFPs, we may hold such single-quasienergy imbalances fixed and instead tune the imbalance between the distributions of couplings at {\it distinct} quasienergies. Transitions driven by such cross-quasienergy imbalances will usually involve an onset or change of TTSB in the bulk or at the boundary, and in this sense describe ``time crystallization''.  In some cases, it may be possible to leverage a Jordan-Wigner mapping in conjunction with these infinite-randomness arguments to arrive at a domain-wall description of the critical/multicritical physics. We anticipate that a variety of Floquet symmetry-breaking/symmetry-protected {topological} phases will be amenable to similar analysis, but we defer an exhaustive study to future work.

\section{Strong-Disorder Renormalization Group for Periodically Driven Systems}

\subsection{Introduction}

The study of periodically driven (Floquet) systems lies at the frontier of our understanding of non-equilibrium quantum physics. Despite the restriction to only {\it discrete} time translation symmetry and the attendant lack of full energy conservation, great strides have been made in recent years in understanding these inherently out-of-equilibrium systems. Results of these investigations include proposals to engineer exotic effective Hamiltonians~\cite{RevModPhys.89.011004,doi:10.1080/00018732.2015.1055918,PhysRevLett.116.205301,0953-4075-49-1-013001,PhysRevX.4.031027,PhysRevB.82.235114,Lindner:2011aa,PhysRevX.3.031005,1367-2630-17-12-125014,PhysRevB.79.081406,Gorg:2018aa,PhysRevLett.116.125301}, efforts to classify driven symmetry-protected topological phases~\cite{PhysRevB.92.125107,PhysRevB.93.245145,PhysRevB.93.245146,PhysRevB.93.201103,potter_classification_2016,PhysRevB.94.125105,PhysRevB.94.214203,PhysRevB.95.195128}, and the discovery of discrete time translation symmetry breaking -- dubbed `time \\ \noindent crystallinity'~\cite{pi-spin-glass,else_floquet_2016,PhysRevX.7.011026,PhysRevLett.118.030401,PhysRevB.96.115127,PhysRevLett.119.010602,Zhang:2017aa,Choi:2017aa,PhysRevB.97.184301,PhysRevLett.120.180603} --- among many others.  Floquet systems have thus been shown to host rich phase structures~\cite{khemani_prl_2016}, both extending fundamental concepts of equilibrium statistical mechanics into the non-equilibrium realm, and admitting new possibilities forbidden in equilibrium. 

Any discussion of the late-time limits of periodically driven closed quantum systems must address the issue of thermalization. Since such systems lack energy conservation, the expectation is that as energy is injected into them at periodic intervals, they will heat up to an infinite-temperature Gibbs state, characterized by a Floquet generalization of the eigenstate thermalization hypothesis (ETH)~\cite{PhysRevA.43.2046,PhysRevE.50.888,0034-4885-81-8-082001}.
The possibility of an exponentially long ``prethermal'' regime notwithstanding~\cite{kuwahara_floquetmagnus_2016,PhysRevLett.115.256803}, the only known generic (i.e., not fine-tuned) exceptions to this scenario are systems that exhibit the phenomenon of many-body localization (MBL)~\cite{BASKO20061126,PhysRevB.75.155111,PalHuse,doi:10.1146/annurev-conmatphys-031214-014726,doi:10.1146/annurev-conmatphys-031214-014701,1742-5468-2016-6-064010,ALET2018,2018arXiv180411065A}. Such systems can  avoid heating and retain  a notion of distinct phases of matter even far away from thermal equilibrium~\cite{PhysRevB.88.014206,BahriMBLSPT,PhysRevB.89.144201}. The presence of quenched randomness therefore allows us to sharply define Floquet phases, and naturally leads to the possibility of transitions between them~\cite{khemani_prl_2016,PhysRevLett.114.140401,PONTE2015196,ABANIN20161,PhysRevLett.115.030402}.  

One of the most potent tools for studying one-dimensional random Hamiltonians is the real space renormalization group (RSRG)~\cite{PhysRevB.22.1305,PhysRevB.51.6411,PhysRevB.50.3799,PhysRevLett.69.534}. Though initially introduced as a technique for finding the ground states of random spin chains, these were subsequently extended to study excited states in MBL systems (RSRG-X)~\cite{PhysRevX.4.011052,PhysRevLett.112.217204}. Such RSRG-X techniques are not only a powerful (though approximate) way to obtain the excited states of interacting many-body systems at strong disorder, but also can be used to characterize the universal critical behavior near dynamical transitions between distinct MBL phases. Remarkably, RSRG often becomes asymptotically exact, since the effective disorder strength controlling the approach grows without bound under renormalization. It is therefore natural to examine whether these Floquet MBL systems might be amenable to a similar RSRG approach, improving our understanding of Floquet phases and the transitions possible between them.

In this work, we introduce a real-space renormalization group method for Floquet systems, which we dub ``Floquet RSRG," and apply it to understand the criticality of a canonical Floquet system: the driven Ising chain~\cite{khemani_prl_2016}. We derive a generalization of the Schrieffer-Wolff transformation to unitary operators which serves as our basic technical workhorse, allowing us to perturbatively construct a renormalized Floquet operator that captures the effective dynamics over ever-increasing length scales as the RG progresses. Working in the Majorana fermion language and tracking the flow of the couplings under renormalization, we identify the criticality in the model, including at the multicritical point. In contrast to previous works, our method is generic to {many periodically-driven systems:\footnote{Formally, our method requires that the Floquet unitary be separable into a `strong' piece $F_0$ and `weak' piece $e^{iV}$, as described in more detail in the main text.}} it only assumes the existence of a Floquet operator $F$, and furthermore is not limited to non-interacting Floquet-Ising~\cite{1742-5468-2017-7-073301} or discrete time crystal~\cite{PhysRevLett.118.030401} models. Even when applied to the driven Ising model, our method does not require all bond terms to correspond to near-0 quasienergy and all field terms to be near 0~\cite{1742-5468-2017-7-073301} or near $\pi$~\cite{PhysRevLett.118.030401}. Crucially, this allows us to analyze the proliferation of domain walls  between $0$ and $\pi$ regions, leading to a full description of the critical lines and multicritical point of the model. 
We find precise agreement with an earlier, intuitive, picture we proposed in terms of topological domain walls~\cite{Berdanier201805796}, {generalizing the picture proposed for quantum groundstates by Damle and Huse~\cite{PhysRevLett.89.277203}}. This work therefore provides an exact microscopic justification for results that may also be deduced on general grounds, thereby affirming their universality. {\it In toto}, these two perspectives give a rather complete description of criticality in a specific example of a periodically driven system, and in particular provide a concrete example of how an effective description in terms of domain walls arises from the underlying microscopic Floquet physics. 

The remainder of this paper is organized as follows. In Section~\ref{sec:theory} we review some basic aspects of Floquet theory, introduce the Floquet Schrieffer-Wolff transformation, and outline the general framework of Floquet RSRG. In Section~\ref{sec:Ising} we exemplify the use of this method on the driven Ising model, working first within the free-fermion limit. In Section~\ref{sec:ints} we show how Floquet RSRG can be used to argue analytically  for the irrelevance of interactions in the strong-disorder limit. Finally, we close with a summary of the work and a discussion of future applications of Floquet RSRG (Sec.~\ref{sec:discussion}).

\subsection{Floquet RSRG via Floquet Schrieffer-Wolff transformations} \label{sec:theory} 

Floquet systems are defined by a time-periodic Hamiltonian $H(t) = H(t+T)$. Equivalently, the Hamiltonian has a discrete time translation symmetry by the drive period $T$. As with Bloch's theorem for Hamiltonians with discrete spatial translation symmetry, the eigenstates of a time-periodic Hamiltonian must satisfy $\ket{\psi_\alpha(t)} = e^{-i E_\alpha t} \ket{\phi_\alpha(t)}$, where $\ket{\phi_\alpha(t+T)} = \ket{\phi_\alpha(t)}$ in units where $\hbar = 1$~\cite{PhysRev.138.B979,PhysRevA.7.2203}. A central object of interest is the single-period time evolution operator, 
\begin{equation}
F \equiv U(T) = \mathcal T { e}^{-i  \int_0^T dt H(t) } = { e}^{-i T H_F},
\end{equation}
where $\mathcal T$ denotes that the exponential is time-ordered, and  the last equality defines the so-called Floquet Hamiltonian. In general, the Floquet Hamiltonian is quite different from $H(t)$ at any $t$, and may in fact be non-local. 
Crucially, $H_F$ has eigenvalues that are constrained to lie on a circle, so $H_F = H_F + 2\pi/ T$. This in general eliminates the notion of a ground state,
requiring us to consider  the entire spectrum of  eigenstates of the Floquet operator $F$.

Our goal is to study such Floquet systems in one dimension in the presence of strong disorder, using a real-space renormalization group (RSRG) approach. Also termed the `strong-disorder renormalization group,' this method was initially introduced as a means of constructing the ground state~\cite{PhysRevB.22.1305,PhysRevB.51.6411,PhysRevB.50.3799,PhysRevLett.69.534} of one-dimensional Hamiltonians with quenched disorder, and later {extended (under the name `RSRG-X') to study} the full spectrum\cite{PhysRevX.4.011052} of such systems. The basic steps involved at any given stage of this RG scheme are (i) to identify the largest coupling in the effective Hamiltonian $H$ at that stage (which sets the characteristic energy scale); (ii) to `solve' the effective local problem $H_0$ defined by turning off all other couplings, assumed to be much weaker by virtue of the broad distribution of couplings; and (iii)  performing perturbation theory to determine the new couplings mediated by virtual fluctuations between eigenstates of $H_0$, thus defining a new effective Hamiltonian that can be fed back into step (i) in the next iteration. Tracking the flows of the distributions of couplings under this RG gives access to various quantities including correlation functions~\cite{PhysRevB.51.6411,PhysRevB.50.3799,PhysRevLett.69.534}, dynamical properties~\cite{PhysRevLett.84.3434,PhysRevB.63.134424}, and the entanglement entropy~\cite{PhysRevLett.93.260602}. Crucially, in many cases this procedure has an asymptotically self-consistent justification: to wit, initially broad distributions of the couplings broaden further with increasing iterations, indicating a flow to an `infinite randomness' fixed point where the RG procedure becomes exact.

At the heart of such RSRG methods is the Schrieffer-Wolff (SW) transformation~\cite{PhysRev.149.491}, a perturbative unitary rotation that eliminates off-diagonal elements of the Hamiltonian $H$ with respect to a ``strong'' piece $H_0$. In particular, one writes $H = H_0 + V$, where $\abs{V} \ll \abs{H_0}$, then seeks a unitary operator $e^{i S}$ such that $[e^{iS} H e^{-iS}, H_0] = 0$ to the desired order in $V$. This gives a self-consistent equation for $S$ to each order, which can be readily solved. Projecting onto an eigenstate subspace of the strong-coupling piece, one finds the overall energy shift and renormalized couplings between the remaining degrees of freedom. Projecting onto the lowest-energy eigenstate in each step then picks out the ground state of the problem, whereas projecting onto different subspaces besides the lowest-energy state allows access to excited states --- this being the key modification involved in RSRG-X.

In a similar spirit, let us imagine decomposing a Floquet operator $F$ into the product of a ``strong''\footnote{Generically, since the eigenvalues of a unitary operator lie on a circle, perturbation theory on a unitary operator is not well defined as the eigenvalues are not well-ordered. We note that care must be taken in identifying a ``strong'' piece such that perturbation theory is well-controlled; this can fail, as it does at the naive decimation of domain walls in the main text.} piece $F_0$ and a ``weak'' piece $V$, where $V \in \mathcal O(\lambda)$ with $\lambda$ a small parameter:
\begin{equation}
F = F_0 e^{i V}.
\end{equation}
The ordering of $F_0$ and $e^{i V}$ is an arbitrary choice of convention. We then seek a unitary operator $e^{i S}$ that transforms $F$ to $\tilde F \equiv e^{iS} F e^{-iS}$ such that $[\tilde F,F_0] = 0$ to the desired order in $V$. 
Expanding, we have  
\begin{align}
\label{eq:F_tilde}
\tilde F =& F_0 + i[S, F_0] + i F_0 V -\frac{1}{2} \{S^2, F_0 \}  
+ [F_0 V, S] + S F_0 S - \frac{1}{2} F_0 V^2 + \ldots .
\end{align}
We now write $S$ as a power series in $\lambda$, $S = \sum_n S_{(n)}$ with $S_{(n)} \in \mathcal O(\lambda^n)$. First, note that $S_{(0)} = 0$. This is because, to $0^{\text{th}}$ order in $V$, $F = F_0 + \mathcal O(V)$ already commutes with $F_0$. Expanding, the above expression becomes
\begin{align*}
\tilde F =& F_0 + (i[S_{(1)}, F_0] + i F_0 V) + ( i[S_{(2)}, F_0]-\frac{1}{2} \{S_{(1)}^2, F_0 \} \\ & +[F_0 V, S_{(1)}] + S_{(1)} F_0 S_{(1)} - \frac{1}{2} F_0 V^2) + \ldots,
\end{align*}
where we have grouped terms according to their order in $V$. Each of these grouped pieces self-consistently defines $S_{(n)}$. In particular, we require that each of them commute with $F_0$ to that order. In general, each self-consistent equation will be of the form 
\begin{equation}
\label{eq:Mn}
[ i [S_{(n)}, F_0] + M_{(n)} , F_0 ] = 0,
\end{equation}
where $M_n$ are the $n$th order terms obtained from expanding the expression for $\tilde F$: $M_{(1)} = i F_0 V$ and $M_{(2)} = -\frac{1}{2} \{S_{(1)}^2, F_0 \} + [F_0 V, S_{(1)}] + S_{(1)} F_0 S_{(1)} - \frac{1}{2} F_0 V^2$. In order to solve equations of the form of Eq. \ref{eq:Mn}, it is convenient to introduce the set of projectors $\{P_\alpha \}$  onto the eigenspaces of $F_0$, where $F_0 P_\alpha = \alpha P_\alpha$, $\alpha \in U(1)$ since $F$ is unitary, and $P_\alpha^2 = P_\alpha$. Then Eq. \ref{eq:Mn} can be readily solved as\footnote{Note that the solution to Eq. \ref{eq:Mn} is only unique up to operators $\mathcal O$ satisfying $[[\mathcal O, F_0],F_0]=0$. Our solution is chosen to give blocks of zeros in $M_n$ corresponding to the eigenbasis of $F_0$.}

\begin{equation}
S_{(n)} = \sum_{\alpha \not= \beta} \frac{1}{i(\alpha - \beta)} P_\alpha M_{(n)} P_\beta.
\end{equation}

Having now described the framework for performing perturbation theory on a Floquet unitary,  the Floquet RSRG method proceeds as follows. First, identify the strongest piece $F_0$ in the Floquet evolution operator. This is similar to identifying the strongest bond $\Omega$ in the usual RSRG method; however, since quasi-energies take values on a circle,  they do not form a well-ordered set, and some care must be taken in identifying $\Omega$; we turn to this in the next section. For the moment, assume that such a  ``strong piece'' $F_0$ has been identified; then the rest of the chain is identified with $e^{iV}$. We then perform a Floquet SW transformation on $F$, truncating at second order in $V$. This generates a virtual coupling mediated by the strong piece $F_0$, giving a renormalized coupling between neighboring degrees of freedom. Iterating this procedure generates a flow of the couplings in the chain in much the same way as is in the usual RSRG method.

\subsection{Application: the driven Ising chain} \label{sec:Ising}

\begin{figure}
\includegraphics[width=\columnwidth]{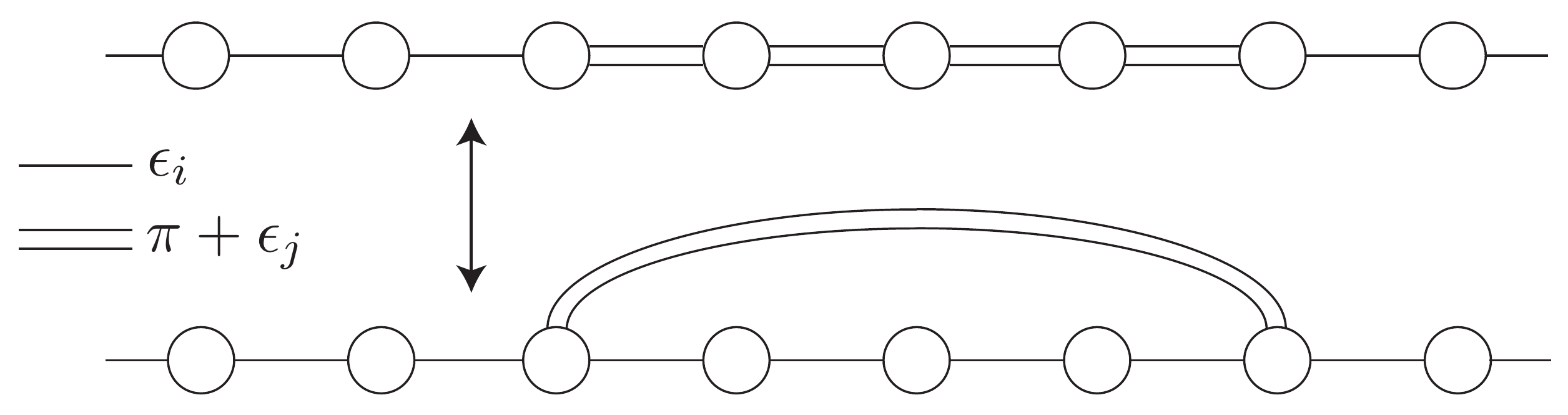}
\caption{\label{fig:factoring} (Top) A generic configuration of the disordered periodically driving Ising chain in the Majorana fermion representation. Circles represent Majorana fermion operators, single lines represent bilinear couplings $\epsilon_i$ with $\epsilon_i \in [-\pi/2,\pi/2]$, and double lines represent bilinear couplings $\pi + \epsilon_j$ with $\epsilon_j$ in the same range. The chain has domains with couplings near $0$ (``0-domains'') and with couplings near $\pi$ (``$\pi$-domains''). (Bottom) We can factor out the exact $\pi$ pulses to form one long-ranged exact $\pi$ pulse spanning the $\pi$-domain, with all couplings now in the range $[-\pi/2, \pi/2]$.}
\end{figure}

In order to demonstrate how Floquet RSRG works in practice, we now turn to a concrete application of the method outlined in the preceding section to a prototypical Floquet system: namely, the periodically driven Ising chain, defined by the sequence
\begin{equation}
H(t) \!=\! \begin{cases}
H_1 \!=\! \sum_i  \!h_i \sigma^x_i  + U^{xx}_i \sigma_i^x \sigma_{i+1}^x, & \!\!0 \leq t \leq T/2 \\
H_2 \!=\! \sum_i \!J_i \sigma^z_i \sigma^z_{i+1} +  U^{zz}_i \sigma_i^z \sigma_{i+2}^z, & \!\!T/2 \leq t \leq T,
\end{cases}
\end{equation}

where $\sigma_i^{x,z}$ are Pauli operators on site $i$, $J_i$ and $h_i$ are uncorrelated random variables, and $U$ corresponds to small interaction terms that respect the Ising $\Z_2$ symmetry generated by $G_{\rm Ising} = \prod_i \sigma_i^x$. For now, we will take $U^{xx}_i = U^{zz}_i =0$ and discuss the role of interactions in Section~\ref{sec:ints}. Applying a Jordan-Wigner transformation to rewrite the chain in terms of Majorana fermions~\cite{sachdev2011quantum}, we see that the Floquet evolution operator is
\begin{equation}
F = \exp( \frac{1}{2}\sum_{i} J_{2i} \gamma_{2i} \gamma_{2i+1}) \exp( \frac{1}{2} \sum_{i} J_{2i-1}\gamma_{2i-1} \gamma_{2i}),
\end{equation}
where we set $T=1$ for convenience. The odd Majorana bonds correspond to field terms for the spins ($h_i \leftrightarrow J_{2i-1}$) while the even Majorana bonds correspond to bond terms for the spins ($J_i \leftrightarrow J_{2i}$). The Majorana operators $\gamma_j$ obey $\gamma_j^\dagger = \gamma_j$, $\gamma_j^2 = 1$, and $\{\gamma_i, \gamma_j\} = 2\delta_{ij}$. We restrict the couplings to fall in a window of width $2\pi$, specifically the range $J_i \in [-\pi/2,3\pi/2)$ which is symmetric about $0$ and $\pi$. All couplings may be brought into this range by noting that $e^{(J_{ij} + 2 \pi n) \gamma_i \gamma_j/2} \propto e^{J_{ij} \gamma_i \gamma_j/2}$ for integer $n$, which will share the same eigenstates and hence the same phase structure.

This model hosts four phases~\cite{khemani_prl_2016}. Two are connected to static counterparts in the $T\to 0$ limit: (1) a trivial paramagnet, which is short-range correlated and does not exhibit spontaneous symmetry breaking (SSB); and (2) a spin glass phase which spontaneously breaks the Ising symmetry and is long-range correlated in time, or equivalently hosts a localized edge Majorana fermion mode at zero quasi-energy. Two phases are unique to the driven setting: (3) a ``$\pi$-spin glass'', which is long-range correlated and spontaneously breaks both the Ising symmetry and time translation symmetry, or equivalently hosts an Majorana edge mode at $\pi$ quasi-energy, and is often referred to as a ``time crystal''; and (4) a ``$0\pi$-paramagnet'', which has short-range bulk correlations, does not break the Ising symmetry, but does break time translation symmetry, and equivalently hosts edge Majorana modes at both 0 and $\pi$ quasi-energy. 

A generic configuration of the chain will have some couplings closer to $0$ and others closer to $\pi$. This leads to two types of domain: domains of couplings nearer to 0 (``$0$ domains''), and domains of couplings nearer to $\pi$ (``$\pi$-domains''). We will assume that in these domains, the couplings are either very close to $0$, or very close to $\pi$: even if this is not initially the case microscopically, we will see that this becomes true self-consistently after running the RG, {\it i.e.}, this is a property of the RG fixed points we are after. Note that Majorana fermions obey a simple evolution equation: $e^{\theta \gamma_i \gamma_j} = \cos\theta + \gamma_i \gamma_j \sin \theta$ for $i\not=j$. Therefore, within a $\pi$-domain we can factor out all of the $\pi$ pulses to simply extract one long-ranged pulse at strength exactly $\pi$: $(\gamma_0 \gamma_1 )( \gamma_1 \gamma_2) \ldots (\gamma_{L-1}\gamma_L ) = \gamma_0 \gamma_L = e^{\frac{\pi}{2} \gamma_0 \gamma_L}$.

We perform this factoring across the entire chain as a first step, as diagrammed in Figure~\ref{fig:factoring}. This is one of the most important steps in our procedure, as what remains are small couplings (controlling perturbation theory) that we will now show how to decimate, along with large non-local couplings that significantly modify the decimation rules from those of similar static Hamiltonians. From here on we use the notation $J_i$ to denote a Majorana coupling in the range $[-\pi/2,\pi/2)$, assuming all larger couplings have been factored out.

\subsubsection{Decimating inside a 0-domain or $\pi$-domain}

\begin{figure}
\includegraphics[width=\columnwidth]{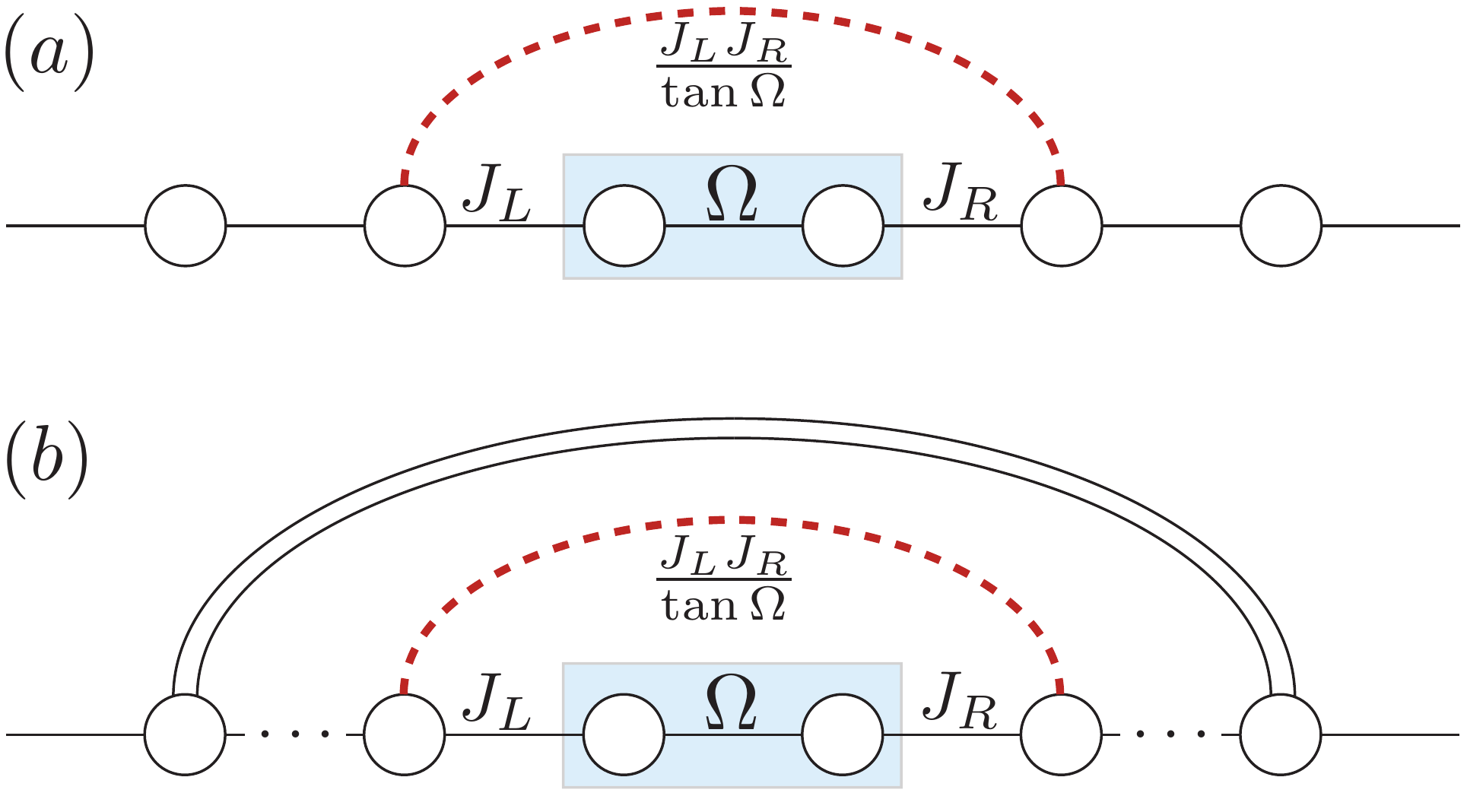}
\caption{\label{fig:domain_decimation} Decimation inside a 0-domain or a $\pi$-domain follows the same rule. (a) Within a 0-domain, a strong bond $\Omega$ is renormalized via the rule $\tilde J = J_L J_R / \tan \Omega$. (b) Within a $\pi$-domain, the long-ranged $\pi$ pulse does not affect the unitary rotation, giving the same virtual coupling. We can view decimation within a $\pi$-domain as decimation towards a $\pi$-bond by replacing the long-ranged $\pi$-pulse with a product of short-ranged ones.}
\end{figure}

As a warm-up exercise, consider decimation deep within a 0-domain or a $\pi$-domain. Assume that the strongest coupling is $J_0 \equiv \Omega \ll 1$, and that the neighboring couplings satisfy $J_L, J_R \ll \Omega \ll 1$, and are not on domain walls. This is actually the generic case deep in {\it both} a 0-domain and a $\pi$-domain; see Figure~\ref{fig:domain_decimation} for a diagram. We then have 
\begin{align}
F_0 &= e^{\Omega \gamma_0 \gamma_1} = \cos \Omega + \gamma_0 \gamma_1 \sin \Omega, \nonumber \\
V &= -i J_L \gamma_{-1} \gamma_0 - i J_R \gamma_0 \gamma_1.
\end{align}
All other terms commute with $F_0$, so we drop them. $F_0$ has two eigenstate projectors: $P_{\pm} = \frac{1}{2}(1 \pm i \gamma_0 \gamma_1)$ which project onto the $i \gamma_0 \gamma_1 = \pm 1$ subspaces, with eigenvalues $\lambda_\pm = e^{\mp i \Omega}$. Solving Eq.~\ref{eq:Mn} for $n=1,2$, we find
\begin{align}
S_{(1)} &= \frac{1}{2} i [ \gamma_{-1} \gamma_1 J_L \cot \Omega + \gamma_{-1} \gamma_0 J_L + J_R (\gamma_1 \gamma_2 - \gamma_0 \gamma_2 \cot \Omega) ], \\
S_{(2)} &= 0.
\end{align}
Computing $P_+ \tilde F P_+$ and setting $i \gamma_0 \gamma_1 = 1$, 
\begin{align}
P_+ \tilde F P_+ = \frac{1}{2} \sin\Omega (\cot \Omega - i) \left[ -i(J_L^2 + J_R^2 ) \cot \Omega + 2 \gamma_{-1} \gamma_2 J_L J_R \cot\Omega + 2 \right].
\end{align}
Now note that this is of form $P_+ \tilde F P_+ = e^{i \theta} e^{ \tilde J \gamma_{-1} \gamma_2} = e^{i \theta} (\cos \tilde J + \gamma_{-1} \gamma_2 \sin \tilde J ) = A + B \gamma_{-1} \gamma_2$. Thus, $\tan \tilde J = B / A$. Looking above to identify $A$ and $B$ and dividing, we find
\begin{equation}
\label{eq:deep_rule}
\tan \tilde J = \frac{J_L J_R}{\tan \Omega - \frac{i}{2} (J_L^2 + J_R^2) } \approx \frac{J_L J_R}{\tan \Omega}.
\end{equation}

That is, the renormalized coupling is related to the {\it tangent} of the strong bond $\Omega$; this reduces to the well-known static rule $\tilde J = J_L J_R / \Omega$ in the limit $\Omega \to 0$. We can similarly project onto the $P_-$ subspace to access other branches of the many-body spectrum, finding the same rule of $\tan \tilde J = J_L J_R / \tan \Omega$ but a different overall quasi-energy shift (Eq.~\ref{eq:quasi-energy}). Though Ref.~\citenum{1742-5468-2017-7-073301} obtains a superficially similar formula, we note that in contrast to Eq.(45) of Ref.~\citenum{1742-5468-2017-7-073301}, Eq.~\ref{eq:deep_rule} is valid when all nearby bonds are near $\pi$ {\it as well as} near $0$, due to the factoring of the $\pi$ pulses.

One crucial difference with the static case is that the ordering of terms is now important; one might wonder if the proposed renormalization procedure even gives back a self-similar evolution operator for this model. Indeed it does, as can be seen by examining $F$. Taking $\Omega$ in the even sublattice for the sake of argument, let us decimate $\Omega \gamma_0 \gamma_1$:
\begin{align}
&F = \ldots e^{J_{-2} \gamma_{-2} \gamma_{-1} + J_{2} \gamma_2 \gamma_3} e^{\Omega \gamma_0 \gamma_1} e^{J_{-1} \gamma_{-1} \gamma_0 + J_{1} \gamma_1 \gamma_2} e^{J_{-3} \gamma_{-3} \gamma_{2} + J_{3} \gamma_3 \gamma_4} \ldots \nonumber \\
&\xrightarrow{\text{RG}} \ldots e^{J_{-2} \gamma_{-2} \gamma_{-1} + J_{2} \gamma_2 \gamma_3} e^{\tilde J \gamma_{-1} \gamma_2 + i \theta} e^{J_{-3} \gamma_{-3} \gamma_{2} + J_{3} \gamma_3 \gamma_4} \ldots \nonumber \\
&= \ldots e^{i \theta} e^{J_2 \gamma_2 \gamma_3 + J_{-2} \gamma_{-2} \gamma_{-1}} e^{J_{-3} \gamma_{-3}\gamma_{-2} + \tilde J \gamma_{-1}\gamma_2 + J_3 \gamma_3 \gamma_4} \ldots ,
\end{align}
where in the first step we factored commuting pieces, and $e^{i\theta}$ is the overall quasi-energy shift. Odd sublattice decimations proceed similarly. Therefore, we see that the decimation has produced a self-similar Floquet operator, and decimating a bond in the even (odd) sublattice produces a renormalized bond in the odd (even) sublattice, as in the static case. 

We can also determine how much the quasi-energy has shifted by isolating $e^{i\theta}$. Since we expect a renormalized Floquet operator of form $e^{i \theta}e^{\phi \tilde{\mathcal O} }$, with $\tilde{\mathcal O} $ some operator, we recover $e^{i\theta}$ by simply taking $\tilde{\mathcal O} \to 0$, or equivalently $\gamma_i \to 0$ for all $i$. Thus,
\begin{equation} \label{eq:quasi-energy}
e^{i \theta} = \frac{(J_L^2 + J_R^2) \cos \Omega + 2i c\sin\Omega}{e^{i 2 c \Omega} - 1},
\end{equation}
where $c = \pm 1$ picks the $P_+$ or $P_-$ branch, respectively. In the limit $\Omega \ll 1$, we can expand this expression and check that it reproduces the energy shift in the static case~\cite{PhysRevX.4.011052}: $\theta \approx c ( \frac{J_L^2 + J_R^2}{2 \Omega} + \Omega ) + \mathcal O (\Omega^2)$, as it should.  \\

Thus, for decimations that do not encounter a domain wall, we obtain a straightforward generalization of the usual RSRG decimation rules for the static Ising chain. Note that, if we are within a $\pi$-domain, we can replace the long-range $\pi$-pulse by a product of short range $\pi$-pulses at any time by just reversing the factoring argument above section A (see Figure~\ref{fig:factoring}). This tells us that after we decimate within a $\pi$-region, we are actually decimating the bond towards $\pi$. Therefore, the domain type is maintained under renormalization.

\subsubsection{Decimating near a domain wall}

\begin{figure}
\includegraphics[width=\columnwidth]{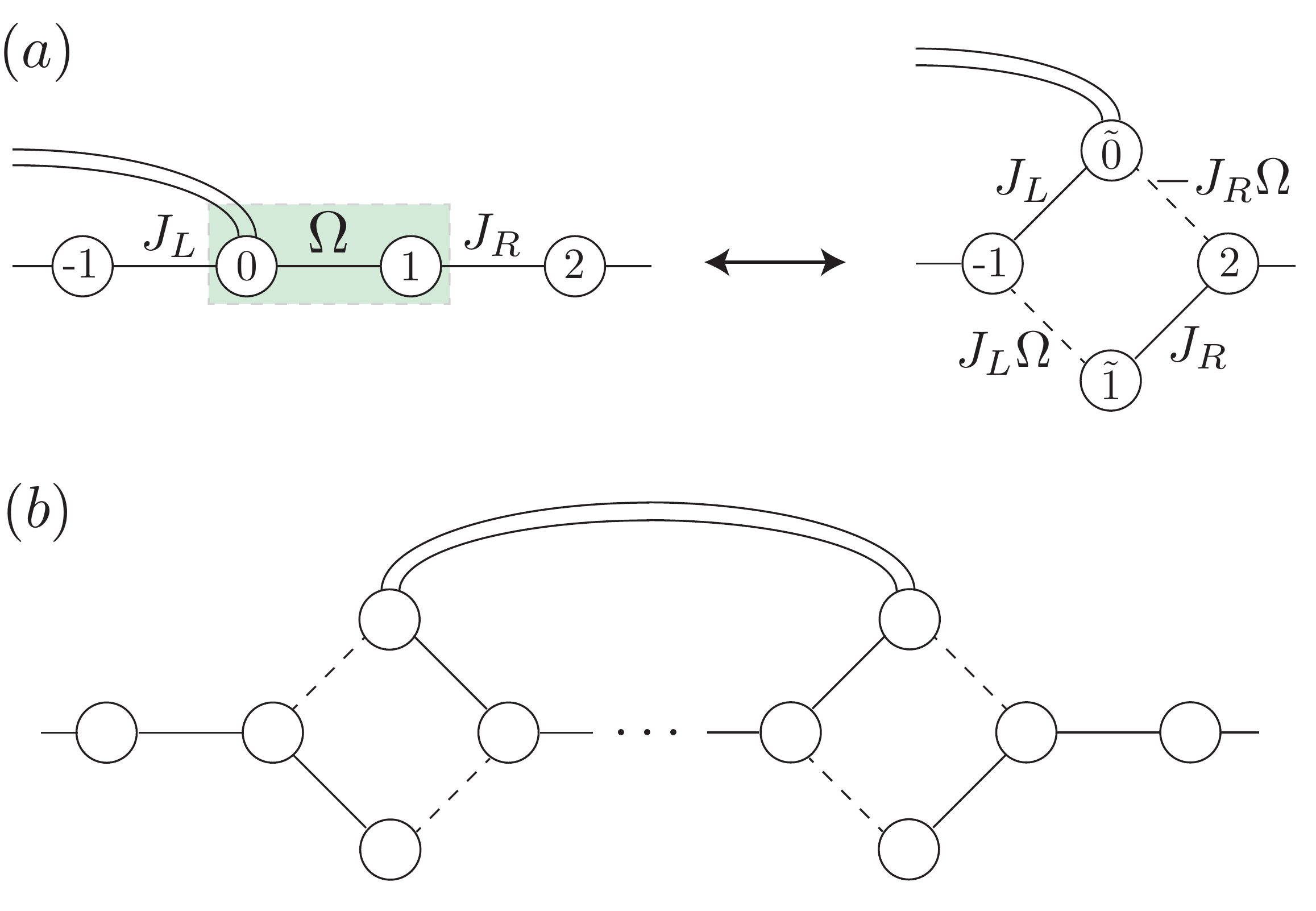} \\
\caption{\label{fig:dw} (a) An exact transformation of each domain wall reveals that we can view the $\pi$ and 0 puddles as being connected by 2nd order terms (see below for details). We will never decimate these because $J_R \Omega < J_R$ and $J_L \Omega < J_L$, so at every step $J_L$ or $J_R$ dominate. (b) The chain after the exact transformations of factoring out $\pi$-pulses and rotating the Majoranas at every domain wall. Approximating $\cos\Omega \approx 1, \sin \Omega \approx \Omega$ shows that the puddles are decoupled at leading order.}
\end{figure}

Now let us consider the interesting case where we want to decimate a bond on the domain wall between a 0-domain and a $\pi$-domain. Let us imagine that the strongest bond in the chain lies at the domain wall between a 0-domain and a $\pi$-domain. That is, let the bonds $J_j$ with $-L<j<0$ be near $\pi/2$, and other bonds near 0. The Floquet operator is
\begin{equation}
F = \ldots e^{\Omega \gamma_0 \gamma_1} \gamma_{-L} \gamma_0 e^{J_L \gamma_{-1}\gamma_0 + J_R \gamma_1 \gamma_2} \ldots 
\end{equation}
where we have pulled out the $\pi$-pulses within the $\pi$-domain. Focusing on the domain wall at site $0$, note that the Floquet operator in this vicinity is $e^{\Omega \gamma_0 \gamma_1} \gamma_{-L} \gamma_0 e^{J_L \gamma_{-1}\gamma_0}$. Let us assume that $\Omega > J_L$; then naively we may wish to identify the ``strong'' piece as $F_0 = e^{\Omega \gamma_0 \gamma_1} \gamma_{-L} \gamma_0$. However, the eigenvalues of this operator actually {\it do not depend} on $\Omega$, so perturbation theory in $\Omega^{-1}$ is not well-controlled. To see this, note that the middle piece can be factored as $ \exp(\Omega \gamma_0 \gamma_1) \gamma_{-L} \gamma_0 =\gamma_{-L} (\gamma_0 \cos\Omega - \gamma_1 \sin \Omega)$. We can define rotated Majorana operators
\begin{equation}
\begin{pmatrix} \tilde \gamma_0 \\
\tilde \gamma_1 \end{pmatrix} = \begin{pmatrix} \cos\Omega & -\sin \Omega \\
\sin \Omega & \cos \Omega \end{pmatrix} \begin{pmatrix} \gamma_0 \\ 
\gamma_1 \end{pmatrix} \approx \begin{pmatrix} 1 & -\Omega \\
\Omega & 1 \end{pmatrix}  \begin{pmatrix} \gamma_0 \\ 
\gamma_1 \end{pmatrix} 
\end{equation}
since $\Omega \ll 1$ (we factored out all $\pi$-pulses). One can verify that this gives a new set of Majorana fermions: $\tilde \gamma_{0,1}^\dagger = \tilde \gamma_{0,1}$, $\tilde \gamma_{0,1}^2 = 1$, $\{\gamma_i, \tilde \gamma_{0,1}\}_{i\not=0,1} = 0$, and $\{\tilde \gamma_0, \tilde \gamma_1\} = 0$. Rewriting in terms of these variables,
\begin{align}
F = \ldots \gamma_{-L} \tilde \gamma_0 \exp(&J_L \gamma_{-1} \tilde \gamma_0 + J_L \Omega \gamma_{-1} \tilde \gamma_1 
 - J_R \Omega \tilde \gamma_0 \gamma_2 + J_R \tilde \gamma_1 \gamma_2) \ldots,
\end{align}
where we have eliminated all couplings of strength $\Omega$ in favor of weaker terms.
This transformed picture is diagrammed in Figure~\ref{fig:dw}(a). We note that we could have just as easily performed a similar rotation based around $J_L \gamma_{-1} \gamma_0$, but find it more convenient to always choose to rotate the bond that is closer to 0 than to $\pi$. In Ref.~\citenum{potter_classification_2016}, a similar rotation argument was given to demonstrate that bilinear couplings between zero- and $\pi$-quasienergy Majorana fermions cannot change their quasienergy.

\begin{figure}
\includegraphics[width=\columnwidth]{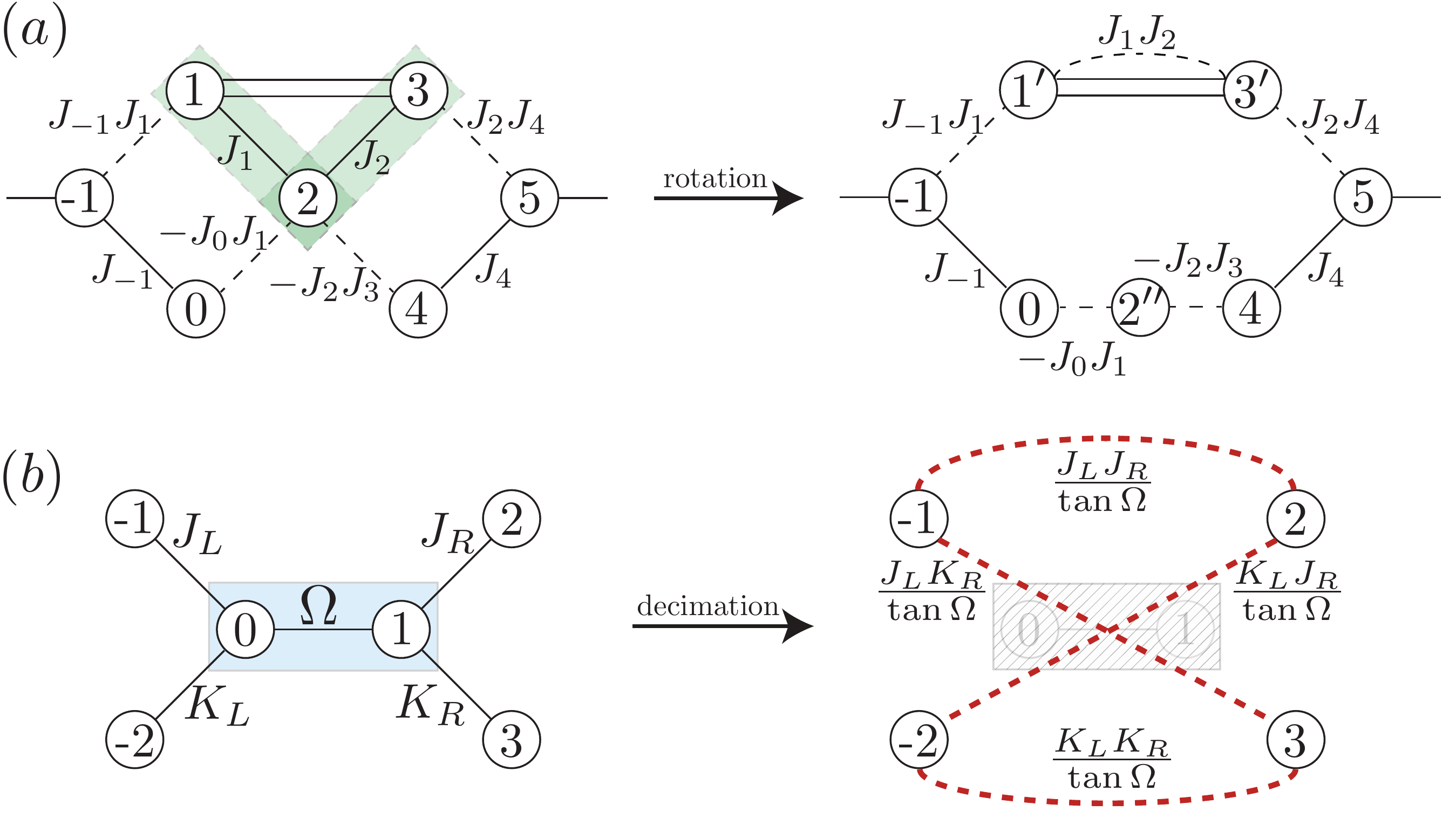}
\caption{\label{fig:pathway} (a) An even-length domain, after it has been decimated down to only two bonds, can be rotated into this two-chain form without decimating. (b) When decimating the central bond in the setup above, every three-step pathway through the decimated bond gets renormalized according to $M_L M_R/\tan \Omega \approx M_L M_R/\Omega$, where $M_{L,R}$ are the bonds to the left and right respectively. This type of decimation occurs with odd-length domains.}
\end{figure}

\begin{figure*}[ht!]
\includegraphics[width=\columnwidth]{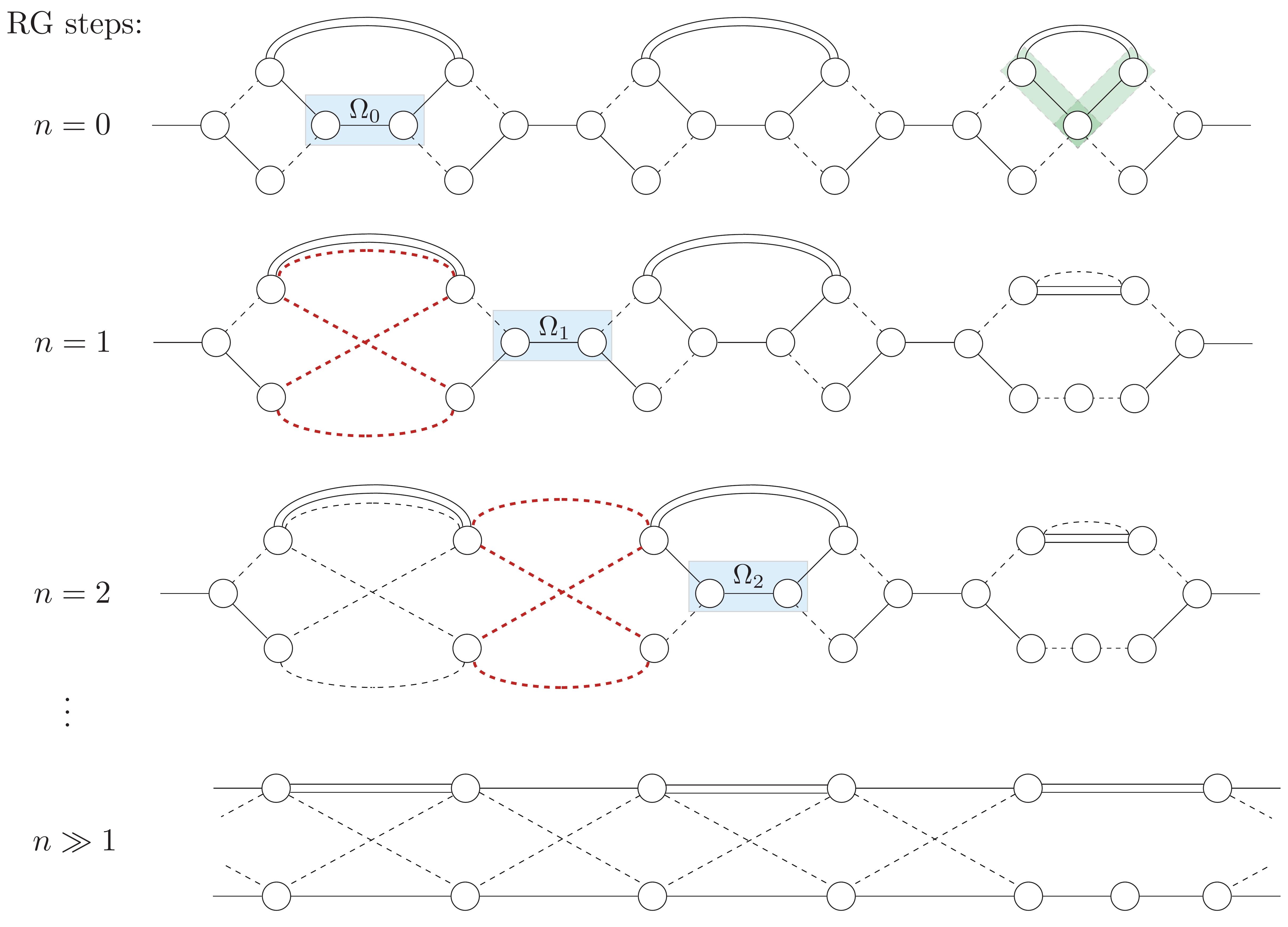}
\caption{\label{fig:two_chains} After the 0 and $\pi$-domains have been decimated, we are left with a configuration of similar form to the chain in (1). As we decimate the central couplings of each puddle, we renormalize to the chain in step (2), and so on. This leads to two linked chains.}
\end{figure*}

The above argument applies equally well to the bond $J_L \gamma_{-1}\gamma_0$, so we find that neither bond touching the domain wall Majorana should be (naively) decimated. This has a physical interpretation: decimating either bond is akin to decimating the topological edge mode that lies at this interface between a 0 region and a $\pi$ region, so to lowest order is at exactly $\pi$ quasi-energy independent of the surrounding coupling strengths. Since these edge modes are at 0 or $\pi$ quasi-energy, perturbation theory is not well-controlled when attempting to decimate them. Since, as previously argued~\cite{Berdanier201805796}, these topological edge modes control the criticality, it would not be sensible to decimate them outright, and the surrounding bonds will only be decimated at higher order, later in the RG.

\subsubsection{Decimations between the $0$- and $\pi$-chains}

Let us run the above RG until all of the 0-domain and $\pi$-domains have been decimated and we are left with many domain walls. Note first that we must distinguish between domains that began with an even number of bonds versus those that began with an odd number. Since a decimation removes two Majoranas, after many decimations an odd-length puddle will be left with three bonds, while an even-length puddle will be left with only two. We find it convenient to first perform a rotation on the two-bond puddles, described below. 

Let us consider a two-bond $\pi$-domain, namely $J_j \gamma_j \gamma_{j+1}$ which has $J_j \approx \pi$ for $j=0,1$ and $J_j$ near 0 otherwise. We first rotate this domain's domain walls as above, constructing $\tilde \gamma_{0,1} = \gamma_{0,1} \pm J_0 \gamma_{1,0}$ and $\gamma_{1,2}' = \gamma_{1,\tilde 2} \mp J_1 \gamma_{\tilde 2,1}$. This gives a domain structure as diagrammed in Fig.~\ref{fig:dw}b and the left part of Fig.~\ref{fig:pathway}(a), dropping tildes and primes for clarity. Then let us define two consecutive rotations, $\gamma_{0}' = \gamma_{0} + J_0 \gamma_{1}$, $\gamma_{1}' = \gamma_{1} - J_0 \gamma_{0}$, followed by $\gamma_{1}'' = \gamma_{1}' - J_1 \gamma_{2}$, $\gamma_{2}'' = \gamma_{2} + J_1 \gamma_{1}'$. Dropping third order terms, this leads to the following bonds: $J_{-1}J_1 i \gamma_{-1} \gamma_1$, $(\pi/2 + J_1 J_2) i \gamma_1' \gamma_3'$, $J_{-1} i \gamma_{-1} \gamma_0$, $-J_0 J_1 i \gamma_0 \gamma_2''$, $-J_2 J_3 \gamma_2'' \gamma_4$, $J_4 i \gamma_4 \gamma_5$, and $J_2 J_4 i \gamma_3' \gamma_5$, as diagrammed in the right half of Fig.~\ref{fig:pathway}(a). Therefore, with a two-bond puddle it is sensible to rotate the bonds so that the puddle is of two-chain form and each bond can be straightforwardly decimated. 

Now consider an odd-length puddle. As we run the RG, we avoid decimating the domain-wall bonds, so the puddle size shrinks until there are just three bonds remaining. These 3-bond puddles will be left with a configuration similar to Figure~\ref{fig:two_chains} at RG step 1. We now want to decimate the central bond of one of the puddles. Let us calculate this rule, diagrammed in Figure~\ref{fig:pathway}(b). The Floquet operator is of the form
\begin{equation}
F = \ldots e^{\Omega \gamma_0 \gamma_1} e^{J_L \gamma_{-1} \gamma_0 + K_L \gamma_{-2} \gamma_0+J_R \gamma_1 \gamma_2 + K_R \gamma_1 \gamma_3} \!\ldots\!\!
\end{equation}
so we identify
\begin{align}
F_0 &= e^{\Omega \gamma_0 \gamma_1} \\
V &= -i J_L \gamma_{-1} \gamma_0 - i K_L \gamma_{-2} \gamma_0 - i J_R \gamma_1 \gamma_2 - i K_R \gamma_1 \gamma_3\nonumber.
\end{align}

Applying the machinery from above, we arrive at a renormalized Floquet operator where every 3-step pathway through the central bond is renormalized as $M_L M_R / \tan\Omega$, where $M_{L,R}$ are the bonds on the left and right of $\Omega$, respectively. That is,
\begin{align}
\tilde F = \ldots \exp \Bigg( &\frac{J_L J_R}{\tan\Omega} \gamma_{-1} \gamma_2 + \frac{J_L K_R}{\tan\Omega} \gamma_{-1} \gamma_3
+ \frac{K_L J_R}{\tan\Omega} \gamma_{-2} \gamma_2 + \frac{K_L K_R}{\tan\Omega} \gamma_{-2} \gamma_3 \Bigg)\ldots
\end{align}
with quasi-energy shift
\begin{equation}
e^{i \theta} = \frac{\cos\Omega (J_L^2 + J_R^2 + K_L^2 + K_R^2) + 2 i c \sin \Omega}{e^{i 2 c \Omega} - 1},
\end{equation}
where again $c=\pm 1$ specifies the branch choice. Applying this decimation rule repeatedly will lead to two connected chains, as shown in Figure~\ref{fig:two_chains}. 

Note that the bottom chain has all couplings near 0 and hence is near 0 quasi-energy, while the top chain has every other coupling near $\pi$, and hence is at $\pi$-quasi-energy.\footnote{This can be verified by comparison with the microscopic phase diagram of the model, as shown in Ref.~\citenum{Berdanier201805796}} We might worry that the couplings connecting the chains will need to be decimated. However, they cannot be decimated for the same reason that we could not directly decimate a bare domain-wall coupling earlier: each of these links has a coupling near $\pi$ to one side and a coupling near 0 to the other. Thus, at leading order the energetics are independent of these couplings, and we can perform a Majorana rotation as above to push them to higher order. Once we do this, we find that the higher order couplings actually {\it still} connect a 0 coupling on one side to a $\pi$ coupling on the other, and can be rotated yet again to push these couplings to an even higher order. Ultimately, if we continue to rotate ad infinitum, we will decouple the chains without doing any decimations. This leaves us with two effectively decoupled chains, one at 0 and one at $\pi$ quasi-energy. 

{
We can write flow equations, using the rules above, for the four distributions in the problem: the distributions of bonds and fields near 0 and $\pi$. Define logarithmic variables $\Gamma\equiv \log(\Omega_I / \Omega)$, which sets the overall RG scale for some arbitrary initial scale $\Omega_I$; $\zeta^{0,\pi} \equiv \log(\Omega/J^{0,\pi})$, where $J$ is from the even sublattice in terms of Majoranas; and $\beta^{0,\pi} \equiv \log(\Omega/h^{0,\pi})$, where $h$ is from the odd sublattice. These last four logarithmic variables have four associated coupling distributions: $P_{\beta,\zeta}^{0,\pi}$. Then the above decimation rules translate to sum rules in terms of logarithmic variables {at strong disorder}: $\tilde \beta^\theta = \zeta^\theta_L + \zeta^\theta_R$, $\tilde \zeta^\theta = \beta^\theta_L + \beta^\theta_R$, with $\theta = 0, \pi$ indexing the two chains. These rules give rise to two coupled RG flow equations for each chain~\cite{PhysRevB.51.6411}
\begin{align}
\pd{P^\theta_\zeta}{\Gamma}(\zeta^\theta) &= \pd{P^\theta_\zeta}{\zeta^\theta}(\zeta^\theta) + P_\beta^\theta(0) \int_0^\infty d\bar\zeta^\theta P_\zeta^\theta(\bar \zeta^\theta) P_\zeta^\theta(\zeta^\theta - \bar \zeta^\theta) + P_\zeta^\theta(\zeta^\theta) [ P_\zeta^\theta(0) - P_\beta^\theta(0) ], \nonumber \\
\pd{P^\theta_\beta}{\Gamma}(\beta^\theta) &= \pd{P^\theta_\beta}{\beta^\theta}(\beta^\theta) + P_\zeta^\theta(0) \int_0^\infty d\bar\beta^\theta P_\beta^\theta(\bar \beta^\theta) P_\beta^\theta(\beta^\theta - \bar \beta^\theta) + P_\beta^\theta(\beta^\theta) [ P_\beta^\theta(0) - P_\zeta^\theta(0) ],
\end{align}
with $\theta = 0,\pi$. These equations give rise to the usual infinite-randomness fixed-point distributions $\tilde P_{\beta}^\theta(\beta^\theta) = \frac{1}{\Gamma} e^{-\beta^\theta / \Gamma}$, $\tilde P_{\zeta}^\theta(\zeta^\theta) = \frac{1}{\Gamma} e^{-\zeta^\theta / \Gamma}$ {at criticality}, showing flow to two IRFPs at 0 and $\pi$ quasienergy.}

\begin{figure}
\centering{
\includegraphics[width=0.6\columnwidth]{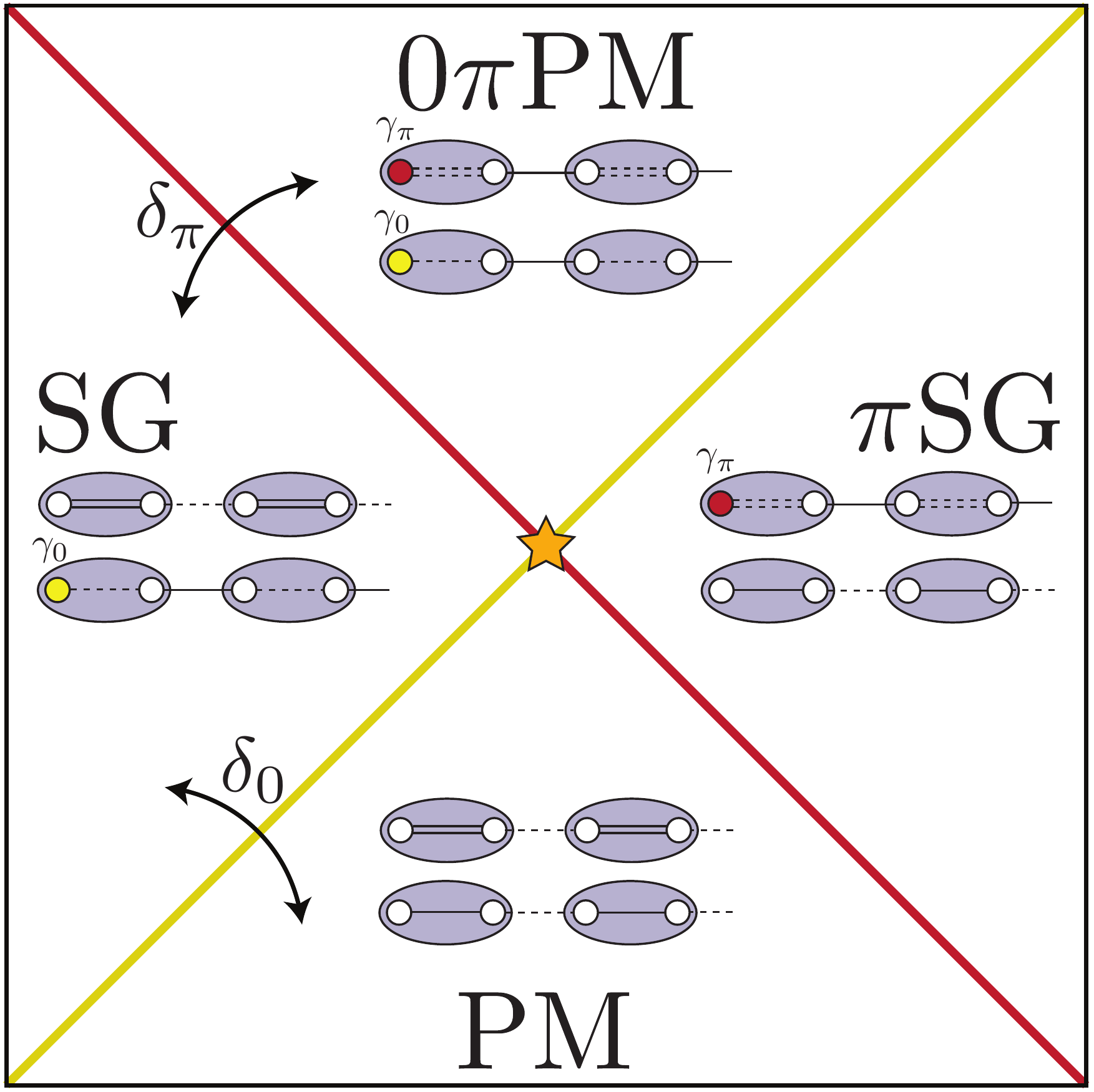}
\caption{\label{fig:phases} Cartoon of the four possible phases and their transitions as deduced from the strong-disorder RG flow to two decoupled Majorana chains near 0 and $\pi$ quasienergy. The phases are controlled by the dimerization of these chains $\delta_{0,\pi}$. The PM has both chains trivially dimerized; the SG has the 0-chain topologically dimerized and the $\pi$-chain trivially dimerized; the $\pi$SG has the 0-chain trivially dimerized and the $\pi$-chain topologically dimerized; and the $\pi$SG has both chains topologically dimerized (insets). Both chains can be independently tuned to Ising-class criticality (yellow and red lines), resulting in a twice-Ising multicritical point (orange star).
}
}
\end{figure}

Now that we have flowed to two decoupled Ising chains, the four possible phases of the model become clear: each chain can be independently dimerized, with a dimerization parameter $\delta_{0,\pi}$ controlling the phases. If the 0 ($\pi$)-chain is dimerized in the trivial pattern $\delta_{0(\pi)} < 0$, there is no edge mode, while if the 0 ($\pi$)-chain is dimerized in the topological pattern $\delta_{0(\pi)} > 0$, there is an edge Majorana mode at 0 ($\pi$) quasienergy. The four phases are thus identified by the sign of the dimerization patterns at $0,\pi$; accordingly we may label the phases by $(\delta_0,\delta_\pi)$ where $\delta_{0\pi} = +,-$. With this convention,  we identify the PM as both chains trivially dimerized $(-,-)$; the SG as the 0 chain topologically dimerized and the $\pi$-chain trivially dimerized, $(+,-)$; the $\pi$SG as the 0-chain trivially dimerized and the $\pi$-chain topologically dimerized $(-,+)$; and the $0\pi$PM as both chains topologically dimerized $(+,+)$. The critical lines are then set by $\delta_{0,\pi} = 0$ with both vanishing at the multicritical point of the model (see Fig.~\ref{fig:phases}). One can tune between these phases microscopically by adjusting the ratio of couplings near 0 and $\pi$.~\cite{Berdanier201805796}

This two-chain structure also elucidates the model's (multi)criticality. Each chain can be tuned to criticality independently, and each will be in the Ising universality class~\cite{PhysRevB.51.6411,PhysRevLett.69.534}. In particular, this implies that the coupling fixed-point distribution will be in the form of a stretched exponential $\sim e^{-1/w}$, where $w$ sets the disorder strength and $w\to \infty$ under renormalization when the system is at criticality ($\delta = 0$). This infinite randomness fixed point has dynamical exponent $z = 1 / 2 \abs{\delta} \to \infty$, characteristic of slow, glassy scaling. As one tunes slightly away from criticality, one finds that the correlation length diverges as $\xi \sim \abs{\delta}^{-\nu}$, with $\nu = 2$ or $\nu = 1$ for average ($\avg{\mathcal O}$) or typical ($e^{\avg{\log\mathcal O}}$) quantities, respectively. Further results include the critical spin-spin correlation function scaling $\avg{ \expectation{\sigma_j^z \sigma_{j+x}^z}} \sim 1/x^{2-\phi}$, with $\phi = (1+\sqrt{5})/2$ the golden ratio, and the entanglement entropy scaling $\overline{S_\ell} = (\tilde c / 6) \log \ell$ with $\tilde c = (1/2) \ln 2$ with open boundary conditions for a cut from the boundary of length $\ell$.~\cite{PhysRevLett.93.260602} At the system's multicritical point, we find that the chain flows under renormalization to two chains, one at 0 and the other at $\pi$ quasi-energy, each in the random Ising universality class.

\subsection{Interactions \label{sec:ints}}

Now that we have addressed the free problem, let us return to adding small but finite interactions $U \ll 1$. As a reminder, we introduced interaction terms such that the first piece of the drive was $H_1 = \sum_i h_i \sigma_i^x + U^{xx}_i \sigma_i^x \sigma_{i+1}^x$ and the second piece was $H_2 = \sum_i J_i \sigma_i^z \sigma_{i+1}^z + U^{zz}_i \sigma_i^z \sigma_{i+2}^z$. In terms of Majoranas, these interactions are statistically\footnote{Absent disorder, the self-duality would be exact, but here it is only true in a statistical sense, i.e. it is to be understood as a duality map of various observables that occurs when the different couplings are drawn from the same distribution.}  {self-dual}
 under the usual Ising bond-field duality that interchanges $\sigma^x_i \leftrightarrow \sigma^z_i \sigma^z_{i+1}$. Therefore bond-field duality exchanges the operator content of $H_1$ and $H_2$, while preserving coefficients. Upon performing a Jordan-Wigner transformation for these interaction terms, and redefining $J_{2i} \equiv h_i$, $J_{2i+1} \equiv J_i$, $U_{2i} \equiv U^{xx}_i$, $U_{2i+1} \equiv U^{zz}_i$, the full Floquet operator reads 
\begin{align}
F = &e^{\sum_j J_{2j} i \gamma_{2j} \gamma_{2j+1} - U_{2j} \gamma_{2j} \gamma_{2j+1} \gamma_{2j+2} \gamma_{2j+3} } \nonumber \\
&\times e^{\sum_j J_{2j+1} i \gamma_{2j+1} \gamma_{2j+2} - U_{2j+1} \gamma_{2j+1} \gamma_{2j+2} \gamma_{2j+3} \gamma_{2j+4} }.
\end{align}

Let us first discuss decimating deep in one of the 0-domain or $\pi$-domains, such that all bonds are near 0. Then let us say that the strongest bond is $J_3$, such that 
\begin{align}
F_0 &= e^{J_3 \gamma_3 \gamma_4}, \\
V &= i J_0 \gamma_0 \gamma_1 + i J_1 \gamma_1 \gamma_2 + i J_4 \gamma_4 \gamma_5 + i J_5 \gamma_5 \gamma_6 + i J_6 \gamma_6 \gamma_7 \nonumber \\
& - U_0 \gamma_{0123} - U_1 \gamma_{1234} - U_2 \gamma_{2345} - U_3 \gamma_{3456} - U_4 \gamma_{4567},
\end{align}
where $\gamma_{ijk\ldots l} \equiv \gamma_i \gamma_j \gamma_k \ldots \gamma_l$ for short. Performing the Floquet Schrieffer-Wolff transformation and projecting onto the subspace $i\gamma_3 \gamma_4 = c$ with $c=\pm 1$, we find, after some algebra, that the renormalized operator is 
\begin{align}
\tilde F = &\ldots e^{i \theta} e^{ \tilde J_0 \gamma_0 \gamma_1 + \tilde J_2 \gamma_2 \gamma_5 + \tilde J_6 \gamma_6 \gamma_7 +  \tilde U_0 i \gamma_{0125} + \tilde U_2 i \gamma_{2567} + \tilde W \gamma_{012567} } \nonumber \\ 
&\times e^{ \tilde J_1 \gamma_1 \gamma_2 + \tilde J_5 \gamma_5 \gamma_6 + \tilde U_1 i \gamma_{1256} } \ldots 
\end{align}
with the following decimation rules:
\begin{align}
&\tilde J_0 = J_0 + c \frac{J_2 U_0}{\tan J_3}, 
~~~ &\tilde J_1 &= J_1 + c U_1, \nonumber \\
&\tilde J_2 = \frac{J_2 J_4}{\tan J_3} + U_2, 
~~~ &\tilde J_5 &= J_5 + c U_3, \nonumber \\
&\tilde J_6 = J_6 + c \frac{J_4 U_4}{\tan J_3}, 
~~~ &\tilde U_0 &= \frac{J_4}{\tan J_3} U_0, \nonumber \\
&\tilde U_1 = 0, 
~~~ &\tilde U_2 &= \frac{J_2}{\tan J_3} U_4, \nonumber \\
&\tilde W = \frac{U_0 U_4}{\tan J_3},
\end{align}
and overall quasi-energy shift
\begin{align}
e^{i \theta} = &\frac{1}{2} e^{-i c J_3} \left[ 2- i c \frac{J_2^2 + J_4^2 + U_0^2 + U_4^2}{\tan J_3} -J_0^2 - (J_1 + U_1)^2- (J_5 + U_3)^2 - J_6^2 - U_2^2 \right].
\end{align}

First, note that in the limit $\tan J_3 \approx J_3$, we recover the decimation rules of the static case,\footnote{These static case rules are derived in the presence of only $\sigma^x_i \sigma^x_{i+1}$ interactions in Appendix B of Ref.~\citenum{PhysRevX.4.011052}, though we note that the published version contains several sign errors.} as we must. A six-fermion term $-i \tilde W \gamma_{012567}$ is generated in the Schrieffer-Wolff transformation that is second order in the interaction strength, which is already taken to be weak. The generation of fermion strings from RSRG methods has already been explored in the static case~\cite{PhysRevLett.112.217204}, where it was found that as the RG progresses, fermion strings of length $2m$ ($\sigma^x$ strings of length $m$) were generated at order $J (J/h)^{m-2}$. Since these coefficients are exponentially decaying with $m$, they can be safely discarded, though one can keep track of them if one wishes. We find that they exponentially decay here as well, so we neglect these strings and keep only fermion bilinears and four-fermion interaction terms. 

Now, consider the role of interaction terms at the domain wall. Let us rotate $\gamma_{2,3}$ to form $\tilde \gamma_2 = \gamma_2 + J_{2} \gamma_3$, $\tilde \gamma_3 = \gamma_3 - J_{2} \gamma_2$. Then a four-fermion interaction term $U_0 \gamma_{0123}$ will rotate to $U_0 \gamma_0 \gamma_1 (\tilde \gamma_2 - J_2 \tilde \gamma_3)(\tilde \gamma_3 + J_2 \tilde \gamma_2) = U_0 \gamma_0 \gamma_1 ( \tilde \gamma_2 \tilde \gamma_3 - J_2 \tilde \gamma_3^2 + J_2 \tilde \gamma_2^2 - J_2^2 \tilde \gamma_3 \tilde \gamma_2 ) = U_0 \gamma_0 \gamma_1 \tilde \gamma_2 \tilde \gamma_3 ( 1 + J_2^2)$. Therefore, the effect of the rotation is simply to renormalize the interaction to second order, as $U_0 (1+J_2^2) \gamma_{0 1 \tilde{2} \tilde{3}} \approx U_0 \gamma_{0 1 \tilde{2} \tilde{3}}$. Thus if the interaction term involves both of the rotated Majorana operators, the interaction is essentially unchanged. If it only involves one, on the other hand, we generate a new longer-ranged interaction term at one higher order: $U_{-1} \gamma_{-1012} = U_{-1}  \gamma_{-101\tilde 2} + U_{-1} J_2 \gamma_{-101\tilde 3}$. Finally, if an interaction term spans two separate rotations we obtain the following: $U_3 \gamma_{3456} = U_3 (\tilde \gamma_3 + J_2 \tilde \gamma_2)\gamma_4 \gamma_5 (\tilde \gamma_6 - J_2 \tilde \gamma_7) = U_3 \gamma_{\tilde{3} 45 \tilde{6}} + U_3 J_2 \gamma_{\tilde{2} 45 \tilde{6}} - U_3 J_6 \gamma_{\tilde{3} 45 \tilde{7}} - J_{2} J_{6} \gamma_{\tilde{2} 45 \tilde{7}}$. That is, we couple the Majoranas involved in both domain walls at leading order $\mathcal O(U)$ for the two nearest, one higher order $\mathcal O(UJ)$ for the next nearest, and one higher order still $\mathcal O(UJ^2)$ for the farthest away.

Having seen that no fundamentally new interactions are generated by this rotation, let us consider a bilinear decimation in the most general case of an odd-length domain, similar to that shown in Fig.~\ref{fig:pathway}(b). Then the Floquet operator will take the form 
\begin{align*}
  F =& \exp( J_0 \gamma_{01} + J_2 \gamma_{23} + \Omega \gamma_{45} + J_6 \gamma_{67} + J_8 \gamma_{89} + \ldots \\
  &~~~-i U_0 \gamma_{0123} -i U_2 \gamma_{2345} -i U_4 \gamma_{4567} -i U_6 \gamma_{6789} + \ldots ) \\ & \times \exp(J_1 \gamma_{12} + J_3 \gamma_{34} + J_5 \gamma_{56} + J_7 \gamma_{78} + \ldots  - i U_1 \gamma_{1234}
  \\ & ~~~- i U_3 \gamma_{3456} + U_5 \gamma_{5678} + U_7 \gamma_{789,10} + \ldots),
\end{align*}
where $\Omega = J_4$ refers to the strongest coupling. After rotating $\gamma_{2,3}$ and $\gamma_{6,7}$, we find
\begin{align*}
  F = & \exp( J_0 \gamma_{01} + \Omega \gamma_{45} + J_8 \gamma_{89} + \ldots -i U_0 \gamma_{01\tilde{2}\tilde{3}} -i U_{2} \gamma_{\tilde{2}\tilde{3}45}-i U_{4}\gamma_{45\tilde{6}\tilde{7}} -i U_{6}\gamma_{\tilde{6}\tilde{7}89} + \ldots ) \\
  \times &\exp(J_1 \gamma_{1\tilde{2}} + J_{1} J_{2} \gamma_{1\tilde{3}} +J_{3} \gamma_{\tilde{3}4} + J_{2} J_{3} \gamma_{\tilde{2}4} + J_{5} \gamma_{5\tilde{6}}+J_{5} J_{6} \gamma_{5\tilde{7}} + J_{7}\gamma_{\tilde{7}8} + J_{7} J_{6} \gamma_{\tilde{6}8} + \ldots
  \\&~ -i U_1 \gamma_{1 \tilde{2}\tilde{3}4} -i U_3 \gamma_{\tilde{3}45\tilde{6}} -i U_3 J_{2} \gamma_{\tilde{2}45\tilde{6}} - i U_3 J_6 \gamma_{\tilde{3}45\tilde{7}} - i U_3 J_2 J_6 \gamma_{\tilde{2}45\tilde{7}} 
  \\&~-i U_5 \gamma_{5\tilde{6}\tilde{7}8} -i U_7 \gamma_{\tilde{7}89,10} - i U_7 J_6 \gamma_{\tilde{6}89,10}+ \ldots).
\end{align*}
We can therefore write the most general case as 
\begin{align}
F_0 &= e^{\Omega_{45} \gamma_4 \gamma_5}, \nonumber \\
iV &= J_{01} \gamma_{01} + J_{12} \gamma_{12} + J_{13} \gamma_{13} + J_{34} \gamma_{34} + J_{24} \gamma_{24} + J_{56} \gamma_{56} \nonumber \\
&+ J_{57} \gamma_{57} + J_{78}\gamma_{78} + J_{68} \gamma_{68} + J_{89} \gamma_{89}+ \ldots  \nonumber \\
&-i U_{0123} \gamma_{0123} -i U_{1234} \gamma_{1234} - i U_{2345} \gamma_{2345} -i U_{3456} \gamma_{3456} \nonumber \\
&-i U_{2456} \gamma_{2456} -i U_{3457} \gamma_{3457} - i U_{2457} \gamma_{2457} - i U_{4567} \gamma_{4567} \nonumber \\
&-i U_{5678} \gamma_{5678} -i U_{6789} \gamma_{6789} -i U_{789,10} \gamma_{789,10} \nonumber \\
&- i U_{689,10} \gamma_{689,10}+ \ldots,
\end{align}
where we have dropped the tilde's and relabelled the couplings more explicitly as $J_{ij} i \gamma_i \gamma_j$ and $U_{ijkl} \gamma_i \gamma_j \gamma_k \gamma_l$. Though tedious, these decimation rules are straightforward to compute. We find
\begin{align}
&\tilde J_{01} = J_{01}, 
 &\tilde J_{12} &= J_{12} + c \frac{J_{34} U_{1234}}{\tan \Omega_{45}}, \nonumber \\
&\tilde J_{13} = J_{13} - c \frac{J_{24} U_{1234}}{\tan \Omega_{45} }, 
~~~ &\tilde J_{23} &= c U_{2345}, \nonumber \\
&\tilde J_{26} = \frac{J_{24} J_{56}}{\tan \Omega_{45}} + c U_{2456},
~~~ &\tilde J_{27} &= \frac{J_{24}J_{57}}{\tan{ \Omega_{45}}} + c U_{2457}, \nonumber \\
&\tilde J_{36} = \frac{J_{34} J_{56}}{\tan \Omega_{45}} + c U_{3456},
~~~ &\tilde J_{37} &= \frac{J_{34} J_{57} }{\tan \Omega_{45}} + c U_{3457}, \nonumber \\ 
&\tilde J_{67} = -c U_{6789},
~~~ &\tilde J_{68} &= J_{68} - c \frac{J_{57} U_{5678}}{\tan \Omega_{45}}, \nonumber \\
&\tilde J_{78} = J_{78} + c \frac{J_{56} U_{5678}}{\tan \Omega_{45}},
~~~ &\tilde J_{89} &= J_{89}, \nonumber \\
~~~ &\tilde J_{9,10} = J_{9,10}, & & \nonumber \\
&\tilde U_{0123} = U_{0123},
~~~ &\tilde U_{1236} &= \frac{J_{56} U_{1234}}{\tan \Omega_{45}}, \nonumber \\
&\tilde U_{1237} = \frac{J_{57} U_{1234}}{\tan \Omega_{45}}, 
~~~ &\tilde U_{2678} &= \frac{J_{24} U_{5678}}{\tan \Omega_{45}}, \nonumber \\
&\tilde U_{3678} = \frac{J_{34} U_{5678}}{\tan \Omega_{45}}, 
~~~ &\tilde U_{689,10} &= U_{689,10},  \nonumber \\
&\tilde U_{789,10} = U_{789,10},
~~~ &\tilde W_{123678} &= \frac{U_{1234} U_{5678}}{\tan \Omega_{45}}.
\end{align}
Assuming $U \ll J$ as before, most of the interaction-induced dressing of the bilinears leads to irrelevant higher-order corrections. Interestingly, the interaction terms generate a new bilinear coupling between the two rotated sites, namely $\tilde J_{23} = c U_{2345}$ and $\tilde J_{67} = -c U_{6789}$, where the new bilinear is entirely due to the interaction term. However, note that this is still simply a bilinear coupling between the $0$ and $\pi$ quasi-energy chains, so again may be rotated away, and cannot affect the universality class of the transition.  

We have now computed how the couplings and quasi-energies are affected by the introduction of interaction terms. We find that the structure of the RG rules is very similar to the static case, except that since $\Omega$ is treated non-perturbatively, the denominators are $\tan \Omega$ instead of $\Omega$. As in the static case, then, interaction terms within each chain are irrelevant~\cite{PhysRevB.51.6411,PhysRevLett.69.534}, since they decay in strength as $\overline{\log(\Omega/U)} \sim \Gamma^{\phi}$ with $\Gamma \equiv \log \Omega_0 / \Omega$ the RG scale and $\phi=(1+\sqrt{5})/2$ the golden ratio, compared with $\overline{\log(J / \Omega) } \sim \Gamma$ for bilinear decimations. Tracking the interactions at the domain wall, we find that some interaction terms are generated that couple the two chains in the late stages of the RG; though by the above argument the intra-chain interactions are irrelevant, it is natural to wonder whether the inter-chain interactions are irrelevant as well. We show this by noting that our model maps explicitly onto a disordered XYZ model, $H_{XYZ} = \sum_i J_i^{XX} \sigma_i^x \sigma_{i+1}^x + J_i^{YY} \sigma_i^y \sigma_{i+1}^y + J_i^{ZZ} \sigma_i^z \sigma_{i+1}^z$. To derive the mapping, first apply a Jordan-Wigner transformation $\sigma_j^x = i \gamma_{2j} \gamma_{2j+1}$, $\sigma_j^y =  (\prod_{l < j} i \gamma_{2l} \gamma_{2l+1}) \gamma_{2j+1}$, $\sigma_j^z = (\prod_{l < j} i \gamma_{2l} \gamma_{2l+1}) \gamma_{2j}$. This implies $\sigma_j^x \sigma_{j+1}^x = - \gamma_{2j} \gamma_{2j+1} \gamma_{2j+2} \gamma_{2j+3}$, $\sigma_j^y \sigma_{j+1}^y = -i \gamma_{2j} \gamma_{2j+3}$, $\sigma_j^z \sigma_{j+1}^z = i \gamma_{2i+1} \gamma_{2i+2}$. This gives $H_{XYZ} = \sum_i J_i^{ZZ} i \gamma_{2i+1} \gamma_{2i+2} - J_i^{YY} i \gamma_{2i} \gamma_{2i+3} - J_i^{XX} \gamma_{2i} \gamma_{2i+1} \gamma_{2i+2} \gamma_{2i+3}$. This is manifestly of the form of two disordered Majorana chains coupled by density-density interactions, i.e. our model. Interactions in the XYZ model were studied and found to be irrelevant by Slagle et. al.~\cite{PhysRevB.94.014205}; hence, we conclude that these interactions also cannot change the universal critical physics of the free model above.

Note that this argument shows that interactions are irrelevant at the infinite-randomness fixed point. For ground states, it is possible to show that even weak disorder ultimately flows to this infinite-randomness fixed point, even in the presence of interactions. For excited states (i.e., RSRG-X) and for Floquet systems, however, there remains the possibility that rare many-body resonances can disrupt the  flow to infinite randomness, resulting ultimately in thermalization~\cite{PhysRevB.95.155129}. Such resonances, which are enabled by interactions, are not captured by the RSRG and therefore not ruled out by our treatment above which relied on proximity to the infinite-randomness fixed point. Therefore, we must always leave open the possibility that the ultimate fate of the critical/multi-critical system on the longest length and time scale is to thermalize to the infinite-temperature Gibbs state. This would be characterized by thermal correlations and volume-law ($\propto \ell$) entanglement scaling rather than the scaling discussed above. Nevertheless, for sufficiently strong disorder the dynamics on all reasonable (i.e., experimentally or numerically accessible) length and time scales will be controlled by the infinite-randomness fixed point, with a crossover to thermalization on exponentially long scales. Indeed, a definitive determination of which of these two scenarios occurs in the thermodynamic limit is outside the capability of current numerical simulations. We note that previous studies~\cite{khemani_prl_2016} of level statistics of the Floquet spectrum at criticality {for small system sizes and a given disorder strength} showed results intermediate between the Poisson statistics characteristic of localized systems and the circular orthogonal ensemble expected of thermalizing systems, {indicating that a systematic analysis of the dependence on disorder strength for larger system sizes would be needed.}

\subsection{Concluding remarks\label{sec:discussion}}

In this paper, we have introduced a general method for performing real space renormalization for Floquet systems based on a generalization of the Schrieffer-Wolff transformation to Floquet unitaries. We have applied it to study the criticality in a paradigmatic model hosting several Floquet MBL phases and phase transitions -- the driven Ising model -- finding agreement with an earlier picture based on topological domain walls~\cite{Berdanier201805796}. Indeed, this calculation can be viewed as providing a microscopic derivation of the topological domain wall argument from a more mathematical perspective. 

This Floquet RSRG procedure can be readily applied to many one-dimensional Floquet MBL systems. For instance, a natural application would be to the periodically driven parafermion chain~\cite{PhysRevB.94.045127}, whose evolution operator has a similar structure to that of the driven Kitaev chain considered in this work. Importantly, this method does not depend on {\it a priori} knowledge of the phase structure of a given model, which can be deduced from the flow of the Floquet operator under renormalization. It is also applicable in cases where relative topological edge modes between different Floquet MBL phases either do not exist or are not known. In this vein, another natural problem to which this method may be fruitfully applied is to periodically driven anyon chains~\cite{PhysRevLett.114.217201}. {Finally, it is not always clear how to pick out a `strong' and `weak' piece of the Floquet unitary operator for every periodically driven system; for instance, when we drive the Ising model sinusoidally instead of in a piecewise fashion, it is no longer obvious how to determine $F_0$. We note that our previous topological arguments suggest that such a separation is indeed possible at late RG times~\cite{Berdanier201805796}. For now, we leave this factorization, as well as further applications and generalizations of our method, to future work. }

\section{Appendices}

\subsection{Numerical precision and convergence}\label{sec:numericalPrecision}

\begin{figure}
	\centering
	\includegraphics[width = \columnwidth]{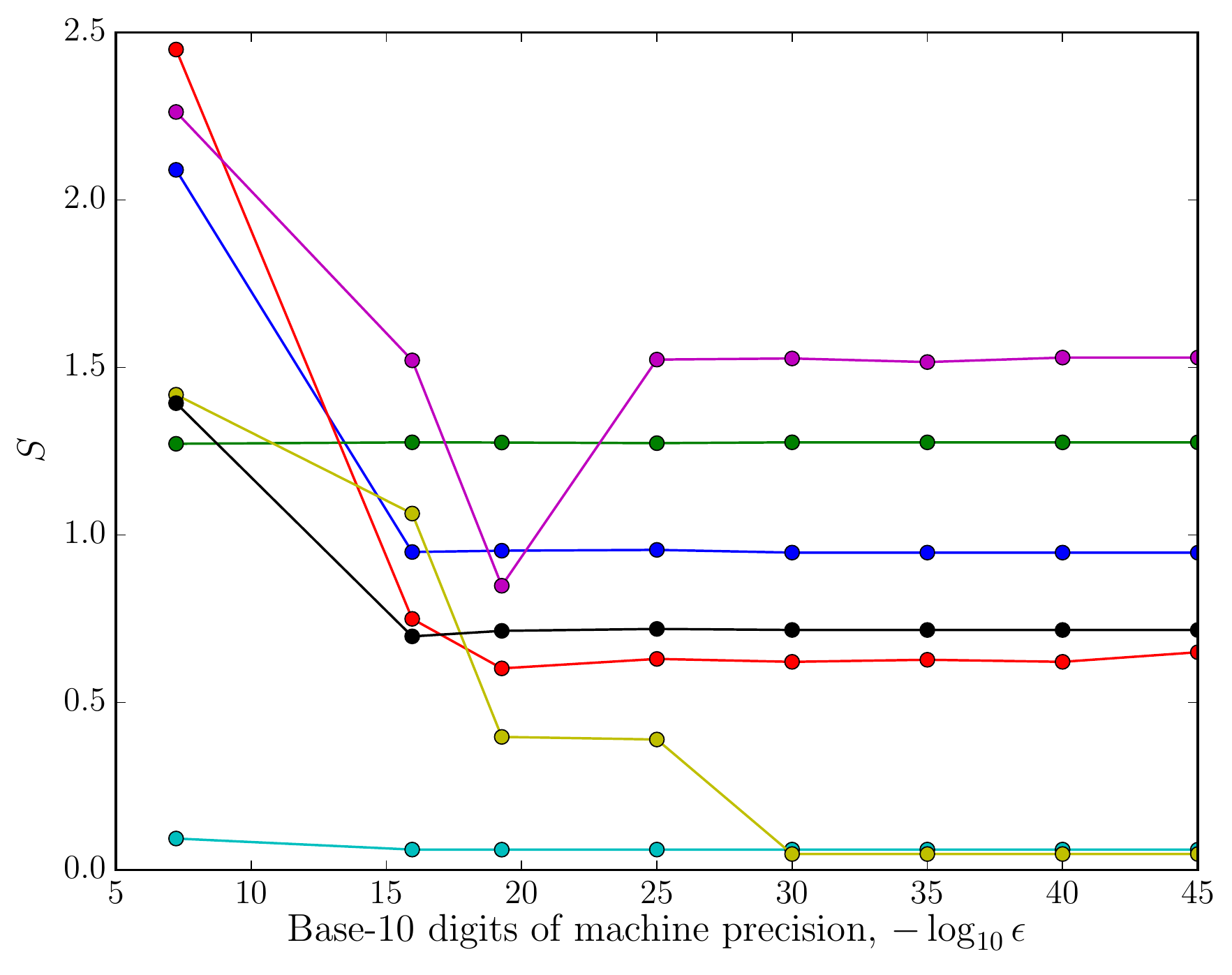}
	\caption{\label{fig:convergence} Entanglement entropy converges with increased machine precision. Several disorder realizations (see main text) at system size $L=400$ are plotted in various colors. Even for this relatively modest system size and box disorder, high precision is needed to accurately calculate the entanglement entropy. For comparison, float has $\epsilon \approx 10^{-8}$ and double has $\epsilon \approx 10^{-16}$.}
\end{figure}

As argued in the main text, the disordered periodically driven Ising model flows to infinite randomness about both $0$ and $\pi$. This leads to an abundance of single-particle modes exponentially near $E_F = 0$ and $E_F = \pi$, worsening with larger system sizes and stronger disorder. These near-degenerate modes give rise to a subtle numerical instability issue that requires going to high numerical precision, especially at strong disorder and/or large system sizes, in order to reliably calculate the correlation function and all derived quantities. \\

To that end, we implemented arbitrary precision\footnote{By `precision' we mean `machine epsilon', namely the smallest number $\epsilon$ such that $1+\epsilon$ is distinct from 1.} diagonalization in C++ using the Eigen\footnote{\url{http://eigen.tuxfamily.org/}} and MPFR\footnote{\url{http://www.holoborodko.com/pavel/mpfr/}} C++ libraries. We then checked convergence of our entropy calculations on a realization-by-realization basis by increasing the precision by hand. In Figure \ref{fig:convergence} we show a sample convergence plot for several disorder realizations at system size $L = 400$ Majoranas. Even at this relatively modest system size, precisions of $\epsilon < 10^{-20}$ are needed to reliably converge most disorder realizations. Data in the main text has been converged accordingly.

\subsection{0$\pi$ Ising duality, emergent symmetry, and infinite randomness structure}\label{sec:sym}

In this appendix, we give a complete account of the 0$\pi$ duality transformation described in the main text. We then calculate the associated emergent symmetry operator $D$ to second order in the weak coupling parameters $\epsilon,\delta \ll 1$, and show that it is simply the product of domain wall Majorana modes dressed by neighboring weak bonds, as argued in the main text. Finally, we detail a calculation of the ratio of the relative number of quasienergies near $0$ to those near $\pi$ based on the infinite randomness domain wall structure from the main text and compare with numerical data. \\

\subsubsection{0$\pi$-Ising duality and bond-field duality}\label{sec:duality}

First, consider the well-known bond-field duality of the static Ising model. This transformation defines dual spins on the bonds of the previous model, mapping $h_j \sigma_j^x \mapsto h_j \sigma_{j-1/2}^z \sigma_{j+1/2}^z$ and $J_j \sigma_j^z \sigma_{j+1}^z \mapsto J_j \sigma_{j+1/2}^x$. In the static case, this produces an Ising Hamiltonian with $h \leftrightarrow J$. Therefore this map does not preserve the spectrum, but does return a self-similar Hamiltonian. In the Floquet context, $F$ is actually not self-similar under this transformation. A bond-field transformation would send 
\begin{align}
&F = {e}^{- \frac{i}{2} \sum_i   J_i \sigma_i^z \sigma_{i+1}^z + U \sigma_i^z \sigma_{i+2}^z} {e}^{-\frac{i}{2} \sum_i h_i \sigma_i^x + U \sigma^x_i \sigma^x_{i+1}} \nonumber \\ &\mapsto F' = {e}^{-\frac{i}{2} \sum_i   J_i \sigma_i^x + U \sigma^x_i \sigma^x_{i+1} } {e}^{-\frac{i}{2} \sum_i h_i \sigma_i^z \sigma_{i+1}^z + U \sigma_i^z \sigma_{i+2}^z},
\end{align}
where $F'$ is not in the same form as $F$ since the order of the field and bond terms is switched. However, $F'$ and $F$ are related by a unitary rotation: since $F$ is the product of two unitaries $F = AB$, conjugating either by $A$ or by $B^\dagger$ will give $F' = BA$. Also note that the transformation sending $h_i \mapsto -h_i$ without modifying $J$ is just a global $\pi$ rotation about the $y$- or $z$-axis, and hence does not affect the spectrum. Bond-field duality thus sends $h_i \mapsto \pm J_i$. \\

Unique to the Floquet case is the self-similar map $J_i \mapsto \pi + J_i$ and its bond-field dual $h_i \mapsto \pi + h_i$. For the first map, the Floquet operator transforms as
\begin{align}
&{e}^{ -\frac{i}{2} \sum_i  J_i \sigma_i^z \sigma_{i+1}^z + U \sigma_i^z \sigma_{i+2}^z} {e}^{-\frac{i}{2}\sum_i h_i \sigma_i^x + U \sigma^x_i \sigma^x_{i+1}} \nonumber \\ &\mapsto (\prod_{\forall i} \sigma^z_i \sigma^z_{i+1} ){e}^{- \frac{i}{2} \sum_i   J_i \sigma_i^z \sigma_{i+1}^z + U \sigma_i^z \sigma_{i+2}^z} {e}^{- \frac{i}{2} \sum_i h_i \sigma_i^x + U \sigma^x_i \sigma^x_{i+1}},
\end{align}
using the identity $e^{i \frac{\pi}{2} \sigma_i^{(x,y,z)}} = i \sigma_i^{(x,y,z)}$, where we have dropped the overall phase factor because we are interested only in the operator content. Under periodic boundary conditions, $\prod_{\forall i} \sigma^z_i \sigma^z_{i+1} = 1$, so $F$ maps precisely onto itself. The dual transformation sends $F$ to $FG_{\rm Ising}$, where $G_{\rm Ising} = \prod_{\forall i} \sigma_i^x$ is the usual Ising symmetry operator. Since $[F,G_{\rm Ising}]=0$, this leaves the eigenstates invariant, but simply shifts their eigenvalues. \\

Composing the spectrum-preserving maps $(J_j,h_j)\mapsto (-J_j,-h_j)\mapsto (\pi - J_j,\pi-h_j)$ with bond-field duality gives the 0$\pi$-Ising duality transformation of $\sigma_j^z \sigma_{j+1}^z \mapsto (\pi-J_j) \sigma_{j+1/2}^x$, $h_j \sigma_j^x \mapsto (\pi - h_j) \sigma_{j-1/2}^z \sigma_{j+1/2}^z$ that maps the $\pi$SG to the PM and the SG to the $0\pi$PM. 

\subsubsection{Emergent symmetry and chain decoupling}\label{sec:decoupling}

Significant insight about this problem can be gleaned by considering not $F$ but rather $F^2$, in the same vein as Yao et al~\cite{PhysRevLett.118.030401}. In particular, let us begin by considering a typical configuration of the model, where some couplings will be near 0 and others will be near $\pi$. As described in the main text, flow towards infinite randomness dictates that the multicritical point should be controlled by such chain configurations. Further, interaction terms are irrelevant, so in the limit of infinite randomness they have reached their fixed point of $U=0$. \\

Let us work in the Majorana fermion picture, related to the spin picture via a Jordan-Wigner transformation, detailed in the Materials and Methods section. For convenience, let us relabel the two Majorana sublattices as $a_j = \gamma_{2j}$, $b_j = \gamma_{2j+1}$. In this language, the natural domains near the multicritical point are regions of Majorana couplings that are near $\pi$ ($J_i = \pi + \epsilon_i, h_i = \pi + \delta_i$) or near $0$ ($J_j = \pi + \epsilon_j, h_j = \pi + \delta_j$), separated by domain walls. Considering a single domain wall, let all bonds to the left of $i=0$ be near $0$, and let all bonds to the right be near $\pi$. The Floquet operator in this case is then
\begin{equation}
F = e^{ \sum_{i<0} \epsilon_i b_i a_{i+1} + \sum_{i\ge 0} (\pi/2 + \epsilon_i) b_i a_{i+1} }e^{ \sum_{i<0} \delta_i a_i b_i + \sum_{i \ge 0} (\pi/2 + \delta_i) b_i a_{i+1}},
\end{equation} 
where we have absorbed factors of $\frac{1}{2}$ into $\epsilon,\delta$. As a reminder, these Majorana operators obey the algebra $\{a_i,b_j\} = 2 \delta_{ab}\delta_{ij}$, and $a^2 = b^2 = 1$. As can be straightforwardly shown, the Majorana fermion evolution operator is just $e^{\theta ab} = \cos\theta + ab \sin\theta$, where in particular, $e^{\frac{\pi}{2} a b} = a b$. We will drop overall phase factors to focus on operator content. Factoring out the $\pi$ pulses, then, 
\begin{align}
F &= (\prod_{i\ge0} b_i a_{i+1}) e^{ \sum_{i} \epsilon_i b_i a_{i+1} }e^{ \sum_{i} \delta_i a_i b_i } (\prod_{i\ge 0} a_i b_i) \notag \\
&= a_0 e^{-\epsilon_{-1} b_{-1} a_0 + \sum_{i\not=-1} \epsilon_i b_i a_{i+1} } e^{\sum_i \delta_i a_i b_i},
\end{align}
where we have used the fact that if operators $A,B$ anticommute, then $e^B A = A e^{-B}$. If we now compute $F^2$, we get 
\begin{align}
F^2 &= a_0 e^{-\epsilon_{-1} b_{-1} a_0 + \sum_{i\not=-1} \epsilon_i b_i a_{i+1} } e^{\sum_i \delta_i a_i b_i}  a_0 e^{-\epsilon_{-1} b_{-1} a_0 + \sum_{i\not=-1} \epsilon_i b_i a_{i+1} } e^{\sum_i \delta_i a_i b_i} \notag\\
&= e^{\sum_i \epsilon_i b_i a_{i+1} } e^{\sum_{i\not=0} \delta_i a_i b_i - \delta_0 a_0 b_0 } e^{\sum_{i\not=-1} \epsilon_i b_i a_{i+1} - \epsilon_{-1} b_{-1} a_0 } e^{\sum_i \delta_i a_i b_i} \notag \\
&= \exp\{ 2 \sum_{i \not=-1} \epsilon_i b_i a_{i+1} + 2 \sum_{i\not=0} \delta_i a_i b_i + 2 \sum_{\forall i} (\epsilon_{i-1} \delta_i b_{i-1} b_i - \epsilon_i \delta_i a_{i-1} a_i) - 4 \epsilon_{-1} \delta_0 b_{-1} b_0  + \mathcal O(\lambda^3)\}
\end{align} 
where $\epsilon_i, \delta_j$ are $\mathcal O(\lambda)$. To lowest order in BCH, both bonds touching $a_0$ cancel, {\it isolating this Majorana}.  Now, as argued in \cite{PhysRevX.7.011026}, $F$ can be brought into the form $F = D e^{-i \tilde H}$, where $\tilde H$ is a quasi-local Hamiltonian, and $D$ satisfies $D^2 = 1$ and $[D,\tilde H] = 0$ (hence $[D,F] = 0$). Now, $D$ both squares to 1 and commutes with $F$, as does $G_{\rm Ising} = \prod_{\forall i} \sigma_i^x = \prod_{\forall j} i a_j b_j$. This lead us, as in the main text, to identify $D$ as a symmetry operator for $F$, with 
\begin{equation}
D = F \sqrt{F^2}^\dagger.
\end{equation}

This will only be well-defined if the branch cut of the square root operator at $-1$ is well-separated from the eigenvalues of $F^2$, which is indeed true in the limit of infinite randomness, since all eigenvalues of $F^2$ will be near $1$. We can calculate $\tilde H$ and $D$ explicitly here using our formulas from above. The quasi-local Hamiltonian $\tilde H$ is
\begin{equation}
\tilde H = i \sum_{i \not=-1} \epsilon_i b_i a_{i+1} + i \sum_{i\not=0} \delta_i a_i b_i + i \sum_{\forall i} (\epsilon_{i-1} \delta_i b_{i-1} b_i - \epsilon_i \delta_i a_{i-1} a_i) - 2 i \epsilon_{-1} \delta_0 b_{-1} b_0  + \mathcal O(\lambda^3),
\end{equation}
and the symmetry operator $D$ is
\begin{align}
D &= a_0 e^{-\epsilon_{-1} b_{-1} a_0 + \sum_{i\not=-1} \epsilon_i b_i a_{i+1} } e^{\sum_i \delta_i a_i b_i} e^{i \tilde H} \notag \\
&= a_0 e^{\delta_0 a_0 b_0 - \epsilon_{-1} b_{-1} a_0 +\epsilon_{-1} \delta_{-1} a_{-1} a_0 - \epsilon_0 \delta_0 a_0 a_1 + \mathcal O(\lambda^3)} \notag \\
&=e^{-\frac{1}{2}(\delta_0 a_0 b_0 - \epsilon_{-1} b_{-1} a_0 +\epsilon_{-1} \delta_{-1} a_{-1} a_0 - \epsilon_0 \delta_0 a_0 a_1+\ldots)} a_0 e^{\frac{1}{2}(\delta_0 a_0 b_0 - \epsilon_{-1} b_{-1} a_0 +\epsilon_{-1} \delta_{-1} a_{-1} a_0 - \epsilon_0 \delta_0 a_0 a_1 + \ldots)} \notag \\
&= \tilde a_0,
\end{align}
where in the penultimate step we have used the fact that $a_0$ anticommutes with all terms in the exponential, and in the final step we have recognized that $\tilde a_0$ is simply $a_0$ dressed by a unitary rotation. Hence, $\tilde a_0$ is still a Majorana fermion operator, satisfying $\tilde a_0^2 = 1$ and $\tilde a_0^\dagger = \tilde a_0$. The exponential pieces on either side of $a_0$ are its exponential (or stretched exponential) tails. We identify $\tilde a_0$ as the emergent $\pi$ Majorana at this $0\pi$ domain wall.\\

For more generic domain wall configuration, we will see that a chain of Majoranas emerges whose dynamics are decoupled from the initial Majoranas. In particular, the emergent symmetry operator $D$ will take the form
\begin{equation}
D = \prod_{i \in \{{\rm DWs}\}} \tilde \gamma_i
\end{equation}
as claimed in the main text. Since domain walls must come in pairs in a chain with finite domain sizes, this will be just 

\begin{equation}
D = \prod_{j} \tilde \sigma^x_j,
\end{equation}
that is, the Ising symmetry operator (parity operator) of the chain formed by these domain wall Majoranas, up to an overall phase factor. These domain wall Majoranas are coupled by exponentially small tunneling terms: for a domain of size $L$, we must go to order $L$ in BCH to see a coupling term appear in the exponent. For $0/\pi$ regions deep in a phase, this coupling is thus exponentially small in $L$. For critical regions, exponential scaling is replaced by stretched exponential scaling, but the above is otherwise the same. \\

Finally, we note that our picture of two decoupled chains can be mapped explicitly onto the disordered $XY$ model, with interactions between the chains promoting this to an $XYZ$ model. Following a Jordan-Wigner transformation, we see that $X_i X_{i+1} = i b_i a_{i+1}$ and $Y_{i} Y_{i+1} = i a_i b_{i+1}$, where $a$ and $b$ are the two Majorana fermion sublattices. The even $J_{XX}$ couplings and odd $J_{YY}$ couplings then form one Majorana chain, and the remaining couplings form the other. If the  $J_{XX}$,$J_{YY}$ couplings are drawn from the same distribution, these two Majorana chains are critical. The flow of interactions under renormalization has been studied in \cite{PhysRevB.94.014205}, where they found the $J_{ZZ}$ interaction terms to be irrelevant.

\subsubsection{Infinite randomness structure and domain wall quasienergies}

\begin{figure}
	\centering
	\includegraphics[width=\columnwidth]{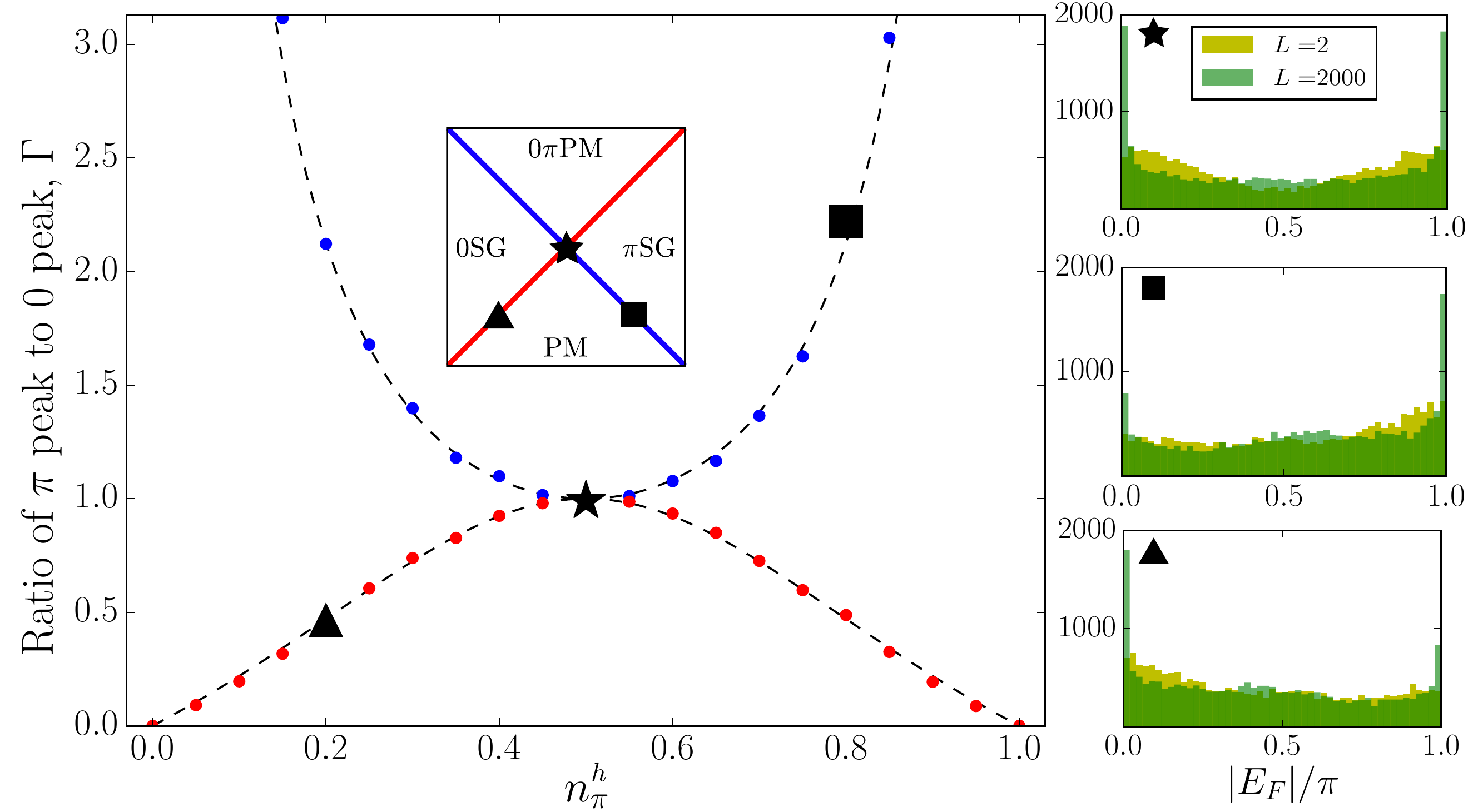}
	\caption{\label{fig:Gamma} Left: Relative number of single-particle quasienergies within 2\% of $\pi$ to those within 2\% of $0$ ($\Gamma$ ratio), tracked numerically along the lines $n_\pi^J = n_\pi^h$ (red) and $n_\pi^J = 1-n_\pi^h$ (blue), at system size $L=2000$ and averaged over $100$ realizations. The inset shows where these lines and the highlighted points (triangle, square, and star) lie on the phase diagram. The dashed line is the prediction from below. Right: Averaged quasienergy histograms showing the development of these peak heights around 0 and $\pi$ with increasing system size. } 
\end{figure}

Here we detail a simple calculation of the relative number of states with single particle quasienergies at $0$ and $\pi$ based on our infinite randomness arguments in the main text, and compare with numerics. We utilize the parameters $n_\pi^{h,J}$ introduced in the main text. Define the ratio of the number of single particle quasienergies within some small number $\epsilon$ of $\pi$ to the number within $\epsilon$ of 0 to be $\Gamma_\epsilon$. First, let us move along the line $n_\pi^h = n_\pi^J$. In our infinite randomness picture, we claim that each domain wall should host a single-particle mode at quasienergy $\pi$, and the other quasienergies should be near 0. The probability of a domain wall is $n_{\rm DW} = 2 n_\pi^h(1-n_\pi^h)$, so we predict that along this line, taking $\epsilon\to 0$, $\Gamma(n_\pi^h) =\frac{ n_{\rm DW} }{1 - n_{\rm DW}} = \frac{1}{2n_\pi^h(n_\pi^h-1) + 1} - 1$. Moving along the other critical line, $n_\pi^h = 1-n_\pi^J$, we expect each domain wall to host a mode at quasienergy $0$, and hence the $\Gamma$ ratio should be $\Gamma(n_\pi^h) = \frac{1-n_{\rm DW}}{n_{\rm DW} } = \frac{1}{2 n_\pi^h} + \frac{1}{2(1-n_\pi^h)} - 1$. These predictions are compared with numerics in Figure \ref{fig:Gamma}, finding good agreement. Therefore, even though we start far from the infinite randomness fixed point, the energetics are nonetheless controlled by the infinite randomness domain wall structure.

\subsection{Microscopic justification for infinite randomness}\label{sec:inf_rand}

\begin{figure}
	\centering
	\includegraphics[width=\columnwidth]{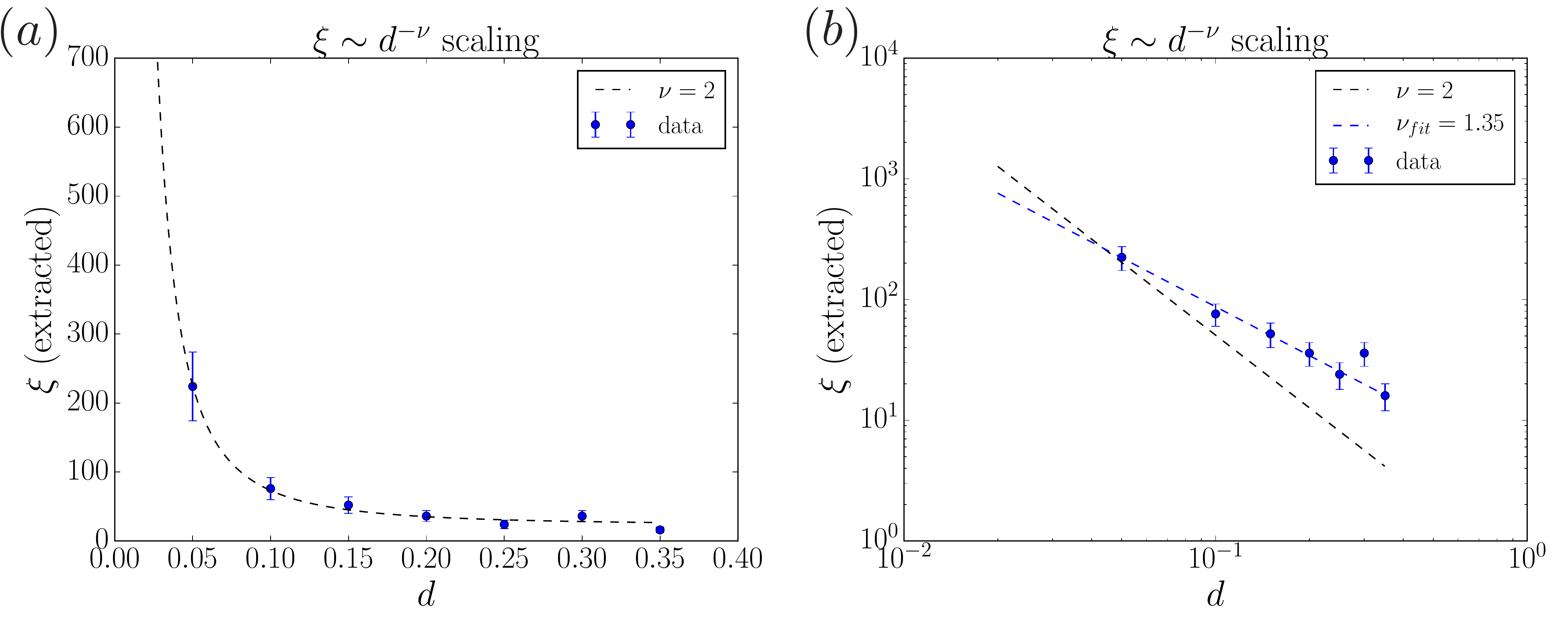}
	\caption{\label{fig:inf_rand} Scaling of the correlation length with distance from criticality. We define the ``correlation length'' $\xi$ as the length at which the disorder-averaged entanglement entropy $\overline{S}(l)$ crosses over from logarithmic growth to a constant (area-law), i.e. the length when $\overline{S}(l)$ saturates to within $\eps = 1.5 \%$ (see text). We take the path $(n^\pi_h,n^\pi_J) = (0.4,0.4 - d)$ here. (a) Scaling of $\xi$ with $d$ (blue dots). The black dashed line is a fit to the expected scaling form $\xi = a \times d^{-\nu} + b$, with $a$ and $b$ fit parameters and $\nu = 2$. (b) The same data on a log-log plot. One can see power-law scaling, with fit exponent $\nu_{\rm fit} \approx 1.35$; given the complexities of the IRFP scaling and disorder averaging, we would expect $\nu_{\rm typ} = 1<\nu_{\rm fit}<2 = \nu_{\rm avg}$ from a coarse measure like entanglement.}
\end{figure}

In this appendix, we provide further evidence for flow to an infinite randomness fixed point. First, we extract a ``correlation length'' from the scaling of the average entanglement, and see how this quantity scales with the distance from criticality. Second, we present results of an upcoming article that gives a microscopic justification of flow to infinite randomness based on an explicit strong-disorder renormalization group procedure~\cite{PhysRevB.98.174203}.

One of the pillars of our argument in the main text is the assumption that critical points between Floquet MBL phases are controlled by infinite randomness fixed points (IRFPs) of a strong-disorder renormalization group. From this assumption, we derived several nontrivial predictions, one of which is the (open boundary condition) entanglement entropy scaling $S(L) \sim (\tilde c / 6) \log L$, with $\tilde c = (1/2) \ln 2$ along the critical lines of the driven disordered Ising model  $\tilde c = \ln 2$ at the multicritical point. Indeed, entanglement entropy scaling is one of the best and most reliable indicators of an infinite randomness fixed point, as it is self-averaging and allows for direct extraction of the disordered central charge, which identifies the universality class of the fixed point~\cite{PhysRevLett.93.260602,PhysRevB.72.140408}. We found good numerical agreement with these predictions in the main text. 

Here we provide further numerical evidence of this IRFP by analyzing the scaling of the entanglement correlation length with distance from criticality. At an IRFP, we generically expect power-law scaling of a correlation length $\xi$ with distance from criticality $d$ as $\xi \sim d^{-\nu}$, where $\nu = 1$ for a typical quantity (i.e, $e^{\overline{\log \mathcal{O}}}$) and $\nu = 2$ for an average quantity (i.e. $\overline{\mathcal{O}}$)~\cite{PhysRevLett.69.534,PhysRevB.51.6411}. Given that our analysis has focused on the entanglement entropy, and that spin-spin correlation functions are difficult to extract in the Majorana basis as they involve arbitrarily-long `string' operators, we extract a correlation length directly from the entanglement data (see Figure~\ref{fig:inf_rand}). In particular, in a one-dimensional MBL phase, we expect the entanglement entropy to cross over from logarithmic growth to a constant, satisfying an entanglement `area-law.' We can numerically estimate the length scale of this crossover as the length $\xi$ at which the disorder-averaged entanglement entropy saturates to within some numerical threshold $\epsilon$. In Figure~\ref{fig:inf_rand}, we set this threshold at $\epsilon = 1.5 \%$, and calculate this crossover length $\xi$ along the line $(n^\pi_h, n^\pi_J) = (0.4, 0.4 - d)$. Since the entanglement is a disorder-averaged quantity, we would expect $\nu > 1$; given the immense length scales required to see the long-distance Griffiths effects that distinguish average from typical quantities, it does not numerically saturate to $\nu = 2$. We instead expect some intermediate value of $2 > \nu > 1$, as has been found in the literature; for instance, Ref.~\citenum{PhysRevLett.118.030401} found $\nu = 1.3$. We find $\nu_{\rm fit} = 1.35$, in accordance with this expectation.

Next, we present some results of an upcoming section giving a microscopic justification of flow to infinite randomness~\cite{PhysRevB.98.174203}. In this work, we introduce a generalization of the Schrieffer-Wolff transformation to Floquet unitary operators, allowing for decimation of `strong' bonds in much the same way as in the static problem, with the strong bond treated non-perturbatively and the rest of the chain a perturbation. In particular, we find that for decimations within a domain of bonds all near 0 or near $\pi/2$ (within a 0 or $\pi$ domain), decimating a strong bond $\Omega i \gamma_1 \gamma_2$ with neighbors $J_L i \gamma_0 \gamma_1$ and $J_R i \gamma_2 \gamma_3$ gives a renormalized coupling $\tilde J i \gamma_0 \gamma_3$ with the rule

\begin{equation}
\tan \tilde J \approx \tilde J = \frac{J_L J_R}{\tan \Omega}.
\end{equation}

Note that this reproduces the static rule of $\tilde J = J_L J_R / \Omega$ when $\Omega \ll 1$, as it must. Within a $\pi$ domain, the renormalized bond is $\pi/2 + \tilde J$. Rewriting this rule in terms of logarithmic variables, $\log \tilde J = \log J_L + \log J_R - \log \tan \Omega$, the distributions flow to an infinite randomness fixed point by analogy with the analysis from Fisher~\cite{PhysRevLett.69.534,PhysRevB.51.6411}. We also find that care must be taken at the domain walls, as suggested by the picture in the main text that these domain walls host topological edge modes. We find that the decimations involving domain wall Majorana operators naturally lead to two decoupled Majorana chains, one near 0 and the other near $\pi$ quasienergy, that can independently be tuned to criticality -- precisely the result from the main text. This analysis provides a microscopic justification for our more coarse-grained arguments, further justifying a flow to infinite randomness.

\subsection{Clean (non-disordered) criticality}\label{sec:clean}

\begin{figure}
	\centering
	\includegraphics[width = \columnwidth]{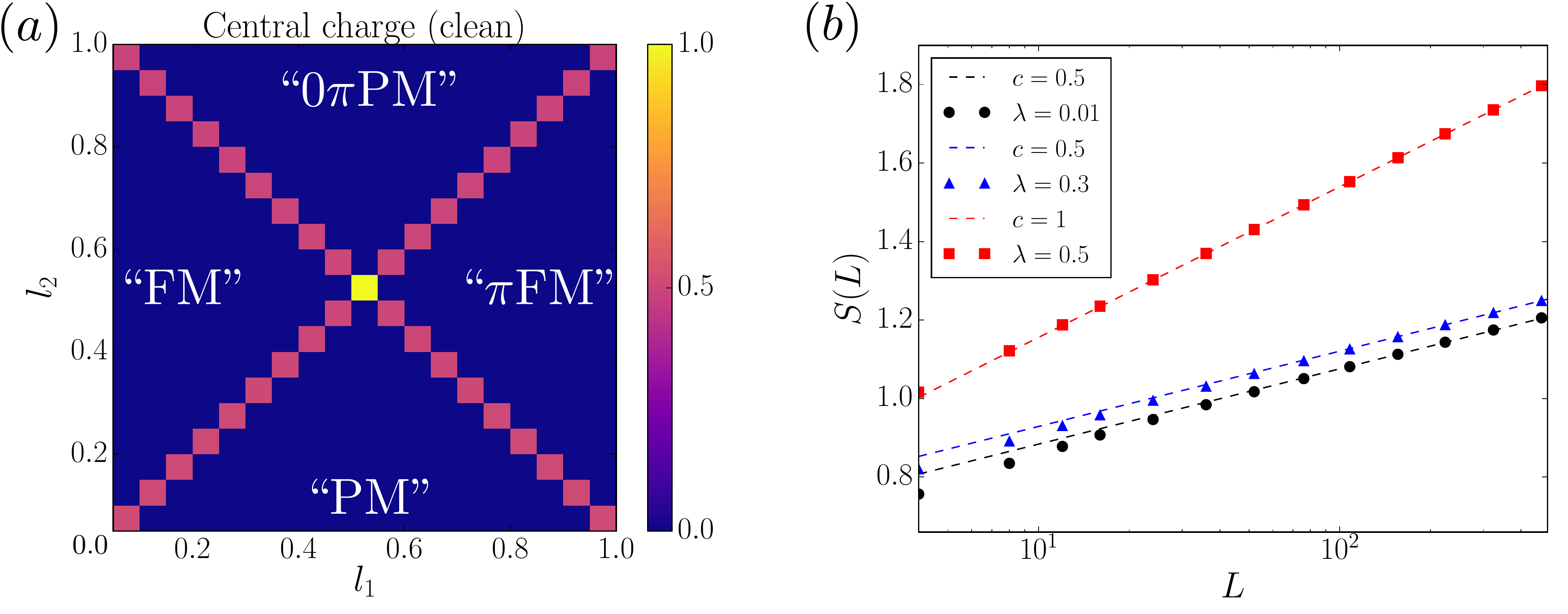}
	\caption{\label{fig:clean} Criticality in the clean (non-disordered) model. (a) Phase diagram of the clean model showing the central charge extracted from entanglement scaling $S\sim \frac{c}{3} \log L$ in the Floquet ground state (all negative quasienergies filled). {Unlike the high temperature spin glass phases in the main text, these phases are more accurately labeled as ferromagnets or paramagnets, as they behave like ground states of the Ising model. We have used quotation marks to indicate that such phases are not robust to heating, unlike their counterparts in the disordered system.} Along the critical lines of the diagram the Floquet ground state is described by a CFT with central charge $c=1/2$, while the multicritical point shows $c=1$. The phases saturate to area-law (constant) entanglement. (b) Entropy slices at the points $l_1 = l_2 = \lambda$ in the diagram to the left. $\lambda = 0$ and $\lambda = 0.3$ both show $c=1/2$, while the multicritical point $\lambda=0.5$ shows $c=1$.} 
\end{figure}

In this appendix, we discuss the criticality of the clean (non-disordered) driven Ising model. In the absence of interactions, this model still has a non-trivial phase structure~\cite{pi-spin-glass,khemani_prl_2016}. For clarity, the clean model is defined by the Floquet evolution operator
\begin{equation}
F = {e}^{-i \frac{\pi}{2}  \sum_i   l_2 \sigma_i^z \sigma_{i+1}^z} { e}^{-i \frac{\pi}{2} \sum_i l_1 \sigma_i^x },
\end{equation}
where $l_1, l_2$ are normalized variables defined modulo 1. In the clean model, we do not expect a flow to infinite randomness, and thus cannot apply the picture of criticality from the main text, which relied on viewing a typical configuration of the chain at criticality as composed of adjoining regions deep in the neighboring phases. Nonetheless, the clean case still displays critical behavior, described instead by a conformal field theory (CFT). This is possible because the clean case, in the absence of interactions, is integrable; with interactions we expect that the model displays only prethermal phases and critical lines, and will eventually thermalize to infinite temperature~\cite{PhysRevLett.115.256803}.  \\

We note that the critical lines of the clean model are set entirely by the dualities mentioned in the main text and explained in Appendix \ref{sec:sym}. Bond-field duality, as in the static clean Ising model, implies that the self-dual line of $l_1=l_2$ will be critical. In the driven case, 0$\pi$ Ising duality maps $l_1 \sigma_i^x \mapsto (1 - l_1) \sigma_i^z \sigma_{i+1}^z $ and $l_2 \sigma_i^z \sigma_{i+1}^z  \mapsto (1 - l_2) \sigma_i^x $. This changes the order of the pulses, but as discussed above, since $F$ is of the form $F=AB$ with $A,B$ unitary operators, $F' = BA$ is related to $F$ by a simple unitary transformation and hence the spectrum is unaffected. A second self-dual and hence critical line is then $l_1 = 1-l_2$. These two self-dual lines cross at the multicritical point $(l_1,l_2) = (0.5,0.5)$. \\

As mentioned in the main text, Floquet systems generally have no notion of a ground state, since quasi-energy is only defined modulo $2\pi/T$ and hence lives on a circle. Nonetheless, for weak enough driving, one state will be adiabatically connected to the ground state -- usually deemed the ``Floquet ground state'' (see, e.g., \cite{0295-5075-115-3-30006}). In our model, this state corresponds to simply filling all single-particle modes of negative quasi-energy. In the weak-driving limit, all quasi-energies will be near 0, and hence the problem will be nearly static; the Floquet ground state is then just the ground state of the static model ($T\to0$). We expect that in the $T \to 0$ limit that the Floquet ground state should be described by a CFT with central charge $c=1/2$,  since this limit recovers a static critical Ising model~\cite{di_francesco_conformal_2011}. \\

In Figure \ref{fig:clean}, we show numerical calculations of the entanglement entropy in the Floquet ground state from which we can extract the central charge $c$: with periodic boundary conditions, we expect the entanglement entropy to satisfy $S\sim \frac{c}{3} \log L$~\cite{1751-8121-42-50-504005}. We find that along the critical lines of the diagram, the Floquet ground state gives $c=1/2$ while the multicritical point shows $c=1$. 
\begin{savequote}
\begin{tabular}{ll}
古池や&The old pond \\
蛙飛び込む &A frog jumped in, \\
水の音 &Kerplunk! \\
\end{tabular}
\qauthor{松尾 芭蕉 \quad Matsuo Bash\=o, trans. Allen Ginsburg}
\end{savequote}

\chapter{Hydrodynamics, Viscosity and Thermal Coulomb Drag}
\label{ch:Hydro}

Hydrodynamics is in some sense the most universal theory in physics. Virtually all systems that reach thermal equilibrium\footnote{And even some that don't, such as integrable systems, obey a generalized form of hydrodynamics.} obey hydrodynamics in the late-time limit, as they approach the global equilibrium state. It is also a theory of the weakest form of non-equilibrium matter, in that we still assume \emph{local} thermodynamic equilibrium, with the whole system a patchwork of local equilibria stitched together.\footnote{This is still non-equilibrium enough to cover turbulence and other complicated dynamical phenomena, though, as the equations of motion are nonlinear.} Quantities such as temperature $T$ or pressure $P$ become functions of space and time, $T(r,t)$ and $P(r,t)$, and the hydrodynamic equations, specifically the Navier-Stokes equations, govern how these quantities change. In tune with their universal nature, very few microscopic parameters need go into the hydrodynamic equations to determine how the system globally evolves. We usually only need to condense a microscopic model into a few length and time scales to get a hydrodynamic theory of the system. 

That said, solids -- or more precisely, electrons in solids -- almost never behave hydrodynamically. This is because hydrodynamics is built on the foundation of kinetic theory, which requires conservation of momentum and energy as essential ingredients. In a solid, though, electrons relax their momenta all the time! Electrons can lose momentum to scattering from impurities, for instance, which are rife in solids. Electrons can also give up their momentum to phonons, the quanta of lattice vibrations, as they bounce off of the ions in their environment. More fundamentally, since in quantum mechanics electrons are waves, their momenta are simply not conserved in a periodic potential like a lattice.\footnote{`Momentum', strictly speaking, is a free-space concept that loses its meaning without translation symmetry. Continuous translation symmetry gives the usual momentum as a continuous eigenvalue of the operator $-i \partial_x$. But once we move to a system with only discrete translation symmetry, momentum is only uniquely defined within one particular \emph{Brillouin zone}, and ceases to be exactly conserved.} This is usually elided into the definition of `crystal momentum', which is only conserved modulo a reciprocal lattice vector $\vec{G}$. Even without external collisions, if two electrons collide, the sum of their old and new momenta may differ by some reciprocal lattice $\vec{G}$, in a process known as \emph{Umklapp scattering}.\footnote{This rather ugly word derives from the German \emph{umklappen}, meaning `to flip over', and was introduced by Peierls during his 1929 crystal lattice studies under Pauli.~\cite{peierlsMemoir} For some reason, it is always capitalized, despite simply being a loan word and not, e.g., a name.} This has the potential to convert two left-moving particles into two right-moving particles, for instance.

Nonetheless, the prospect of the hydrodynamic flow of electrons has excited theorists for a long time, dating back to at least the 1960s~\cite{gurzhi}. The possibility of novel types of fluids arising from the quantum mechanical nature of electrons was, and remains, particularly tantalizing. It was thus quite exciting when the hydrodynamic flow of electrons was observed experimentally, first in two-dimensional electron gases (2DEGs) AlAs and GaAs~\cite{PhysRevB.51.13389} in 1995,\footnote{This is probably Laurens Molenkamp's most famous work from \emph{before} he discovered topological insulators in the mid 2000s.} then later in graphene~\cite{Bandurin1055} and palladium cobaltate~\cite{Moll1061} in 2016. In particular, in graphene a novel form of `relativistic' fluid, dubbed the `Dirac fluid', may form, with unusual properties such as a strong violation of the Wiedemann-Franz law~\cite{Crossno1058}. These systems behave hydrodynamically because they are ultra-pure, so nearly all collisions conserve momentum, and because Umklapp scattering is not important. 

The hydrodynamic equations, i.e. the Navier-Stokes (NS) equations, can often be straightforwardly numerically solved.\footnote{That said, mathematically proving properties about them is immensely challenging. Proving even the \emph{existence} of a smooth, well-behaved solution to the three-dimensional NS equations -- let alone solving them! -- is worth a one million dollar prize from the Clay Mathematics Foundation (as of 2020). Interestingly, global existence and smoothness has been proven in two dimensions since the 1960s~\cite{navierStokesBook}, and many results have been proven for ``weak'' solutions (replacing quantities by their averages in the equations), as well as other partial results. Terence Tao recently proved~\cite{Tao_2015} that a slightly modified version of the equations has a smooth solution that blows up in finite time, casting some doubt on the NS smoothness problem. Nevertheless, the majority view is that the three-dimensional NS equations likely do have smooth and well-behaved solutions.} Their derivation, though, requires a brief detour into kinetic theory. We first give an introduction to this formalism, highlighting the importance of the scattering time approximation and its use in semi-classical calculations (used in the rest of the chapter). We then discuss the phenomenon of Coulomb drag and its relation to viscosity, before finally delving into the author's work.

\section{Hydrodynamics and Drag}

The subject of hydrodynamics is old, dating back to the 1800s. Consequently, there are many excellent places to learn the material. In particular, the two books by Kardar~\cite{kardar_fields,kardar_particles} are modern classics, though the old fogies and/or Russians will undoubtedly steer you towards Landau and Lifshitz~\cite{lifshitz-kinetics}.\footnote{This is the last volume in the Landau-Lifshitz series, and was published decades after Landau's death. Tragically, Landau died on 1 April 1968, aged 60, from complications sustained in a car accident 6 years prior. That accident had left him comatose for two months, and he never fully regained his physics capabilities. Showing his humor, though, after awakening from the coma he remarked ``I'm afraid my brain is just not the same as it was. I'll never again be able to do physics like Landau. Maybe I could do physics like Zeldovich.'' (Zeldovich was one of Landau's colleagues, often the butt of his jokes; the quote is from recollections of Gell-Mann, \url{https://www.youtube.com/watch?v=fhioDOi2g4E}.)} Many classic condensed matter books, such as Chaikin-Lubensky and Altland-Simons, contain linear response chapters. The best book the author has found, particularly in its treatment of Kubo formulas and the like, is by Pottier~\cite{pottier}. As far as short and sweet notes go, though, none can beat David Tong's,\footnote{\url{http://www.damtp.cam.ac.uk/user/tong/kinetic.html}} from which this section is heavily borrowed.\footnote{It is a general feature of Tong's notes, on every subject, that they are good. Though the author regrets not having had him as a lecturer at Cambridge, his excellent General Relativity and Quantum Field Theory notes, among others, got him through the rather milquetoast Part III lectures.}

\subsection{Kinetic theory: from phase space to the Quantum Boltzmann Equation}

The starting point for classical hydrodynamics is, as with all classical mechanics, phase space. Say we have a system of many, many (identical) particles -- perhaps $\mathcal O(10^{23})$. Then while we could write down the Hamiltonian, namely 

\begin{equation}
H = \frac{1}{2 m} \sum_i p_i^2 + \sum_i V(r_i) + \sum_{i<j} U(r_i - r_j),
\end{equation} 

and its resultant equations of motion, we'd have no hope whatsoever of solving them. The first approximation we make, then, is to admit our ignorance of the full, detailed solution in phase space and introduce a \emph{probability density function} over phase space. That is, we seek $f$ such that $$\int \prod_{i} d^3 p_i d^3 r_i \ f(\{r_j, p_j\}_{j=1}^N ) = 1$$ that reproduces the statistical properties of the collection of particles, if not their exact dynamics. This is still a tremendously complicated function living in an $\mathcal O (10^{23})$-dimensional space, so at first it may seem like we haven't simplified things much -- but bear with us. 

Hamilton's equations of motion on the raw variables $r_i, p_i$, coupled with the conservation of probability (continuity equation), imply that the distribution $f$ must satisfy what is known as \emph{Liouville's equation}. Rather simply, this is $\frac{df}{dt} = 0$, which has the physical meaning that probability does not change as we move along flows in phase space. Using Hamilton's equations $\partial_t p_i = -\partial_{r_i} H$, $\partial_t r_i = \partial_{p_i} H$, this implies

\begin{equation}
\pd{f}{t} = \pd{f}{p_i} \cdot \pd{H}{r_i} - \pd{f}{p_i} \cdot \pd{H}{r_i} = \{H, f\},
\end{equation}

where in the last equation we defined the \emph{Poisson bracket} between $H$ and $f$. Now, equilibrium distributions have $\pd{f}{t} = 0$, so if $\{H, f\}$ is small, we are only weakly out of equilibrium. So far, though, we have made no assumption about how far we are away from the equilibrium $f$; our only approximation has been to use $f$ in place of the full dynamics. 

The next key approximation is to admit further ignorance and simplify $f$, in particular, by considering the one-particle distribution function rather than the full $N$-particle distribution function. This is gotten by averaging out all the other $N-1$ particles from $f$:

$$
f_1(r_1,p_1;t) \equiv N \int \prod_{i=2}^N  d^3 r_i d^3 p_i \ f(\{r_j, p_j\}_{j=1}^N ; t) .
$$

This is a powerful quantity; many macroscopic physical properties of the system can be written in terms of $f_1$ alone. For instance, the density is $n= \int d^3 p_1 \ f_1$, with similar expressions for the mean velocity and mean energy flux. It turns out that $f_1$ satisfies a Liouville-like equation, namely

\begin{equation}
\pd{f_1}{t} = \{H_1, f_1\} + N \int \prod_{i=2}^N  d^3 r_i d^3 p_i \sum_{k=2}^N \pd{U(r_1-r_k)}{r_1} \cdot \pd{f}{p_1},
\end{equation}

where we define the \emph{one-particle Hamiltonian} $H_1 = p_1^2 / 2m + V(r_1)$.\footnote{$V$ here is the external potential; the interactions are all in the $U$ term.} The first term on the right, $\{H_1, f_1\}$, is called the \emph{streaming term}. The second is the \emph{collision integral}, usually written as 

$$
 N \int \prod_{i=2}^N  d^3 r_i d^3 p_i \sum_{k=2}^N \pd{U(r_1-r_k)}{r_1} \cdot \pd{f}{p_1} \equiv \left( \pd{f_1}{t}\right)_{\text{coll}}.
$$

Crucially, the collision integral cannot be expressed solely in terms of $f_1$, and needs the full $f$. This makes sense; the collision integral is supposed to capture the interactions between particles, after all. We can, though, expand it, making use of higher-body distribution functions. These are defined by averaging over the remaining particles, 

\begin{equation}
f_n(r_1, \ldots, r_n, p_1, \ldots, p_n; t) = \frac{N!}{(N-n)!} \int \prod_{i=n+1}^N d^3 p_i d^3 r_i \  f(\{r_j, p_j\}_{j=1}^N;t).
\end{equation}

As a first step, note that we can express $f_1$'s collision integral in terms of $f_2$, as 

$$
\left( \pd{f_1}{t}\right)_{\text{coll}} = \int d^3 p_2 d^3 r_2 \pd{U(r_1-r_2)}{r_1} \cdot \pd{f_2}{p_1}.
$$
 
This gives a closed Liouville equation for $f_1$ in terms of $f_1$ and $f_2$. We can generalize this to all $f_n$, it turns out, in an expansion known as the \emph{BBGKY hierarchy}:\footnote{Rather a mouthful, this is for Bogoliubov, Born, Green, Kirkwood and Yvon.}

\begin{equation}
\pd{f_n}{t} = \{H_n, f_n\} + \sum_{i=1}^n \int d^3 r_{n+1} d^3 p_{n+1} \pd{U(r_i - r_{n+1})}{r_i} \cdot \pd{f_{n+1}}{p_i}.
\end{equation}

So far we still haven't made much progress; one equation in $10^{23}$ variables has been replaced by $10^{23}$ coupled equations. Now, though, we make an enormous approximation, and truncate the BBGKY hierarchy. This is better than truncating our original expression, because these are in some sense the `right' variables; $f_1$ and the low $f_n$ contain all of the interesting macroscopic content, and so by truncating higher $f_n$, we are naturally moving to a zoomed-out, universal description.\footnote{We can further justify the truncation by comparing the scales of the $f_n$ terms. Roughly, $f_3$ is smaller than $f_2$ by a factor $N d^3 / V$, with $d^3$ the typical particle (atomic) volume. For a dilute gas at room temperature, this is around $10^{-3}$ to $10^{-4}$, so we can safely ignore the higher $f_n$.}

There are many ways to truncate this hierarchy, but the usual one is via the \emph{scattering time approximation}. We assume that there are only two important timescales in the system, namely the scattering time $\tau$, which is the time between collisions, and the collision time $\tau_{\text{coll}}$, which is the time for collisions to occur. We assume that $\tau \gg \tau_{\text{coll}}$, so collisions happen almost instantaneously, as with billiard balls on a table. Collisions happen occasionally and suddenly, which we assert must be reflected in the collision integral. Namely, we have that the differential rate of scattering events is

$$
d\Gamma = Q_{p_1 p_2}^{p_1' p_2'} f_2(r,r,p_1, p_2) d^3 p_2 d^3 p_1' d^3 p_2',
$$

where $p_1, p_2$ are incoming momenta, $p_1', p_2'$ are outgoing momenta, $f_2(r,r,p_1, p_2) $ is the likelihood of two particles with those incoming momenta being in the same place at the same time, and $Q_{p_1 p_2}^{p_1' p_2'}$ is a function weighting how important that process is.\footnote{Q is related to the differential scattering cross-section, and can be computed from the interaction $U$.} We can then write the collision integral by integrating this differential rate:

$$
\left(\pd{f_1}{t}\right)_{\text{coll}} = \int d^3 p_2 d^3 p_1' p_2' \left[ Q_{p_1 p_2}^{p_1' p_2'} f_2(r,r,p_1, p_2)  - Q_{p_1' p_2'}^{p_1 p_2} f_2(r,r,p_1', p_2') \right].
$$ 

The first term on the right hand side represents the forward scattering process, and the second term the backwards process. Note that the two-particle distribution function $f_2$ enters into the rate, so we have not expressed the collision integral solely in terms of $f_1$ to close the BBGKY hierarchy. 

We can finish this task by making two further assumptions. First, if the interaction $U$ obeys parity invariance, which sends $r, p \to -r, -p$, as well as time-reversal invariance, which sends $r, p \to r, -p$, then the scattering factor $Q$ must obey $Q_{p_1 p_2}^{p_1' p_2'} = Q_{p_1' p_2'}^{p_1 p_2}$. That is, $Q$ is invariant under exchanging incoming and outgoing momenta. Second, we assume that the velocities of the two particles involved in the scattering are initially \emph{uncorrelated}, which means that the two-particle distribution function can be written in terms of products of one-particle distribution functions as

\begin{equation}
f_2(r, r, p_1, p_2) = f_1(r, p_1) f_1(r, p_2).
\end{equation}

This sometimes goes by the name of the assumption of \emph{molecular chaos}.\footnote{This is quite an amazing assumption, actually! Though I won't give much detail here, this assumption is ultimately what leads to the ``arrow of time''. One of Boltzmann's seminal results was his $H$-theorem, which is something of a precursor to the second law of thermodynamics (in fact, $S$$=$$-k_B H$). The theorem states that the quantity $H$ is monotonically decreasing with time. But this seems impossible; the laws we started with were time-reversal symmetric, so how could any quantity only \emph{decrease} with time? If we reversed time, wouldn't that quantity now increase, contradicting the theorem? (This point was first made by Loschmidt soon after Boltzmann discovered his $H$-theorem.) It turns out that this assumption is what violates time reversal, though it was highly non-obvious at first. By assuming that velocities are uncorrelated going into the collision, but not necessarily coming out of the collision, we have introduced an asymmetry with respect to time evolution. Had we assumed the opposite -- that velocities were uncorrelated after the collision but not before -- we would get that entropy always decreases, rather than increases. Interestingly, even in systems that don't have an $H$-theorem since they violate molecular chaos, there is still a form of the second law of thermodynamics (!) -- for an example, see Ref.~\citenum{PhysRevA.4.747}.} Finally, with these assumptions in hand, we arrive at the closed-form equation for the one-particle scattering integral. This is the (classical) \emph{Boltzmann equation},

\begin{equation}
\label{eq:Boltzmann}
\pd{f_1}{t} = \{H_1, f_1\} + \int d^3 p_2 d^3 p_1' d^3 p_2' \ Q_{p_1 p_2}^{p_1' p_2'} \left[ f_1(r, p_1') f_1(r, p_2') - f_1(r, p_1) f_1(r,p_2) \right].
\end{equation}

Note that this equation must change when we work with (electronic) quantum systems; we have to ensure that the states we are scattering \emph{into} are unoccupied due to the Pauli exclusion principle. This means that we have

\begin{align}
\label{eq:QBoltzmann}
\pd{f_1}{t} = \{H_1, f_1\}  + \int d^3 p_2 d^3 p_1' &d^3 p_2' \ Q_{p_1 p_2}^{p_1' p_2'} \Big[ f_1(r, p_1') f_1(r, p_2') (1-f_1(r, p_1))(1-f_1(r,p_2)) \nonumber \\ &-f_1(r, p_1) f_1(r,p_2) (1-f_1(r,p_1'))(1-f_1(r,p_2')) \Big].
\end{align}

This goes by the name of the \emph{Quantum Boltzmann Equation}, or QBE, and will be used extensively later in this chapter. 

The equilibrium form of the single-particle distribution function is just the Maxwell-Boltzmann distribution; call this $f_1^{\text{MB}} = n(r,t) (2\pi m k_B T)^{-3/2} \exp( - m (v - u)^2 / 2 k_B T)$, where the temperature $T$ and drift velocity field $u$ are space and time dependent. But it turns out that this doesn't solve the Boltzmann equation. In particular, it fails to solve the streaming term, $\{H_1, f_1\}$ (though it does solve the collision term, $\left( \pd{f_1^{\text{MB}}}{t}\right)_{\text{coll}} =0$). To find the solution, then, we perturbatively write $f_1 = f_1^{\text{MB}} + \delta f_1$. The heart of the scattering time approximation, also called the \emph{relaxation time approximation}, is to assume

\begin{equation}
\left( \pd{f_1}{t}\right)_{\text{coll}} = - \frac{\delta f_1}{\tau}.
\end{equation}

This is also called the Bhatnagar-Gross-Krook (BGK) operator~\cite{PhysRev.94.511}. In principle $\tau$ may be a function of momentum, $\tau(q)$, but often a constant suffices. 

The consequences of this seemingly benign approximation are profound, particularly for transport properties. Under this approximation, we have that the thermal conductivity $\kappa$ is proportional to this scattering time, with $\kappa = (5/2) \tau n k_B^2 T$, and similarly the shear viscosity is also proportional to $\tau$ as $\eta = \tau n k_B T$. In general, any transport quantity with a Kubo formula should have $\tau$ appear. The Kubo formula, explained in more detail later on, gives the linear-response coefficient for a given driving current (i.e. the conductivity $\sigma$ for charge currents, $\kappa$ for heat currents, etc.) in terms of the two-point fluctuations of the driving current. Using the fluctuation-dissipation theorem, the two-point fluctuations must decay at the same rate $\tau$ that the current dissipates, giving the result that generally transport coefficients are proportional to the scattering time. 

We are often interested in so-called `semi-classical' calculations. This means that, though the underlying system is quantum, the only effect of quantum mechanics is on the scattering time $\tau \to \tau_Q$. This $\tau_Q$ is calculated from first principles in the quantum system, either using perturbation theory (Feynman diagrams)\footnote{Sometimes the transport coefficients $\kappa$ or $\sigma$ are calculated first using diagrams, then $\tau_Q$ is extracted from that, but the results are the same.} or Fermi's Golden Rule. From that point, though, we simply insert $\tau_Q$ into the \emph{classical} kinetic theory equations, treating everything else as a classical system. Despite the somewhat shoehorned nature of this treatment, it is remarkably accurate. Though prefactors are often incorrect, the dependence of transport quantities with temperature and other thermodynamic variables is often precisely right. This is quite remarkable given the drastic nature of this approximation -- an essentially classical picture turns out to be right. Transport formulas for quantum systems involving $\tau_Q$ are often referred to as \emph{Drude formulas}, since the classical results are due to Drude. A central goal of transport calculations, then, is to get $\tau_Q$; once we have it, we can grab all the transport quantities in one fell swoop. This strongly motivates our calculation of $\tau_Q$ in the context of thermal Coulomb drag later in the chapter. 

For completeness, let us finally state the hydrodynamic equations of motion -- the famous Navier-Stokes (NS) equations. The NS equations deal with coarse-grained quantities, namely density $\rho(r,t)$, pressure $P(r,t)$, temperature $T(r,t)$, external forces $F(r,t)$, and the drift velocity $u(r,t)$. The density obeys a continuity equation, due to the conservation of mass: $\partial_t \rho + \nabla \cdot (\rho u) = 0$. We further assume that the viscosity is roughly constant, $\nabla \eta = 0$. These conditions, in conjunction with the Boltzmann equation above (Eq.~\ref{eq:Boltzmann}), give the \emph{Navier-Stokes equations}

\begin{align}
\label{eq:NS}
\rho \left( \partial_t + u\cdot \nabla \right) T &= \frac{2}{3} \kappa \nabla^2 T - \frac{2m}{3} P \nabla \cdot u \\
\rho \left( \partial_t + u\cdot \nabla \right) u &= \frac{\rho}{m} F - \nabla P + \eta \nabla^2 u + \left( \frac{\eta}{3} + \zeta\right) \nabla (\nabla \cdot u).
\end{align}

The first equation is sometimes called the heat conduction equation, with the second the canonical Navier-Stokes equation. We have generalized slightly to include the possibility of a \emph{bulk viscosity} $\zeta$, which vanishes for dilute gases, in addition to the shear viscosity $\eta$. 

\subsection{Coulomb drag and viscosity}

When two plates are brought together in quantum mechanics, strange things may happen. Even two plates in vacuum feel a small, but nonzero, force of attraction, called the \emph{Casimir force}. This is due to the fact that there are more modes outside the plates -- infinitely many, in fact -- than there are between the plates, of which there are only discretely many. The quantum fluctuations, therefore, do not cancel between the inside and the outside, leading to a small inward push that gets stronger as the distance between the plates decreases. 

Much less exotically, simply running a current through one plate can `drag along' a current in the second plate, due to the quantum interactions between the plates. This cannot happen in classical systems in general,\footnote{Simply, a current-carrying wire/plate only creates magnetic fields, not electric ones, and hence pulls no current in a parallel wire/plate. Since both plates are at charge neutrality, the enclosed charge is 0 for any Gaussian surface, so by Gauss's law we have no electric field. This picture is slightly complicated by the discrete nature of charge carriers and the electro-dynamical aspects of drag, but the absence of Coulomb drag remains in classical, or high-$T$ quantum, systems.} due to symmetry considerations, but the higher-order fluctuations of quantum systems can again yield a non-zero drag current. Concretely, charge carriers may tunnel from the bottom plate to the top, interact, then tunnel back, which would be impossible in classical systems. This (generally order $U^2$) process is what is known as \emph{Coulomb drag}. First discussed by Pogrebinskii~\cite{pogrebinskii}, it has proved to be a sensitive probe of quantum transport. Depending on the underlying charge carriers and their interactions, the \emph{Coulomb drag conductivity}, defined as $\sigma_D\equiv J^{(1)} / E^{(2)}$ where $ J^{(1)}$ is the response in layer 1 to a drive electric field $E^{(2)}$ in layer 2, displays multiple regimes with different scalings with temperature. The drag conductivity goes to zero at both small and large temperatures (recovering the classical limit), while it peaks at some intermediate temperature scale. A measurement of this quantity can thus distinguish between different competing theories of quantum systems' quasiparticle content and shed light on the relevant energy scales where crossovers occur.

Within the context of hydrodynamics, the drag conductivity may be thought of as a kind of shear viscosity. The usual `two-plate' definition of shear viscosity considers two plates mechanically connected by intervening fluid. If we set plate 1 in motion at drift velocity $u^{(1)}$, then the bottom plate would be dragged by the velocity field in the fluid, requiring a force $F^{(2)}$ in the $-\hat{u}^{(1)}$ direction to hold it still. We'd then have 

$$
F^{(2)} = \eta \frac{A}{d}u^{(1)},
$$

with $A$ the area of the plates and $d$ their separation. For an applied electric field $E^{(2)}$ we have a force $F^{(2)} = q E^{(2)}$, where the total charge is $q = -e n A$, with $n$ the electronic charge density. Similarly the current is related to the velocity by $j = - e n u$. We then have $-e n A  E^{(2)} = -\eta A j^{(1)} / d n e$. Equating terms yields

\begin{equation}
\eta = \frac{d e^2 n^2}{\sigma_D} = d e^2 n^2\rho_D,
\end{equation}

with $\rho_D$ the drag resistivity. 

With all this in mind, we seek to investigate Coulomb drag as the quantum analog of the shear viscosity. Nearly all studies of drag in the past 40 years have focused on the above setup of charge drag, but we may just as well ask what happens when the drive current is a \emph{heat} current rather than a charge current. We now give our investigations into the thermal analog of the Coulomb drag, and its differences with traditional charge drag. 

\section{Energy Drag in Particle-Hole Symmetric Systems as a Quantum Quench}

\subsection{Introduction}

Since its inception~\cite{pogrebinskii}, the Coulomb drag phenomenon -- whereby a charge current in one layer pulls a reciprocal current in another through Coulomb interactions alone -- has shed light on the special role of interaction effects in quantum transport~\cite{RevModPhys.88.025003}. Coulomb drag measurements have been instrumental in studying the microscopic structure of systems as diverse as double-quantum well structures~\cite{PhysRevLett.66.1216,EISENSTEIN1992107}, excitons in electron-hole bilayers~\cite{doi:10.1063/1.2132071,PhysRevLett.101.246801,PhysRevLett.102.026804,PhysRevLett.106.236807}, quantum Hall states~\cite{PhysRevB.49.11484,PhysRevLett.88.126804,PhysRevLett.93.036801,PhysRevLett.84.5808,doi:10.1116/1.3319260}, Luttinger liquids~\cite{Debray_2001,Laroche631}, spin currents in two-dimensional electron gases~\cite{PhysRevB.62.4853,Weber:2005aa}, and bilayer \\ \noindent graphene~\cite{KIM20121283,PhysRevB.83.161401,Gorbachev:2012aa,PhysRevB.76.081401,PhysRevLett.109.236602,PhysRevLett.110.026601,PhysRevLett.111.166601}, among others. From the theoretical point of view, the Coulomb drag conductivity generally shows a rich dependence with temperature, with each regime dominated by different microscopic processes, and has been generalized in many directions~\cite{RevModPhys.88.025003}. Given the recent interest in the hydrodynamic behavior of electrons in solids~\cite{PhysRevLett.94.111601,PhysRevLett.118.226601,Mackenzie_2017}, an analogy can also be made between the Coulomb drag and the shear viscosity, two processes leading to the equalization of currents in neighboring layers.

In light of this history, it stands to reason that the Coulomb drag effect between {\it thermal} currents, first studied to our knowledge in Ref.~\citenum{PhysRevB.98.035415} -- in which a thermal current in one layer may drag along a reciprocal thermal current in the other through Coulomb interactions -- could elucidate the microscopic structure of quantum systems as well. In fact, in one particularly interesting class of quantum systems -- those having particle-hole symmetry -- Coulomb charge drag effects are known to vanish at leading order~\cite{PhysRevB.76.081401}.  Momentum is transferred between the layers at this order, but it cannot result in a charge current~\cite{Rojo_1999}.  This is not a straightforward effect of symmetry, which would lead to vanishing at all orders; rather the leading process in perturbation theory is independent of the sign of the scattering potential, as with the Born approximation, so that the currents induced by particle-particle and particle-hole scattering cancel.  Such systems are prime candidates for the study of thermal drag, as thermal drag need not vanish under particle-hole symmetry, and we find it to be the dominant form of drag in such systems. Examples include the Hubbard model at half filling, graphene near the Dirac point, and superconductors probed at low energy, among others.

\begin{figure}
    \centering
    \includegraphics[width=0.7\columnwidth]{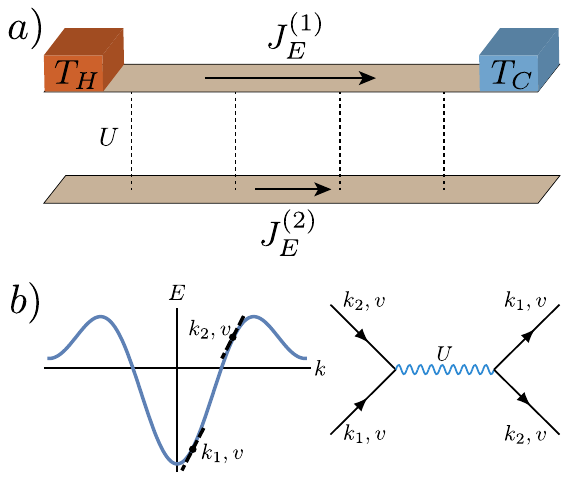}
    \caption{\label{fig:cartoon} (a) The thermal Coulomb drag geometry considered in this paper. A conducting quantum system's top layer is held at a temperature gradient by connecting it to two reservoirs at temperatures $T_H > T_C$, causing a thermal current to flow; through quantum interactions $U$, a thermal current is dragged in the bottom layer. (b) The source of the divergent scattering process leading to the breakdown of the usual Fermi's Golden Rule in one-dimensional systems, namely when all incoming and outgoing particles have the same velocity $v$ but differ in energy.}
\end{figure}

In this Letter, we focus on thermal drag between particle-hole symmetric quantum systems, viewed through the lens of a quantum quench of the inter-layer interactions in a bilayer system. We find that thermal drag does indeed dominate drag physics in these systems and, in sharp contrast to charge drag, suffers from a scattering singularity generic to one-dimensional band structures. This singularity leads to a violation of the na\"ive Fermi's Golden Rule, where the rate of change of the thermal current is {\it logarithmic} in time rather than constant, in the thermodynamic limit. This implies that a simple scattering rate analysis is generally incorrect, and more sophisticated perturbation theory analysis must be used; in particular, the approximation of linearizing the spectrum cannot be used when dealing with thermal currents without some method of regulation.

\subsection{A quench and a Kubo formula} 

To study the thermal drag, let us consider the paradigmatic one-dimensional Hubbard model,

\begin{equation}
    H = -t \sum_{\langle i j \rangle,\sigma} c_{i,\sigma}^\dagger c_{j,\sigma}+ U \sum_i \left(n_{i,\uparrow}-\frac{1}{2} \right) \left(n_{i,\downarrow} - \frac{1}{2}\right)
\end{equation}

where $\{c_{i,\sigma}^\dagger,c_{j,\sigma'}\} = \delta_{ij} \delta_{\sigma \sigma'}$. Let us view the two spin species as each forming separate quantum wires, with on-site interactions coupling them. {We note that the limit of on-site interactions can be physically motivated as originating from a screened Coulomb potential with small screening length.} Initialize one species, say spin-down, in a thermal state at temperature $T$ with some small initial energy current, and initialize the other spin species in a thermal state with no energy current (with $U=0$). Explicitly, since the free fermion chain may be diagonalized by a simple Fourier transform with energies $E_k = -2t \cos k$ and velocities $v_k = 2t \sin k$ (assuming periodic boundary conditions), such a state is given by 

\begin{equation}
\langle n_k^{\sigma} \rangle = \frac{1}{1+\exp(\beta(-2 t \cos k - \mu))} - \delta_{\sigma \downarrow}\epsilon \sin(2k),
\end{equation}

with $\epsilon$ a small parameter {ensuring the validity of linear response}. The charge and thermal current operators carried by the $\sigma$ spin species are given respectively by $J^\sigma = L^{-1} \sum_k v_k n_k^\sigma$ and $J_E^\sigma = L^{-1}\sum_k E_k v_k n_k^\sigma$, hence this initial state has $\langle J_E^\sigma \rangle = \epsilon\delta_{\sigma \downarrow}$ and $\langle J^\sigma \rangle=0$ (diagrammed in Fig.~\ref{fig:cartoon}(a)). In this setup, the spin-down channel is the ``drive'' layer and the spin-up channel is the ``response'' layer in the usual terminology of Coulomb drag, with the caveat that the ``drive'' current is allowed to relax (which does not change the short-time dynamics). We note that, while somewhat unorthodox, this quench interpretation of the Coulomb drag problem is physically reasonable and allows for the use of techniques from scattering theory and integrability that would be inapplicable in an equilibrium description. To avoid confusion, from now on, we set the Hubbard hopping parameter $t=1$.

At time $t=0$, let us quench on the interaction term $U$. We are interested in the change over time of the heat current in the spin-up channel. From the perspective of linear response, one would expect that an initial thermal current in the spin-down channel would drag along a thermal current in the spin-up channel, leading to the development of a temperature gradient for the spin-up species that is proportional to the initial energy current. This would give a thermal drag conductivity of 

\begin{equation}
\kappa_D = \frac{J_E^{(1)} }{\nabla T^{(2)} }
\end{equation}

where $J_E^{(1)}$ is taken at time $t=0$, and here $(1)$ refers to spin-up and $(2)$ to spin-down. Now, generally speaking, there is no perturbing Hamiltonian for a temperature gradient, so there is no straightforward method of deriving a Kubo formula for thermal conductivities. One may argue, however, based on entropy production in the system, that there exists an effective perturbing Hamiltonian and from this derive a Kubo formula~\cite{pottier}. Adapting this method, we arrive at a Kubo formula for the thermal drag conductivity~\cite{PhysRevB.98.035415},\footnote{See Sec.~\ref{sec:kubo} for a derivation of the thermal drag Kubo formula, which includes Refs.~\citenum{doi:10.1143/JPSJ.12.1203,PhysRev.135.A1505,pottier,PhysRevB.78.205407,ziman_book,PhysRevB.78.205407,PhysRevB.73.165104}.}

\begin{equation}
\label{eq:Kubo}
\kappa_{ab}^{\sigma\sigma'}(q,\omega) = \frac{1}{VT} \int_0^\infty dt e^{(i\omega - 0^+)t} \int_0^\beta d\lambda \langle J_{Q,b}^\sigma(-q,-i\lambda) J_{Q,a}^{\sigma'}(q,t)\rangle
\end{equation}
with $V$ the system size, $\sigma$ and $\sigma'$ layer indices, $q$ the wavevector, $J_Q$ the heat current, and $a$ and $b$ spatial indices (in the case of higher dimensional systems). 

With this Kubo formula in hand, we can connect our quench picture to the thermal drag conductivity by the following argument: if the initial rate of change of the energy current in the spin-down species is some rate $\partial_t \langle J_E^\uparrow \rangle = \Gamma$, then by the fluctuation-dissipation theorem~\cite{Kubo_1966} we should expect that the two-point function is exponentially decaying with the same rate $\Gamma$. This would give $\kappa_D \sim \int_0^\infty dt e^{i\omega t} e^{-\Gamma t} = 1/(\Gamma - i \omega)$, which, identifying $\Gamma = 1/\tau$ with $\tau$ a scattering time, would reproduce the usual Drude relation. We caution that in this case, however, a na\"ive Drude analysis will fail due to the complicated behavior of the energy current post-quench, which we examine below.

To calculate $\Gamma$, we seek the quantity $\partial_t n_k^\uparrow$, under the perturbation of the Hubbard interaction. To lowest (second) order in $U$,

\begin{equation}
\label{eq:dtnk}
\partial_t n_k^\uparrow = U^2 \sum_{k_2,k_3,k_4}S_{k k_2}^{k_3 k_4} \frac{\sin(t \Delta E)}{\Delta E} \delta(\Delta k),
\end{equation}

where $S_{k k_2}^{k_3 k_4} = (1-n_k^\uparrow) (1-n_{k_2}^\downarrow)n_{k_3}^\uparrow n_{k_4}^\downarrow - n_k^\uparrow n_{k_2}^\downarrow (1-n_{k_3}^\uparrow)(1-n_{k_4}^\downarrow)$ is the net Fermi factor for the inward and outward scattering processes, $\Delta k = k + k_2 - k_3 - k_4$ and $\Delta E = E_{k} + E_{k_2} - E_{k_3} - E_{k_4}$. In the usual Fermi's Golden Rule, one takes the limit of large $t$, which sends $\sin(t \Delta E)/\Delta E \to \pi \delta(\Delta E)$ provided that the quantity being integrated against does not diverge at $\Delta E = 0$. This is the case for Coulomb drag of charge currents, which is well-behaved; however, this is {not} the case for the energy current, as we shall see, and we must deal with the divergence carefully.

Imposing momentum conservation, the energy current grows as

\begin{equation}
\label{eq:dtJE}
\partial_t J_E^\uparrow = \frac{2}{L} \sum_k \sin(2k) \partial_t n_k^\uparrow.
\end{equation}

{We can usefully rewrite this expression by moving the sum on $k$ to an integral in energy space of a quantity $G(E)$, integrating against a kind of ``density of states''.\footnote{See Sec.~\ref{sec:log} for a full derivation of the logarithmic divergence encountered in perturbation theory.} Focusing on half-filling $\mu=0$, the function $G(E)$ contains the essential divergence of the response energy current, namely

\begin{align}
G(E)& \propto  \int dk_1 dk_3 \sum_{\nu = 1,2}  \frac{F(k_1,k_{2,\nu},k_3)}{\abs{v(k_1 + k_{2,\nu} - k_3) - v(k_{2,\nu})}}
\end{align}

with $v(k) = \partial_k E(k) \propto \sin k$ the group velocity, the function $F$ does not diverge, and $\nu$ indexes the solutions to $\Delta E - E = 0$. Clearly, the source of the divergence is the difference of velocities in the denominator, corresponding to a resonance of points in $k$ space with different energies but the same velocity. Physically, this shows that the energy current operator diverges at small energies $\Delta E \approx 0$ which are directly probed by the $\sinc(t\Delta E )$ term in perturbation theory, and it is because of this singular behavior that Fermi's Golden Rule breaks down.}

There are two conditions under which the denominator diverges: the trivial case of $k_1=k_3$, and the nontrivial second solution. In the first instance, one can readily see that the numerator also vanishes, and hence there is no divergence. For the second solution, which occurs here at $k_1 + k_2 - k_3 = \pi - k_2$ but must occur somewhere in a generic one-dimensional band structure, one finds that the numerator also vanishes for a charge current -- and hence, it is well-behaved -- while it does not for the energy current. The divergence is point-like, in the sense that for every incoming $k$ there is a finite set of partners $\{k'\}$ with the same velocity. That there must be at least one partner is a consequence of the lattice, i.e. the periodicity of the band structure (see Figure~\ref{fig:cartoon}(b)).

{At small but finite $E$, we can regularize the denominator, ultimately leading to a logarithmic divergence. A careful accounting yields }

\begin{equation}
g(E) = \epsilon \frac{4 U^2}{(2\pi)^3} \int_{-\pi}^\pi dk \frac{f(k)}{\abs{\sin k/2}} \log E
\end{equation}

where $g(E) =(G(E) + G(-E))/2$ is the symmetric part of $G(E)$, \\ \noindent $f(k) = -2 \sin^2(k) n(E_k) n(-E_k)$, and $n(E)$ is the Fermi-Dirac distribution. Finally, using $\int_{-\infty}^\infty dx \ \log(x) \sinc(x t) = - \pi (\gamma + \log t) / t$, with $\gamma$ the Euler-Mascheroni constant, and keeping only the dominant term in the large $t$ limit, we arrive at the result

\begin{equation}
\frac{\partial_t J_E^\uparrow}{J_E^\downarrow(t=0)} = \alpha \log t + \mathcal{O}(1),
\end{equation}
with 
\begin{equation}
\label{eq:alpha}
\alpha(T) = \frac{U^2}{\pi^2} \int_{0}^{\pi} dk \abs{u(k)}^2 \frac{\sin^2 k \csc(k/2)}{1 + \cosh(2 \beta \sin(k/2))},
\end{equation}

 {where for generality, we have allowed for $k$-dependent interactions, $U(k) = U u(k)$, and in the Hubbard model with onsite interactions $u(k) = 1$. We remark that this logarithmic behavior is quite general: we expect it for any lattice band structure in 1D, as such band structures must generically have points where $v(k) = v(k')$ but $E(k) \not = E(k')$. Further, other kinds of interactions only modify the prefactor of the log growth.} This integral cannot be computed analytically, but for Hubbard, the low- and high-temperature limits are readily analyzed. First, at low temperatures, the denominator is a strongly peaked function about $k=0$; expanding the numerator in Taylor series and performing the integration yields

\begin{equation}
\alpha(T) \underset{T \ll 1} {\approx} \frac{4 U^2\log 2}{\pi^2} T^2,
\end{equation}

in units of Hubbard hopping $t=1$ and $k_B=1$. In the high temperature limit the denominator is approximately constant, yielding

\begin{equation}
\alpha(T)  \underset{T \gg 1} {\approx} 4 U^2 / 3 \pi^2.
\end{equation}

\begin{figure}
	\centering
	\includegraphics[width=\columnwidth]{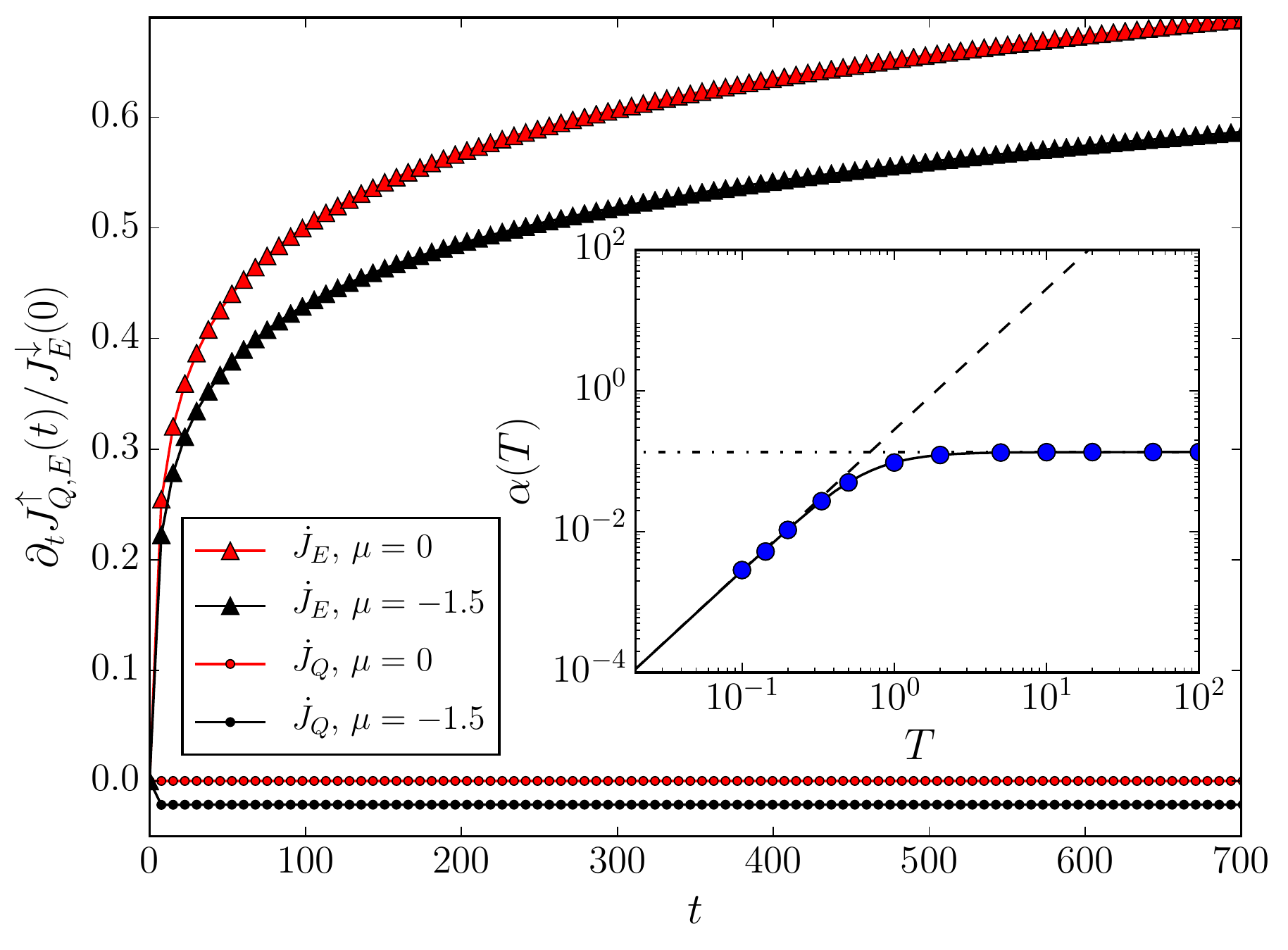}
	\caption{\label{fig:data} The growth of the heat and energy currents in the bottom layer due to the Coulomb drag, to $\mathcal O(U^2)$ in perturbation theory. At half filling $\mu=0$, no charge drag occurs due to particle-hole symmetry (red dots); this is no longer true away from half filling (black dots). In both cases, thermal drag is nonzero and the rate of change grows logarithmically in time as $\alpha(T) \log t$ (red and black triangles), rather than saturating to a constant as would be na\"ively expected. Inset: the prefactor for this log growth $\alpha(T)$ as a function of temperature. Agreement with the analytical formula of Eq.~\ref{eq:alpha} is excellent (solid line); the asymptotics are $\alpha(T) = 4 U^2 T^2 \log 2 / \pi^2$ for small $T$ (dashed line) and $\alpha(T) = 4 U^2 / 3 \pi^2$ for large $T$ (dotted line).}
\end{figure}

We have numerically checked this expression by exactly summing Eq.~\ref{eq:dtnk} on system sizes of $L>3000$ and calculating $\partial_t J_E$ and $\partial_t J$. The results are shown in Fig.~\ref{fig:data}; the logarithmic growth of the energy current is clear both at half-filling ($\mu=0$) and away from half-filling ($\mu=-1.5$). We recover the result that, as expected, there is no charge drag at half filling, confirming that thermal drag dominates in this regime, while we do notice a drag thermopower effect away from half filling. Finally, the observed dependence on temperature of the prefactor of the log, obtained by fitting at various temperatures, is in excellent agreement with Eq.~\ref{eq:alpha}, which we integrate numerically and whose asymptotics we plot. This confirms that the processes considered in this section indeed dominate the thermal drag to an excellent approximation.

A few remarks are now in order. First, the breakdown of Fermi's Golden Rule for the energy current is generic to one-dimensional systems, as any band structure will display the same kind of divergence. Second, due to the divergence, the widespread technique of linearizing the spectrum~\cite{Giamarchi:743140} will fail badly in analyses of thermal drag. {In this case, band curvature effects may be included directly in the field theory and treated perturbatively~\cite{RevModPhys.84.1253}.} Third, the timescale for the validity of perturbation theory is parametrically reduced for thermal drag calculations: perturbation theory holds only up to a timescale $t_*^{-1} \sim U^2 \log U$. {Finally, one may consider the effects of adding a small magnetic field: to lowest order, the field would simply shift the chemical potential in the two species in opposite directions,\footnote{Excluding the effects of the field on the hopping, which are expected to be small.} effectively breaking particle-hole symmetry. In that case, we no longer expect a vanishing charge drag. However, the logarithmic growth of the response heat current would remain, as it is present for any chemical potential, being a consequence of the band structure.}

To access longer times, we make the approximation of a linear spectrum (Luttinger liquid) and regulate the breakdown of Fermi's Golden Rule.\footnote{See Sec.~\ref{sec:FGR}, where we analyze the ``generalized Golden Rule'' trick in more detail.} Linearizing the spectrum produces a left- and a right-moving mode, described by wavevector $q_{L/R} = k \pm k_F$ with dispersion relation $E(q_{L/R}) = \mp v_F q_{L/R}$. We must then consider 8 possible scattering channels: two forward scattering channels, two Umklapp channels, and four backward scattering channels. For simplicity, we slightly modify the setup such that one spin species is kept at a temperature gradient with $k<0$ at $T_L$ and $k>0$ at $T_R$, with the other species in the ground state ($T=0$). 

Analyzing these possible scattering channels, we find that, while the Umklapp and backscattering channels give a finite rate, the forward scattering channel leads to a divergence with system size{, a one-dimensional incarnation of the well-known ``collinear scattering singularity'' in Dirac-dispersing systems~\cite{doi:10.1002/andp.201700043,PhysRevB.83.155441,RevModPhys.88.025003}}. This is due to the fact that, for the forward scattering channel, conservation of energy and momentum become the same constraint, leading to a delta function squared appearing under the scattering integral. This type of divergence was noted in Ref.~\citenum{PhysRevB.78.205407} in the case of Coulomb drag for spinful Luttinger liquids. To recover a finite answer, it was proposed that one go past lowest order perturbation theory, inserting the RPA propagator in place of the bare propagator in the scattering integral (dubbed the ``generalized Fermi's golden rule''). In our case, it amounts to taking the incoming particles to have velocity $v_F$ while the outgoing particles have velocity $u$, the Luttinger velocity, which is interaction dependent. Under this prescription, we find a heat current growth rate that is actually {\it first-order} in the interaction $U$,

\begin{equation}
    \partial_t J_E^{\uparrow} \sim U \frac{2\pi^4 \log 2}{3\hbar v_F} k_B^3 (T_R^3 - T_L^3),
\end{equation}

due to the interaction-renormalized outgoing velocity canceling a power of $U$. In sum, due to the unique divergences of heat drag as opposed to charge drag, we expect a logarithmic heat current growth rate at the shortest times that is second order in $U$, followed by a longer regime of heat current growth rate that is constant in time and first order in $U$. We emphasize that the charge drag in particle-hole symmetric systems vanishes to lowest order, and only enters at order $U^3$ (if at all); hence thermal drag is the dominant form of drag physics in this broad class of systems. 

\subsection{Long-time limit and higher dimensions}
 
Generally speaking, the long-time limit of this quench is outside the realm of validity of perturbation theory, and therefore inaccessible. However, here we may exploit the integrability of the one-dimensional Hubbard model to make progress~\cite{esslerbook}. In particular, due to its integrability, the one-dimensional Hubbard model hosts a tower of conserved quantities, the number of which is extensive in system size. One such quantity, known as $Q_3$, differs from the total energy current operator only by a term of order $U$; that is,

$$
J_E = t^2 \sum_{l,\sigma} i (c^\dagger_{l+1,\sigma} c_{l-1,\sigma} - c^\dagger_{l-1,\sigma} c_{l+1,\sigma}) - \frac{U t}{2}\sum_{l,\sigma} (j_{l-1,\sigma}+j_{l,\sigma}) (n_{l {\bar \sigma}} - 1/2),
$$

which takes the same form as $Q_3$ except for a factor of 2 in the term proportional to $U$~\cite{PhysRevLett.117.116401}. This implies that in the limit of small $U$, $J_E \approx Q_3$ and is hence conserved. (We note that even in the limit of stronger $U$, the overlap of $J_E$ with $Q_3$ will be conserved, leaving some energy current in the final state.) Under the assumption of approach to a generalized Gibbs ensemble final state~\cite{Vidmar_2016} with this same value of $Q_3$, we expect that the energy current will be equally divided between the two wires. That is,

\begin{equation}
J_E^{\uparrow}(t\to \infty) = J_E^{\downarrow}(t\to \infty) = \frac{J_E^{\downarrow}(t=0)}{2}.
\end{equation} 

The conservation of the energy current is likely a special feature due to the integrability of the Hubbard model, but we remark that in this case it leads to an intriguing hydrodynamic transport of energy current reminiscent of the Dirac fluid~\cite{Crossno1058}.

Since the source of the divergent heat drag is related to special properties of scattering in 1D, we do not expect the same divergence to appear generically for higher dimensional systems.
As a check, we have considered the Hubbard model on the square lattice with nearest-neighbor hopping.\footnote{See Sec.~\ref{sec:higherD}, where we show data for the two-dimensional Hubbard model thermal quench.}
We have numerically explored this model for various values of the chemical potential and temperature on system sizes of up to $L_x=L_y=100$. We find that the thermal drag indeed dominates near half-filling, and it does not appear to be divergent. We defer an exhaustive analysis of the two-dimensional case to future work.

\subsection{Discussion} 

We have analyzed a thermal analogue of the Coulomb drag in interacting quantum systems with particle-hole symmetry via a quantum quench in the Hubbard model. We have found that, due to the vanishing of the charge Coulomb drag, the thermal drag effect dominates. In one dimension, its growth is drastically different than the charge drag due to the structure of the energy current operator: the short-time limit shows logarithmic non-Fermi's golden rule growth, followed by a longer regime of linear growth given by a generalized Golden rule, with the late-time limit in this case obtained from integrability arguments. 

We expect these conclusions to apply to a broad range of experimentally realizable systems, including perhaps most prominently graphene near charge neutrality. It is an interesting question whether some components of the thermal Coulomb drag may be topologically quantized in certain systems, especially in light of recent experiments on the thermal Hall effect at non-chiral Hall edges~\cite{Banerjee:2018aa}. We emphasize that, despite the vast literature on the charge Coulomb drag, the thermal drag effect is largely unexplored,\footnote{We note for completeness that another form of thermal drag was recently studied in Ref.~\citenum{PhysRevB.99.201406} but there the drag was mediated by thermal photons rather than the direct Coulomb interaction between charge carriers.}, and is ripe for further study.

\section{Appendices}

\subsection{Thermal drag conductivity Kubo formula} \label{sec:kubo}

In this appendix, we provide a derivation of the thermal drag conductivity Kubo formula presented in the main text.

Generally speaking, since there is no external perturbation (nor perturbing Hamiltonian) associated to thermal gradients and heat currents, there is no straightforward way to apply a Kubo-type formalism to heat transport. However, it is nonetheless possible to relate the presence of a temperature gradient to an ``effective'' perturbing Hamiltonian that would produce the same entropy growth in the system~\cite{doi:10.1143/JPSJ.12.1203,PhysRev.135.A1505,pottier}, and this leads to a Green-Kubo formula for the thermal conductivity. Here we apply this argument to the thermal drag conductivity. 

Consider a bilayer system, labeled by $\sigma = 1,2$ with unperturbed Hamiltonian $H_0$. Usually, to derive a Green-Kubo formula we would write a perturbing Hamiltonian $H_{\rm pert} = \int dr \phi(r,t) \rho(r)$, with $\phi(r,t)$ the applied potential and $\rho(r)$ the degree of freedom of the system that couples to the potential. The time derivative is $\adot H_{\rm pert} = \int dr \phi(r,t) \adot \rho(r)$, with $\adot \rho(r)$ corresponding to the unperturbed evolution of $\rho(r)$ via $i \adot \rho = [\rho,H_0]$ (in units of $\hbar=1$). Now, one can show that $\adot H_{\rm pert}$ balances the entropy production of the system: $\adot H_{\rm pert} = -T \int dr \sigma_S$, with $\sigma_S$ the entropy source; we will use this expression to write a ``perturbing Hamiltonian'' even in the absence of a perturbing potential $\phi$. 

Let us assume that each layer (with layer index $\sigma$) is in a situation of local equilibrium, where the local temperature is $T^\sigma(r,t) = \tilde T^\sigma + \delta T^\sigma(r,t)$, the local energy density is $\epsilon^\sigma(r,t) = \tilde \epsilon^\sigma + \delta \epsilon^\sigma(r,t)$, and the local particle density is $n^\sigma(r,t) = \tilde n^\sigma + \delta n^\sigma(r,t)$. Let us introduce the perturbing Hamiltonian

\begin{equation}
H_{\rm pert}^\sigma = \int dr \frac{\delta T^\sigma(r,t)}{\tilde T^\sigma} \left(\epsilon^\sigma(r,t) - \frac{\tilde \epsilon^\sigma + \tilde P^\sigma}{\tilde n^\sigma} n^\sigma(r,t)\right)
\end{equation}

where $\tilde P^\sigma$ is the equilibrium pressure. Appealing to the relation $TdS = dE + PdV$ and the condition $dN=0$, we see that $TdS = d\epsilon - dn (\tilde \epsilon + P)/\tilde n$ and hence the right hand side can be interpreted as a local density of thermal energy. 

Now consider the time derivative of $H_{\rm pert}$. We assume that we are in a hydrodynamic regime $\omega \tau \ll 1$, $q l \ll 1$, with $\tau$ the typical time between collisions and $l$ the mean free path. This allows us to use hydrodynamic equations, which in linearized form are $\partial_t n(r,t) + \tilde n \nabla \cdot u(r,t) = 0$ and $\partial_t \epsilon(r,t) + \nabla \cdot J_E(r,t) = 0$, with $u(r,t)$ the velocity of the fluid (suppressing layer indices). The linearized energy flux is $J_E(r,t) = (\tilde \epsilon + \tilde P)u(r,t) + J_Q(r,t)$. Finally, this allows us to write the time derivative of the perturbing Hamiltonian in terms of the heat current:

\begin{equation}
\adot H_{\rm pert}^\sigma = - \int dr \frac{\delta T^\sigma(r,t)}{\tilde T^\sigma} \nabla \cdot J_Q^\sigma = -\tilde T^\sigma \int dr J_Q^\sigma \cdot \nabla \frac{1}{T^\sigma}
\end{equation}

where in the last equality we have integrated by parts, and used the fact that, to lowest order in $\delta T$, we have $\nabla \delta T = - \tilde T^2 \nabla(1/T)$. 

With this perturbing Hamiltonian in hand, we can now turn the crank of linear response and produce a Green-Kubo formula for the drag thermal conductivity. Write

\begin{align}
\langle J_{Q,a}^{\sigma'}(r,t) \rangle &= -\frac{1}{\tilde T^\sigma} \int dr' \int_{-\infty}^\infty dt' \chi_{BA}(r-r',t-t') \nabla_b \delta T^\sigma(r',t')  \nonumber \\
&= -\frac{1}{\tilde T^\sigma} \int dr' \int_{-\infty}^t dt' \int_0^\beta d\lambda \langle J_{Q,b}^\sigma(r',-i\lambda) J_{Q,a}^{\sigma'}(r,t-t')\rangle \nabla_b \delta T^\sigma(r',t')
\end{align}

with $a,b$ spatial indices, and $\chi_{BA}$ the canonical Kubo correlation function with $\adot A(r) = J_{Q,b}^\sigma$ and $B(r) = J_{Q,a}^{\sigma'}$. Fourier transforming gives the definition of the drag thermal conductivity,

\begin{equation}
\langle J_{Q,a}^{\sigma'}(q,\omega)  \rangle = - \kappa_{ab}^{\sigma \sigma'}(q,\omega) [\nabla_b \delta T^\sigma](q,\omega).
\end{equation}

Finally, comparison gives the Kubo formula for the drag thermal conductivity tensor,

\begin{equation}
\kappa_{ab}^{\sigma\sigma'}(q,\omega) = \frac{1}{VT} \lim_{\eta \to 0^+} \int_0^\infty dt e^{(i\omega - \eta)t} \int_0^\beta d\lambda \langle J_{Q,b}^\sigma(-q,-i\lambda) J_{Q,a}^{\sigma'}(q,t)\rangle,
\end{equation}

recalling that $a,b$ are spatial indices and $\sigma,\sigma'$ are layer indices.

\subsection{Logarithmic divergence of heat drag}\label{sec:log}

In this appendix, we show that lowest-order time-dependent perturbation theory does not predict a Fermi's Golden Rule for the heat drag, but rather predicts a logarithmic divergence with time. 

We treat charge ($a=1$) and heat drag ($a=2$) on the same footing. At $t=0$, the up spins are thermal at temperature $T$, and the down spins have a distribution $n(k)$ given by thermal (at the same $T$) plus a small $\sin(a k)$ component:

\begin{align}
n_{k,\downarrow}^0 &= \frac1{1+e^{-2\beta_{\downarrow} \cos(k)}} + \eta \sin( a k) \\
n_{k,\uparrow}^0 &= \frac1{1+e^{-2\beta_{\uparrow} \cos(k)}}
\end{align}

The initial current is

\begin{align}
J_{b,\downarrow}(t=0) &= \frac2N \sum_k \sin( b k) n_{k,\downarrow}^0 = \eta \frac1{\pi} \int_{-\pi}^{\pi} \sin(a k)\sin(b k) = \eta \delta_{ab}.
\end{align}

To leading order in $U$, one finds for the time derivative of the current in the up-spins (writing $J = J^\uparrow_a$):

\begin{align}
\partial_t J &= \frac2N \sum_{k_3} \sin(a k_3) \partial_t \left\langle n_{k_3,\uparrow}(t) \right\rangle \nonumber \\ &= -4 t \eta \frac{1}{N^3} \sum_{k_1,k_2,k_3} F(k_1,k_2,k_3) \abs{U(k_3 - k_1)}^2 \mathrm{sinc}\left(t E\right),
\end{align}

where $F$ is a well-behaved function (no divergence), given by a sums of product of Fermi-Dirac terms and sines, and where $E = (\epsilon_{k_1} - \epsilon_{k_3} + \epsilon_{k_2} - \epsilon_{k_1+k_2-k_3})$. $k_3 - k_1$ is the net momentum transferred between the two layers, and in the Hubbard model, we have $U(q) = U$ since the interaction is on-site. 

It is instructive to rewrite this as

\begin{align}
\partial_t J = t \int dE \ G(E) \ \mathrm{sinc}\left(t E\right)
\end{align}

with

\begin{align}
&G(E) = -4 \eta \frac{1}{N^3} \sum_{k_1,k_2,k_3} F(k_1,k_2,k_3) \abs{U(k_3 - k_1)}^2 \delta(\epsilon_{k_1}  + \epsilon_{k_2}- \epsilon_{k_3} - \epsilon_{k_1+k_2-k_3} -E) \nonumber \\
&= -4 \eta \frac{1}{(2 \pi)^3} \int dk_1 dk_2 dk_3 F(k_1,k_2,k_3) \abs{U(k_3 - k_1)}^2 \delta(\epsilon_{k_1}  + \epsilon_{k_2}- \epsilon_{k_3} - \epsilon_{k_1+k_2-k_3} -E) \nonumber \\
&= - 2\eta \frac{1}{(2 \pi)^3} \int dk_1 dk_3 \sum_{\mu=1,2} \frac{ F(k_1,k_{2,\mu},k_3)}{|\sin(k_1+k_{2,\mu}-k_3) - \sin(k_{2,\mu})|} \abs{U(k_3 - k_1)}^2 ,
\end{align}

where $\mu$ indexes the solutions of $\epsilon_{k_1}  + \epsilon_{k_2}- \epsilon_{k_3} - \epsilon_{k_1+k_2-k_3} -E =0$, of which there are generically two for a given $k_1, k_3$.
We are only interested in the $E$-even part of $G$, so let us define

\begin{equation}
g(E) = \frac{1}{2} (G(E)+G(-E))
\end{equation}

with

\begin{align}
&g(E) = - 2\eta U^2 \frac1{(2 \pi)^3} \int dk_1 dk_3 \sum_{\mu=1,2}  \frac{f(k_1,k_{2,\mu},k_3)}{|\sin(k_1+k_{2,\mu}-k_3) - \sin(k_{2,\mu})|} \abs{U(k_3 - k_1)}^2, \nonumber \\
&f(k_1,k_{2,\mu},k_3) = \frac12 (F(k_1,k_{2,\mu},k_3) + F(k_3,k_1+k_3-k_{2,\mu},k_1).
\end{align}

Writing $f$ explicitly leads to

\begin{align}
2 f(k_1,k_{2},k_3) =  
\sin(a k_3) 
\Big[ &\sin(a (k_1+k_2-k_3)) n_{k_3,\uparrow}^0  (1-n_{k_1,\uparrow}^0) (1-n_{k_2,\downarrow}^0) \nonumber \\
&-\sin(a k_2) n_{k_3,\uparrow}^0 n_{k_1+k_2-k_3,\downarrow}^0 (1-n_{k_1,\uparrow}^0)  \nonumber \\
& + \sin(a(k_1+k_2-k_3)) (1-n_{k_3,\uparrow}^0) n_{k_1,\uparrow}^0  n_{k_2,\downarrow}^0 ) \nonumber \\ &- \sin(a k_2) (1-n_{k_3,\uparrow}^0) (1-n_{k_1+k_2-k_3,\downarrow}^0) n_{k_1,\uparrow}^0  \Big] \nonumber \\
&\hspace{-20mm}+\sin(a k_1)  \Big[ \sin(a (k_2)) n_{k_1,\uparrow}^0 (1-n_{k_3,\uparrow}^0) (1-n_{k_1+k_2-k_3,\downarrow}^0) \\ &-\sin(a (k_1+k_2-k_3)) n_{k_1,\uparrow}^0 n_{k_2,\downarrow}^0 (1-n_{k_3,\uparrow}^0) \nonumber \\
&+ \sin(a k_2) (1-n_{k_1,\uparrow}^0) n_{k_3,\uparrow}^0  n_{k_1+k_2-k_3,\downarrow}^0 ) \nonumber \\&- \sin(a (k_1+k_2-k_3)) (1-n_{k_1,\uparrow}^0) (1-n_{k_2,\downarrow}^0) n_{k_3,\uparrow}^0 \Big], 
\end{align}

which can be rearranged as

\begin{align*}
&2 f(k_1, k_2, k_3) =  \Big(\sin(a k_3) - \sin(a k_1)\Big) \nonumber \\ &\times \Big[-\sin(a k_2) \Big\{ (1-n_{k_3,\uparrow}^0) (1-n_{k_1+k_2-k_3,\downarrow}^0) n_{k_1,\uparrow}^0 + n_{k_3,\uparrow}^0 n_{k_1+k_2-k_3,\downarrow}^0 (1-n_{k_1,\uparrow}^0) \Big\} \nonumber \\
&\qquad + \sin(a(k_1+k_2-k_3)) \Big\{ (1-n_{k_3,\uparrow}^0)  n_{k_1,\uparrow}^0  n_{k_2,\downarrow}^0 + n_{k_3,\uparrow}^0 (1-n_{k_1,\uparrow}^0) (1-n_{k_2,\downarrow}^0) \Big\}  \Big].
\end{align*}

Clearly, the only source of divergence is the factor $\frac1{|\sin(k_1+k_{2,\mu}-k_3) - \sin(k_{2,\mu})|}$. 
This term diverges in two cases, $k_1=k_3$, and $k_1+k_2-k_3 = \pi - k_2$.
In the case of $k_1=k_3$, one can see that $f$ always vanishes, and the divergence is therefore cured.

The case of $k_1+k_2-k_3 = \pi - k_2$ is more tricky. Plugging this in $f$ leads to

\begin{align*}
&2 f_\text{div}(k_1,k_{2},k_3) =  \Big(\sin(a k_3) - \sin(a k_1)\Big) \\ &\times\Big[n_{k_3,\uparrow}^0 (1-n_{k_1,\uparrow}^0) (1-n_{k_2,\downarrow}^0) \sin(a (\pi - k_2)) - n_{k_3,\uparrow}^0 n_{\pi - k_2,\downarrow}^0 (1-n_{k_1,\uparrow}^0) \sin(a k_2)  \\
&\qquad + (1-n_{k_3,\uparrow}^0)  n_{k_1,\uparrow}^0  n_{k_2,\downarrow}^0 \sin(a(\pi - k_2)) - (1-n_{k_3,\uparrow}^0) (1-n_{\pi - k_2,\downarrow}^0) n_{k_1,\uparrow}^0  \sin(a k_2) \Big].
\end{align*}

Focusing on half-filling, one has the property that $(1-n^0_{k})=n^0_{\pi-k}$. This leads to

\begin{align}
2 f_\text{div}(k_1,k_{2},k_3) =  &\Big(\sin(a (\pi - k_2)) - \sin(a k_2)\Big) \Big(\sin(a k_3) - \sin(a k_1)\Big) \nonumber \\ &\times \Big[ n_{k_3,\uparrow}^0 (1-n_{k_1,\uparrow}^0) (1-n_{k_2,\downarrow}^0)  + (1-n_{k_3,\uparrow}^0)  n_{k_1,\uparrow}^0  n_{k_2,\downarrow}^0 \Big].
\end{align}

The first factor vanishes for odd $a$, but is finite for even $a$, and will remain finite unless we fine-tune $\abs{U(k_3 - k_1)}^2$. 
This is why charge drag is not divergent (and actually zero at half-filling, but could be finite away from half-filling), while heat drag is divergent.

\subsubsection{Taking care of the divergence at $k_1+k_2-k_3=\pi - k_2$}

Let us first try naively at $E=0$. In that case, the solutions for $k_2$ are $k_{2,\nu=1} = \pi - k_1$ and $k_{2,\nu=2}=k_3$.
Plugging this in $k_1+k_2-k_3=\pi - k_2$ leads to a line of divergences at $k_1 + k_3 = \pi$,  in both cases.

At small but finite $E$, we can regularize the $\frac1{|\sin(k_1+k_{2,\mu}-k_3) - \sin(k_{2,\mu})|}$ factor as 

\begin{align}
\frac1{|\sin(k_1+k_{2,\mu}-k_3) - \sin(k_{2,\mu})|} \mapsto \mathrm{Re}\left[\frac{1}{\sqrt{(\sin(k_3) - \sin(k_1))^2 - 2 E (\cos(k_3)-\cos(k_1))}}\right].
\end{align}

We finally find

\begin{align}
g(E) = &-\frac{2 \eta}{(2 \pi)^3} \int dk_1 dk_3 \abs{U(k_3 - k_1)}^2 \nonumber \\&\times \sum_{\mu=1,2} f(k_1,k_{2,\mu},k_3) \mathrm{Re}\left[\frac{1}{\sqrt{(\sin(k_3) - \sin(k_1))^2 - 2 E (\cos(k_3)-\cos(k_1))}}\right].
\end{align}

We change variables first, using $k_\pm = k_1 \pm k_3$, obtaining

\begin{align}
g(E) = &-\frac{\eta}{(2 \pi)^3} \int dk_+ dk_- \abs{U(-k_-)}^2 \nonumber \\&\times \sum_{\mu=1,2}  f_\mu(k_+,k_-) \mathrm{Re}\left[\frac{1}{\sqrt{4 \cos(k_+/2)^2 \sin(k_-/2)^2 - 4 E \sin(k_+/2) \sin(k_-/2)}}\right].
\end{align}

On physical grounds, for time-reversal symmetric interactions, we expect $U(-q) = U(q)$, so from now on let us assume that $U$ is even. Since the integral will be dominated by the near-divergence of the denominator close to $k_+=\pi$, we can approximate $f$ to take its value on that line, which is the same for the two solutions:

\begin{align}
g(E) = &- \frac{2 \eta U^2}{(2 \pi)^3} \int dk_+ dk_-  \abs{U(k_-)}^2 \nonumber \\&\times f(k_+=\pi,k_-) \mathrm{Re}\left[\frac{1}{\sqrt{4 \cos(k_+/2)^2 \sin(k_-/2)^2 - 4 E \sin(k_+/2) \sin(k_-/2)}}\right].
\end{align}

We can now perform the integral over $k_+$. Writing $k_+ = \pi + \epsilon$, one finds

\begin{align}
g(E) &= - \frac{4 \eta }{(2 \pi)^3} \int_{-\pi}^{\pi} dk_- \abs{U(k_-)}^2 \int_{-K}^{+K} d\epsilon   f(k_+=\pi,k_-) \mathrm{Re}\left[\frac{1}{\sqrt{\epsilon^2 \sin(k_-/2)^2 - 4 E  \sin(k_-/2)}}\right] 
\end{align}

where we added a factor of 2 because we restricted $k_1$ and $k_3$ to lie in the first quadrant (which leads to $k_-$ running from $-\pi$ to $\pi$). $K$ is a large momentum cutoff, which will lead to a non-divergent piece that will be discarded.

Using

\begin{equation}
\int_{-K}^{+K} dx \mathrm{Re}\left[\frac1{\sqrt{x^2-b}}\right] \rightarrow - \log |b| 
\end{equation}

in the small $b$ limit, one finds

\begin{align}
g(E) &= \frac{4 \eta U^2 }{(2 \pi)^3} \int_{-\pi}^{\pi} dk_- \abs{U(k_-)}^2 \frac{f(k_+=\pi,k_-)}{|\sin(k_-/2)|}  \log\left(\frac{4 |E|}{|\sin(k_-/2)|} \right).
\end{align}

Focusing on the divergent piece, we finally find

\begin{align}
g(E) &= \alpha \log|E|, \qquad \qquad 
\alpha = 4 \eta U^2  \frac1{(2 \pi)^3} \int_{-\pi}^{\pi} dk_- \abs{U(k_-)}^2    \frac{f(k_+=\pi,k_-)}{|\sin(k_-/2)|},
\end{align}

where, at half-filling and $k_+=\pi$, one has

\begin{align}
2 f(k_1,k_{2},k_3) = &\Big(\sin(a (\pi - k_2)) - \sin(a k_2)\Big) \Big(\sin(a k_3) - \sin(a k_1)\Big) \nonumber \\&\times\Big[ n_{k_3,\uparrow}^0 (1-n_{k_1,\uparrow}^0) (1-n_{k_2,\downarrow}^0)  + (1-n_{k_3,\uparrow}^0)  n_{k_1,\uparrow}^0  n_{k_2,\downarrow}^0 \Big].
\end{align}

Focusing on heat drag ($a=2$), one finds

\begin{align}
2 f(k_1,k_3) &=  - 4 \sin(k_-)^2  \Big( (n_{k_3,\uparrow}^0)^2  n_{k_1,\downarrow}^0  + (n_{k_1,\uparrow}^0)^2  n_{k_3,\downarrow}^0 \Big) \nonumber \\&= - 4 \sin(k_-)^2  \Big( (n_\uparrow(\epsilon)^2  n_\downarrow(-\epsilon)  + (n_\uparrow(-\epsilon)^2 n_\downarrow(\epsilon)\Big),
\end{align}

where $\epsilon(k_-)=-2\sin(k_-/2)$ and $n_{\uparrow,\downarrow}(\epsilon) = 1/(1+e^{\beta_{\uparrow,\downarrow} \epsilon})$.

Let us focus on $T_\uparrow = T_\downarrow$ for now, leading to

\begin{align}
2 f(k_1,k_3) &=  - 4 \sin(k_-)^2  \Big( (n(\epsilon)^2  n(-\epsilon)  + (n(-\epsilon)^2 n(\epsilon)\Big) = - 4 \sin(k_-)^2   n(\epsilon)  n(-\epsilon).
\end{align}

We finally do the integral over $E$, by using

\begin{equation}
\int_{-\infty}^\infty dx \log(x)\ \mathrm{sinc}(xt) = - \pi \frac{\gamma + \log(t)}t,
\end{equation}

where $\gamma$ is the Euler-Mascheroni constant.
Keeping only the dominant term, this leads to

\begin{align}
\partial_t J &= t \alpha  \int dE \ \log(E) \ \mathrm{sinc}\left(t E\right) = \alpha' \log(t) + \mathcal{O}(1),
\end{align}

where

\begin{align}
\alpha' = -\pi \alpha &= \frac{\eta U^2 }{\pi^2} \int_{-\pi}^{\pi} dk_-  \abs{U(k_-)}^2  \frac{\sin(k_-)^2}{|\sin(k_-/2)|}      n(2\sin(k_-/2))  n(-2\sin(k_-/2)) \nonumber \\&\equiv    \frac{\eta U^2}{\pi^2} I(T),
\end{align}

where we write the interaction has a having a characteristic scale $U$ such that $U(q) = U u(q)$, and

\begin{equation}
I(T) \equiv \int_{-\pi}^{\pi} dk_-  \abs{u(k_-)}^2  \frac{\sin(k_-)^2}{|\sin(k_-/2)|}      n(2\sin(k_-/2))  n(-2\sin(k_-/2)).
\end{equation}

Normalizing by the initial current in the down spins leads to

\begin{equation}
\frac{\partial_t J_\uparrow}{J_\downarrow(t=0)}  =  \frac{U^2}{\pi^2} I(T) \log(t)
\end{equation}

as shown in the main text. Finally, we can massage the expression for $I(T)$, using the fact that the integrand is even, to obtain

\begin{align}
	I(T) &= 2\int_{0}^{\pi} dk  \abs{u(k)}^2  \frac{\sin^2 k}{\sin(k/2)}  \frac{1}{1+\exp(-2t\beta \sin(k/2))} \frac{1}{1+\exp(2t\beta \sin(k/2))} \nonumber \\
	&= \int_0^{\pi} dk \abs{u(k)}^2 \frac{\sin^2(k) \csc(k/2)}{1 + \cosh(2 t\beta \sin (k/2))},
\end{align}

where $t$ is the Hubbard hopping parameter (not time).\footnote{Don't blame me; I think this is a terrible notation, but we are bound by historical convention here.} For the particular case $U(k) = U$, i.e. $u(k) = 1$, we have

\begin{equation}
I(T) =  \int_0^{\pi} dk \frac{\sin^2(k) \csc(k/2)}{1 + \cosh(2 t\beta \sin (k/2))} .
\end{equation}

We can analyze its limits as follows: for small $\beta \ll 1$, the denominator of the integrand is approximately constant, so 

\begin{equation}
I(T) \approx  \frac{1}{2} \int_0^{\pi} dk \sin^2(k) \csc(k/2) = 4/3.
\end{equation}

The other limit of $\beta \gg 1$ requires a bit more work, but the key is that the denominator is a sharply peaked function. Rewriting in terms of $\epsilon = 2 t \sin(k/2)$ with $t=1$ for convenience, 

\begin{equation}
I(T) = \int_{-2}^2 d\epsilon \frac{\abs{\sin(2 \arcsin(\epsilon/2))}}{1 + \cosh(\beta \epsilon)}.
\end{equation}

Noting that the integrand is strongly peaked around $\epsilon = 0$ when $\beta \gg 1$, we can extend the limits of integration and Taylor expand the numerator in $\epsilon$, yielding

\begin{equation}
I(T) \approx \int_{-\infty}^\infty d\epsilon \frac{\abs{\epsilon}}{1 + \cosh(\beta \epsilon)}. 
\end{equation}

Finally, we change variables again to $\epsilon' = \beta \epsilon$, yielding 

\begin{equation}
I(T) \approx \frac{2}{\beta^2} \int_0^\infty d\epsilon'  \frac{\epsilon'}{1 + \cosh(\epsilon')} = \frac{4 \ln 2}{\beta^2}.
\end{equation}

A similar analysis may be made for other forms of the interaction $U(q)$. 

\subsubsection{Validity of perturbation theory}

Perturbation theory relies on the occupation numbers (and therefore the currents) being close to their initial values, i.e. $|n_k(t)-n_k(0)| \ll 1$.

The current at time $t$ is given by the integral of the rate,

\begin{align}
\frac{J_\uparrow(t)}{J_\downarrow(t=0)} &= \int_0^t dt' \frac{\partial_t J_\uparrow}{J_\downarrow(t=0)}
= 2 \frac{U^2}{\pi^2} I(T) \ t \log(t) + \mathcal{O}(t)
\end{align}

where we have neglected an $\mathcal{O}(1)$ term in the rate, which leads to an $\mathcal{O}(t)$ term in the current.

Perturbation theory should be accurate as long as 

\begin{equation}
\frac{J_\uparrow(t)}{J_\downarrow(t=0)}  \ll 1,
\end{equation}

which leads to

\begin{equation}
t \log(t) \ll \frac{\pi^2}{2 U^2 I(T)}.
\end{equation}

To a good approximation, this is equivalent to

\begin{equation}
t \ll t^* \sim \frac1{U^2 \log \abs{U}}.
\end{equation}

Remarkably, this is parametrically shorter than the usual time for FGR to break, which is $\sim 1/U^2$.

\subsection{Generalized Fermi's Golden Rule}\label{sec:FGR}

In this appendix, we demonstrate that Fermi's Golden Rule applied to thermal drag in the spinful Luttinger liquid produces a divergence, and detail a method to correct this behavior known as the ``generalized Golden Rule''~\cite{PhysRevB.78.205407}. 

Let us do an explicit calculation of $J_E(t)$ at short times and small $U$ using Fermi's Golden Rule. We set up the problem as follows. Consider the one-dimensional Hubbard model. Prepare the spin-up channel at temperature $T$, such that we have the initial distribution function

\begin{equation}
n_{\up}^0(k) = f_T(k)
\end{equation}

where we define 

\begin{equation}
f_T(k) = \frac{1}{1 + \exp(- \frac{1}{k_B T} 2 t_h \cos k)}
\end{equation}

letting the hopping parameter be $t_h$ to avoid confusion with time $t$. Prepare the spin-down channel with a bath at temperature $T_L$ for the left-movers, $T_R$ for the right-movers (so the left side is at $T_R$ and the right side at $T_L$). Then the distribution function is 

\begin{equation}
n_{\down}^0(k) = \begin{cases}
f_{T_L}(k) & k<0 \\
f_{T_R}(k) & k>0
\end{cases}
\end{equation}

which should be normalized already since the Fermi-Dirac distribution at half filling is symmetric about $k=0$. Note that it is discontinuous at $k=0$, though. Now, the Fermi's Golden Rule transition rate between the states $\ket{k_1\up, k_2\down} \to \ket{k_3\up, k_4\down}$ is~\cite{ziman_book}

\begin{align}
Q_{k_1\up, k_2\down}^{k_3\up, k_4\down} &= \frac{2\pi}{\hbar} \abs{ \bra{k_1\up, k_2\down}\hat U \ket{k_3\up, k_4\down} }^2 \delta(E_{k_1} + E_{k_2} - E_{k_3} - E_{k_4}) \nonumber \\ &=  \frac{2\pi}{\hbar} U^2 \delta(k_1 + k_2 - k_3 - k_4)\delta(E_{k_1} + E_{k_2} - E_{k_3} - E_{k_4})
\end{align}

which explicitly conserves momentum and energy. Since the scattering is reversible, $Q_{k_1\up, k_2\down}^{k_3\up, k_4\down} = Q^{k_1\up, k_2\down}_{k_3\up, k_4\down}$. Hence the rate of change of $n_{k_1}^\up$ is 

\begin{align}
\partial_t n_{k_1}^\up &= \int dk_2 dk_3 dk_4 [n_{k_3}^\up n_{k_4}^\down (1-n_{k_1}^\up)(1-n_{k_2}^\down) - n_{k_1}^\up n_{k_2}^\down (1-n^\up_{k_3} )(1-n_{k_4}^\down) ] Q_{k_1\up, k_2\down}^{k_3\up, k_4\down} \nonumber \\
&=  \frac{2\pi}{\hbar} U^2 \int dk_2 dk_3 dk_4 [n_{k_3}^\up n_{k_4}^\down (1-n_{k_1}^\up)(1-n_{k_2}^\down) - n_{k_1}^\up n_{k_2}^\down (1-n^\up_{k_3} )(1-n_{k_4}^\down) ]  \nonumber \\ 
& \qquad\qquad \times \delta(k_1 + k_2 - k_3 - k_4)\delta(E_{k_1} + E_{k_2} - E_{k_3} - E_{k_4})
\end{align}

with $E_k = -2 t_h \cos k$. Now let's linearize the spectrum. Write

\begin{equation}
\begin{cases}
k = k_F + q_R & k>0 \\
k = -k_F + q_L & k<0
\end{cases}
\end{equation}

and approximate cosine spectrum by a pair of lines:

\begin{equation}
\begin{cases} 
E_k = E_{k_F} + v_F q^R & k > 0 \\
E_k = E_{-k_F} - v_F q^L & k < 0 
\end{cases}
\end{equation}

We take the left and right movers $q^{R,L}$ to be defined in some bandwidth about the Fermi wavevector: $q \in [-\Lambda,\Lambda]$, with $k$ outside this range not contributing to the low-energy physics. We then take the limit $\Lambda \to \infty$, assuming that large $q$ does not contribute very much. 

At half-filling, $E_{k_F} = E_{-k_F} = 0$, $k_F = \pi/2a$ (where the lattice spacing $a=1$) and $v_F = 2t_h$. Hence $E_k = \pm v_F q^{R,L}$.This means that 

\begin{equation}
\int_{BZ} dk = \int_{k<0} dk + \int_{k>0} dk = \int_{-\infty}^{\infty} dq^L + \int_{-\infty}^\infty dq^R
\end{equation}

hence

\begin{align}
&\int dk_2 dk_3 dk_4 f(n_{k_i}) \delta(\sum k) \delta(\sum E_k) = \\ &\left(\int dq_2^L + \int dq_2^R\right) \left(\int dq_3^L + \int dq_3^R\right) \left(\int dq_4^L + \int dq_4^R \right)f(n_{k_i})  \delta(\sum k) \delta(\sum E_k) \nonumber
\end{align}

so we've now turned one integral into 8 integrals (and we integrate over $q_1^{R,L}$ as allowed by momentum conservation). The following processes then contribute to $\partial_t n^\up_k$:

\begin{table}
	\caption{Scattering processes in FGR calculation.}
	\begin{center}
		\begin{tabular}{ c|c|c|c|c } 
			1$\up$ & 2$\down$ & 3$\up$ & 4$\down$ & process \\
			\hline
			L & L & L & L & forward \\
			R & R & R & R & \\
			R & R & L & L & Umklapp \\
			L & L & R & R & \\ 
			R & L & R & L & backscattering \\
			R & L & L & R & \\
			L & R & R & L & \\
			L & R & L & R & 
		\end{tabular}
	\end{center}
\end{table}

\subsubsection{Forward scattering}

Let's look at just one of these, say the LLLL channel. Recalling that the energy current is 

\begin{equation}
\partial_t J_E^\up = \int dk_1 v_{k_1} E_{k_1} \partial_t n_{k_1}^\up
\end{equation}

with $v_{k_1} = -v_F$ and $E_{k_1} = -v_F q_1^L$, the LLLL channel contribution to $\partial_t J_E^\up$ is

\begin{align}
&\frac{2\pi}{\hbar} U^2 \int dq_1^L dq_2^L dq_3^L dq_4^L \delta(q_3^L + q_4^L - q_2^L - q_1^L) \delta(v_F(q_3^L + q_4^L - q_2^L - q_1^L))  \\&\times [ f_T(q_3^L) f_{T_L}(q_4^L)(1-f_T(q_1^L)) (1-f_{T_L}(q_2^L)) - f_T(q_1^L) f_{T_L}(q_2^L)(1-f_T(q_3^L)) (1-f_{T_L}(q_4^L))] v_F^2 q_1^L. \nonumber
\end{align}

Using the delta function identity $\delta(ax) = \delta(x) / \abs{a}$ to pull out a factor $1/v_F$, and integrating $dq_2^L$, such that $q_2^L = q_3^L + q_4^L - q_1^L$, we get

\begin{align}
\frac{2\pi}{\hbar} U^2 v_F \delta(0) \int& dq_1^L dq_3^L dq_4^L \Bigg[ f_T(q_3^L) f_{T_L}(q_4^L)(1-f_T(q_1^L)) (1-f_{T_L}(q_3^L + q_4^L - q_1^L)) \nonumber \\&- f_T(q_1^L) f_{T_L}(q_3^L + q_4^L - q_1^L)(1-f_T(q_3^L)) (1-f_{T_L}(q_4^L)) \Bigg]q_1^L. 
\end{align}

This $\delta(0)$ factor we can take to be regularized as $\delta(0) = L$; it comes from conservation of momentum and energy being the same constraint for a linear spectrum. We can also now use the Fermi-Dirac distribution identity $1-f_{T_L}(q_3^L + q_4^L - q_1^L) = f_{T_L}(q_1^L - q_3^L - q_4^L)$. 

Now let's approximate $T = 0$ so that we only have one temperature scale ($T_L$) in the problem. Then $f_T(q^L) = \Theta(-v_F q^L)$, where $\Theta$ is the Heaviside step function, $$ \Theta(\epsilon) = \begin{cases} 0 & \epsilon<0 \\ 1 & \epsilon>0 \end{cases}.$$ We use this $\Theta$ function to fix the limits of integration. Plugging in the FD distributions:

\begin{align}
&\frac{2\pi}{\hbar} U^2 v_F L \Bigg\{ \int_{-\infty}^0 dq_1^L \int_0^\infty dq_3^L \int_{-\infty}^\infty dq_4^L  \frac{q_1^L}{1+\exp(-\beta_L v_F q_4^L)} \frac{1}{1+\exp(\beta_L v_F (q_4^L - (q_1^L - q_3^L)))}   \nonumber \\&- \int_{0}^\infty dq_1^L \int_{-\infty}^0 dq_3^L \int_{-\infty}^\infty dq_4^L \frac{q_1^L}{1+\exp(\beta_L v_F q_4^L)} \frac{1}{1+\exp(-\beta_L v_F (q_4^L - (q_1^L - q_3^L)))}   \Bigg\}.
\end{align}

Focusing on the first integral,

\begin{align}
\int_{-\infty}^0 dq_1^L &\int_0^\infty dq_3^L \int_{-\infty}^\infty dq_4^L  \frac{q_1^L}{1+\exp(-\beta_L v_F q_4^L)} \frac{1}{1+\exp(\beta_L v_F (q_4^L - (q_1^L - q_3^L)))} \nonumber \\ 
&= \int_{-\infty}^0 dq_1^L \int_0^\infty dq_3^L \frac{1}{\beta_L v_F} \frac{ q_1^L(q_3^L - q_1^L)}{1 - \exp(\beta_L v_F (q_3^L - q_1^L))} \nonumber \\
&= \int_{-\infty}^0 dq_1^L \left[ \frac{1}{(\beta_L v_F)^2} \Li_2(e^{\beta_L v_F q_1^L}) + \frac{1}{\beta_L v_F} q_1^L \log(1-e^{\beta_L v_F q_1^L}) \right]q_1^L \nonumber \\
&= -\frac{\pi^4}{30} \frac{1}{(\beta_L v_F)^4}
\end{align}

with $\Li_2(z)$ the (order 2) polylogarithm function. Similarly the second integral contributes

\begin{align}
-\int_{0}^\infty dq_1^L &\int_{-\infty}^0 dq_3^L \int_{-\infty}^\infty dq_4^L \frac{q_1^L}{1+\exp(\beta_L v_F q_4^L)} \frac{1}{1+\exp(-\beta_L v_F (q_4^L - (q_1^L - q_3^L)))} \nonumber \\&= -\frac{\pi^4}{30} \frac{1}{(\beta_L v_F)^4}.
\end{align}

Thus, the LLLL channel contributes

\begin{equation}
-\frac{2\pi}{\hbar} U^2 v_F L \frac{\pi^4}{15} \frac{1}{(\beta_L v_F)^4} 
\end{equation}

to $\partial_t J_E(t=0)$. Now, by the same logic the RRRR channel contributes

\begin{equation}
+\frac{2\pi}{\hbar} U^2 v_F L \frac{\pi^4}{15} \frac{1}{(\beta_R v_F)^4} .
\end{equation}

Hence the total contribution from forward scattering is, using $v_F = 2t_h$ at half filling,

\begin{equation}
\frac{\pi^5}{60 \hbar} \frac{U^2}{t_h^3} L k_B^4 \left( T_R^4 - T_L^4 \right).
\end{equation}

For ease of reference, $\pi^5/60 \approx 5.1$. 

\subsubsection{Umklapp}

Now consider the Umklapp channels, RRLL and LLRR. First focus on RRLL. Conservation of momentum reads $\sum k = (k_F + q_1) + (k_F + q_2) - (-k_F + q_3) - (-k_F + q_4) = 4k_F + q_1 + q_2 - q_3 - q_4 = 0$. Since we are at half filling, $4 k_F = 4(\pi/2) = 2\pi = 0 \mod 2\pi$. Hence conservation of momentum is simply $q_1 + q_2 - q_3 - q_4 = 0$. Conservation of energy reads $\sum E_k = 0 = v_F q_1 + v_F q_2 - (-v_F q_3 - v_F q_4) = v_F (q_1 + q_2 + q_3 + q_4)$. Hence, the contribution to $\partial_t J_E$ is 

\begin{align}
\frac{2\pi}{\hbar} v_F^2 U^2 \Bigg\{ &\int_{-\infty}^0 dq_1 \int_{-\infty}^\infty dq_2 \int_{-\infty}^0 dq_3 \int_{-\infty}^{\infty} dq_4  f_{T_R}(-q_2) f_{T_L}(q_4) \nonumber \\ &\hspace{20mm}\times \delta(q_1 + q_2 - q_3 - q_4)\delta(  v_F (q_1 + q_2 + q_3 + q_4)) \nonumber \\ 
&-\int_{0}^\infty dq_1 \int_{-\infty}^\infty dq_2 \int_{0}^\infty dq_3 \int_{-\infty}^{\infty} dq_4  f_{T_R}(q_2) f_{T_L}(-q_4) \nonumber \\ &\hspace{20mm}\times \delta(q_1 + q_2 - q_3 - q_4)\delta(  v_F (q_1 + q_2 + q_3 + q_4)) \Bigg\}.
\end{align}

The integral over $q_2$ sets $q_2 = q_3 + q_4 - q_1$, and then conservation of energy reads 

$$
\delta(v_F(q_1 + (q_3 + q_4 - q_1) + q_3 + q_4) = \frac{1}{2v_F} \delta(q_3 + q_4).
$$

Integrating over $q_4$ then yields

\begin{align}
\frac{\pi}{\hbar v_F}U^2 &\Bigg\{ \int_0^\infty dq_1 q_1 \int_0^\infty dq_3 f_{T_R}(q_1) f_{T_L}(-q_3) - \int_{-\infty}^0 dq_1 q_1 \int_{-\infty}^0 dq_3 f_{T_R}(-q_1) f_{T_L}(q_3) \Bigg \} \nonumber \\&= \frac{\pi^3 \log 2}{6 \hbar} \frac{U^2}{v_F^2} \frac{1}{\beta_R^2 \beta_L}.
\end{align}

Similarly, LLRR yields $$-\frac{\pi^3 \log 2}{6 \hbar} \frac{U^2}{v_F^2} \frac{1}{\beta_L^2 \beta_R}.$$ Hence, using $v_F = 2t_h$, the total contribution from Umklapp scattering is 

\begin{equation}
\frac{\pi^3 \log 2}{24\hbar} \frac{U^2}{t_h^2} k_B^3 \left( T_R^2 T_L - T_L^2 T_R \right).
\end{equation}

\subsubsection{Backscattering}

Finally, consider the four backscattering channels: RLRL, LRLR, RLLR, LRRL. Taking RLRL, the integrals we want to compute are

\begin{align}
&v_F^2 \frac{2\pi}{\hbar} U^2 \Bigg\{ \int_0^\infty dq_1 q_1 \int_{-\infty}^\infty dq_2 \int_{-\infty}^0 dq_3 \int_{-\infty}^\infty dq_4 f_{T_L}(-q_2) f_{T_L}(q_4) \delta(\sum k) \delta(\sum E_k)  \nonumber \\ 
&-\int_{-\infty}^0 dq_1 q_1 \int_{-\infty}^\infty dq_2 \int_0^\infty dq_3 \int_{-\infty}^\infty dq_4 f_{T_L}(q_2) f_{T_L}(-q_4) \delta(\sum k) \delta(\sum E_k)  \Bigg\}
\end{align}

with $\sum k = q_1 + q_2 - q_3 - q_4$ and $\sum E_k = v_F (q_1 - q_3 - q_2 + q_4)$. Then when we integrate $dq_2$, conservation of momentum will set $q_2 = q_3 + q_4 - q_1$ and hence the conservation of energy delta function will read 

$$
\delta(2v_F(q_1-q_3)) = \frac{1}{2v_F} \delta(q_1 - q_3)
$$

and will pick out $q_1 = q_3$. However, $q_1 = q_3$ means both are 0 in the above integrals! This picks out just one point under the integral sign (i.e. a set of measure 0), so the contribution from this process {\it vanishes}. Similarly, LRLR vanishes. This is intuitively reasonable, since if you draw out the setup in RLRL and LRLR, the only way to conserve energy and momentum is for the particles to not scatter at all. 

Now consider the true backscattering processes, LRRL and RLLR. Focusing on LRRL:

\begin{align}
&v_F^2 \frac{2\pi}{\hbar} U^2 \Bigg\{ \int_{-\infty}^0 dq_1 q_1 \int_{-\infty}^\infty dq_2 \int_{-\infty}^0 dq_3 \int_{-\infty}^\infty dq_4 f_{T_R}(-q_2) f_{T_L}(q_4) \delta(\sum k) \delta(\sum E_k)  \nonumber \\ 
&-\int_0^\infty dq_1 q_1 \int_{-\infty}^\infty dq_2 \int_0^\infty dq_3 \int_{-\infty}^\infty dq_4 f_{T_R}(q_2) f_{T_L}(-q_4) \delta(\sum k) \delta(\sum E_k) \Bigg \}
\end{align}

with $\sum k = q_1 + q_2 - q_3 - q_4$ and $\sum E_k = v_F (q_2 - q_1 + q_4 - q_3)$. Performing the integral over $q_2$ sets $q_2 = q_3 + q_4 - q_1$ hence conservation of energy becomes

$$
\delta(2v_F (q_4 - q_1)) = \frac{1}{2v_F} \delta(q_4 - q_1).
$$

Then performing the integral over $q_4$ gives

\begin{align}
\frac{\pi v_F}{\hbar} U^2 \Bigg\{ &\int_{-\infty}^0 dq_1 q_1 \int_{-\infty}^0 dq_3 f_{T_R}(-q_3) f_{T_L}(q_1) -  \int_0^\infty dq_1 q_1 \int_0^\infty dq_3 f_{T_R}(q_3) f_{T_L}(-q_1)  \Bigg\} \nonumber \\&= - \frac{\pi^3 \log 2}{6 \hbar} {U^2}{v_F^2} \frac{1}{\beta_L^2 \beta_R}.
\end{align}

The total contribution from from backscattering is then 

\begin{equation}
\frac{\pi^3 \log 2}{24\hbar} \frac{U^2}{t_h^2} k_B^3 \left( T_R^2 T_L - T_L^2 T_R \right).
\end{equation}

For ease of reference, ${\pi^3 \log 2} / {24} \approx 0.9$.

\subsubsection{``Generalized Golden Rule'' trick and forward scattering}

We found that the forward scattering process leads to a divergence in the case of the (spinful) Hubbard model, and generally between two wires. As described in ~\cite{PhysRevB.78.205407}, the forward scattering integral is finite with spinless fermions but diverges when we have multiple spin species. They use a trick (referred to as the `generalized Fermi's golden rule') to fix this divergence, which we adapt here. 

The issue with the forward scattering channel is that conservation of momentum and conservation of energy are the same constraint due to the linear spectrum, so we get a delta function squared under the integral. The intuitive picture for why it is inconsistent to use $v_F$ for the energy conservation constraint in a Fermi's golden rule calculation is that the Fermi velocity is renormalized by interactions at first order in $U$. The patch that Yashenkin et al. prescribe is to replace the usual energy constraint $v_F q_{in} = v_F q_{out}$ with $v_F q_{in} = u q_{out}$, with $u$ the Luttinger liquid velocity. That is, we replace

\begin{align}
\delta(q_1 + q_2 - q_3 - q_4) &\delta(v_F(q_1 + q_2) - v_F(q_3 + q_4)) \nonumber \\&\mapsto \delta(q_1 + q_2 - q_3 - q_4) \delta(v_F(q_1 + q_2) - u(q_3 + q_4)),
\end{align}

with Luttinger velocity (using $g_4 = g_2 = U$)

\begin{equation}
u = v_F \sqrt{1 + U/\pi v_F}.
\end{equation}

For completeness, the $K$ parameter is $K =1/ \sqrt{1+U/\pi v_F}$ (note that for Hubbard, since all $g$ are equal we get $uK = v_F$). Now, using this new delta function, the LLLL contribution to $\partial_t J_E(t=0)$ is

\begin{align}
&\frac{2\pi}{\hbar} U^2 v_F^2 \int_{\mathbb{R}^4} dq_{1234} \delta(q_1 + q_2 - q_3 - q_4) \delta(v_F(q_1 + q_2) - u(q_3 + q_4)) \nonumber \\ \times &q_1 \{ f_T(q_3) f_{T_L}(q_4) f_T(-q_1) f_{T_L}(-q_2) - f_T(q_1) f_{T_L}(q_2) f_T(-q_3) f_{T_L}(-q_4)  \},
\end{align}

using the fact that $f_T(-k) = 1-f_T(k)$ for the Fermi-Dirac distribution. Integrate over $q_2$, which enforces the first delta function, setting $q_2 = q_3 + q_4 - q_1$, and setting the second delta function to $\delta((u-v_F)(q_3+q_4) = \delta(q_3+q_4) / \abs{u-v_F}$. We get

\begin{align}
\frac{2\pi}{\hbar} U^2 \frac{v_F^2}{\abs{u-v_F}} \int_{\mathbb{R}^3}& dq_{134} \delta(q_3 + q_4)  \nonumber \\ \times &q_1 \Bigg\{ f_T(q_3) f_{T_L}(q_4) f_T(-q_1) f_{T_L}(-(q_3 + q_4 - q_1)) \nonumber \\ &- f_T(q_1) f_{T_L}(q_3 + q_4 - q_1) f_T(-q_3) f_{T_L}(-q_4)  \Bigg\}.
\end{align}

Now integrate over $q_4$, setting $q_4 = -q_3$:

\begin{align}
\frac{2\pi}{\hbar} U^2 \frac{v_F^2}{\abs{u-v_F}} \int_{\mathbb{R}^2}& dq_{13} q_1 \Bigg\{ f_T(q_3) f_{T_L}(-q_3) f_T(-q_1) f_{T_L}(q_1) \nonumber \\&- f_T(q_1) f_{T_L}(- q_1) f_T(-q_3) f_{T_L}(q_3)  \Bigg\}.
\end{align}

Assume for simplicity that $T = 0$. Then we have 

\begin{align}
\frac{2\pi}{\hbar} U^2 \frac{v_F^2}{\abs{u-v_F}} &\Bigg\{ \int_{-\infty}^{0} dq_1 \int_0^\infty dq_3 \frac{q_1}{1 + \exp(\beta_L v_F q_3)} \frac{1}{1+\exp(-\beta_L v_F q_1)} \nonumber\\&- \int_0^\infty dq_1 \int_{-\infty}^0 dq_3 \frac{q_1}{1+\exp(-\beta_L v_F q_3)} \frac{1}{1+\exp(\beta_L v_F q_1)} \Bigg\}. 
\end{align}

We can do these integrals straightforwardly, giving

\begin{equation}
- \frac{2\pi}{\hbar} U^2 \frac{v_F^2}{\abs{u-v_F}} \frac{\pi^2 \log 2}{6 \beta_L^3 v_F^3}.
\end{equation}

Expanding the velocity prefactor, we have 

\begin{equation}
\frac{v_F^2}{\abs{u-v}} = \frac{v_F}{\sqrt{1 + U/\pi v_F} - 1} \approx 2\frac{\pi v_F^2}{U}.
\end{equation} 

Thus our answer is actually first order in $U$, as found by Yashenkin et al as well. The final answer is then (to first order in $U$)

\begin{equation}
\partial_t J_E(t=0) \sim U  \frac{2\pi^4 \log 2}{3 \hbar v_F}k_B^3 \left(T_R^3 - T_L^3 \right),
\end{equation}

giving a finite Fermi's Golden Rule rate $\Gamma$. For ease of reference, $2(\pi^4 \log 2) / 3 \approx 45$.

\subsection{Higher dimensions and lack of divergence}\label{sec:higherD}

\begin{figure}[ht!]
	\includegraphics[width = \columnwidth]{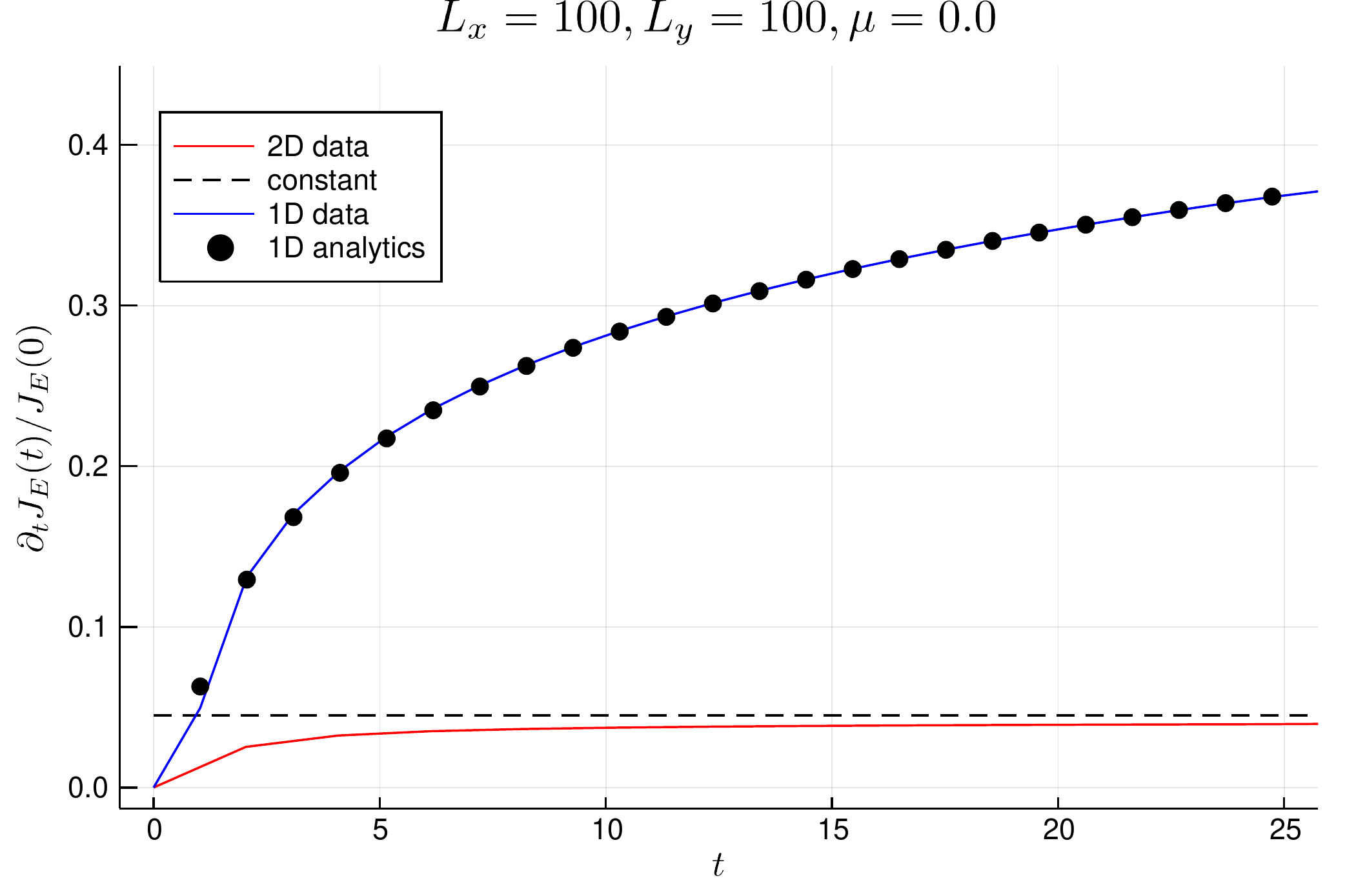}
	\caption{\label{fig:2D} Comparison of the 2D Hubbard model quench with $L_x = L_y = 100$, $t_x = t_y = t = 1.0$ (red line), to the 1D case with $L = 101$ (blue line). All data was taken at temperature $\beta_\uparrow = \beta_{\downarrow} = 1.0$, $U = 1.0$ and chemical potential $\mu = 0.0$ (the particle-hole symmetric point). We see that the 2D quench does not appear to be divergent, in accordance with expectations.}
\end{figure}

In this appendix, we detail a two-dimensional version of the 1D Hubbard model quench presented in the main text, showing numerical evidence that it does not suffer from the same divergence. 

Consider the two-dimensional Hubbard model, $$H = \sum_{\langle ij \rangle, \sigma} t_{ij} c_{i,\sigma}^\dagger c_{j,\sigma} + h.c. + U \sum_i n_{i,\uparrow} n_{i,\downarrow}.$$ When $U=0$ this model can be diagonalized by Fourier transform, giving energies

\begin{equation}
E = -2 t_x \cos k_x - 2 t_y \cos k_y,
\end{equation}

where we have set $t_{ij} = t_x$ or $t_y$ for horizontal or vertical bonds. In what follows let us further simplify to the case $t_x = t_y = t$. We again initialize the problem with an initial $J_E^{\downarrow}$ as in the one-dimensional case, and at $t=0$ quench on the interactions $U$. We compute $\partial_t J_E^{\uparrow}$ numerically using lowest-order perturbation theory as in the main text. We note that we do not expect the same divergence in the 2D case as the 1D one due to fundamental differences in the band structure scattering processes. Our results are summarized in Fig.~\ref{fig:2D}; we do not observe the same divergence in 2D as in the 1D case at the largest accessible system sizes, and the Fermi's Golden Rule limit of a constant rate appears to be valid. 

\subsection{Generic interactions}\label{sec:ints}

In the main text, we considered Hubbard contact interactions, which are $k$-independent. In this appendix we show that the divergence associated to the response energy current occurs for generic interactions.

Let us first consider Coulomb interactions. We expect~\cite{PhysRevB.73.165104}

\begin{equation}
U_{\mathrm{Coulomb}}(x) = U \frac{1}{\sqrt{x^2 + d^2}}
\end{equation}

with $d$ the distance between the layers (or a cutoff for the spin-spin interactions). The Fourier transform of this is a modified Bessel function of the second kind (up to a constant prefactor),

\begin{equation}
U_{\mathrm{Coulomb}}(q) = U K_0(\abs{q}d).
\end{equation}

This function is logarithmically singular near $(d \abs{k}) \to 0$ and falls off exponentially as $e^{-d \abs{k}} / \sqrt{\abs{k}}$ at large $d \abs{k}$. Following the analysis in Appendix~\ref{sec:log}, we expect a log growth of the energy drag with prefactor

\begin{equation}
I(T) =  \int_0^{\pi} dk K_0(k d)^2\frac{\sin^2(k) \csc(k/2)}{1 + \cosh(2 t\beta \sin (k/2))},
\end{equation}

which may be integrated numerically. Finally, let us consider short-ranged interactions with decay length $\xi$, i.e.

\begin{equation}
U_{\mathrm{exp}}(x) = U \frac{e^{-\abs{x} / \xi}}{2 \xi},
\end{equation}

which is normalized such that $\xi\to 0$ recovers the contact interaction. This has Fourier transform

\begin{equation}
U_{\mathrm{exp}}(q) = U \frac{1}{1+\xi^2 q^2},
\end{equation}

entailing a prefactor of the log growth of 

\begin{equation}
I(T) =  \int_0^{\pi} dk \frac{1}{(1+ \xi^2 k^2)^2}\frac{\sin^2(k) \csc(k/2)}{1 + \cosh(2 t\beta \sin (k/2))}.
\end{equation}

We have checked these predictions numerically (see Fig.~\ref{fig:ints}) and find excellent agreement.

\begin{figure}
\includegraphics[width=0.5\columnwidth]{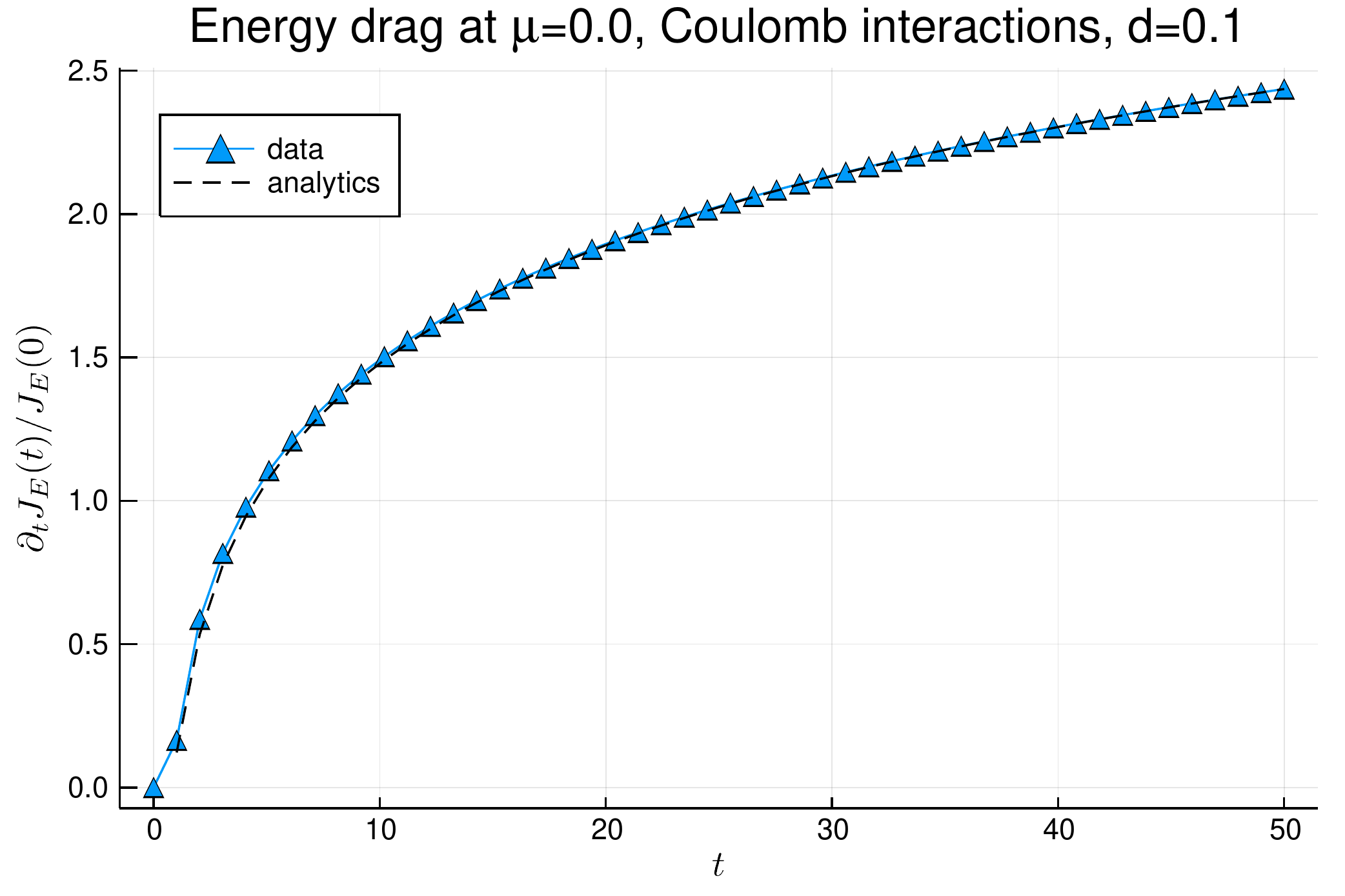}   \includegraphics[width=0.5\columnwidth]{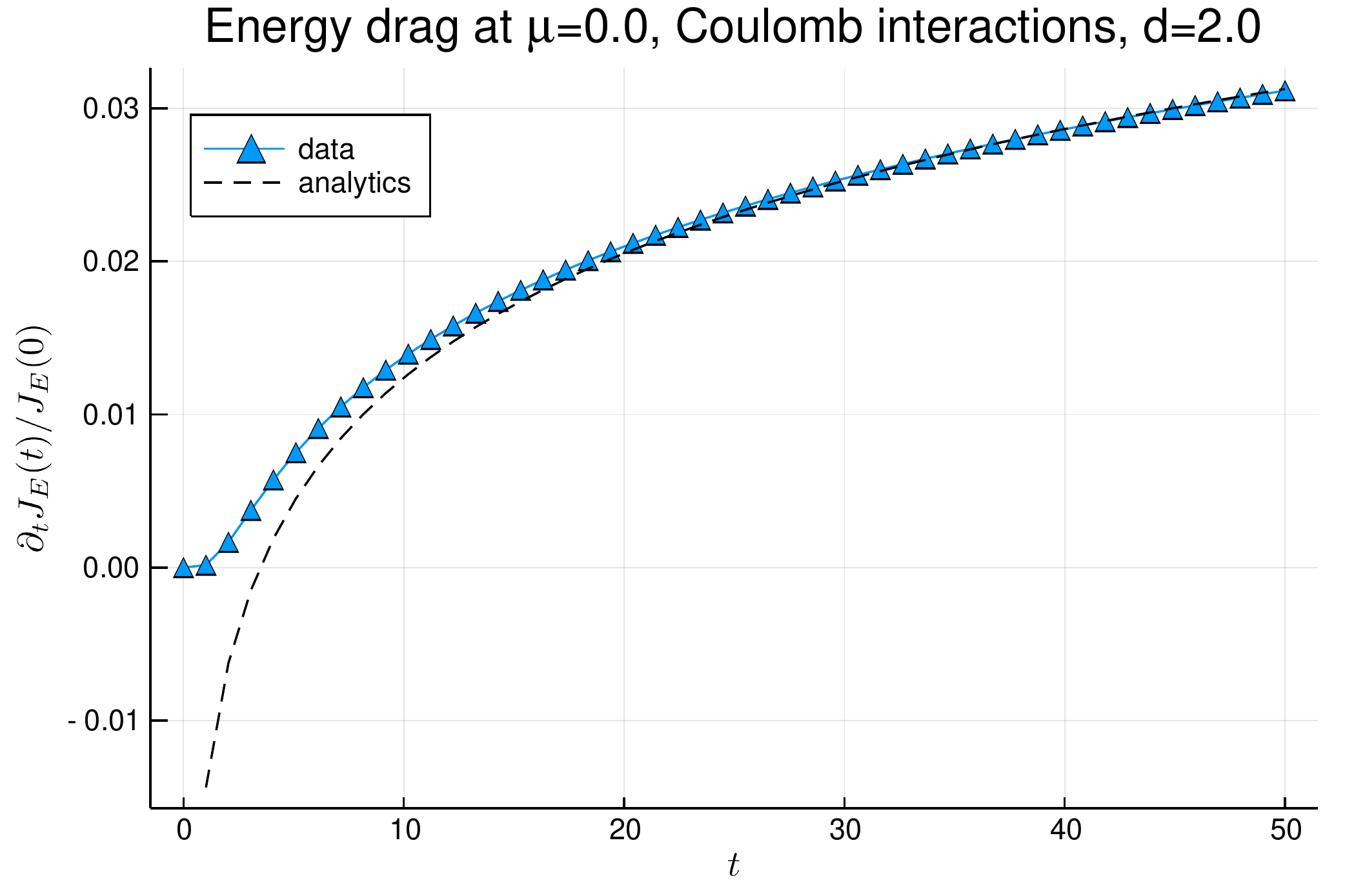} \\
\includegraphics[width=0.5\columnwidth]{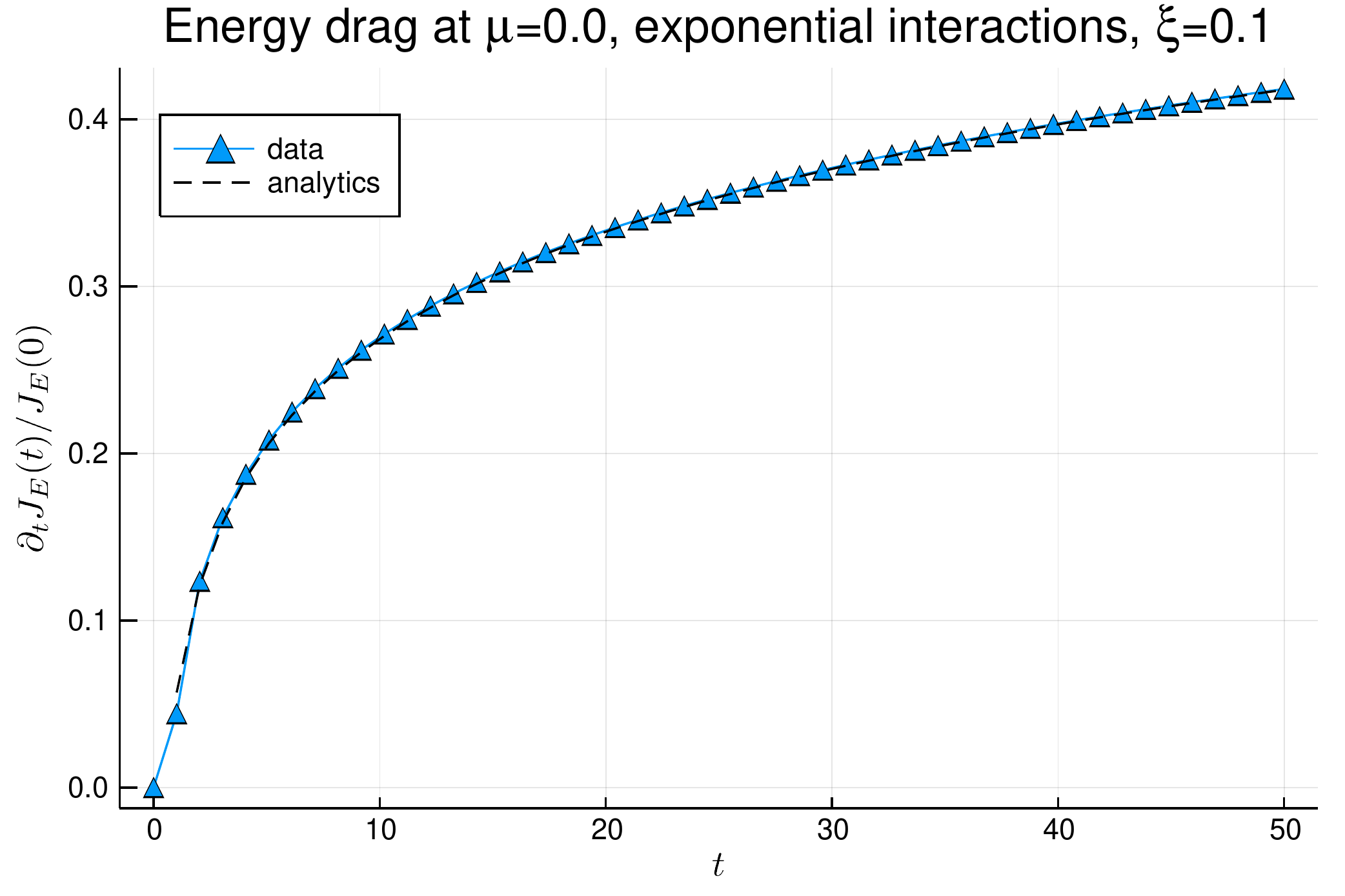} 
\includegraphics[width=0.5\columnwidth]{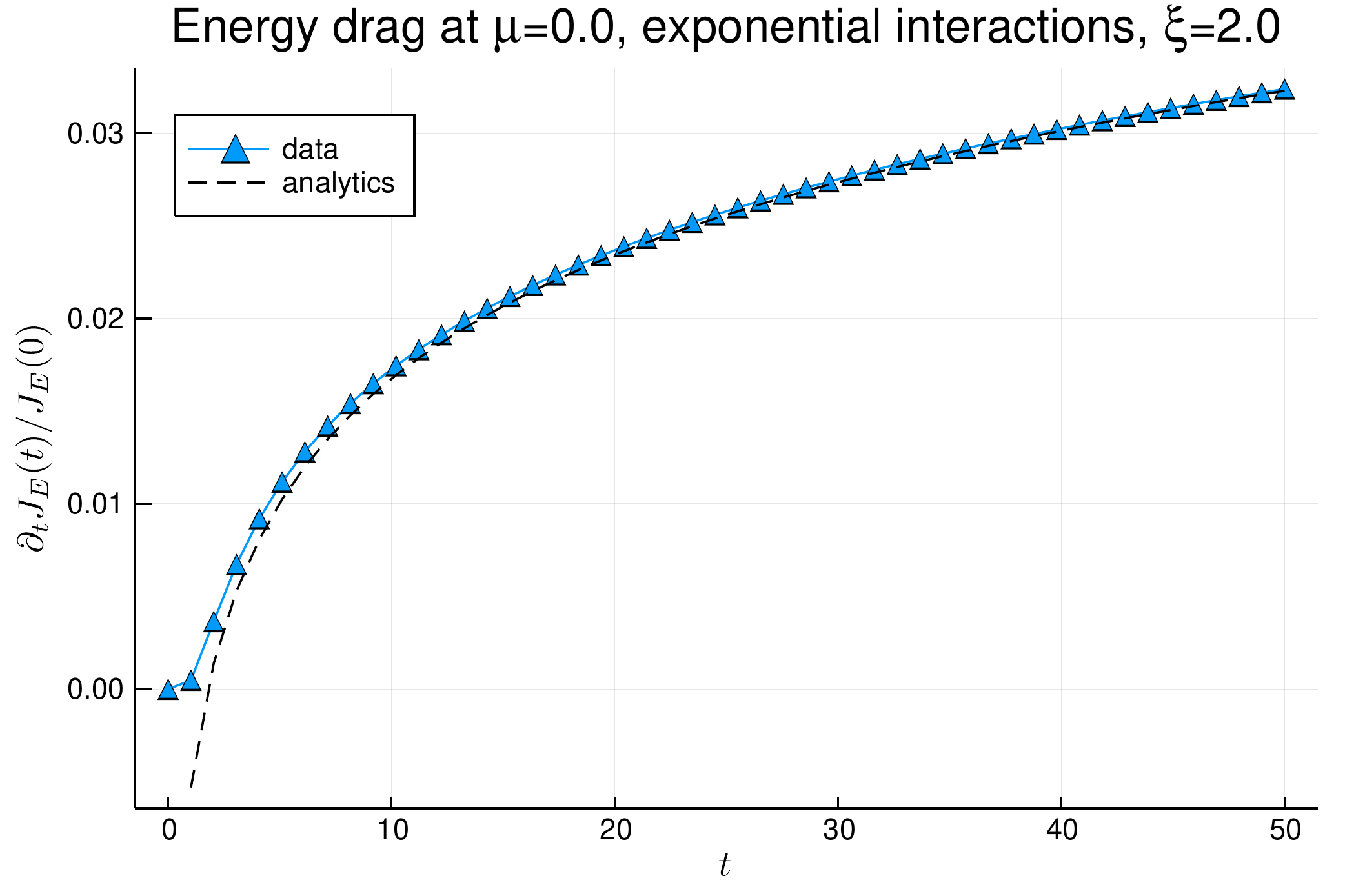} 
\caption{\label{fig:ints} Logarithmic growth of the response energy current for Coulomb interactions with $d = 0.1,2.0$ and for short-ranged interactions with decay length $\xi = 0.1, 2.0$. All data was taken with $\mu = 0.0$, $\beta = 1.0$ with hopping parameter $t=1.0$, on system size $L=301$ (in the thermodynamic limit).}
\end{figure}


\begin{savequote}
Then world behind and home ahead, \\
We'll wander back and home to bed. \\
Mist and twilight, cloud and shade, \\
Away shall fade! Away shall fade! 
\qauthor{J. R. R. Tolkein, \textit{The Fellowship of the Ring}}
\end{savequote}

\chapter{Conclusion}
\label{conclusion}

We have seen that, despite the apparent reliance on thermodynamic equilibrium of tools like the renormalization group, universality is possible even in non-equilibrium quantum systems. In Chapter~\ref{ch:CFT}, we analyzed initially equilibrium quantum critical systems driven at their boundaries. Using the tools of conformal field theory, we saw that the dynamics generally inherit universality from the bulk, both under periodic (Floquet) driving and under stochastic driving. In particular, the Loschmidt echo was shown to be a universal scaling function, and many more complicated quantities should universally collapse as well. We also saw that some dynamical phenomena, like Kibble-Zurek scaling, manifest as we drive these critical points, smoothly changing the critical exponents. In Chapter~\ref{ch:RSRG}, we saw that strongly disordered, periodically driven systems can also show universality. Transitions are generally controlled by infinite randomness fixed points, as shown with our renormalization group procedure, and the periodic driving allows access to classes not possible without the drive. For instance, the disordered driven Ising chain can exhibit universal scaling with $\tilde{c}=\ln 2$, in contrast to the undriven case of $\tilde{c}=\ln 2 / 2$. These disordered systems are robust to heating due to the phenomenon of many-body localization, though this may break down at criticality. Finally, in Chapter~\ref{ch:Hydro}, we analyzed a quantum limit of the shear viscosity -- namely the Coulomb drag -- within the context of the universal, hydrodynamic flow of electrons in solids. We found that drag between heat currents shows striking differences with conventional charge drag, with new divergences in one-dimension leading to the failure of Fermi's Golden Rule. This was due to the periodicity of the band structure, a fundamental aspect of all one-dimensional systems; hence, this divergence would be seen in virtually all thermal drag scenarios. 

Perhaps the most obvious shortcoming of these results is that they lack a general framework. This is unfortunately par for the course of non-equilibrium systems, as we can only make inroads using a limited toolkit. A grand challenge for the future, then, is to develop such a generalized framework that could tackle, if not all non-equilibrium systems, then at least a large swath in one go. This seems out of reach at the moment, but work on tensor network approaches lends some hope to this task. Tensor networks provide a way to succinctly represent a quantum state, and though non-equilibrium states tend to be volume-law entangled and hence numerically difficult for this representation, analytic treatments may make headway. There has also been exciting progress in using machine-learning-inspired frameworks, such as the Restricted Boltzmann Machine ansatz, to study non-equilibrium systems. With the impetus of recent experimental advances in cold-atom systems, hopefully the future holds similar breakthroughs in our abilities to simulate and solve such non-equilibrium quantum systems. 

A further critique is that the universality demonstrated here is not so different from equilibrium universality, in that the universality classes are just equilibrium classes in disguise. For instance, the scaling seen in driven CFTs is simply that of the equilibrium universality class, and transitions in the disordered Floquet case are controlled by equilibrium universality classes as well. I would first caution that the ability to observe universality in these non-equilibrium settings at all is remarkable, as we have good reason to think \textit{a priori} that heating should destroy the delicate quantum coherences at critical points. That said, it is true that these classes are not completely novel. In fact, it is extremely challenging to find truly novel universality classes out of equilibrium, and almost all other works in this vein have found the same. For instance, the seminal work of Cardy and Calabrese on quantum quenches in CFTs also found universal responses controlled by the \emph{equilibrium} quantum critical point. More recently, studies of random unitary quantum circuits interspersed with projective measurements have found universal scaling out of equilibrium, resulting from competition between the scrambling of random Hamiltonian evolution and the freezing of quantum measurements~\cite{PhysRevB.98.205136, PhysRevX.9.031009}.\footnote{In a hand-waving way, we can see this as follows: the random Hamiltonian evolution leads (almost surely) to a volume-law-entangled state at late times. However, measurements project onto a single quantum state, so if we measure every underlying qudit, we form a product state that has no entanglement. If we measure a fraction $p$ of the qudits at every time step, then for some critical $p_c$ we transition from volume-law entanglement to area-law, and precisely at this point we have logarithmic growth of entanglement (in one dimension) -- a kind of critical state.} While there are hints that the universality class here may be novel, it is not completely understood, and our best theories show that it is a kind of percolation transition (i.e., another known non-equilibrium class, at least in a certain limit)~\cite{PhysRevB.101.104301, PhysRevB.101.104302}. A similar transition seen when tuning the bond dimension of random tensor networks also points to percolation~\cite{PhysRevB.100.134203}; we know at least that it has a CFT description with $c=0$, though these theories (`logarithmic CFTs') are themselves notoriously poorly understood. Famously, the MBL-ETH transition, which is the onset of many-body localization as a function of disorder strength, is a novel non-equilibrium universality class. But this is quite a subtle subject, as it is not even clear which aspects of the transition are universal. Recent work has argued that the transition exhibits Kosterlitz-Thouless (KT) scaling~\cite{PhysRevB.99.094205}, again an equilibrium class, though competitors have refined their arguments to propose a possibly new class with many similarities to KT~\cite{morningstar2020manybody}. Suffice it to say, truly novel universality classes are few and far between. 

Nonetheless, the prospects for understanding universality out of equilibrium, even if in a  piecemeal fashion, are bright. The tools of CFT and disordered systems have shone great light on the dynamics of quantum systems, and no doubt will continue to be used to great effect. Though not considered in this thesis, great strides have been made in understanding open quantum systems, which may dissipate to their environments. The interplay between driving and dissipation (aptly, `driven-dissipative systems') holds promise for realizing intriguing non-equilibrium steady states that may display unique forms of universality. Physics has historically been an experimentally driven science, and the advent of precision measurements in cold atomic systems of fully non-equilibrium systems, including their time-resolved dynamics, bolsters the prospects of our reaching, someday, a full understanding of quantum non-equilibrium universality. 

\setstretch{\dnormalspacing}

\backmatter

\end{document}